\documentclass[
    reprint,
    superscriptaddress,
    nofootinbib,
    amsmath,amssymb,
    aps,
    pre,
]{revtex4-2}

\usepackage{cmap}
\usepackage[T1]{fontenc}
\usepackage{newtxtext}
\usepackage{newtxmath}
\usepackage{microtype}
\usepackage{graphicx}
\graphicspath{ {figures} }
\usepackage{dcolumn}
\usepackage{bm}
\usepackage{hyperref}
\hypersetup{colorlinks=true,allcolors=blue}
\usepackage{mathtools}
\usepackage{upgreek}
\usepackage{orcidlink}

\begin{document}

\title{Orthogonal projections of hypercubes}

\author{Yoshiaki Horiike\,\orcidlink{0009-0000-2010-2598}}
\email{yoshi.h@nagoya-u.jp}
\affiliation{
    Department of Applied Physics,
    \href{https://ror.org/04chrp450}{Nagoya University},
    Nagoya, Japan
}
\affiliation{
    Department of Neuroscience,
    \href{https://ror.org/035b05819}{University of Copenhagen},
    Copenhagen, Denmark
}

\author{Shin Fujishiro\,\orcidlink{0000-0002-0127-0761}}
\affiliation{
    Fukui Institute for Fundamental Chemistry,
    \href{https://ror.org/02kpeqv85}{Kyoto University},
    Kyoto, Japan
}

\date{August 19, 2025}

\begin{abstract}
    Projections of hypercubes have been applied to visualize
    high-dimensional binary state spaces in various scientific fields.
    Conventional methods for projecting hypercubes, however,
    face practical difficulties.
    Manual methods require nontrivial adjustments of the projection basis,
    while optimization-based algorithms limit the interpretability and
    reproducibility of the resulting plots.
    These limitations motivate us to explore theoretically analyzable
    projection algorithms such as principal component analysis (PCA).
    Here, we investigate the mathematical properties of
    PCA-projected hypercubes.
    Our numerical and analytical results show that PCA effectively captures
    polarized distributions within the hypercubic state space.
    This property enables the assessment of the asymptotic distribution of
    projected vertices and error bounds,
    which characterize the performance of PCA in the projected space.
    We demonstrate the application of PCA to visualize the hypercubic energy
    landscapes of Ising spin systems,
    specifically finite artificial spin-ice systems,
    including those with geometric frustration.
    By adding projected hypercubic edges,
    these visualizations reveal pathways of correlated spin flips.
    We confirm that the time-integrated probability flux exhibits patterns
    consistent with the pathways identified in the projected
    hypercubic energy landscapes.
    Using the mean-field model, we show that dominant state transition pathways
    tend to emerge around the periphery of the projected hypercubes.
    Our work provides a better understanding of how PCA discovers
    hidden patterns in high-dimensional binary data.
\end{abstract}

\maketitle


\section{Introduction}
By generalizing the idea of a three-dimensional cube to higher dimensions,
one obtains a high-dimensional cube or a hypercube~\cite{Coxeter1973}
(see Appendix~\ref{sec:constructing-hypercube} for details
on the construction of a hypercube).
Although a hypercube is purely a geometric
concept~\cite{Stringham1880, BooleStott1900, Coxeter1973, Gardner1975},
it has applications not only in
physics~\cite{Gamow1988, Rucker1977, Rucker2014},
electrical engineering~\cite{Gray1953, Gardner1972, Harary1988,
Saad1988, Kumar1992, Malluhi1994},
graph theory~\cite{Gilbert1958, Gardner1972},
and recreational mathematics~\cite{Gardner1972, Gardner1975, Gardner1987},
but also in a wide range of interdisciplinary arts~\cite{sm}.
Since the pioneering work of almost 150 years ago~\cite{Stringham1880} and
subsequent contributions~\cite{Hinton1980, Coxeter1973},
interest in high-dimensional geometry, especially hypercubes, has been
continuously growing.

\begin{figure}[tb]
    \includegraphics{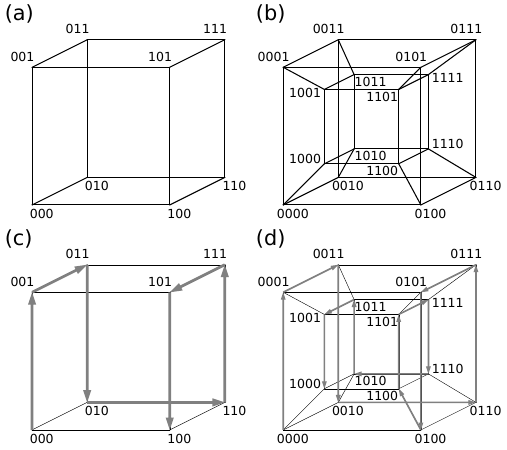}
    \caption{
        A gallery of hypercubes and Hamiltonian (directed) paths on them.
        (a) A cube, or three-dimensional hypercube,
        with three-dimensional coordinates of vertices.
        (b) A tesseract, or four-dimensional hypercube,
        with four-dimensional coordinates of vertices.
        (c) A Hamiltonian (directed) path on a cube.
        (d) A Hamiltonian (directed) path on a tesseract.
        In (c) and (d), the arrow indicates the direction of the path.
        One obtains a Hamiltonian path
        by converting the decimals to Gray code~\cite{Gray1953, Gardner1975} and
        following them in ascending order.
        See Tables~\ref{tab:decimal-binary-gray-3d}
        and~\ref{tab:decimal-binary-gray-4d}
        for Gray codes used to visualize Hamiltonian paths on
        three-dimensional and four-dimensional hypercubes.
    }
    \label{fig:hypercube-hamiltonian-pathway}
\end{figure}

In the sciences,
hypercubes are used to visualize binary state\footnote{
    Also called microstate, configuration, or phase.
}
spaces or state transitions in a wide range of fields,
from physics to biology.\footnote{
    See, for instance,
    evolutionary landscapes~\cite{Wright1932, Kauffman1987, Kauffman1991,
        Tan2011, DeVisser2014, Kondrashov2015, Mira2015, Johnston2016,
    Zagorski2016, Greenbury2020, Moen2023, Weaver2024},
    epistasis~\cite{Wright1932, Esteban2020, Eble2023},
    chemical reaction networks~\cite{Glass1975, Glass1975a},
    learning in neural networks~\cite{Rujan1989},
    genetic code space~\cite{Jimenez-Montano1996, Klump2020, Greenbury2020},
    allostery~\cite{Weiss1996, Weiss1996a, Weiss1996b, Hall2000, Biddle2021,
    Schirmeyer2021},
    quantum states~\cite{Zurek2001, Howard2019},
    quantum walks~\cite{Shenvi2003},
    language space~\cite{Matsen2004},
    data visualization~\cite{Acar2010, Peng2015},
    probability currents~\cite{Szabo2010},
    protein folding~\cite{Prinz2011, Olsson2017},
    energy landscapes~\cite{Farhan2013},
    gene regulation~\cite{Estrada2016, Nam2022, Owen2023, Martinez-Corral2024,
    Mahdavi2024},
    quantum many-body scar states~\cite{Turner2018, Turner2018a, Turner2021},
    quantum many-body
    localization~\cite{Roy2019,Roy2019a,Roy2020,Roy2020a,DeTomasi2021,Roy2021,Roy2025,Creed2023},
    morphology~\cite{Gerber2019, Budd2021},
    disease progression~\cite{Greenbury2020, Moen2023},
    gene regulatory networks~\cite{Rozum2021},
    gene expression~\cite{Zhu2022},
    and
    neural spike dynamics~\cite{Lynn2022, Lynn2022a}.
}
This is because hypercubic vertices correspond to binary states of a system
[Figs.~\ref{fig:hypercube-hamiltonian-pathway}(a)
and~\ref{fig:hypercube-hamiltonian-pathway}(b)],
and hypercubic edges represent transitions\footnote{
    Strictly speaking,
    only \emph{one} of the dimensions, components, or elements
    of the system must change in the transition on the hypercubic edge.
    This kind of dynamics is called asynchronous dynamics
    (update)~\cite{Glauber1963, Hopfield1982, Cornforth2005, Rozum2021}.
    It is also called multipartite dynamics~\cite{Horowitz2014,
    Wolpert2020, Lynn2022, Lynn2022a}
    in the context of stochastic thermodynamics.
    On the other hand,
    transitions with several changes of elements are called
    synchronous dynamics (update)~\cite{McCulloch1943,
        Caianiello1961, Little1974,
    Kauffman1969, Amari1972, Wolfram1983, Holland1992}.
}
between states
[Figs.~\ref{fig:hypercube-hamiltonian-pathway}(c)
and~\ref{fig:hypercube-hamiltonian-pathway}(d)].
This can be naturally applied to illustrate
the high-dimensional binary state space or state transition diagram.
We refer to such a binary state space as a \emph{hypercubic state space},
and such a state transition diagram as a
\emph{hypercubic state transition diagram}.
As an example of transitions between the vertices,
we visualize the Hamiltonian path on hypercubes
[Figs.~\ref{fig:hypercube-hamiltonian-pathway}(c)
    and~\ref{fig:hypercube-hamiltonian-pathway}(d),
    Tables~\ref{tab:decimal-binary-gray-3d}
and~\ref{tab:decimal-binary-gray-4d}].
The advantage of visualizing hypercubes is that, in addition to  displaying
the state space as hypercubic vertices,
it also illustrates state transitions as hypercubic edges.
Such representations provide valuable insight into the dynamics of
many-body systems,
particularly by illustrating correlated state transition pathways.

By using these useful two-dimensional illustrations of hypercubes,
scientists have visualized high-dimensional hypercubic state spaces
to intuitively understand the state space structure and
state transition dynamics of systems of interest.
Despite the practical usefulness of hypercubes, in general,
it is ``disappointingly difficult'' to visualize hypercubes,
even for low-dimensional ones~\cite{Abramson2003}.
Thus, studies on projecting the hypercube have targeted relatively
lower-dimensional systems,
and the procedure to project hypercubic state spaces is still unclear.

One method to project a hypercubic state space,
or a hypercube, is orthogonal projection~\cite{Coxeter1973},
which is reproducible.
In orthogonal projection,
by casting a shadow perpendicular to the two-dimensional plane with
a distant light source,
the parallels and lengths are preserved between the edges
representing the same dimensions.
Despite this useful property,
the weakness of orthogonal projection is that
one must manually determine the alignment of
the object to the targeted plane:
the projection of the unit vector of each dimension is manually determined.
For projections of higher-dimensional hypercubes,
it is nontrivial and impractical to manually determine
the projection of the unit vectors.
How can one decide the projection of the unit vector of each dimension
to create informative projections of hypercubes?

\begin{table}[tb]
    \begin{tabular}{ccc}
        \hline
        \hline
        Decimal & Binary & Gray \\
        \hline
        000     & 000    & 000  \\
        001     & 001    & 001  \\
        002     & 010    & 011  \\
        003     & 011    & 010  \\
        004     & 100    & 110  \\
        005     & 101    & 111  \\
        006     & 110    & 101  \\
        007     & 111    & 100  \\
        \hline
        \hline
    \end{tabular}
    \caption{
        Decimal identifiers of three-dimensional hypercubic vertices,
        and corresponding binary and Gray codes~\cite{Gray1953, Gardner1972}.
        The first digit (left-most digit) of the Gray code is the same as
        the binary code.
        One obtains the digit of the Gray code
        by performing an exclusive-\textsc{or} (\textsc{xor}) operation
        on the corresponding digit of the binary code with its left digit.
    }
    \label{tab:decimal-binary-gray-3d}
\end{table}

\begin{table}[tb]
    \begin{tabular}{ccc}
        \hline
        \hline
        Decimal & Binary & Gray \\
        \hline
        0000    & 0000   & 0000 \\
        0001    & 0001   & 0001 \\
        0002    & 0010   & 0011 \\
        0003    & 0011   & 0010 \\
        0004    & 0100   & 0110 \\
        0005    & 0101   & 0111 \\
        0006    & 0110   & 0101 \\
        0007    & 0111   & 0100 \\
        0008    & 1000   & 1100 \\
        0009    & 1001   & 1101 \\
        0010    & 1010   & 1111 \\
        0011    & 1011   & 1110 \\
        0012    & 1100   & 1010 \\
        0013    & 1101   & 1011 \\
        0014    & 1110   & 1001 \\
        0015    & 1111   & 1000 \\
        \hline
        \hline
    \end{tabular}
    \caption{
        Decimal identifiers of four-dimensional hypercubic vertices,
        and corresponding binary and Gray codes~\cite{Gray1953, Gardner1972}.
        The first digit (left-most digit) of the Gray code is the same as
        the binary code.
        One obtains the digit of the Gray code
        by performing an exclusive-\textsc{or} (\textsc{xor}) operation
        on the corresponding digit of the binary code with its left digit.
    }
    \label{tab:decimal-binary-gray-4d}
\end{table}

An alternative method to project a hypercube is
optimizing the coordinates of projected vertices following
a predefined error function~\cite{Abramson2003}.
While this method projects hypercubes automatically,
it has several limitations.
The first limitation is reproducibility.
Because of the stochastic nature of the optimization process,
projections generated from the same vertices are not guaranteed to be the same
(unless all relevant parameters are fixed).
Interpretability is another weakness of projecting hypercubes
by optimization.
Because of its nonlinear nature,
what is indicated from the resulting plot
is not obvious (see, however, Refs.~\cite{Gower1988, Gower1992, Faust2019}).
How can one create reproducible and interpretable
projections of hypercubes?

Here, we reveal that principal component analysis
(PCA)~\cite{Pearson1901, Hotelling1933}
provides reproducible, interpretable, and automatic projections of hypercubes.
PCA is a linear dimensionality reduction method frequently performed in
statistics and machine learning~\cite{Jolliffe2002, Bishop2006}.
By interpreting the principal component (PC) loading
as the basis for the projection of a hypercube,
we show that one can draw the edges of a hypercube
on a projected two-dimensional plane.
This idea is closely related to the biplot~\cite{Gabriel1971},
a method to visualize loadings (eigenvectors) with data points
to assist in the interpretation of resulting plots.
By combining ideas from geometry and statistics (or machine learning),
we achieve informative projections of hypercubes~\cite{Gerber2019}:
we can analytically obtain the properties of the projection,
and the resulting plots are examined with such prior knowledge.
Such prior knowledge enhances
our interpretation of the resulting plots,
enabling a deeper understanding of high-dimensional systems.
As an application of these informative projections of hypercubes,
we visualize the state space of the Ising spin system and demonstrate that
the dynamical behavior of the system can be inferred from the resulting plots.

This paper is organized as follows.
In Sec.~\ref{sec:orthogonal-projection-hypercubes},
we present the Hamming and fractal projections with
introducing manual orthogonal projections of hypercubes.
Then, in Sec.~\ref{sec:interpreting-pca},
we show how biplots enable us to interpret the resulting plots of PCA on
hypercubic vertices.
We provide some examples of orthogonal projections of hypercubes using PCA
in Sec.~\ref{sec:hypercubic-pca}.
Through analytical and numerical investigations
in Sec.~\ref{sec:pc-loading-corresponding},
we show that principal components are informative for weighted vertices.
In Sec.~\ref{sec:quality-centrality-upper-bound},
using the inner-product error,
we show that the vertices around the origin of the projected space
less accurately preserve the original distances between them.
We apply our method to visualize the hypercubic energy landscapes and
probability flux of Ising spin systems in Sec.~\ref{sec:applications}
with analytical understanding through the mean-field model,
and conclude this paper in Sec.~\ref{sec:conclusions}.

\section{Orthogonal Projections of Hypercubes}
\label{sec:orthogonal-projection-hypercubes}
Orthogonal projection is a linear method to project a high-dimensional object.
We introduce the concept of the contribution vector,
which is the projected unit vector of each dimension.
We then illustrate isometric, Hamming, and fractal projections of hypercubes
by manually varying the contribution basis.

\subsection{Dimensionality reduction and orthogonal projection}
To visualize hypercubic vertices or high-dimensional binary data,
one needs to reduce the dimensionality of the data.
Dimensionality reduction methods are roughly divided into two categories:
nonlinear and linear methods.
The former often employ optimization of target functions
to determine the projected coordinates,
while the latter corresponds to linear projection.
Modern nonlinear dimensionality reduction methods, such as
t-distributed stochastic neighbor embedding (t-SNE)~\cite{vanderMaaten2008}
and
uniform manifold approximation and projection (UMAP)~\cite{McInnes2018},
project high-dimensional objects by optimizing target functions
with an emphasis on patterns or clusters in the data.
Although the resulting plots preserve
the neighboring relationships between vertices well,
it is nontrivial to interpret the meaning of the projected coordinates
or to extract information from the resulting plots.
On the other hand, resulting plots of linear methods
are more intuitive because of their linearity:
the projection is a shadow of the object.

A linear method to project a high-dimensional object or polytope is suggested
in high-dimensional geometry~\cite{Coxeter1973}.
Parallel projection illustrates polytopes (or high-dimensional polyhedra)
in lower-dimensional space
by moving the vertices of the object parallel to the selected direction until
they reach the desired lower-dimensional space.
Orthogonal projection is a special form of parallel projection
where the selected direction is orthogonal to the lower-dimensional space.
It is known that orthogonal projection can be derived from
concentrically overlapping the cross-sections (sections) of a polytope
and connecting the pairs of vertices of edges,
with the condition that the cross-sections include the vertices and
are parallel to the targeted lower-dimensional space of the object.
For projections of hypercubes,
orthogonal projection is a natural choice:
edges that are parallel before the projection remain parallel,
and edges that are parallel to each other share the same length after
the projection~\cite{Segerman2013},
which assists in the interpretation of the resulting plots.

The coordinates
$
\bm{r}
\coloneqq
\begin{bsmallmatrix}
    r_1 & r_2
\end{bsmallmatrix}^\top
\in
\mathbb{R}^2
$
of an orthogonally projected hypercubic vertex in two-dimensional space
are given as a linear transformation of binary coordinates $\bm{b}$:
\begin{equation}
    \begin{bmatrix}
        \vrule \\
        \bm{r} \\
        \vrule
    \end{bmatrix}
    =
    \begin{bmatrix}
        \text{---} & \bm{v}_1^\top & \text{---} \\
        \text{---} & \bm{v}_2^\top & \text{---}
    \end{bmatrix}
    \begin{bmatrix}
        \vrule \\
        \bm{b} \\
        \vrule
    \end{bmatrix}
    ,
    \label{eq:orthogonal-projection}
\end{equation}
where
$
\bm{b}
\coloneqq
\begin{bsmallmatrix}
    b_1 & \cdots & b_N
\end{bsmallmatrix}^\top
\in \left\{1, 0\right\}^N
$
is an $N$-dimensional coordinate of the hypercubic vertices.
The reduced original dimensions are represented by
two $N$-dimensional vectors,
$\bm{v}_1, \bm{v}_2 \in \mathbb{R}^N$.
We can write Eq.~\eqref{eq:orthogonal-projection} using
the transformation matrix
$
\bm{V}
\coloneqq
\begin{bsmallmatrix}
    \bm{v}_1 & \bm{v}_2
\end{bsmallmatrix}
\in \mathbb{R}^{N \times 2}
$,
\begin{equation}
    \bm{r}
    =
    \bm{V}^\top
    \bm{b}
    .
    \label{eq:orthogonal-projection-matrix}
\end{equation}

\subsection{Contribution vector}
To understand the resulting projection,
we rewrite~\cite{Coxeter1973} Eq.~\eqref{eq:orthogonal-projection} using the
\emph{contribution basis},
\begin{equation}
    \bm{r}
    =
    \sum_{i=1}^{N}
    b_i
    \tilde{\bm{e}}_i
    \label{eq:contribution-vector-to-projection}
    ,
\end{equation}
where the contribution vector,
\begin{equation}
    \tilde{\bm{e}}_i
    \coloneqq
    \begin{bmatrix}
        v_{i;1} \\
        v_{i;2}
    \end{bmatrix}
    ,
\end{equation}
is the $i$th column of the projection matrix
$
\begin{bsmallmatrix}
    \bm{v}_1 & \bm{v}_2
\end{bsmallmatrix}^\top
$,
\begin{equation}
    \begin{bmatrix}
        \vrule           &        & \vrule           \\
        \tilde{\bm{e}}_1 & \cdots & \tilde{\bm{e}}_N \\
        \vrule           &        & \vrule
    \end{bmatrix}
    \coloneqq
    \begin{bmatrix}
        \text{---} & \bm{v}_1^\top & \text{---} \\
        \text{---} & \bm{v}_2^\top & \text{---}
    \end{bmatrix}
    .
\end{equation}
Here, $v_{i;j}$ is the $i$th element of $\bm{v}_j$.

The interpretation of the contribution vector
in Eq.~\eqref{eq:contribution-vector-to-projection}
is the contribution from the $i$th dimension to the projected space.
Because $b_i \in \left\{1, 0\right\}$,
the contribution vector corresponds to the projected unit vector of
each dimension.
Indeed, the alternative definition of contribution vector is
\begin{equation}
    \tilde{\bm{e}}_i
    \coloneqq
    \bm{V}^\top
    \bm{e}_i
\end{equation}
where
$
\bm{e}_i
\coloneqq
\begin{bsmallmatrix}
    0 & \cdots & 0 & 1 & 0 & \cdots & 0
\end{bsmallmatrix}^\top
\in
\left\{1, 0\right\}^N
$
is the $i$th standard unit vector in the original space.
Thus, by introducing the contribution basis $\left\{\bm{e}_i\right\}_{i=1}^N$,
we can decompose the resulting projection into
the contributions from each dimension as shown in
Eq.~\eqref{eq:contribution-vector-to-projection}.
The remaining question is how to determine the contribution basis
$\left\{\tilde{\bm{e}}_i\right\}_{i=1}^N$.

\begin{figure*}[p]
    \includegraphics{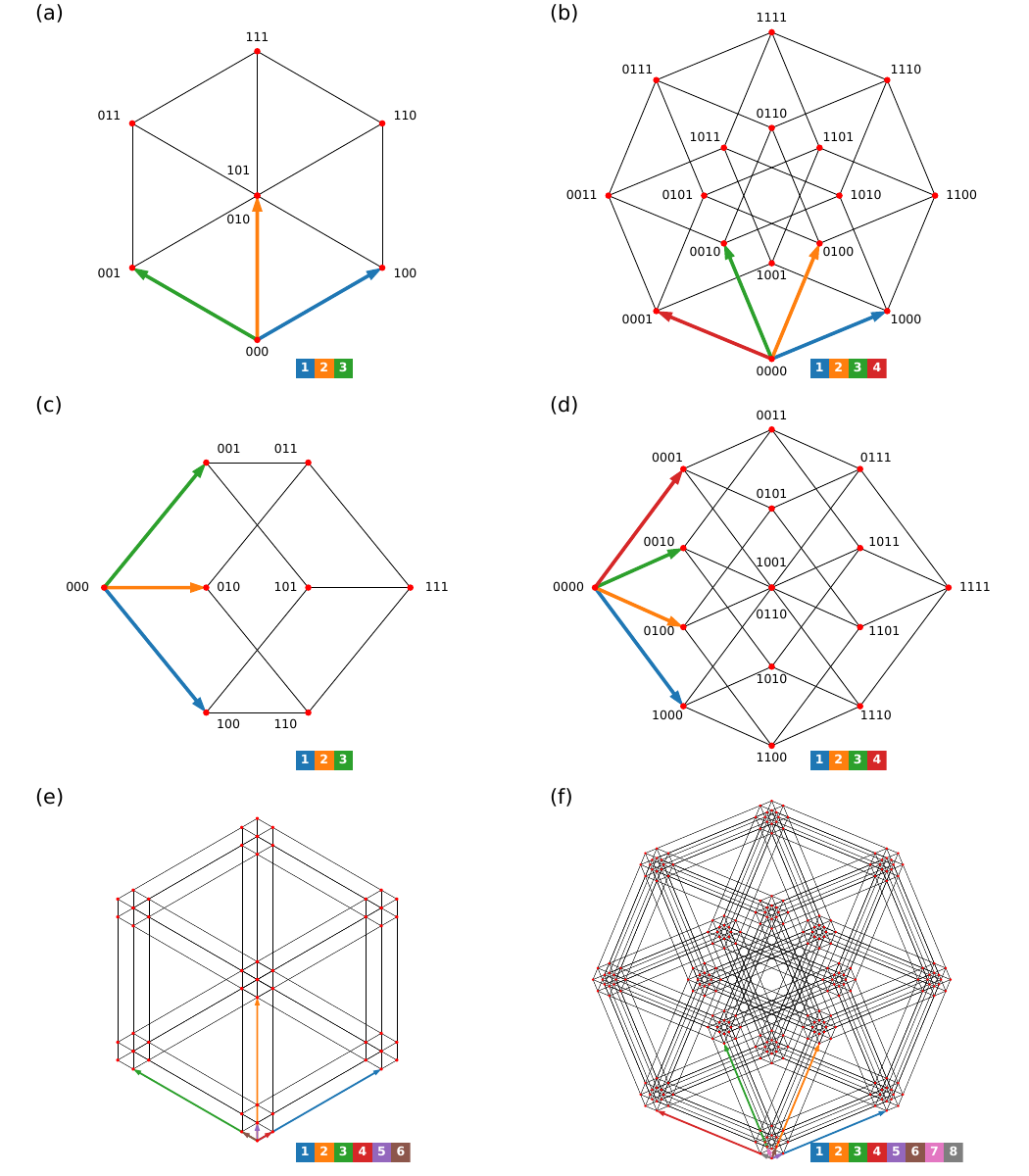}
    \caption{
        A gallery of orthogonal projections of hypercubes.
        Colored arrows represent the contribution vectors corresponding to
        the original dimensions.
        The boxes on the bottom right indicate
        the correspondence between the colors and the original dimensions.
        (a) An isometric projection of a cube.
        Notice that
        $
        \begin{bsmallmatrix}
            1 & 0 & 1
        \end{bsmallmatrix}^\top
        $
        and
        $
        \begin{bsmallmatrix}
            0 & 1 & 0
        \end{bsmallmatrix}^\top
        $
        are overlapped.
        (b) An isometric projection of a tesseract.
        (c) A Hamming projection of a cube.
        (d) A Hamming projection of a tesseract.
        Notice that
        $
        \begin{bsmallmatrix}
            1 & 0 & 0 & 1
        \end{bsmallmatrix}^\top
        $
        and
        $
        \begin{bsmallmatrix}
            0 & 1 & 1 & 0
        \end{bsmallmatrix}^\top
        $
        are overlapped.
        In Hamming projections (c) and (d),
        the contribution vector of each dimension has
        the same horizontal contribution.
        (e) A fractal projection of a six-dimensional hypercube.
        (f) A fractal projection of an eight-dimensional hypercube.
        In fractal projections (e) and (f),
        the contribution vectors of the first half (left half) of the code are
        ten times longer than the rest.
        More projections of hypercubes are available in
        the Supplemental Material~\cite{sm}.
    }
    \label{fig:isometric-hamming-fractal}
\end{figure*}

\subsection{Isometric projection}
One particularly regular and symmetric projection is called
isometric projection,
where a Petrie polygon is projected as
a regular polygon located at the periphery
of the projection of a polytope~\cite{Coxeter1973}.
Petrie polygons are equatorial polygons,
lying in planes crossing the center of the object
and inscribed in great circles of the circumsphere of the object.
To achieve isometric projection, one needs
the projected unit vector of the $i$th dimension
to form consecutive edges of the projected Petrie polygon, i.e.,
$\bm{e}_i$ form a Petrie polygon together with a reversed vector $-\bm{e}_i$.
To project the Petrie polygon as a regular polygon,
the contribution basis are determined as
$
\bm{e}_i
=
\begin{bsmallmatrix}
    \cos \left(\theta_i\right) &
    \sin \left(\theta_i\right)
\end{bsmallmatrix}^\top
$,
where
$\theta_i = \frac{\uppi}{N}\left(i-1\right) + \phi$,
with a constant $\phi$.
The resulting isometric projections and vectors $\bm{e}_i$ used to create them
are shown in Figs.~\ref{fig:isometric-hamming-fractal}(a)
and~\ref{fig:isometric-hamming-fractal}(b),
where Petrie polygons are projected as a regular hexagon
[Fig.~\ref{fig:isometric-hamming-fractal}(a)] and
octagon [Fig.~\ref{fig:isometric-hamming-fractal}(b)].
We determine the angle $\phi = \frac{\uppi}{2N}$ for symmetry of
the projections.
In isometric projections,
all edges of the hypercube are drawn with the same length.

\subsection{Hamming and fractal projections}
By modifying the contribution basis
$\left\{\tilde{\bm{e}}_i\right\}_{i=1}^N$
for projection,
one can \emph{view} a high-dimensional hypercube from various angles.
Several methods reflecting the cross-section of the hypercube
have already been suggested~\cite{Coxeter1973}.
Here, we introduce two other orthogonal projections of hypercubes
used in sciences.

A Hamming projection
[Figs.~\ref{fig:isometric-hamming-fractal}(c)
and~\ref{fig:isometric-hamming-fractal}(d)],
we call,
is a type of projection
that plots vertices according to the Hamming distance~\cite{Hamming1950}
from a selected vertex.
To achieve this, an example of a contribution vector is
$
\tilde{\bm{e}}_i
=
\begin{bsmallmatrix}
    1 &
    \left(i - \frac{N+1}{2}\right)
    \frac{2}{N-1}
    \sin \left(\frac{\varphi}{2}\right)
\end{bsmallmatrix}^\top
$,
where
$
\varphi
\coloneqq
\arccos
\left(
    \frac
    {\tilde{\bm{e}}_1^\top \tilde{\bm{e}}_N}
    {\sqrt{\tilde{\bm{e}}_1^\top \tilde{\bm{e}}_1}
    \sqrt{\tilde{\bm{e}}_N^\top \tilde{\bm{e}}_N}}
\right)
$
is the angle between the first and last contribution vector, but it can be any.
The first element of $\tilde{\bm{e}}_i$ is the same for all $i$.
The second element can be any,
but we determine them to increase linearly with $i$.
This type of projection is found
not only in research in
physics\footnote{
    Examples are found in quantum walk~\cite{Shenvi2003},
    unsupervised learning of states of Ising spin
    system~\cite{Wang2016,Hu2017,Kiwata2019},
    quantum many-body scar states~\cite{Turner2018, Turner2018a, Turner2021},
    and quantum many-body
    localization~\cite{Roy2019,Roy2020,Roy2020a,DeTomasi2021,Roy2021,Roy2025,Creed2023}.
}
but also in
biology.\footnote{
    Examples include evolutionary landscapes~\cite{Wright1932,
        DeVisser2014, Kondrashov2015, Mira2015, Johnston2016,
    Zagorski2016, Moen2023, Weaver2024},
    epistasis~\cite{Wright1932, Esteban2020},
    genetic code space~\cite{Greenbury2020},
    data visualization~\cite{Acar2010, Peng2015},
    protein folding~\cite{Prinz2011, Olsson2017},
    gene regulation~\cite{Estrada2016},
    disease progression~\cite{Greenbury2020, Moen2023},
    and allostery~\cite{Biddle2021}.
}
The horizontal distance between
the Hamming projected hypercubic vertices corresponds to
the Hamming distance from a reference vertex on the rightmost or leftmost side
of the projected space.
In Figs.~\ref{fig:isometric-hamming-fractal}(c)
and~\ref{fig:isometric-hamming-fractal}(d),
the horizontal axis corresponds to the Hamming distance from the origin
$
\begin{bsmallmatrix}
    0 & 0 & 0
\end{bsmallmatrix}^\top
$
and
$
\begin{bsmallmatrix}
    0 & 0 & 0 & 0
\end{bsmallmatrix}^\top
$, respectively.
Notice that this projection corresponds to
the Hasse diagram~\cite{Birkhoff1967}.

A fractal projection
[Figs.~\ref{fig:isometric-hamming-fractal}(e)
and~\ref{fig:isometric-hamming-fractal}(f)],
we name,
has coordinates of vertices in a fractal pattern,
i.e., if one magnifies one of the clusters of vertices,
one finds a structure similar to the whole.
We determine the contribution vector as
$
\tilde{\bm{e}}_i
=
\begin{bsmallmatrix}
    \cos \left(\theta_i\right) &
    \sin \left(\theta_i\right)
\end{bsmallmatrix}^\top
$
for
$
i
\in
\left\{
    i \in \mathbb{Z}
    \, \middle| \,
    1 \leq i \leq \frac{N}{2}
\right\}
$,
and
$
\tilde{\bm{e}}_i
=
\zeta
\begin{bsmallmatrix}
    \cos \left(\theta_i\right) &
    \sin \left(\theta_i\right)
\end{bsmallmatrix}^\top
$
for
$
i
\in
\left\{
    i \in \mathbb{Z}
    \, \middle| \,
    \frac{N}{2} + 1 \leq i \leq N
\right\}
$,
to create Figs.~\ref{fig:isometric-hamming-fractal}(e)
and~\ref{fig:isometric-hamming-fractal}(f) as examples.
Here, $\zeta \sim 0.1 \ll 1$ is the length of the smaller edge,
and $N$ is restricted to be even in our example.
This projection is found in electrical engineering~\cite{Kumar1992, Malluhi1994}
and genetic code visualization~\cite{Jimenez-Montano1996}.
This projection method provides a way to visualize the clusters of vertices
in a high-dimensional space.
Generalization to odd-dimensional hypercubes is possible
by repeating fractal structures while ignoring a single contribution vector.

\subsection{Exchanging the labels of vertices}
One can exchange the labels or original coordinates of projected vertices
after performing the projection.
Suppose we have vectors
$\bm{b}^\prime, \bm{b}^{\prime\prime} \in \left\{1, 0\right\}^N$
with binary elements representing vertices of a hypercube,
and we want to swap their labels and exchange other labels accordingly.
One can obtain swapped labels by performing the conversion~\cite{Abramson2003},
$
\bm{b}
\leftarrow
\left(
    \bm{b}
    \oplus
    \bm{b}^\prime
\right)
\oplus
\bm{b}^{\prime\prime}
$,
where $\oplus$ is the bitwise exclusive-\textsc{or} (\textsc{xor}) operation.

\begin{figure*}[tb]
    \includegraphics{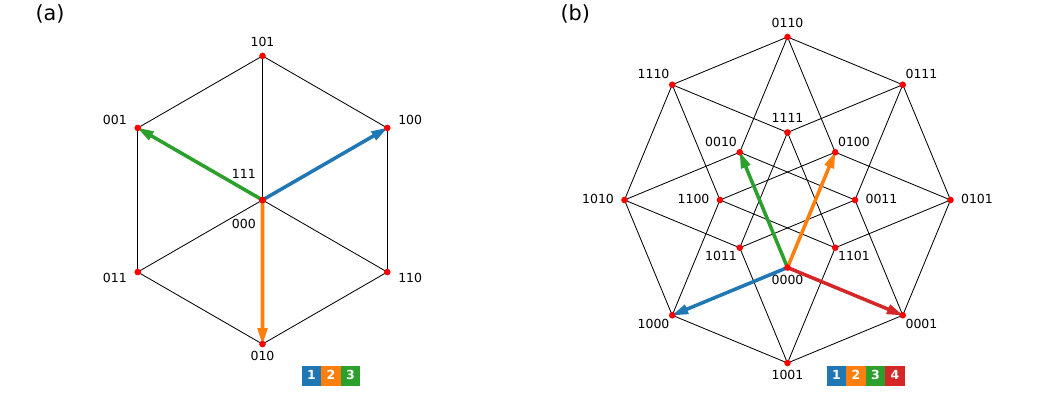}
    \caption{
        Exchanging the labels of vertices
        by reversing the direction of
        the contribution vectors of corresponding digits.
        The boxes on the bottom right indicate
        the correspondence between the colors and the dimensions.
        (a) By reversing the contribution vector for the second digit,
        one can exchange the cubic labels of pairs of vertices,
        $
        \begin{bsmallmatrix} 0 & 0 & 0
        \end{bsmallmatrix}^\top$
        and
        $
        \begin{bsmallmatrix} 0 & 1 & 0
        \end{bsmallmatrix}^\top$,
        $
        \begin{bsmallmatrix} 1 & 0 & 0
        \end{bsmallmatrix}^\top$
        and
        $
        \begin{bsmallmatrix} 1 & 1 & 0
        \end{bsmallmatrix}^\top$,
        $
        \begin{bsmallmatrix} 0 & 0 & 1
        \end{bsmallmatrix}^\top$
        and
        $
        \begin{bsmallmatrix} 0 & 1 & 1
        \end{bsmallmatrix}^\top$,
        and
        $
        \begin{bsmallmatrix} 1 & 0 & 1
        \end{bsmallmatrix}^\top$
        and
        $
        \begin{bsmallmatrix} 1 & 1 & 1
        \end{bsmallmatrix}^\top$.
        Compare this with Fig.~\ref{fig:isometric-hamming-fractal}(a).
        (b) Obtaining different labels for a tesseract
        by swapping two contribution vectors.
        The first and fourth unit vectors are reversed,
        cf. Fig.~\ref{fig:isometric-hamming-fractal}(b).
    }
    \label{fig:swapping-vector}
\end{figure*}

Here, we show a visual understanding of this conversion
through contribution vectors.
By reversing the direction of the contribution vector,
one can swap the projected coordinates of the rest,
cf.
Figs.~\ref{fig:isometric-hamming-fractal}(a)
and~\ref{fig:isometric-hamming-fractal}(b),
or Figs.~\ref{fig:swapping-vector}(a)
and~\ref{fig:swapping-vector}(b).
The dimension of the reversed contribution vector corresponds to
the dimension where there is a $1$ in
$\bm{b}^\prime \oplus \bm{b}^{\prime\prime}$.
Visualizing the projected vector provides
a way to understand exchanging the labels via
reversing the direction of unit vectors.

\subsection{The limitation of manual orthogonal projections}
Although one can create reproducible and interpretable
(and even visually label-exchangeable)
two-dimensional projections
of hypercubes in the abovementioned way,
the disadvantage is that one must manually determine the contribution basis of
each dimension.
For low-dimensional hypercubes,
it is manageable to adjust the contribution basis oneself
to make the projections easy to interpret.
Still, it is nontrivial to decide on contribution basis for
high-dimensional hypercubes.
Thus, automatic and unsupervised projection methods are demanded,
especially for projections of high-dimensional hypercubes.
How can one project a hypercube with such a strategy?

\section{Orthogonal projections of hypercubes using PCA}
\label{sec:interpreting-pca}
To answer the question raised in the previous section---how can one create
a linear but automatic and unsupervised projection of hypercubes---we suggest
using a linear dimensionality reduction method: PCA\@.
In this section, we derive PCA from minimizing an error function,
but how can one interpret the projection of a hypercube using PCA\@?
We show that biplot enable us to interpret PCA as an orthogonal projection of
a hypercube.

\subsection{PCA by minimizing inner-product error}
\label{sec:pca-minimizing-inner-product-error}
Unsupervised and automatic dimensionality reduction is possible by
minimizing an error function to create projections of high-dimensional objects.
Unlike the previous study~\cite{Abramson2003},
where the error function is based on the difference between
the pairwise Euclidean distance in low-dimensional projected space and
the Hamming distance in high-dimensional original space,
we suggest minimizing the difference of the
\emph{inner product} between vertices in the projected and original space.
The inner product measures the similarity between two vectors, thus
minimizing the difference between them results in a projection that
preserves the original similarity between the vertices.

We define the error as the difference of the inner product
in the original space and the projected space:
\begin{equation}
    \varepsilon\left(\bm{s}, \bm{s}^\prime\right)
    \coloneqq
    \bm{s}^\top
    \bm{s}^\prime
    -
    \bm{r}^\top \left(\bm{s}\right)
    \bm{r} \left(\bm{s}^\prime\right)
    ,
    \label{eq:inner-product-error}
\end{equation}
where,
$
\bm{s}
\coloneqq
\begin{bsmallmatrix}
    s_1 & \cdots & s_N
\end{bsmallmatrix}^\top
\in \left\{+1, -1\right\}^N
$
is a coordinate of the hypercubic vertices in the original space but with
the Ising spin variable~\cite{Ising1925, Newell1953, Brush1967, Cipra1987}.
Instead of binary variables $b_i \in \left\{ 1, 0\right\}$,
we introduce Ising spin variables $s_i = 2b_i - 1 \in \left\{+1, -1\right\}$
to calculate the inner product
because the inner product among the Ising variables reflects
the similarity or overlap between the vertices.
We need to find the vectors in the projection matrix
$
\begin{bsmallmatrix}
    \bm{v}_1 & \bm{v}_2
\end{bsmallmatrix}^\top
$
of Eq.~\eqref{eq:orthogonal-projection} that minimizes the mean squared error
\begin{equation}
    \left<\varepsilon^2\right>
    \coloneqq
    \sum_{\bm{s}, \bm{s}^\prime}
    p\left(\bm{s}\right)
    p\left(\bm{s}^\prime\right)
    \varepsilon^2\left(\bm{s}, \bm{s}^\prime\right)
    \label{eq:mean-squared-inner-product-error}
    ,
\end{equation}
where $p\left(\bm{s}\right) \in \left[0, 1\right]$
is the normalized weight, i.e., the probability\footnote{
    We use the notation
    $p\left(\bm{s}\right)$
    as the probability mass function
    (discrete probability distribution).
}
of finding the Ising coordinate $\bm{s}$,
satisfying $\sum_{\bm{s}} p\left(\bm{s}\right) = 1$.

Minimizing $\left<\varepsilon^2\right>$
with the normalization constraint
$\left\{\bm{v}_i^\top \bm{v}_i = 1\right\}_{i=1}^2$
corresponds to classical multidimensional scaling
(MDS)~\cite{Young1938, Torgerson1952}.
Classical MDS\footnote{
    Also called
    classical scaling,
    Torgerson scaling,
    Torgerson--Gower scaling,
    or principal coordinates analysis.
}
with squared Euclidean (Pythagorean) distance
provides lower-dimensional scaled coordinates preserving
the original distance between the vertices.
It is shown that classical MDS with squared Euclidean distance is equivalent to
PCA~\cite{Gower1966},
which is a \emph{linear} dimensionality reduction method.
See relevant studies~\cite{Cox2001,Borg2005,Ghojogh2023}
for the equivalence of classical MDS and PCA\@.
Interested readers can obtain the derivation of PCA from
the minimum inner-product error formulation and other formulations
(maximum projection variance formulation
and minimum reconstruction error formulation)
in Appendix~\ref{sec:formulations-pca}.

The minimum inner-product error formulation of PCA has two advantages.
First, unlike other formulations of PCA,
the minimum inner-product error formulation minimizes a function that sums
a term involving pairs of vertices
[Eq.~\eqref{eq:mean-squared-inner-product-error}]
rather than minimizing (or maximizing) a function
summing a term involving a single vertex
[Eqs.~\eqref{eq:variance-func} and~\eqref{eq:reconstruction-error-func}].
Thus, we can examine which distortions between pairs of vertices contribute to
the error.
We numerically and analytically investigate the quality of the projection of
hypercubes through the inner-product error
in Sec.~\ref{sec:quality-centrality-upper-bound} by taking advantages of
this feature.
Second, this formulation is useful for understanding
the projection of hypercubes
because we can investigate where the error arises in the projection.
The minimum inner-product error formulation of PCA can provide
a visual understanding of distortion in the projection of hypercubes.
Unlike the variance, reconstruction error or distance, the inner-product error
can be readily estimated from the resulting plot.
Through the minimum inner-product error formulation of PCA,
we can visually estimate the quality of the projection of hypercubes.

As we show in the following
Secs.~\ref{sec:pc-loading-corresponding-analytical},~\ref{sec:number-of-vertices-along-pc1}
and~\ref{sec:upper-bound},
the projection of hypercubes using PCA is informative because the properties of
the projection can be analytically obtained.
In addition to the biplot-assisted interpretation of PCA
in the next Sec.~\ref{sec:pca-and-biplot},
we can analytically obtain the properties of the projection of hypercubes
for special cases.
We analytically reveal both the strengths and weaknesses of PCA
in projecting hypercubes in
Secs.~\ref{sec:pc-loading-corresponding-analytical},~\ref{sec:number-of-vertices-along-pc1}
and~\ref{sec:upper-bound}.
With such prior knowledge of the projection of hypercubes,
we can examine the resulting plots more deeply,
even if the knowledge is from special cases.

\subsection{PCA and biplot}
\label{sec:pca-and-biplot}
Considering that PCA is an unsupervised and linear method for projection,
we suggest interpreting PCA on binary vertices
as a reproducible, interpretable, and automatic projection of a hypercube.
PCA~\cite{Pearson1901, Hotelling1933},
a method of statistics and unsupervised machine learning,
is an essential technique for analyzing high-dimensional
data~\cite{Jolliffe2002, Bishop2006, Jolliffe2016, Greenacre2022}.
By calculating eigenvalues (explained variance)
and eigenvectors (PC loadings) of the covariance matrix,
PCA finds a set of vectors that maximize the variance of the data\footnote{
    See Appendix~\ref{sec:maximum-projection-variance-formulation}
    for the derivation of PCA by the maximum projection variance formulation.
}
and is often employed to perform linear dimensionality reduction.\footnote{
    In practical applications of PCA,
    the probability distribution $p\left(\bm{s}\right)$ is approximated by
    the empirical distribution of the data points.
}
To perform PCA, one needs to calculate the covariance matrix of the data points.
Specifically, in the context of hypercubic vertices, the covariance matrix
$\bm{\varSigma} \in \mathbb{R}^{N\times N}$ is defined as
\begin{align}
    \bm{\varSigma}
    &\coloneqq
    \left<
    \left(
        \bm{s}
        -
        \left<\bm{s}\right>
    \right)
    \left(
        \bm{s}
        -
        \left<\bm{s}\right>
    \right)^\top
    \right>
    \nonumber
    \\&=
    \sum_{\bm{s}}
    p\left(\bm{s}\right)
    \left(
        \bm{s}
        -
        \left<\bm{s}\right>
    \right)
    \left(
        \bm{s}
        -
        \left<\bm{s}\right>
    \right)^\top
    \label{eq:covariance-matrix}
    ,
\end{align}
where
$
\left<\bm{s}\right>
\coloneqq
\sum_{\bm{s}}p\left(\bm{s}\right)\bm{s}
$
is the mean vector of the Ising coordinate.

Obtaining the eigenvalues $\left\{\lambda_i\right\}_{i=1}^N$ and
eigenvectors $\left\{\bm{u}_i\right\}_{i=1}^N$
of the covariance matrix corresponds to
finding the PCs of the hypercubic vertices.
The eigenvalue equation is
\begin{equation}
    \bm{\varSigma}
    \bm{u}_i
    =
    \lambda_i
    \bm{u}_i
    ,
\end{equation}
where the eigenvalues are sorted in descending order,
$\lambda_1  \geq \cdots \geq \lambda_N \geq 0$,
and the fraction of explained variance by the $i$th eigenvector is defined as
normalized explained variance
$
\tilde{\lambda}_i
\coloneqq
\frac{
    \lambda_i
}{
    \sum_{i=1}^{N} \lambda_i
}
$.
The magnitude of the eigenvalues indicates the explained variance,
i.e.,
the importance of the corresponding eigenvectors in the projection.
By assigning the eigenvectors with the largest eigenvalues as
the direction of the projection in Eq.~\eqref{eq:orthogonal-projection}, i.e.,
$\bm{v}_1 = \bm{u}_1$,
one can perform a linear projection of the hypercube
where the projected coordinate corresponds to the PC1 score
$
r_1 \left(\bm{s}\right)
=
\bm{u}_1^\top \bm{s}
$.
In general, one obtains the projected coordinate by PC$i$ as
\begin{equation}
    r_i \left(\bm{s}\right)
    =
    \bm{u}_i^\top \bm{s}
    .
\end{equation}
Hence, PCA provides high-dimensional projection vectors preserving
the original distance, similarity, or variance as much as possible.
Although one can have the basis of the orthogonal projection using PCA,
how can one interpret the resulting plot as a projection of hypercubes?

It has been known that biplots~\cite{Gabriel1971}
assist in the interpretation of data points in the projected space
by plotting both data points transformed using PCA
and arrows indicating the contribution of each original dimension to
PCs~\cite{Gower1996, Gower2011}.
For example, when we plot the data points by the first two PCs,
an arrow that has the $i$th element of $\bm{u}_1$ and $\bm{u}_2$,
i.e.,
$
\begin{bsmallmatrix}
    u_{i;1} & u_{i;2}
\end{bsmallmatrix}^\top
$
is plotted as the contribution from the $i$th dimension to the plot.
Here, we suggest interpreting the arrow of a biplot
as the contribution vector of each dimension of a hypercube,
namely loading contribution vector of PC$j$ and $k$ as
\begin{equation}
    \tilde{\bm{e}}_i
    =
    2
    \begin{bmatrix}
        u_{i;j} \\
        u_{i;k}
    \end{bmatrix}
    \label{eq:biplot-vector}
    .
\end{equation}
For PCA, we call contribution vector
$\tilde{\bm{e}}_i$
\emph{biplot vector}.
Notice that,
when the data points are binary vertices
$\bm{b} \in \left\{1, 0\right\}^N$ of a system,
the biplot basis exactly match the basis of
the orthogonal projection of a hypercube,
while when the data points are Ising vertices
$\bm{s} \in \left\{+1, -1\right\}^N$,
the biplot basis with doubled magnitudes exactly match the contribution basis.
This difference arises from the difference in the length of hypercubic edges:
they are one for binary vertices and two for Ising vertices.
This is why we add factor $2$ in Eq.~\eqref{eq:biplot-vector}.
By performing PCA and plotting biplot vectors or biplot basis,
one can obtain
a reproducible, interpretable, and automatic projection of a hypercube.

\section{Hypercubic PCA}
\label{sec:hypercubic-pca}
In the previous Sec.~\ref{sec:interpreting-pca},
we introduced PCA as a method to project hypercubes
and presented the interpretation of the projection using biplot vectors.
PCA, in practice, requires the probability distribution $p\left(\bm{s}\right)$
of the vertices $\bm{s}$ to calculate the covariance matrix.
How does $p\left(\bm{s}\right)$ affect the projection of hypercubes?
How do the resulting coordinates of the vertices change
when $p\left(\bm{s}\right)$ is varied?
In this section, we project hypercubes using
PCA while varying the probability distribution $p\left(\bm{s}\right)$
of the vertices.
Throughout the examples, we reveal the trends of the resulting projections:
the leading PC corresponds to the vertices with higher probability,
and PCA distorts the distance between the vertices
around the origin of the projected space.

\begin{figure*}[p]
    \includegraphics{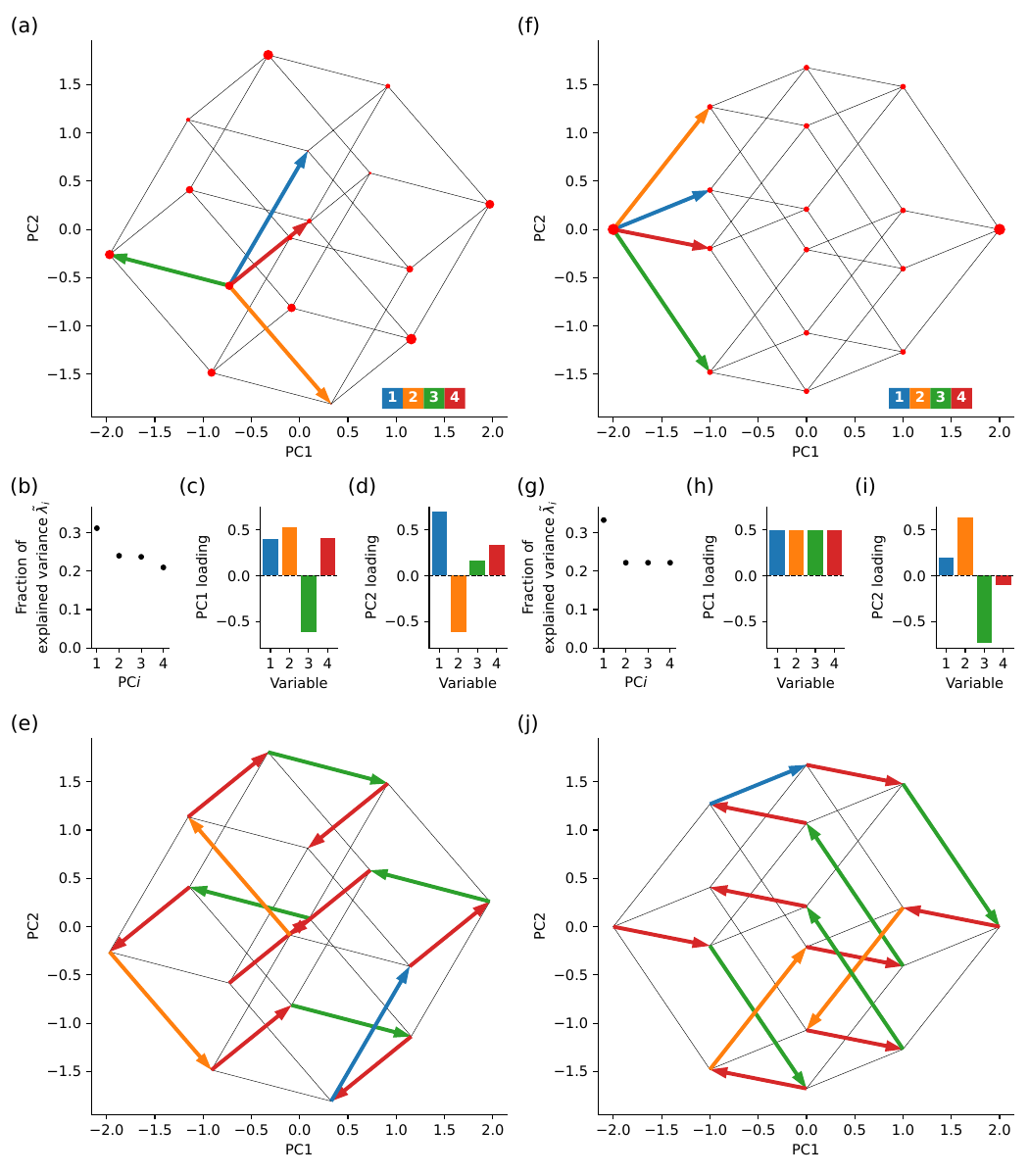}
    \caption{
        Orthogonal projections of four-dimensional hypercubic vertices
        using PCA\@.
        (a) A projection of a four-dimensional hypercube where vertices are
        weighted randomly by uniform random numbers in $\left[0, z\right)$.
        Red filled circles are the vertices and lines are the edges of
        the hypercube.
        The magnitude of weight is proportional to the area of the vertex.
        Arrows are biplot vectors originating from
        $
        \begin{bsmallmatrix}
            - & - & - & -
        \end{bsmallmatrix}^\top
        $
        and the boxes on the bottom right indicate
        the correspondence between the colors of the arrows and
        the original dimensions.
        (b) Fraction of explained variance by each PC of (a).
        (c) PC1 loading,
        and (d) PC2 loading
        of random weighted hypercubic vertices of (a).
        (e) Hamiltonian path on a four-dimensional hypercube in (a).
        For a different realization of random weight, see the Supplemental
        Material~\cite{sm}.
        (f)--(j) Same as (a)--(e) but with bipolar distribution.
        Two of the vertices,
        $
        \begin{bsmallmatrix} - & - & - & -
        \end{bsmallmatrix}^\top$
        and
        $
        \begin{bsmallmatrix} + & + & + & +
        \end{bsmallmatrix}^\top$,
        are two-times more weighted than the others.
    }
    \label{fig:pca-hypercubes-random-pair-weighted}
\end{figure*}

\subsection{Random distribution}
\label{sec:random-distribution}
To begin with,
we consider the random probability distribution of the vertices.
In Fig.~\ref{fig:pca-hypercubes-random-pair-weighted}(a),
we perform PCA on a four-dimensional hypercube
with randomly weighted vertices.
The probability distribution of the vertices is drawn randomly from
a uniform distribution.
The vertices with higher probabilities,
which contribute more to the variance than the others,
tend to lie on the outer part of the projection.
Nevertheless, some less weighted vertices also lie on the outer part of
the projection.
We find that these vertices have counterparts with larger weights across
the origin of the projected space.
For example, in Fig.~\ref{fig:pca-hypercubes-random-pair-weighted}(a),
the vertex around
$
\begin{bsmallmatrix}
    0.4 & -1.7
\end{bsmallmatrix}^\top
$
has a small weight but lies on the outer part of the projection.
Across the origin
(around
    $
    \begin{bsmallmatrix}
        -0.4 & 1.7
    \end{bsmallmatrix}^\top
$),
it has a vertex with a large weight.
Because PCA tries to preserve the variance of the data,
the heavily weighted vertices tend to lie on the outer part of the projection,
and sometimes less weighted vertices are also projected to the outer part of
the projection.

The biplot vectors in Fig.~\ref{fig:pca-hypercubes-random-pair-weighted}(a)
are drawn as arrows from
$
\begin{bsmallmatrix}
    - & - & - & -
\end{bsmallmatrix}^\top
$,
assisting us in estimating the original binary coordinates of the vertices.
Here, we abbreviate $+1$ to $+$ and $-1$ to $-$.
For example,
the projected vertex around
$
\begin{bsmallmatrix}
    -2.0 & -0.25
\end{bsmallmatrix}^\top
$
corresponds to the vertex
$
\begin{bsmallmatrix}
    - & - & + & -
\end{bsmallmatrix}^\top
$
and the projected vertex around
$
\begin{bsmallmatrix}
    -0.4 & 1.7
\end{bsmallmatrix}^\top
$
corresponds to the vertex
$
\begin{bsmallmatrix}
    + & - & + & +
\end{bsmallmatrix}^\top
$.
Equivalently, one can follow the hypercubic edge from
$
\begin{bsmallmatrix}
    - & - & - & -
\end{bsmallmatrix}^\top
$
(the origin of the biplot vectors)
to the vertex of interest by changing the corresponding elements indicated by
the biplot vectors to know the original Ising coordinate.
Notice that it is not necessary to start from
$
\begin{bsmallmatrix}
    - & - & - & -
\end{bsmallmatrix}^\top
$
to infer the original Ising coordinate:
if one knows the original coordinates of any vertex,
one can infer the rest from the biplot vectors,
cf. Eq.~\eqref{eq:contribution-vector-to-projection}.

We then examine the fraction of explained variance and loading of
PCA in Fig.~\ref{fig:pca-hypercubes-random-pair-weighted}(a).
We show the fraction of explained variance by each PC
in Fig.~\ref{fig:pca-hypercubes-random-pair-weighted}(b).
PC1 explains approximately 30\% of the variance, followed by
PC2 and PC3, which explain around 22\% each,
and PC4 explains just above 20\%.
Therefore, less than 60\% of the variance is explained by the first two PCs.
With a randomly weighted hypercube,
the projection by the first two PCs does not explain
a large fraction of the variance.

To understand the projection of the hypercube using PCA,
we show the loading of PC1 and PC2 in
Figs.~\ref{fig:pca-hypercubes-random-pair-weighted}(c)
and~\ref{fig:pca-hypercubes-random-pair-weighted}(d).
The PC1 loading [Fig.~\ref{fig:pca-hypercubes-random-pair-weighted}(c)] shows
that variable 3 contributes negatively, but the rest contribute positively,
and their absolute values are almost the same.
We see that the sign of the PC1 loading corresponds to the vertices with
the highest PC1 score,
$
\operatorname{sgn}
\left(
    \bm{u}_1
\right)
=
\begin{bsmallmatrix}
    + & + & - & +
\end{bsmallmatrix}^\top
$,
indicating that the weighted vertices correspond to the PC1 loading.
Here,
$\operatorname{sgn}$ is the element-wise sign function.
Unlike the PC1 loading,
the PC2 loading [Fig.~\ref{fig:pca-hypercubes-random-pair-weighted}(d)] shows
a different pattern of contribution for each variable.
Although, similar to the PC1 loading,
the sign of the PC2 loading corresponds to the vertices with
the highest PC2 score,
$
\operatorname{sgn}
\left(
    \bm{u}_2
\right)
=
\begin{bsmallmatrix}
    + & - & + & +
\end{bsmallmatrix}^\top
$.
Leading PC captures the weighted vertices,
and the rest of the PCs capture the vertices with lower weights.

We then present the usage of the projection of the hypercube using PCA\@.
In Fig.~\ref{fig:pca-hypercubes-random-pair-weighted}(e),
the Hamiltonian path on the hypercube is shown following the biplot.
Starting from the vertex
$
\begin{bsmallmatrix}
    - & - & - & -
\end{bsmallmatrix}^\top
$,
one can follow the path to the vertex
$
\begin{bsmallmatrix}
    + & + & + & +
\end{bsmallmatrix}^\top
$
in the original high-dimensional space,
by knowing which digit changed by following the biplot vector.
These properties may be useful for understanding
pathways of state transition of the target system.
Notice, however, that the arrows indicating the Hamiltonian path overlap
around the origin of the projected space.
This overlap makes it difficult to follow the path.
We address this issue in Sec.~\ref{sec:quality-centrality-upper-bound}.

\subsection{Bipolar distribution}
\label{sec:bipolar-distribution}
Considering the results of PCA on randomly weighted hypercubic vertices, which
indicate that weighted pairs of vertices play an important role
in the projection,
we perform PCA on hypercubic vertices with a bipolar weight distribution.
In Fig.~\ref{fig:pca-hypercubes-random-pair-weighted}(f),
we show the result of PCA on a four-dimensional hypercube where
all vertices are weighted equally except for two of them.
Two of the vertices
$
\begin{bsmallmatrix}
    - & - & - & -
\end{bsmallmatrix}^\top
$
and
$
\begin{bsmallmatrix}
    + & + & + & +
\end{bsmallmatrix}^\top
$,
which are the most distant from each other in the original
four-dimensional space, are weighted more than the others.
These weighted vertices are projected to have larger magnitudes of PC1 scores,
but the rest of the vertices are projected in the order of Hamming distance from
the weighted vertices along PC1:
the resulting projection is the Hamming projection,
cf. Fig.~\ref{fig:isometric-hamming-fractal}(c)
and~\ref{fig:isometric-hamming-fractal}(d).

We show the fraction of explained variance by each PC
in Fig.~\ref{fig:pca-hypercubes-random-pair-weighted}(g)
to validate the projection
of Fig.~\ref{fig:pca-hypercubes-random-pair-weighted}(f).
Similar to the random weighted case
in Fig.~\ref{fig:pca-hypercubes-random-pair-weighted}(b),
PC1 explains approximately 30\% of the variance.
Unlike the random weighted case
in Fig.~\ref{fig:pca-hypercubes-random-pair-weighted}(b),
PC2 to PC4 have the same explained variance because of
the uniform probability except for two vertices.
The first two PCs explain a comparable proportion of variance
(approximately 50\%)
to Fig.~\ref{fig:pca-hypercubes-random-pair-weighted}(b).

We arrive at the PC loadings of PC1 and PC2 in
Figs.~\ref{fig:pca-hypercubes-random-pair-weighted}(h)
and~\ref{fig:pca-hypercubes-random-pair-weighted}(i),
where the former is expected to correspond to the weighted vertices.
The PC1 loading in Fig.~\ref{fig:pca-hypercubes-random-pair-weighted}(h) shows
that all variables contribute equally, contrary to the randomly weighted case in
Fig.~\ref{fig:pca-hypercubes-random-pair-weighted}(c).
This uniform PC1 loading in
Fig.~\ref{fig:pca-hypercubes-random-pair-weighted}(h)
supports that Fig.~\ref{fig:pca-hypercubes-random-pair-weighted}(f) is
the Hamming projection.
As expected, the most weighted vertices are projected
to have the largest magnitude of PC1 scores,
and the element-wise sign of the PC1 loading corresponds to
the most weighted vertices,
$
\operatorname{sgn}
\left(
    \bm{u}_1
\right)
=
\begin{bsmallmatrix}
    + & + & + & +
\end{bsmallmatrix}^\top
$.
Similar to the randomly weighted PCA in previous
Sec.~\ref{sec:random-distribution},
the element-wise sign of the PC2 loading in
Fig.~\ref{fig:pca-hypercubes-random-pair-weighted}(i)
corresponds to the vertices with the highest PC2 scores,
$
\begin{bsmallmatrix}
    + & + & - & -
\end{bsmallmatrix}^\top
$.
While the PC1 loading relates to the weighted vertex pair,
the PC2 loading seems to be randomly chosen due to
the uniform probability distribution.

The projection of the Hamiltonian path shown in
Fig.~\ref{fig:pca-hypercubes-random-pair-weighted}(j)
is an example usage of the resulting Hamming projection.
In addition to the traceability---visualized original dimension---of
the Hamiltonian path as in
Fig.~\ref{fig:pca-hypercubes-random-pair-weighted}(e),
one can see how the transition on the hypercubic edge relates to
the Hamming distance from the vertices
$
\begin{bsmallmatrix}
    - & - & - & -
\end{bsmallmatrix}^\top
$
and
$
\begin{bsmallmatrix}
    + & + & + & +
\end{bsmallmatrix}^\top
$.

\begin{figure*}[tb]
    \includegraphics{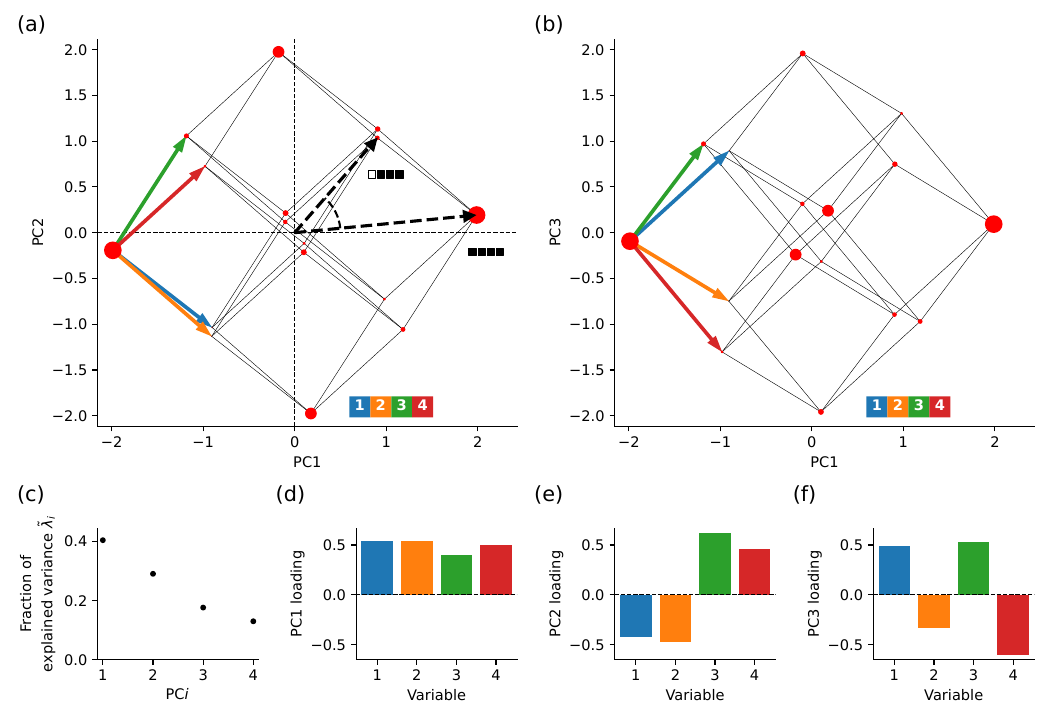}
    \caption{
        Orthogonal projections of four-dimensional hypercubic vertices
        using PCA\@.
        $
        \begin{bsmallmatrix} - & - & - & -
        \end{bsmallmatrix}^\top$ and
        $
        \begin{bsmallmatrix} + & + & + & +
        \end{bsmallmatrix}^\top$
        are the most weighted vertices with weight $3z$,
        $
        \begin{bsmallmatrix} - & - & + & +
        \end{bsmallmatrix}^\top$ and
        $
        \begin{bsmallmatrix} + & + & - & -
        \end{bsmallmatrix}^\top$
        are the second most weighted vertices with weight $2z$,
        $
        \begin{bsmallmatrix} - & + & - & +
        \end{bsmallmatrix}^\top$ and
        $
        \begin{bsmallmatrix} + & - & + & -
        \end{bsmallmatrix}^\top$
        are the third most weighted vertices with weight $z$,
        and the rest of them are weighted randomly by uniform random numbers in
        $\left[0, z\right)$.
        (a) Projection of the hypercube by PC1 and PC2.
        Red filled circles are the vertices and lines are
        the edges of the hypercube.
        The magnitude of weight is proportional to the area of the vertices.
        Arrows are biplot vectors originating from
        $
        \begin{bsmallmatrix}
            - & - & - & -
        \end{bsmallmatrix}^\top
        $
        and the boxes on the bottom right indicate
        the correspondence between the colors of the arrows and
        the original dimensions.
        Dashed arrows correspond to the projection of
        original high-dimensional vectors,
        $
        \begin{bsmallmatrix} - & + & + & +
        \end{bsmallmatrix}^\top$ and
        $
        \begin{bsmallmatrix} + & + & + & +
        \end{bsmallmatrix}^\top$.
        The original coordinate is indicated around the lower right of
        the arrowhead as
        an array of filled $\blacksquare$ (indicates $+$) or
        empty $\square$ (indicates $-$) boxes.
        Horizontal and vertical dashed lines crossing the origin are
        for visual aid.
        (b) Same as (a) but by PC1 and PC3.
        (c) Fraction of explained variance by each PC\@.
        (d) PC1 loading, (e) PC2 loading, and (f) PC3 loading.
    }
    \label{fig:pca-hypercubes-multiple-weighted}
\end{figure*}

\subsection{Sexapolar distribution}
Motivated by the results of PCA on bipolar weighted hypercubic vertices,
we expect that the projection of the hypercube by
PCA reflects the weighted vertices.
In Fig.~\ref{fig:pca-hypercubes-multiple-weighted},
we perform PCA on a four-dimensional hypercube
where three pairs of the most distant vertices are weighted more,
and the rest are weighted randomly.
Each of the three pairs of vertices is weighted differently so that
the most weighted vertices are
$
\begin{bsmallmatrix}
    - & - & - & -
\end{bsmallmatrix}^\top
$
and
$
\begin{bsmallmatrix}
    + & + & + & +
\end{bsmallmatrix}^\top
$,
the second most weighted vertices are
$
\begin{bsmallmatrix}
    - & - & + & +
\end{bsmallmatrix}^\top
$
and
$
\begin{bsmallmatrix}
    + & + & - & -
\end{bsmallmatrix}^\top
$,
and the third most weighted vertices are
$
\begin{bsmallmatrix}
    - & + & - & +
\end{bsmallmatrix}^\top
$
and
$
\begin{bsmallmatrix}
    + & - & + & -
\end{bsmallmatrix}^\top
$.
These weighted pairs are selected to be perpendicular to each other.

We show the projection by the first two PCs
in Fig.~\ref{fig:pca-hypercubes-multiple-weighted}(a),
finding a similar trend as the bipolar weighted PCA
in previous Sec.~\ref{sec:bipolar-distribution}.
The most weighted vertex pair is projected to have a larger magnitude of
PC1 score as in
Fig.~\ref{fig:pca-hypercubes-random-pair-weighted}(f).
The second most weighted vertex pair is projected to have a larger magnitude of
PC2 score.
Notice that the third most weighted vertex pair is projected around the origin,
one is around
$
\begin{bsmallmatrix}
    -0.1 & 0.25
\end{bsmallmatrix}^\top
$
and the other is around
$
\begin{bsmallmatrix}
    0.1 & -0.25
\end{bsmallmatrix}^\top
$,
despite being the most distant from each other in the original space.
This can be understood by the minimum inner-product-error formulation of PCA\@.
In this formulation,
each hypercubic vertex is projected so that the inner product
between the vertices is preserved.
Thus, a vertex should be projected far from the most distant vertex
across the origin.
That is why the most weighted vertex pair is projected to
have a larger magnitude
of the PC1 (or PC2 for the second weighted vertex pair) score.
Then the two PCs are already used to locate
the two most weighted vertex pairs,
and the third most weighted vertices,
which are perpendicular to the most and second most weighted vertices
in our example,
are both projected around the origin
even though they are far from each other in the original high-dimensional space.
Moreover, neighbors in the original high-dimensional space are projected
to be distant from each other in the projected space.
For instance, we emphasize the two vertices,
$
\begin{bsmallmatrix}
    - & + & + & +
\end{bsmallmatrix}^\top
$
and
$
\begin{bsmallmatrix}
    + & + & + & +
\end{bsmallmatrix}^\top
$,
in Fig.~\ref{fig:pca-hypercubes-multiple-weighted}(a).
These vertices are projected to be far from each other in the projected space
though they are neighboring vertices in the original space.
Similarly, the vertices
$
\begin{bsmallmatrix}
    - & + & + & +
\end{bsmallmatrix}^\top
$
and
$
\begin{bsmallmatrix}
    + & - & + & +
\end{bsmallmatrix}^\top
$
are located close to each other around
$
\begin{bsmallmatrix}
    1 & 1
\end{bsmallmatrix}^\top
$,
but they are not neighboring vertices in the original space
(the Hamming distance between them is two).
These results show that the neighboring relationships in the original space
are not necessarily preserved in the projected space:
the distances between the vertices in the projected space can be misleading.
This is the reason why distance-preserving formulation of PCA,
i.e., the idea behind classical MDS, is not suitable to interpret
the projections of hypercubes.

When we plot the projection by PC1 and PC3
in Fig.~\ref{fig:pca-hypercubes-multiple-weighted}(b)
instead of by PC1 and PC2,
we see that the third most weighted vertex pair is projected
to have a larger magnitude of PC3 score
but the second most weighted vertex pair is projected around the origin.
Our results indicate that each PC loading represents
the weighted vertex pair and the PC score corresponds to the similarity to
or distance from the weighted vertex pair.

We validate the projection of
Figs.~\ref{fig:pca-hypercubes-multiple-weighted}(a)
and~\ref{fig:pca-hypercubes-multiple-weighted}(b)
by the explained variance of each PC\@.
We show, in Fig.~\ref{fig:pca-hypercubes-multiple-weighted}(c),
the fraction of explained variance by each PC\@.
The fraction gradually decreases as the PC number increases, unlike
the random weighted case
[Fig.~\ref{fig:pca-hypercubes-random-pair-weighted}(b)]
and
the bipolar weighted case
[Fig.~\ref{fig:pca-hypercubes-random-pair-weighted}(g)].
The explained variance by the first two PCs (more than 60\%)
slightly increases more than that of
the random weighted case [Fig.~\ref{fig:pca-hypercubes-random-pair-weighted}(b)]
and the bipolar weighted case
[Fig.~\ref{fig:pca-hypercubes-random-pair-weighted}(g)].

We show the loading of PC1, PC2, and PC3
in
Figs.~\ref{fig:pca-hypercubes-multiple-weighted}(d),~\ref{fig:pca-hypercubes-multiple-weighted}(e),
and~\ref{fig:pca-hypercubes-multiple-weighted}(f), respectively,
to examine how the PCs represent the weighted vertices.
Similar to the bipolar weighted case,
the element-wise sign of PC1 loading corresponds to the most weighted vertices:
$
\operatorname{sgn}
\left(
    \bm{u}_1
\right)
=
\begin{bsmallmatrix}
    + & + & + & +
\end{bsmallmatrix}^\top
$.
In addition, the element-wise sign of PC2 (PC3) loading also corresponds to
the second (third) most weighted vertices.
Due to the noise from the randomly weighted vertices,
PC1 loading is not completely uniform,
and the projection is not exactly the Hamming projection,
but we find a similar trend.

\subsection{Brief review and preview}
In all orthogonal projections
[Figs.~\ref{fig:pca-hypercubes-random-pair-weighted}(a),~\ref{fig:pca-hypercubes-random-pair-weighted}(f),~\ref{fig:pca-hypercubes-multiple-weighted}(a),
and~\ref{fig:pca-hypercubes-multiple-weighted}(b)]
of hypercubes using PCA, the weighted vertices have high PC1 scores,
indicating that PC1 is associated with the weighted vertices.
This observation is discussed further in
Sec.~\ref{sec:pc-loading-corresponding}.
In the Supplemental Material~\cite{sm},
we show the projections of hypercubes with different realizations of random
weight, finding that the vertices does not necessarily have significantly larger
weight to have larger PC scores.
This result seems to be counterintuitive,
but we address this in Sec.~\ref{sec:pc-loading-corresponding} and
Appendix~\ref{sec:pca-with-perturbation}.

We have observed that some neighboring relationships between the vertices are
distorted:
the distances between the vertices in the projected space can be misleading.
When the weighted vertex pairs are perpendicular to each other
[Fig.~\ref{fig:pca-hypercubes-multiple-weighted}(a)
and~\ref{fig:pca-hypercubes-multiple-weighted}(b)],
the second most weighted vertices are projected to have high PC2 scores,
and the third most weighted vertices are projected to have high PC3 scores.
We discuss this further in Sec.~\ref{sec:quality}
from the perspective of projection quality and the inner-product error.

When two PC loadings cannot capture all the weighted vertices, the missed
weighted vertices can be placed near the origin of the projected space.
This centrality, often seen in studies of Ising spin
system~\cite{Wang2016,Hu2017,Kiwata2019},
of the projected vertices is discussed further in
Sec.~\ref{sec:number-of-vertices-along-pc1}
and its effect on the inner-product error is discussed in
Sec.~\ref{sec:upper-bound}.

\section{PC1 loading and weighted vertices}
\label{sec:pc-loading-corresponding}
In Sec.~\ref{sec:hypercubic-pca}, we find that the element-wise sign of
the PC1 loading corresponds to the vertices with the highest PC scores—the
weighted vertices.
The results indicate that the PC1 loading is related to the weighted vertices,
and the PC1 score corresponds to the similarity to or distance from
the weighted vertices.
To understand this, in this section,
we analytically and numerically examine the properties of PCA,
particularly the correspondence between the leading PC loading and
the weighted vertices.
We first consider the ideal probability distribution,
then expand our consideration to more general distributions.
We also numerically validate the analytical results.

\subsection{Analytical investigation}
\label{sec:pc-loading-corresponding-analytical}
We begin with an ideal situation where the distribution is concentrated on
a few hypercubic vertices.
Consider, for example,
a low-temperature canonical ensemble for an Ising spin system with $N$ spins,
where the vertex distribution is dominated by
a weighted vertex (or ground state)
$\bm{\xi} \in \left\{ +1, -1 \right\}^N$
and its globally spin-flipped vertex (state) $ -\bm{\xi} $.
Suppose the distribution is idealized as a bipolar one as in
Fig.~\ref{fig:pca-hypercubes-random-pair-weighted}(f),
\begin{equation}
    p\left(\bm{s}\right)
    \approx
    \frac{1}{2}
    \left(
        \delta_{+\bm{\xi}, \bm{s}}
        +
        \delta_{-\bm{\xi}, \bm{s}}
    \right)
    .
    \label{eq:bipolar-distribution}
\end{equation}
Here, $ \delta_{\bm{y}, \bm{x}} $ is the Kronecker delta function for vectors,
which is equal to $1$ when $\bm{x} = \bm{y}$ and $0$ otherwise.
Given that the mean vector is the zero vector,
\begin{equation}
    \left<\bm{s}\right>
    =
    \sum_{\bm{s}}
    \frac{1}{2}
    \left(
        \delta_{+\bm{\xi}, \bm{s}}
        +
        \delta_{-\bm{\xi}, \bm{s}}
    \right)
    \bm{s}
    =
    \bm{0}
    ,
\end{equation}
the covariance matrix [Eq.~\eqref{eq:covariance-matrix}] becomes
\begin{equation}
    \bm{\varSigma}
    =
    \sum_{\bm{s}}
    \frac{1}{2}
    \left(
        \delta_{+\bm{\xi}, \bm{s}}
        +
        \delta_{-\bm{\xi}, \bm{s}}
    \right)
    \left(
        \bm{s} - \bm{0}
    \right)
    \left(
        \bm{s} - \bm{0}
    \right)^\top
    =
    \bm{\xi} \bm{\xi}^\top
    \label{eq:covariance-matrix-bipolar}
    ,
\end{equation}
which is already diagonalized with the eigenvector
\begin{equation}
    \bm{u}_1
    =
    \frac{1}{\sqrt{N}}
    \bm{\xi}
    \label{eq:PC1-loading}
    ,
\end{equation}
and the corresponding eigenvalue
\begin{equation}
    \lambda_1 = N
    ,
\end{equation}
i.e.,
$
\frac{1}{\sqrt{N}}
\bm{\varSigma}
\bm{\xi}
=
\frac{1}{\sqrt{N}}
\bm{\xi} \bm{\xi}^\top \bm{\xi}
=
\frac{1}{\sqrt{N}}
N \bm{\xi}
$,
Notice the normalization factor
$
\frac{1}{\sqrt{N}}
=
\frac{1}{\sqrt{\bm{\xi}^\top \bm{\xi}}}
$.
Also, the element-wise sign of the PC1 loading is the weighted vertex vector,
$\operatorname{sgn}\left(\bm{u}_1\right) = \bm{\xi}$.
This explains why we observe that the element-wise sign of the PC loading is
the same as the weighted state.
If we remove the normalization factor,
the PC loading is exactly same as the weighted state.

Next, we consider a finite-temperature canonical ensemble with ground states.
Here, the ground states $+\bm{\xi}$ and $-\bm{\xi}$ are weighted more
and the other probability of states are approximated as uniform.
We assume bipolar distribution with uniform background,
\begin{equation}
    p\left(\bm{s}\right)
    \approx
    c
    \frac{1}{2}
    \left(
        \delta_{+\bm{\xi}, \bm{s}}
        +
        \delta_{-\bm{\xi}, \bm{s}}
    \right)
    +
    \left(1 - c\right)
    \frac{1}{2^N}
    \label{eq:bipolar-background-distribution}
    ,
\end{equation}
where $c \in \left[0, 1\right]$ is a parameter that controls the
distribution.
In the limit $c \to 0$, corresponding to the high-temperature regime,
the distribution approaches the uniform distribution,
$p\left(\bm{s}\right) \to \frac{1}{2^N}$.
All states are found randomly in this regime.
In the opposite limit, $c \to 1$, the distribution reduces to the bipolar form
$
p\left(\bm{s}\right) \to \frac{1}{2}
\left(
    \delta_{+\bm{\xi}, \bm{s}}
    +
    \delta_{-\bm{\xi}, \bm{s}}
\right)
$
[Eq.~\eqref{eq:bipolar-distribution}].
Between these two extremes,
the distribution can resemble a finite-temperature canonical ensemble,
where the ground states are more heavily weighted than the others,
but other states also have nonzero probability.
For such a distribution,
the mean vector remains the zero vector as in the bipolar case,
and the covariance matrix becomes
\begin{equation}
    \bm{\varSigma}
    =
    c
    \bm{\xi}
    \bm{\xi}^\top
    +
    \left(1 - c\right)
    \frac{1}{2^N}
    \bm{I}
    \label{eq:covariance-matrix-bipolar-background}
    ,
\end{equation}
which has the eigenvector
$
\bm{u}_1
=
\frac{1}{\sqrt{N}}
\bm{\xi}
$
and the eigenvalue
$
\lambda_1
=
N c + 1 - c
$.
Thus, as long as $c \neq 0$, the PC1 loading corresponds to that of the
bipolar distribution but with a reduced eigenvalue.
If $c = 0$, the covariance matrix becomes the identity matrix,
$\bm{\varSigma} = \bm{I}$,
and the eigenvectors can be any set of orthonormal vectors.
Therefore, we expect that the PC1 loading remains close to the weighted state
as long as the probability distribution retains a bipolar-like structure,
i.e., $c > 0$.
We further discuss this distribution in the next
Sec.~\ref{sec:pc-loading-correspondence-numerical}.

With bipolar or bipolar distribution with background,
we move on to the projected coordinates.
The projected coordinates on PC1 (PC1 score) become
\begin{equation}
    r_1
    \left(
        \bm{s}
    \right)
    =
    \bm{u}_1^\top
    \bm{s}
    =
    \frac{1}{\sqrt{N}}
    \bm{\xi}^\top
    \bm{s}
    =
    \sqrt{N}
    Q\left(\bm{\xi}, \bm{s}\right)
    \label{eq:PC1-score}
    ,
\end{equation}
where we introduce the overlap (or cosine similarity)
$
Q\left(\bm{\xi}, \bm{s}\right)
\coloneqq
\frac{1}{N}
\bm{\xi}^\top
\bm{s}
\in
\left\{
    \frac{-N + 2i}{N}
    \, \mid \,
    i \in \mathbb{Z}, \,
    0 \leq i \leq N
\right\}
$,
i.e., the normalized inner product between state $ \bm{\xi}$ and
state $\bm{s}$.
Thus, the PC1 score is proportional to the
overlap measure with the weighted state.
For the Ising spin systems with ferromagnetic interactions,
the weighted state is all-spin-up or all-spin-down state,
$\bm{\xi} = \bm{1}$ or $\bm{\xi} = -\bm{1}$,
and the overlap measure is equivalent to the magnetization,
$
m\left(\bm{s}\right)
\coloneqq
\frac{1}{N}
\bm{1}^\top \bm{s}
=
Q\left(\bm{1}, \bm{s}\right)
$.
Thus, the PC1 score is proportional to the magnetization or order parameter,
\begin{equation}
    r_1
    \left(
        \bm{s}
    \right)
    =
    \sqrt{N}
    m\left(\bm{s}\right)
    \label{eq:PC1-score-order-parameter}
    ,
\end{equation}
as numerically pointed out in previous
studies~\cite{Wang2016,Hu2017,Kiwata2019}.

The PC1 score is also the distance from the weighted vertices.
We introduce the Hamming distance
$
D_\mathrm{H}\left(\bm{s}, \bm{s}^\prime\right)
\in
\left\{
    D_\mathrm{H} \in \mathbb{Z}
    \, \mid \,
    0 \leq D_\mathrm{H} \leq N
\right\}
$
which is defined as the number of unmatched elements in the binary vectors
$\bm{b} $ and $ \bm{b}^\prime$,
\begin{equation}
    D_\mathrm{H}\left( \bm{b}, \bm{b}^\prime \right)
    \coloneqq
    \sum_{i=1}^{N}
    \left(1 - \delta_{b_i,b_i^\prime}\right)
    .
\end{equation}
Here, $\delta_{b_i,b_i^\prime}$ is the Kronecker delta function for scalars.
Because $s_i s^\prime_i \in \left\{+1, -1\right\}$
and $\delta_{b_i,b_i^\prime} = \frac{1}{2}\left(1+s_i s^\prime_i\right)$,
the Hamming distance
$D_\mathrm{H}\left( \bm{s}, \bm{s}^\prime \right)$
between two Ising state vectors $\bm{s}$ and $\bm{s}^\prime$ is
\begin{align}
    D_\mathrm{H}\left( \bm{s}, \bm{s}^\prime \right)
    &=
    \sum_{i=1}^{N}
    \left(
        1 - \frac{1 + s_i s^\prime_i}{2}
    \right)
    \nonumber
    \\&=
    \frac{N - \bm{s}^\top \bm{s}^\prime}{2}
    =
    N
    \frac{1-Q\left(\bm{s}, \bm{s}^\prime\right)}{2}
    \label{eq:Hamming-distance}
    .
\end{align}
Using this Eq.~\eqref{eq:Hamming-distance}, we have
$
Q\left(\bm{\xi}, \bm{s}\right)
=
\frac
{
    N - 2D_\mathrm{H}\left(\bm{\xi}, \bm{s}\right)
}{
    N
}
$.
Then, the PC1 score of Eq.~\eqref{eq:PC1-score} is written as
\begin{equation}
    r_1
    \left(
        \bm{s}
    \right)
    =
    \frac{
        +
        N
        -
        2
        D_\mathrm{H}\left(+\bm{\xi}, \bm{s}\right)
    }{
        \sqrt{N}
    }
    \label{eq:PC1-score-Hamming-plus}
\end{equation}
or using
$
r_1\left(\bm{s}\right)
=
-
\left(
    -
    \frac{1}{\sqrt{N}}
    \bm{\xi}
\right)
\bm{s}
=
-
\sqrt{N}
Q\left(-\bm{\xi}, \bm{s}\right)
$
and
$
Q\left(-\bm{\xi}, \bm{s}\right)
=
\frac
{
    N - 2D_\mathrm{H}\left(-\bm{\xi}, \bm{s}\right)
}{
    N
}
$,
\begin{equation}
    r_1
    \left(
        \bm{s}
    \right)
    =
    \frac{
        -
        N
        +
        2
        D_\mathrm{H}\left(-\bm{\xi}, \bm{s}\right)
    }{
        \sqrt{N}
    }
    \label{eq:PC1-score-Hamming-minus}
    .
\end{equation}
The PC1 score of any sample state $ \bm{s} $ is then linearly
equivalent to the
Hamming distance between a state $\bm{s}$ and
the weighted state $\bm{\xi}$ (or $-\bm{\xi}$).
Therefore, the distribution of
Eq.~\eqref{eq:bipolar-distribution} guarantees
the Hamming projection.

For the sexapolar distribution, as considered in
Fig.~\ref{fig:pca-hypercubes-multiple-weighted},
the probability distribution can be approximated as
\begin{equation}
    p\left(\bm{s}\right)
    \approx
    \sum_{\mu=1}^{3}
    \frac{c_\mu}{2}
    \left(
        \delta_{+\bm{\xi}_\mu, \bm{s}}
        +
        \delta_{-\bm{\xi}_\mu, \bm{s}}
    \right)
    ,
\end{equation}
where
$c_1, c_2, c_3 \in \left[0, 1\right]$
are the weights of the selected states,
sorted in descending order,
$c_1 \geq c_2 \geq c_3 \geq 0$,
and satisfying
$\sum_{\bm{s}} p\left(\bm{s}\right) = \sum_{\mu=1}^{3} c_\mu = 1$.
Assuming that the selected states are mutually orthogonal,
$\bm{\xi}_\mu^\top \bm{\xi}_\nu = N \delta_{\mu, \nu}$,
the mean vector becomes
\begin{equation}
    \left<\bm{s}\right>
    =
    \sum_{\mu=1}^{3}
    \frac{c_\mu}{2}
    \left(
        \delta_{+\bm{\xi}_\mu, \bm{s}}
        +
        \delta_{-\bm{\xi}_\mu, \bm{s}}
    \right)
    \bm{s}
    =
    \bm{0}
    ,
\end{equation}
and the covariance matrix becomes
\begin{equation}
    \bm{\varSigma}
    =
    \sum_{\mu=1}^{3}
    \frac{c_\mu}{2}
    \left(
        \delta_{+\bm{\xi}_\mu, \bm{s}}
        +
        \delta_{-\bm{\xi}_\mu, \bm{s}}
    \right)
    \left(
        \bm{s} - \bm{0}
    \right)
    \left(
        \bm{s} - \bm{0}
    \right)^\top
    =
    \sum_{\mu=1}^{3}
    c_\mu
    \bm{\xi}_\mu
    \bm{\xi}_\mu^\top
    \label{eq:covariance-matrix-sexapolar}
    .
\end{equation}
Then, the eigenvectors and corresponding eigenvalues of the
covariance matrix are
\begin{equation}
    \lambda_i = N c_i
\end{equation}
and
\begin{equation}
    \bm{u}_i
    =
    \frac{1}{\sqrt{N}}
    \bm{\xi}_i
    ,
\end{equation}
as already illustrated in Fig.~\ref{fig:pca-hypercubes-multiple-weighted}.

Next, we consider a generalized bipolar distribution:
\begin{equation}
    p\left(\bm{s}\right)
    \approx
    c_+
    \delta_{+\bm{\xi}, \bm{s}}
    +
    c_-
    \delta_{-\bm{\xi}, \bm{s}}
    ,
\end{equation}
where
$c_+, c_- \in [0, 1]$ and $c_+ + c_- = 1$.
For convenience, we define the mean coefficient as
$
\overline{c}
\coloneqq
\frac{1}{2}
\left(
    c_+ + c_-
\right)
$
and the difference as
$
\Delta c
\coloneqq
\frac{1}{2}
\left(
    c_+ - c_-
\right)
$,
so that
$
c_+ = \overline{c} + \Delta c
$
and
$
c_- = \overline{c} - \Delta c
$.
The mean vector is given by
\begin{equation}
    \left<\bm{s}\right>
    =
    \left(c_+ - c_-\right)
    \bm{\xi}
    =
    \left(
        2 \Delta c
    \right)
    \bm{\xi}
    ,
\end{equation}
and the covariance matrix is given by
\begin{align}
    \bm{\varSigma}
    &=
    c_+
    \left[
        +\bm{\xi} - \left(2 \Delta c\right) \bm{\xi}
    \right]
    \left[
        +\bm{\xi} - \left(2 \Delta c\right) \bm{\xi}
    \right]^\top
    \nonumber
    \\&\quad+
    c_-
    \left[
        -\bm{\xi} - \left(2 \Delta c\right) \bm{\xi}
    \right]
    \left[
        -\bm{\xi} - \left(2 \Delta c\right) \bm{\xi}
    \right]^\top
    \nonumber
    \\&=
    \left(
        \overline{c} + \Delta c
    \right)
    \left[
        \bm{\xi} \bm{\xi}^\top
        -
        2 \left(2\Delta c\right) \bm{\xi} \bm{\xi}^\top
        +
        \left(2 \Delta c\right)^2
        \bm{\xi} \bm{\xi}^\top
    \right]
    \nonumber
    \\&\quad+
    \left(
        \overline{c} - \Delta c
    \right)
    \left[
        \bm{\xi} \bm{\xi}^\top
        +
        2 \left(2\Delta c\right) \bm{\xi} \bm{\xi}^\top
        +
        \left(2 \Delta c\right)^2
        \bm{\xi} \bm{\xi}^\top
    \right]
    \nonumber
    \\&=
    2
    \overline{c}
    \left[
        1 + \left(2 \Delta c\right)^2
    \right]
    \bm{\xi} \bm{\xi}^\top
    +
    2
    \Delta c
    \left(
        - 4 \Delta c
    \right)
    \bm{\xi} \bm{\xi}^\top
    \nonumber
    \\&=
    \left[
        2
        \overline{c}
        +
        2
        \overline{c}
        \left(
            2 \Delta c
        \right)^2
        -
        2
        \left(
            2 \Delta c
        \right)^2
    \right]
    \bm{\xi} \bm{\xi}^\top
    \nonumber
    \\&=
    \left[
        1
        -
        \left(
            2 \Delta c
        \right)^2
    \right]
    \bm{\xi} \bm{\xi}^\top
    \label{eq:covariance-matrix-bipolar-generalized}
    .
\end{align}
Note that
$\overline{c} = \frac{1}{2}$
from the normalization condition
$\sum_{\bm{s}} p\left(\bm{s}\right) = c_+ + c_- = 1$.
The covariance matrix is proportional to the outer product of the
weighted vertex,
$\bm{\varSigma} \propto \bm{\xi} \bm{\xi}^\top$.
Thus, the eigenvector is the same as Eq.~\eqref{eq:PC1-loading},
but the eigenvalue decreases as
$\Delta c$ deviates from zero,
$
\lambda_1 = \left[1 - \left(2 \Delta c\right)^2\right] N
$.
As the weight of the pair deviates from $\frac{1}{2}$,
the covariance matrix approaches the zero matrix.
If the distribution is idealized as a unipolar one,
$\Delta c \to \frac{1}{2}$,
the covariance matrix approaches the zero matrix,
$
\bm{\varSigma}
\rightarrow
\bm{0} \bm{0}^\top
$,
which results in a zero eigenvalue and arbitrary orthonormal eigenvectors.
This result indicates that the difference in weights, $\Delta c$,
reduces the contribution of the weighted vertex to the covariance matrix.

Finally, we consider the most general polar distribution:
\begin{equation}
    p\left(\bm{s}\right)
    =
    \sum_{\mu=1}^{M}
    \left(
        c_{+, \mu}
        \delta_{+\bm{\xi}_\mu, \bm{s}}
        +
        c_{-, \mu}
        \delta_{-\bm{\xi}_\mu, \bm{s}}
    \right)
    ,
\end{equation}
where
$M \in \left\{M \in \mathbb{Z} \,\mid\, 1 \leq M \leq 2^{N-1}\right\}$
is the number of weighted vertex pairs,
$c_{+, \mu}, c_{-, \mu} \in [0, 1]$ are the coefficients (weights) of
the selected states, and these coefficients satisfy
$
\sum_{\bm{s}} p\left(\bm{s}\right)
=
\sum_{\mu=1}^{M} \left(c_{+, \mu} + c_{-, \mu}\right)
=
1
$.
If $M = 2^{N-1}$, all vertices are selected.
We define the mean coefficient
$\overline{c}_\mu \coloneqq \frac{1}{2}\left(c_{+, \mu} + c_{-, \mu}\right)$
and the coefficient difference
$\Delta c_\mu \coloneqq \frac{1}{2}\left(c_{+, \mu} - c_{-, \mu}\right)$
for each $\mu$.
Then, the mean vector becomes
\begin{equation}
    \left<\bm{s}\right>
    =
    \sum_{\mu=1}^{M}
    \left(
        c_{+, \mu} - c_{-, \mu}
    \right)
    \bm{\xi}_\mu
    =
    \sum_{\mu=1}^{M}
    \left(
        2
        \Delta c_\mu
    \right)
    \bm{\xi}_\mu
    .
\end{equation}
After some algebra\footnote{
    See the Supplemental Material~\cite{sm} for the detailed derivation.
} similar to derive
Eq.~\eqref{eq:covariance-matrix-bipolar-generalized},
the covariance matrix becomes
\begin{align}
    \bm{\varSigma}
    &=
    \sum_{\mu=1}^{M}
    \left[
        2
        \overline{c}_\mu
        -
        \left(
            2
            \Delta c_\mu
        \right)^2
    \right]
    \bm{\xi}_\mu \bm{\xi}_\mu^\top
    \nonumber
    \\&\quad-
    \sum_{\mu=1}^{M}
    \sum_{\nu=\mu+1}^{M}
    \left(
        2
        \Delta c_\mu
    \right)
    \left(
        2
        \Delta c_\nu
    \right)
    \left(
        \bm{\xi}_\mu \bm{\xi}_\nu^\top
        +
        \bm{\xi}_\nu \bm{\xi}_\mu^\top
    \right)
    \label{eq:covariance-matrix-generalized}
    .
\end{align}
If the coefficient differences are zero, $\Delta c_\mu = 0$ for all $\mu$,
the covariance matrix simplifies to
\begin{equation}
    \bm{\varSigma}
    =
    \sum_{\mu=1}^{M}
    \left(
        2
        \overline{c}_\mu
    \right)
    \bm{\xi}_\mu \bm{\xi}_\mu^\top
    ,
\end{equation}
and, assuming that the selected states are mutually orthogonal,
$\bm{\xi}_\mu^\top \bm{\xi}_\nu = N \delta_{\mu, \nu}$,
the eigenvectors and eigenvalues are the same as in
Eq.~\eqref{eq:covariance-matrix-sexapolar}.

To understand the qualitative behavior,
let us consider $M=2$ with $\Delta c_1 = 0$ and $\Delta c_2 \neq 0$.
In this case, the covariance matrix becomes
\begin{equation}
    \bm{\varSigma}
    =
    2
    \overline{c}_1
    \bm{\xi}_1 \bm{\xi}_1^\top
    +
    \left[
        2
        \overline{c}_2
        -
        \left(
            2 \Delta c_2
        \right)^2
    \right]
    \bm{\xi}_2 \bm{\xi}_2^\top
    \label{eq:covariance-matrix-two-vertices-generalized}
    .
\end{equation}
If $\bm{\xi}_1^\top \bm{\xi}_2 = 0$,
the eigenvalues are
$
\lambda
\in
\left\{
    2 \overline{c}_1 N, \,
    \left[
        2 \overline{c}_2 - \left(2 \Delta c_2\right)^2
    \right] N
\right\}
$
and the corresponding eigenvectors are
$
\bm{u}
\in
\left\{
    \frac{1}{\sqrt{N}} \bm{\xi}_1, \,
    \frac{1}{\sqrt{N}} \bm{\xi}_2
\right\}
$.
Note that even if $\overline{c}_1 < \overline{c}_2$,
depending on the value of $\Delta c_2$,
the eigenvalue corresponding to $\bm{\xi}_2$ can be larger than
the eigenvalue corresponding to $\bm{\xi}_1$.
When $\Delta c_1 \neq 0$ and $\Delta c_2 \neq 0$,
the covariance matrix becomes
\begin{align}
    \bm{\varSigma}
    &=
    \left[
        2 \overline{c}_1 - \left(2 \Delta c_1\right)^2
    \right]
    \bm{\xi}_1 \bm{\xi}_1^\top
    +
    \left[
        2 \overline{c}_2 - \left(2 \Delta c_2\right)^2
    \right]
    \bm{\xi}_2 \bm{\xi}_2^\top
    \nonumber
    \\&\quad-
    \left(
        2
        \Delta c_1
    \right)
    \left(
        2
        \Delta c_2
    \right)
    \left(
        \bm{\xi}_1 \bm{\xi}_2^\top
        +
        \bm{\xi}_2 \bm{\xi}_1^\top
    \right)
    ,
\end{align}
which makes it challenging to analytically find the eigenvectors and
eigenvalues due to the presence of the symmetric outer product matrix
$
\bm{\xi}_1 \bm{\xi}_2^\top + \bm{\xi}_2 \bm{\xi}_1^\top
$.
If $\bm{\xi}_1$ and $\bm{\xi}_2$ are orthogonal,
the eigenvector of this symmetric outer product matrix is
$
\frac{1}{\sqrt{2N}}
\left(
    \bm{\xi}_1 + \bm{\xi}_2
\right)
$
with eigenvalue
$
N
$.
Thus, the eigenvector of the symmetric outer product matrix is
linearly dependent on the weighted vertices,
and an analytical derivation of the eigenvectors and eigenvalues
remains challenging.
This difficulty arises from the nonzero centering by the mean vector.

Our consideration of the generalized bipolar distribution shows that
centering with a nonzero mean vector can complicate analytical understanding.
It is, however, important to note that the center of the hypercube is
a meaningful point of reference for our visualization purposes---if PCA rotates
to project the hypercubic vertices around a point other than the center,
the resulting leading PC loading becomes biased toward
the center of the hypercube.
This can be misleading for visualization and should be avoided.
We expect that centering at the zero vector is the most appropriate choice
for PCA\@.
See also Refs.~\cite{Jolliffe2002,Jolliffe2016} for
a review of centering in PCA\@.

We further extend our analysis to more general distributions with
non-orthonormal weighted vertices,
i.e., $\bm{\xi}_\mu^\top \bm{\xi}_\nu \neq 0$,
using perturbation theory from quantum mechanics.
When the distribution is approximately quadripolar, that is,
\begin{equation}
    p \left(\bm{s}\right)
    \approx
    \frac{1-\epsilon}{2}
    \left(
        \delta_{+\bm{\xi}_1, \bm{s}}
        +
        \delta_{-\bm{\xi}_1, \bm{s}}
    \right)
    +
    \frac{\epsilon}{2}
    \left(
        \delta_{+\bm{\xi}_2, \bm{s}}
        +
        \delta_{-\bm{\xi}_2, \bm{s}}
    \right)
    \label{eq:quadripolar-distribution}
    ,
\end{equation}
with perturbation parameter $0 \leq \epsilon \ll 1$.
Using perturbation theory, we obtain the PC1 loading as
\begin{equation}
    \bm{u}_1 \propto \bm{\xi}_1
    \label{eq:quadripolar-PC1-loading}
\end{equation}
and the PC2 loading as
\begin{equation}
    \bm{u}_2
    \propto
    \bm{\xi}_2
    -
    \frac{1}{N}
    \bm{\xi}_1^\top \bm{\xi}_2 \bm{\xi}_1
    \label{eq:quadripolar-PC2-loading}
    .
\end{equation}
When the two weighted vertices are perpendicular to each other,
$ \bm{\xi}_1^\top \bm{\xi}_2 = 0 $,
the PC2 score depends linearly on the Hamming distance from
the second most weighted vertex, similar to the PC1 score.
The derivation is given in Appendix~\ref{sec:pca-with-perturbation}.

\begin{figure}[tb]
    \includegraphics{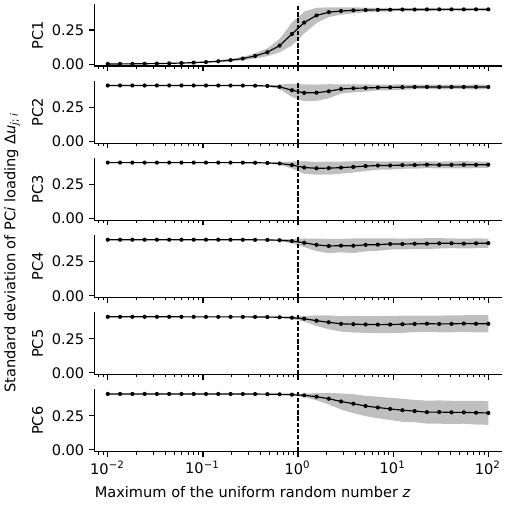}
    \caption{
        The standard deviation of PC loading, $\Delta u_{j;i}$, as a function of
        the weight parameter $z$.
        We performed PCA on the vertices of a six-dimensional hypercube,
        varying the weight $z$ assigned to the selected state $\bm{\xi}$.
        The selected vertices are
        $
        \bm{\xi}
        =
        \begin{bsmallmatrix}
            + & + & + & + & + & +
        \end{bsmallmatrix}^\top
        $
        and
        $
        -\bm{\xi}
        =
        \begin{bsmallmatrix}
            - & - & - & - & - & -
        \end{bsmallmatrix}^\top
        $.
        For each value of $z$,
        we sampled 1024 realizations of random weights
        $\left\{c_{\bm{\xi}^\prime}\left(z\right)\right\}$
        from the uniform distribution, and the shaded area indicates
        the sample standard deviation.
        Vertical dashed lines indicate $z=1$,
        where the selected vertices lose their distinguished weight.
    }
    \label{fig:pc-standard-deviation}
\end{figure}

\subsection{Numerical validation}
\label{sec:pc-loading-correspondence-numerical}
With the ideal bipolar distribution of Eq.~\eqref{eq:bipolar-distribution},
we have shown that the PC loading corresponds to the weighted vertices.
This argument is expected to hold for distributions that are approximately
similar to Eq.~\eqref{eq:bipolar-distribution}.
Indeed, in the previous Sec.~\ref{sec:pc-loading-corresponding-analytical},
a bipolar distribution with background
[Eq.~\eqref{eq:bipolar-background-distribution}] was introduced,
and it was shown that the PC1 loading aligns with the weighted vertices
as long as the distribution retains a bipolar-like structure.
Nevertheless, how robust is the assumption of a bipolar distribution
with background?

To check the validity of the bipolar distribution with background,
we perform PCA on all vertices of the six-dimensional hypercube ($N=6$),
varying the symmetric weights
$p\left(\bm{\xi}\right) = p\left(-\bm{\xi}\right)$
of the selected vertices away from the ideal distribution of
Eq.~\eqref{eq:bipolar-background-distribution}.
We define the probability distribution as
$
p\left(\bm{s}\right)
\propto
\delta_{+\bm{\xi}, \bm{s}}
+
\delta_{-\bm{\xi}, \bm{s}}
+
\sum_{\bm{\xi}^\prime \notin \left\{+\bm{\xi}, -\bm{\xi}\right\}}
c_{\bm{\xi}^\prime} \left(z\right)
\delta_{\bm{\xi}^\prime, \bm{s}}
$,
where
$
+\bm{\xi}
=
\begin{bsmallmatrix}
    + & + & + & + & + & +
\end{bsmallmatrix}^\top
$
and
$
-\bm{\xi}
=
\begin{bsmallmatrix}
    - & - & - & - & - & -
\end{bsmallmatrix}^\top
$
are the selected vertices.
For the remaining vertices
$
\left\{
    \bm{\xi}^\prime
    \notin
    \left\{
        +\bm{\xi}, -\bm{\xi}
    \right\}
\right\}
$,
$
c_{+\bm{\xi}^\prime} \left(z\right)
=
c_{-\bm{\xi}^\prime} \left(z\right)
\in
\left[0, z\right)
$
is a random weight sampled from the uniform distribution,
with $z$ as the upper bound.
According to Eq.~\eqref{eq:PC1-loading}, in the small $z$ limit,
the PC1 loading is proportional to the weighted state,
$
\bm{u}_1
\propto
\bm{\xi}
=
\begin{bsmallmatrix}
    + & + & + & + & + & +
\end{bsmallmatrix}^\top
$,
which has uniform elements in this example.
To quantify the alignment of the PC1 loading with $\bm{\xi}$,
we examine the standard deviation of all PC$i$ loadings:
\begin{equation}
    \Delta u_{j;i}
    \coloneqq
    \sqrt{
        \frac{1}{N}
        \sum_{j=1}^{N}
        \left(
            {u_{j;i}}
            -
            \frac{1}{N}
            \sum_{j=1}^{N}
            u_{j;i}
        \right)^2
    }
    ,
\end{equation}
where $u_{j;i}$ denotes the $j$th element of the $i$th PC loading.
If all elements of a PC loading are identical, the standard deviation is zero.

In Fig.~\ref{fig:pc-standard-deviation},
we show the dependence of the standard deviation on the weight parameter $z$.
At $z \approx 1$, where some randomly weighted vertices start to overwhelm
the of weighted vertices,
the standard deviation of the PC1 loading exhibits a nonlinear increase,
indicating that PC1 is no longer aligned with
the selected state $+\bm{\xi}$ (or $-\bm{\xi}$).
The qualitative equivalence of the covariance matrix
[Eq.~\eqref{eq:covariance-matrix-bipolar-background}]
for the bipolar distribution with background
[Eq.~\eqref{eq:bipolar-background-distribution}]
to the covariance matrix of the bipolar distribution
[Eq.~\eqref{eq:covariance-matrix-bipolar}]
holds for the uniform random weight distribution
as long as the maximum random weight $z$ is approximately less than that of
the selected vertices.

\section{Dependency of inner-product error on projected coordinates and
its inevitability}
\label{sec:quality-centrality-upper-bound}
We have introduced several methods to project hypercubes
and shown that PCA has several advantages.
The remaining question is the quality of these projections.
In this section, we compare the quality of the projections
we have introduced so far.
Through the investigation of quality, we observe that the error
of the projection arises from the vertices located around the center of the
projected space.
We then theoretically explain the reason for this tendency by examining
the number of vertices along PC1 and the upper bound of the
inner-product error.

\begin{figure*}[tb]
    \includegraphics{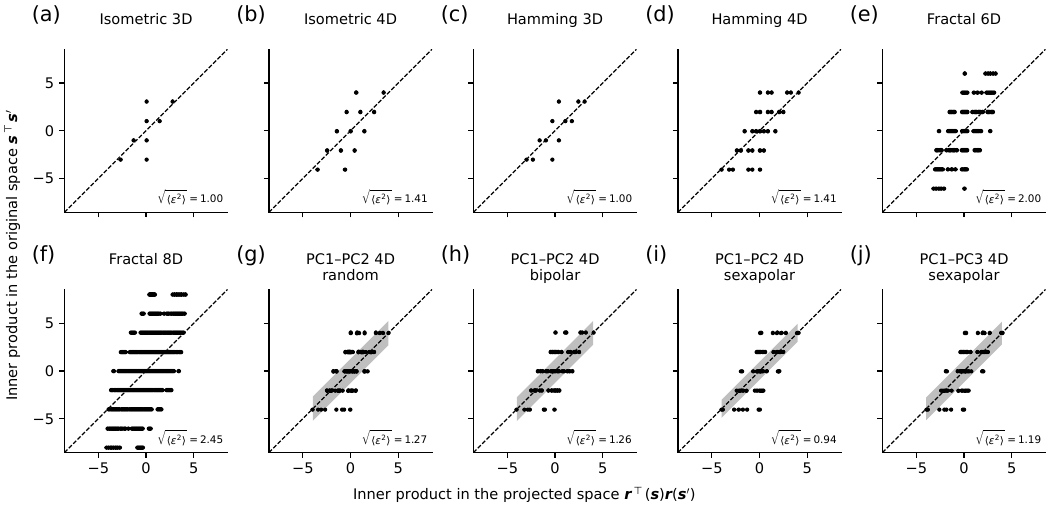}
    \caption{
        Quality of various projections of hypercubes.
        The inner product in the original space $\bm{s}^\top \bm{s}^\prime$
        is plotted as a function of the inner product in the projected space
        $\bm{r}^\top\left(\bm{s}\right) \bm{r}\left(\bm{s}^\prime\right)$.
        (a) The isometric projection of a cube in
        Fig.~\ref{fig:isometric-hamming-fractal}(a).
        (b) The isometric projection of a tesseract in
        Fig.~\ref{fig:isometric-hamming-fractal}(b).
        (c) The Hamming projection of a cube in
        Fig.~\ref{fig:isometric-hamming-fractal}(c).
        (d) The Hamming projection of a tesseract in
        Fig.~\ref{fig:isometric-hamming-fractal}(d).
        (e) The fractal projection of a six-dimensional hypercube
        in Fig.~\ref{fig:isometric-hamming-fractal}(e).
        (f) The fractal projection of an eight-dimensional hypercube
        in Fig.~\ref{fig:isometric-hamming-fractal}(f).
        (g) The projection by PC1 and PC2 of a randomly weighted tesseract
        in Fig.~\ref{fig:pca-hypercubes-random-pair-weighted}(a).
        (h) The projection by PC1 and PC2 of a bipolarly weighted tesseract
        in Fig.~\ref{fig:pca-hypercubes-random-pair-weighted}(f).
        (i) The projection by PC1 and PC2 of a sexapolarly
        weighted tesseract
        in Fig.~\ref{fig:pca-hypercubes-multiple-weighted}(a).
        (j) The projection by PC1 and PC3 of a sexapolarly
        weighted tesseract
        in Fig.~\ref{fig:pca-hypercubes-multiple-weighted}(b).
        The dashed line in each panel crosses the origin with slope one.
        In (g)--(j), we show the squared sum of unexplained variance
        as a shade for the confidence interval.
        The width of the shade is
        $
        2\sum_{i \notin \left\{1, 2\right\}} \lambda_i
        $
        for (g)--(i), and
        $
        2\sum_{i \notin \left\{1, 3\right\}} \lambda_i
        $
        for (j).
        The root mean squared inner-product error is shown at the
        bottom right.
        Note that
        before the calculation of the inner product,
        we centered all variables to the origin, i.e.,
        $\bm{s} \leftarrow \bm{s} - \left<\bm{s}\right>$
        and
        $
        \bm{r}\left(\bm{s}\right)
        \leftarrow
        \bm{r}\left(\bm{s}\right) - \left<\bm{r}\left(\bm{s}\right)\right>
        $.
    }
    \label{fig:quality}
\end{figure*}

\subsection{Quality of the projections}
\label{sec:quality}
To evaluate the quality of orthogonal projections of hypercubes,
we investigate the inner-product error [Eq.~\eqref{eq:inner-product-error}]
between the original space and the projected space
for all possible pairs of vertices of hypercubes.
As mentioned in Sec.~\ref{sec:pca-minimizing-inner-product-error},
inner-product error is minimized in PCA\@.
It indicates which pairs of vertices are responsible for errors.
Inner-product error can also measure the quality of the
projection in general
because it indicates how the similarity between
original and projected vertices is preserved.

In Fig.~\ref{fig:quality},
we show the inner products of all possible pairs of vertices
in the original space
as a function of those in the projected space.
In general,
the difference between the inner product in the original space and
those in the projected space
becomes larger as the inner product in the projected space approaches zero,
even if the projection method is not PCA\@.
Exceptions are found in the fractal projections
[Figs.~\ref{fig:isometric-hamming-fractal}(e)
and~\ref{fig:isometric-hamming-fractal}(f)],
where the inner products in the original space and projected
space do not match
even if the inner product in the projected space is relatively larger.
All the projections show a similar trend that the inner-product error and
the error created by fractal projections is larger than the others.
The inner product becomes zero in two conditions:
when two vectors are orthogonal, or when one or both of them are
zero vectors.
As Fig.~\ref{fig:pca-hypercubes-multiple-weighted}(b) shows,
the latter contributes more than the former to the inner-product
error because
if both vertices of a pair have large probability,
the inner-product error becomes large.
In fact, the inner-product error increases
[cf.\ inset of Figs.~\ref{fig:quality}(i) and~\ref{fig:quality}(j)]
when the weighted vertices are
projected around the origin
[Fig.~\ref{fig:pca-hypercubes-multiple-weighted}(b)]
compared to when they are projected far from the origin
[Figs.~\ref{fig:pca-hypercubes-multiple-weighted}(a)
and~\ref{fig:quality}(i)].

Because of this general trend of the inner-product error,
orthogonal projections of hypercubes can be misleading
for the vertices located around the center of the projected space.
The pairwise inner product between the vertices both located
around the center of the projected space might be
the most distant pair in the original space.
For example, as we mentioned earlier, regarding
Fig.~\ref{fig:pca-hypercubes-multiple-weighted}
in Sec.~\ref{sec:hypercubic-pca},
the orthogonal projection of a hypercube using PCA
locates the third (or second) weighted vertex pair around the origin,
but their original Ising coordinates are the most distant pair.
In the same way, inspecting other types of orthogonal projections in
Figs.~\ref{fig:isometric-hamming-fractal} and~\ref{fig:swapping-vector},
we find that the vertices located and overlapped around the center of
the projected space can be the most distant in the original space.
Why is this trend ubiquitous in the orthogonal projection of hypercubes?
We answer this question
in Sec.~\ref{sec:number-of-vertices-along-pc1} and
Sec.~\ref{sec:upper-bound}.

\subsection{Centrality of the projection}
\label{sec:number-of-vertices-along-pc1}
Several projections of hypercubes, such as
Figs.~\ref{fig:isometric-hamming-fractal}(a),~\ref{fig:isometric-hamming-fractal}(d),~\ref{fig:isometric-hamming-fractal}(e),~\ref{fig:swapping-vector}(a),~\ref{fig:pca-hypercubes-random-pair-weighted}(a),~\ref{fig:pca-hypercubes-random-pair-weighted}(f),~\ref{fig:pca-hypercubes-multiple-weighted}(a),
and~\ref{fig:pca-hypercubes-multiple-weighted}(b),
show that multiple vertices are located around the origin of
the projected space.
In Fig.~\ref{fig:quality}, we find that the pairs of vertices
including those around the origin contribute to the inner-product error.
It seems ubiquitous that a number of vertices projected around the origin
in the orthogonal projection of hypercubes is significant.
Why are some vertices projected to accumulate around the origin?
To answer this question, we consider the number of vertices along
the horizontal axis of the Hamming projection.
We show that at large $N$,
the normalized number of vertices having Hamming distance $D_\mathrm{H}$
is approximated by the zero-mean Gaussian (normal) distribution
with appropriate centering.
In other words, we show that the distribution of the PC1 score of
\emph{unweighted} hypercubic vertices
using the PC1 loading of Eq.~\eqref{eq:PC1-loading} is
the Gaussian distribution.

We first consider the number of vertices along the Hamming-distance axis
of the Hamming projection.
The projected coordinates with Hamming distance
$D_\mathrm{H} = D_\mathrm{H}\left(-\bm{\xi}, \bm{s}\right)$ are given by
Eq.~\eqref{eq:PC1-score-Hamming-minus},
$
r\left(D_\mathrm{H}\right)
\in
\left\{
    -\sqrt{N} + \frac{2}{\sqrt{N}} D_\mathrm{H}
\right\}_{D_\mathrm{H}=0}^{N}
$.
Because the Hamming distance is the number of unmatched elements
in the binary vectors, the binomial coefficient
$
\binom{N}{D_\mathrm{H}}
\coloneqq
\frac{N!}{D_\mathrm{H}! \left(N-D_\mathrm{H}\right)!}
$
gives the number of vertices at each possible $r\left(D_\mathrm{H}\right)$.
The normalized number of vertices with the Hamming distance
$D_\mathrm{H}$ is
$
\varrho\left(D_\mathrm{H}\right)
=
\binom{N}{D_\mathrm{H}}
\frac{1}{2^N}
=
\binom{N}{D_\mathrm{H}}
\left(
    \frac{1}{2}
\right)^{D_\mathrm{H}}
\left(
    1
    -
    \frac{1}{2}
\right)^{N-D_\mathrm{H}}
$,
i.e., the binomial distribution.
With large $N$, de Moivre--Laplace theorem states that
$\varrho\left(D_\mathrm{H}\right)$ is asymptotically
a Gaussian distribution with mean $\frac{N}{2}$ and variance $\frac{N}{4}$,
\begin{equation}
    \varrho\left(D_\mathrm{H}\right)
    \simeq
    \frac{1}{\sqrt{2\uppi \frac{N}{4}}}
    \exp
    \left[
        -
        \frac{1}{2}
        \frac{
            \left(
                D_\mathrm{H} - \frac{N}{2}
            \right)^2
        }{\frac{N}{4}}
    \right]
    \label{eq:PC1-score-distribution-Hamming}
    .
\end{equation}
By changing the variable using Eq.~\eqref{eq:PC1-score-Hamming-minus},
the distribution as a function of the projected coordinate $r$,
$\varrho\left(r\right)$, is obtained:
\begin{equation}
    \varrho\left(r\right)
    =
    \frac{\mathrm{d}D_\mathrm{H}}{\mathrm{d}r}
    \varrho\left(D_\mathrm{H}\right)
    \simeq
    \frac{1}{\sqrt{2\uppi}}
    \exp
    \left(
        -
        \frac{1}{2}
        r^2
    \right)
    \label{eq:PC1-score-distribution}
    ,
\end{equation}
which is the standard Gaussian distribution.
Thus, as the dimension of the hypercube $N$ increases,
a larger number of vertices are projected around the origin
in the Hamming projection.

For the Hamming projection, we find that the distribution of
projected coordinates follows the Gaussian distribution.
To what extent is this result valid?
Our numerical results [Fig.~\ref{fig:pc-standard-deviation}] in
Sec.~\ref{sec:pc-loading-correspondence-numerical}
indicate that as long as the distribution of the vertices is qualitatively
similar to Eq.~\eqref{eq:bipolar-distribution},
the Hamming projection is guaranteed.
Thus, the number of vertices along the Hamming-distance axis (PC1 loading)
is approximated by a distribution close to the Gaussian distribution
if the distribution of the vertices has bipolarity.
This is the reason why the number of vertices along PC1 loading
tends to follow
the Gaussian distribution
even if the distribution is not ideally bipolar~\cite{Gerber2019}.

In general, the distribution of hypercubic vertices along a
linear projection
axis can be shown to be Gaussian under fairly weak assumptions.
First, note that any transformation vector $\bm{v}$ can be represented as
the normalization
$\bm{v} = \frac{\bm{q}}{\left| \bm{q} \right|}$ of a weighted
superposition of
non-overlapping binary sub-states,
\begin{equation}
    \bm{q}
    =
    \sum_{g = 1}^{M_\mathrm{s}} \frac{a_g}{\sqrt{n_g}} \bm{\xi}_g .
\end{equation}
Here, $M_\mathrm{s}$ is the number of sub-states, reflecting the
complexity of
the original transformation vector,
and $\bm{\xi}_g \in \left\{+1, -1, 0\right\}^N$
is the $g$th sub-state with $n_g$ non-zero elements and a weight $a_g$.
Note that, as $\left| \bm{\xi}_g \right| = \sqrt{n_g}$ by definition and the
sub-states do not overlap with each other,
$
\bm{\xi}_g^\top
\bm{\xi}_{g^\prime}
=
n_g \delta_{g, g^\prime}
$,
the squared norm of $\bm{q}$ is the sum of the squared weights:
$\left| \bm{q} \right|^2 = \sum_{g = 1}^{M_\mathrm{s}} a_g^2$.
The projected coordinate for a vertex $\bm{s}$ then is expressed
as a weighted
superposition of sub-coordinates,
\begin{equation}
    r
    =
    \frac{1}
    {
        \left|
        \bm{q}
        \right|
    }
    \bm{q}^\top \bm{s}
    =
    \frac{1}{\left| \bm{q} \right|}
    \sum_{g=1}^{M_\mathrm{s}}
    a_g r_g
    ,
\end{equation}
where $r_g$ is the contribution from the $g$th sub-state,
\begin{equation}
    r_g
    =
    \frac{1}{\sqrt{n_g}} \bm{\xi}_g^\top \bm{s}
    =
    \frac{
        -n_g + 2 D_{\mathrm{H}}^{\left(g\right)}
    }{
        \sqrt{n_g}
    }
    \label{eq:PC1-score-group}
    .
\end{equation}
$D_{\mathrm{H}}^{\left(g\right)}$ is the Hamming distance between
the non-zero
dimensions of the $g$th sub-state $-\bm{\xi}_g$ and the
corresponding dimensions
of a vertex $\bm{s}$, as in Eq.~\eqref{eq:PC1-score-Hamming-minus}.
If the non-zero dimension $n_g$ of all sub-states is sufficiently large, the
distribution of the projected sub-coordinate $r_g$ asymptotically
becomes the
standard Gaussian with the same procedure leading to
Eqs.~\eqref{eq:PC1-score-distribution-Hamming}
and~\eqref{eq:PC1-score-distribution}.
Therefore, the projected coordinate $r$,
which is the weighted sum of the sub-coordinates,
is also Gaussian with zero mean, and its variance is unity:
\begin{equation}
    \left<
    r^2
    \right>_{\varrho\left(r\right)}
    =
    \frac{1}{\left| \bm{q} \right|^2}
    \sum_{g = 1}^{M_\mathrm{s}}
    a_g^2
    \left<
    r_g^2
    \right>_{\varrho\left(r_g\right)}
    =
    1
    .
\end{equation}

What we show here is the reason that many vertices are projected around
the origin in the orthogonal projection---whether by PCA or
not---of hypercubes.
Upon linear projection, hypercubic vertices concentrate near the origin of
the projected space roughly following Gaussian shape.

\subsection{Inner-product error bounds of projections}
\label{sec:upper-bound}
\begin{figure}[tb]
    \includegraphics{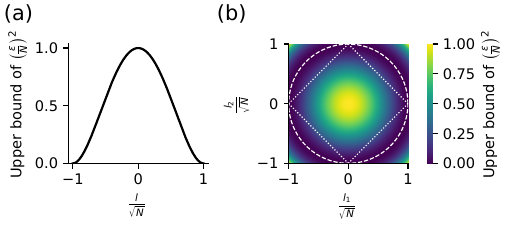}
    \caption{
        The upper bound of the squared inner-product error of
        Hamming projections.
        (a) The upper bound
        [Eq.~\eqref{eq:inner-product-abs-bound-normalized}]
        of the squared inner-product error between
        the vertices sharing the same projected coordinate $l$.
        (b) The upper bound
        [Eq.~\eqref{eq:inner-product-abs-bound-normalized-2D}]
        of the squared inner-product error between
        the vertices sharing the same projected coordinates $\bm{l}$.
        The hypercubic vertices are projected inside the dashed circle
        [Eq.~\eqref{eq:possible-region-general}] in general,
        and in case of Hamming projection,
        inside the dotted square
        [Eqs.~\eqref{eq:bound-r1plusr2} and~\eqref{eq:bound-r1minusr2}].
    }
    \label{fig:inner-product-error-bounds}
\end{figure}

We then evaluate the dependency of the inner-product error on
the projected coordinates.
Suppose the hypercubic vertices are projected to form a Hamming projection,
and the projected coordinates are given as
in Eq.~\eqref{eq:PC1-score-Hamming-plus}.
The Hamming distance between the two vertices $\bm{s}$ and $\bm{s}^\prime$
satisfies the triangular inequalities:
\begin{align}
    D_\mathrm{H}\left(\bm{s}, \bm{s}^\prime\right)
    &\leq
    D_\mathrm{H}\left(+\bm{\xi}, \bm{s}\right)
    +
    D_\mathrm{H}\left(+\bm{\xi}, \bm{s}^\prime\right)
    \\&=
    N
    -
    \frac{\sqrt{N}}{2}
    \left[
        r_1\left(\bm{s}\right)
        +
        r_1\left(\bm{s}^\prime\right)
    \right]
    \label{eq:triangular-inequality-1}
    ,
\end{align}
and
\begin{align}
    D_\mathrm{H}\left(\bm{s}, \bm{s}^\prime\right)
    &\leq
    D_\mathrm{H}\left(-\bm{\xi}, \bm{s}\right)
    +
    D_\mathrm{H}\left(-\bm{\xi}, \bm{s}^\prime\right)
    \\&=
    N
    +
    \frac{\sqrt{N}}{2}
    \left[
        r_1\left(\bm{s}\right)
        +
        r_1\left(\bm{s}^\prime\right)
    \right]
    \label{eq:triangular-inequality-2}
    .
\end{align}
Notice that from Eqs.~\eqref{eq:PC1-score-Hamming-plus}
and~\eqref{eq:PC1-score-Hamming-minus},
$
D_\mathrm{H}\left(\pm\bm{\xi}, \bm{s}\right)
=
\frac{N \mp \sqrt{N}r_1\left(\bm{s}\right)}{2}
$.
Combining Eqs.~\eqref{eq:triangular-inequality-1}
and~\eqref{eq:triangular-inequality-2}, we obtain
\begin{equation}
    D_\mathrm{H}\left(\bm{s}, \bm{s}^\prime\right)
    \leq
    N
    -
    \frac{\sqrt{N}}{2}
    \left|
    r_1\left(\bm{s}\right)
    +
    r_1\left(\bm{s}^\prime\right)
    \right|
    \label{eq:Hamming-distance-inequalities}
    .
\end{equation}
Using
Eqs.~\eqref{eq:Hamming-distance},~\eqref{eq:PC1-score-Hamming-plus},
and~\eqref{eq:Hamming-distance-inequalities},
the inner-product error of Eq.~\eqref{eq:inner-product-error} satisfies
\begin{align}
    \varepsilon(\bm{s}, \bm{s}^\prime)
    &=
    N
    -
    2 D_\mathrm{H}\left(\bm{s}, \bm{s}^\prime\right)
    -
    r_1\left(\bm{s}\right) r_1\left(\bm{s}^\prime\right)
    \nonumber
    \\&\geq
    -
    N
    +
    \sqrt{N}
    \left|
    r_1\left(\bm{s}\right)
    +
    r_1\left(\bm{s}^\prime\right)
    \right|
    -
    r_1\left(\bm{s}\right) r_1\left(\bm{s}^\prime\right)
    \label{eq:inner-product-lower-bound}
    ,
\end{align}
which is the lower bound of the inner-product error.
Because the projected coordinate has linearity under the
reflection of $\bm{s}$,
i.e.,
$
r_1\left(-\bm{s}\right)
=
- r_1\left(\bm{s}\right)
$,
the inner-product error of Eq.~\eqref{eq:inner-product-error} has
bilinearity,
$
\varepsilon\left(-\bm{s}, \bm{s}^\prime\right)
=
\varepsilon\left(\bm{s}, -\bm{s}^\prime\right)
=
-
\varepsilon\left(\bm{s}, \bm{s}^\prime\right)
$.
Thus, from Eq.~\eqref{eq:inner-product-lower-bound},
$
-\varepsilon\left(\bm{s}, \bm{s}^\prime\right)
=
\varepsilon\left(\bm{s}, -\bm{s}^\prime\right)
\geq
-
N
+
\sqrt{N}
\left|
r_1\left(\bm{s}\right)
+
r_1\left(-\bm{s}^\prime\right)
\right|
+
r_1\left(\bm{s}\right) r_1\left(-\bm{s}^\prime\right)
$,
which is equivalent to the upper bound of the inner-product error,
\begin{equation}
    \varepsilon(\bm{s}, \bm{s}^\prime)
    \leq
    +N
    -
    \sqrt{N}
    \left|
    r_1\left(\bm{s}\right)
    -
    r_1\left(\bm{s}^\prime\right)
    \right|
    -
    r_1\left(\bm{s}\right) r_1\left(\bm{s}^\prime\right)
    \label{eq:inner-product-upper-bound}
    .
\end{equation}
We investigate the error between the vertices which are projected to be
overlapped.
If $r_1\left(\bm{s}\right) = r_1\left(\bm{s}^\prime\right) = l$,
the inner-product error satisfies
\begin{equation}
    - N + 2\sqrt{N} \left| l \right| - l^2
    \leq
    \varepsilon
    \leq
    N - l^2
    \label{eq:inner-product-bound-l-1}
    .
\end{equation}
Thus, the squared inner-product error is then bounded by
\begin{equation}
    0
    \leq
    \varepsilon^2
    \leq
    \left(
        N - l^2
    \right)^2
    \label{eq:inner-product-abs-bound}
    .
\end{equation}
We normalize Eq.~\eqref{eq:inner-product-abs-bound} by $N^2$, resulting in
\begin{equation}
    0
    \leq
    \left(
        \frac{
            \varepsilon
        }{
            N
        }
    \right)^2
    \leq
    \left(
        1
        -
        \frac{l^2}{N}
    \right)^2
    \label{eq:inner-product-abs-bound-normalized}
    .
\end{equation}
We plot the upper bound of
Eq.~\eqref{eq:inner-product-abs-bound-normalized} in
Fig.~\ref{fig:inner-product-error-bounds}(a).
We indeed find that the error can deviate most from zero when
the two vertices are projected around the origin $l=0$.

We extend our consideration to the two-dimensional projection by
PC1 and PC2,
with orthogonal weighted states $\bm{\xi}_1$ and $\bm{\xi}_2$.
Assume the probability distribution of
Eq.~\eqref{eq:quadripolar-distribution},
suppose we have a Hamming projection which is projected by
Eqs.~\eqref{eq:quadripolar-PC1-loading}
and~\eqref{eq:quadripolar-PC2-loading}.
For each dimension $r_i$ of the projected coordinates
[Eq.~\eqref{eq:PC1-score}], the inequality of
Eq.~\eqref{eq:Hamming-distance-inequalities} is satisfied.
We then combine them as a single inequality,
\begin{align}
    & 2
    D_\mathrm{H}\left(\bm{s}, \bm{s}^\prime\right)
    \nonumber
    \\&\leq
    2N
    -
    \frac{\sqrt{N}}{2}
    \left[
        \left|
        r_1\left(\bm{s}\right)
        +
        r_1\left(\bm{s}^\prime\right)
        \right|
        +
        \left|
        r_2\left(\bm{s}\right)
        +
        r_2\left(\bm{s}^\prime\right)
        \right|
    \right]
    \nonumber
    \\ &=
    2N
    -
    \frac{\sqrt{N}}{2}
    \left|
    \bm{r}\left(\bm{s}\right)
    +
    \bm{r}\left(\bm{s}^\prime\right)
    \right|_1
    \label{eq:Hamming-distance-inequalities-2D}
    .
\end{align}
Using the same procedure to derive Eqs.~\eqref{eq:inner-product-lower-bound}
and~\eqref{eq:inner-product-upper-bound},
the inner-product error of Eq.~\eqref{eq:inner-product-error} satisfies
\begin{align}
    & -N
    +
    \sqrt{N}
    \left|
    \bm{r}\left(\bm{s}\right)
    +
    \bm{r}\left(\bm{s}^\prime\right)
    \right|_1
    -
    \bm{r}^\top\left(\bm{s}\right) \bm{r}\left(\bm{s}^\prime\right)
    \nonumber
    \\&\leq
    \varepsilon\left(\bm{s}, \bm{s}^\prime\right)
    \nonumber
    \\&\leq
    +N
    -
    \sqrt{N}
    \left|
    \bm{r}\left(\bm{s}\right)
    -
    \bm{r}\left(\bm{s}^\prime\right)
    \right|_1
    -
    \bm{r}^\top\left(\bm{s}\right) \bm{r}\left(\bm{s}^\prime\right)
    \label{eq:inner-product-lower-bound-2D}
    .
\end{align}
Our interest is the inner-product error between vertices sharing the same
projected coordinates.
If
$
\bm{r}\left(\bm{s}\right)
=
\bm{r}\left(\bm{s}^\prime\right)
=
\bm{l}
\coloneqq
\begin{bsmallmatrix}
    l_1 & l_2
\end{bsmallmatrix}^\top
$,
Eq.~\eqref{eq:inner-product-lower-bound-2D} becomes
\begin{equation}
    -N
    +
    2\sqrt{N} \left| \bm{l} \right|_1
    -
    \bm{l}^\top \bm{l}
    \leq
    \varepsilon
    \leq
    N
    -
    \bm{l}^\top \bm{l}
    \label{eq:inner-product-bound-l-2D}
    .
\end{equation}
Thus, the squared inner-product error is bounded by
\begin{equation}
    0
    \leq
    \varepsilon^2
    \leq
    \left(
        N
        -
        \bm{l}^\top \bm{l}
    \right)^2
    \label{eq:inner-product-abs-bound-2D}
    ,
\end{equation}
which is the extension of Eq.~\eqref{eq:inner-product-abs-bound} to
the two-dimensional projection.
We normalize Eq.~\eqref{eq:inner-product-abs-bound-2D} by $N^2$, obtaining
\begin{equation}
    0
    \leq
    \left(
        \frac{
            \varepsilon
        }{
            N
        }
    \right)^2
    \leq
    \left(
        1
        -
        \frac{
            \bm{l}^\top \bm{l}
        }{N}
    \right)^2
    \label{eq:inner-product-abs-bound-normalized-2D}
    .
\end{equation}
We plot Eq.~\eqref{eq:inner-product-abs-bound-normalized-2D} in
Fig.~\ref{fig:inner-product-error-bounds}(b) and find that the error can be
the largest when the two vertices are projected around
the origin $\bm{l} = \bm{0}$.

Notice that the corners of Fig.~\ref{fig:inner-product-error-bounds}(b) have
a slightly higher bound, but here we show that such regions cannot be used
as projected coordinates.
Assume we have a Hamming projection with two weighted states.
From Eq.~\eqref{eq:PC1-score},
the possible range of a single-dimensional projected coordinate is
$-\sqrt{N} \leq r_i\left(\bm{s}\right) \leq \sqrt{N}$.
In Appendix~\ref{sec:possible-region}, however,
we show that a two-dimensional projected coordinate is limited to
a specific region of
\begin{equation}
    \left|\bm{r} \left(\bm{s}\right)\right|
    \leq
    \sqrt{N}
    \label{eq:possible-region-general}
    .
\end{equation}
We plotted the boundary of the possible region with a white dashed line
in Fig.~\ref{fig:inner-product-error-bounds}(b).
Intuitively, this corresponds to the fact that all hypercubic vertices
are on the surface of the $N$-dimensional sphere with radius $\sqrt{N}$,
and the projection of the particular slice of the sphere is the
possible region.
Furthermore, the Hamming projection with two perpendicular weighted states
has a tighter bound on the possible region,
\begin{equation}
    \left|
    r_1\left(\bm{s}\right)
    +
    r_2\left(\bm{s}\right)
    \right|
    \leq
    \sqrt{N}
    \label{eq:bound-r1plusr2}
\end{equation}
and
\begin{equation}
    \left|
    r_1\left(\bm{s}\right)
    -
    r_2\left(\bm{s}\right)
    \right|
    \leq
    \sqrt{N}
    \label{eq:bound-r1minusr2}
    ,
\end{equation}
which is drawn with a white dotted line
in Fig.~\ref{fig:inner-product-error-bounds}(b).
The square shape of the possible region is due to the fact that
four vertices are chosen to form a square in the Hamming projection
as an idealization.
We derive Eqs.~\eqref{eq:bound-r1plusr2} and~\eqref{eq:bound-r1minusr2}
in Appendix~\ref{sec:possible-region}.
Thus, only a limited region of the
$r_1\left(\bm{s}\right)$--$r_2\left(\bm{s}\right)$ plane in
the Hamming projection can be the projected coordinates.

This Sec.~\ref{sec:upper-bound} explains why the inner-product error gets
a dominant contribution from the vertices located around the center of
the projected space.
Indeed, the upper bound of the inner-product error is the largest
when the vertices are projected around the origin.
This tendency is consistent with the numerical results in the
previous work~\cite{Abramson2003}.
Together with the results in Sec.~\ref{sec:number-of-vertices-along-pc1},
we conclude that the vertices projected around the center of the
projected space
contribute most to the inner-product error.

Our discussion in Secs.~\ref{sec:number-of-vertices-along-pc1}
and~\ref{sec:upper-bound} entirely relies on the assumption that
the hypercubic
vertices are projected to form a Hamming projection,
which is a rough approximation and might not be the case in general.
The full properties of such non-Hamming projections cannot be revealed
analytically, but as we show
in Sec.~\ref{sec:pc-loading-correspondence-numerical},
the qualitative aspects of such projections are expected to be similar to
those of the Hamming projection.
Thus, the quality of the projection of hypercubes is typically worse around
the center of the projected space.

\section{Applications}
\label{sec:applications}
So far, we have examined the orthogonal projection of hypercubes by
various methods with particular emphasis on PCA\@.
Through our investigation, we have obtained insights into the projection
using PCA, which enables us to interpret the resulting projections.
Here, employing several Ising spin systems,
we apply the orthogonal projection of hypercubes using PCA,
aiming to obtain the \emph{physical} interpretation of them.

\subsection{Related studies and background}
Several previous works have applied PCA to models in statistical mechanics,
particularly in the study of phase transitions.
Studies~\cite{Wang2016,Hu2017,Kiwata2019} demonstrate that PCA can successfully
identify the order parameter of spin systems from sampled states.
Moreover, PCA has been shown to detect order parameters in, for example,
off-lattice systems~\cite{Jadrich2018},
active matter systems~\cite{McDermott2023}, and
directed percolation~\cite{Shen2022,Muzzi2024},
highlighting its applicability across a broad range of many-body systems.
Thus, applying PCA enables the discovery of a system's
order parameter---even when the underlying Hamiltonian is unknown---which
is valuable for the unsupervised detection of phase transitions.

Similar to previous studies~\cite{Wang2016,Hu2017,Kiwata2019}, in
this Sec.~\ref{sec:applications}, we perform PCA on the states of
Ising spin systems in a canonical ensemble.
Our aim is, however, not only to visualize the states (hypercubic vertices)
but also to capture state transitions (hypercubic edges).
We seek to provide an alternative interpretation of PCA,
focusing on state transition dynamics through correlated spin flips,
rather than merely extracting meaningful parameters that describe
the phase of the system.
We begin by visualizing the hypercubic energy landscapes of Ising spin systems,
which offer insights into the
expected state transition dynamics at low temperatures---dynamics
that are experimentally observable~\cite{Farhan2013}.
Using these hypercubic energy landscapes,
we then directly investigate state transition dynamics by introducing the
time dependence of the probability distribution,
revealing state transition pathways as constrained probability fluxes
that emerge from correlated spin flips.

\begin{figure}[tb]
    \includegraphics{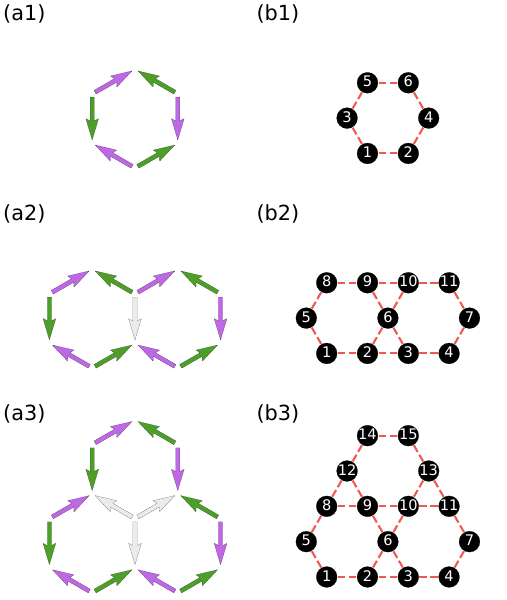}
    \caption{
        The finite artificial kagome spin-ice systems
        and corresponding Ising spin interaction networks.
        (a) The finite artificial spin-ice systems on the kagome lattice;
        the one-ring (a1), the two-ring (a2), and the three-ring
        (a3) systems.
        The arrow indicates the magnetization of the island.
        The color of an arrow is purple (green)
        when it aligns with the clockwise (counterclockwise) direction.
        The color becomes white when the arrow is in the center
        of the system.
        (b) The corresponding Ising spin interaction network for (a);
        one-ring (b1), two-ring (b2), and three-ring (b3) systems.
        The black (white, not shown here) vertex represents
        the up, $\bullet=\,\uparrow\,\coloneqq +1$
        (down $\circ=\,\downarrow\,\coloneqq -1$), Ising spin state.
        Labels of spins are drawn on each vertex.
        The red dashed edge represents the
        antiferromagnetic interaction between spins.
        In (a) and (b), as an example,
        we illustrate the state with the highest
        energy, i.e., the all-spin-up state.
        The Ising spin in (b) is in the up state
        when the magnetization of the island in (a) is directed toward
        the upper triangle of the kagome lattice.
    }
    \label{fig:spin-ice}
\end{figure}

\subsection{The finite artificial kagome spin-ice system}
To demonstrate the usage of the orthogonal projection of
hypercubes using PCA,
we apply our method to statistical mechanical models.
Specifically, we employ the hypercubic energy landscape~\cite{Farhan2013}
of the finite artificial spin-ice systems~\cite{Mengotti2008}.
For this purpose,
we consider the paradigmatic Ising spin system on the kagome lattice because
the kagome spin-ice systems can be mapped to
the antiferromagnetic kagome Ising spin system~\cite{Chioar2016}.
In Fig.~\ref{fig:spin-ice},
we illustrate the finite artificial spin-ice systems
and corresponding Ising spin interaction networks
to be considered:
one-ring [Figs.~\ref{fig:spin-ice}(a1) and~\ref{fig:spin-ice}(b1)],
two-ring [Figs.~\ref{fig:spin-ice}(a2) and~\ref{fig:spin-ice}(b2)],
and
three-ring [Figs.~\ref{fig:spin-ice}(a3)
and~\ref{fig:spin-ice}(b3)] systems.
These systems were experimentally realized~\cite{Mengotti2008,
Farhan2013}, and
the observed dynamics were analyzed through
hypercubic energy landscapes~\cite{Farhan2013} but not with PCA\@.

The Hamiltonian of the Ising spin system is defined as
\begin{align}
    \mathcal{H}\left(\bm{s}\right)
    &\coloneqq
    -
    \frac{1}{2}
    \sum_{i=1}^{N}
    \sum_{j=1}^{N}
    s_i
    J_{i,j}
    s_j
    -
    \sum_{i=1}^{N}
    s_i
    h_i
    \\&=
    -
    \frac{1}{2}
    \bm{s}^\top
    \bm{J}
    \bm{s}
    -
    \bm{s}^\top
    \bm{h}
    \label{eq:ising-spin-hamiltonian}
    ,
\end{align}
where the state vector is defined as
$
\bm{s}
\coloneqq
\begin{bsmallmatrix}
    s_1 & \cdots & s_N
\end{bsmallmatrix}^\top
$
with the element as the Ising spin variable
(up $\uparrow$ or down $\downarrow$),
$
s_i \in
\left\{
    \bullet=\,\uparrow\,\coloneqq +1, \circ=\,\downarrow\,\coloneqq -1
\right\}
$.
A state $\bm{s}$ is a vertex of the $N$-dimensional hypercube,
and a state transition with a single spin flip corresponds to
hypercubic edge.
The element of the interaction matrix
$\bm{J} \in \left\{-1, 0\right\}^{N \times N}$ is
$J_{i,j} = J_{j,i} = -1$
when spin $i$ and $j$ antiferromagnetically interact
(antialignment of spin $i$ and $j$ decreases energy)
on the kagome lattice of
Figs.~\ref{fig:spin-ice}(b1)--\ref{fig:spin-ice}(b3),
and $J_{i,j} = J_{j,i} = 0$ otherwise.
Self-interactions do not exist ($J_{i,i}=0$, $\forall i$).
The external magnetic field
$
\bm{h}
\coloneqq
\begin{bsmallmatrix}
    h_1 & \cdots & h_N
\end{bsmallmatrix}^\top
\in
\mathbb{R}^N
$
is assigned to be zero $\bm{0}$ for projection of the hypercubic
energy landscape, but later we will consider the case with a non-zero field.

Due to the (geometrical) frustration~\cite{Toulouse1977} of
the interaction network and associated degeneracy of ground states,
the energy landscape determines the dynamics of the systems:
one can obtain insights into the dynamics
by visualizing the complexity of the hypercubic energy landscape.
The frustration of the interaction network is quantified by
the frustration function
$
\varPhi
\left(\mathcal{C}\right)
\coloneqq
\operatorname{sgn}
\left(
    \prod_{J_{i, j}\in\mathcal{C}} J_{i, j}
\right)
$
for a closed undirected cycle $\mathcal{C}$
in the interaction network~\cite{Toulouse1977}.
If $\varPhi\left(\mathcal{C}\right) < 0$, the cycle $\mathcal{C}$
is frustrated:
any state cannot satisfy all the interactions in the cycle.

We examine the spin-ice systems through the interaction network
and associated frustration.
The interaction network of the one-ring system
[Fig.~\ref{fig:spin-ice}(b1)] is
a ring without frustration, $\varPhi\left(\mathcal{C}\right) \nless 0$,
and there are two ground states: a ground state
$
\begin{bsmallmatrix}
    \circ & \bullet & \bullet & \circ & \circ & \bullet
\end{bsmallmatrix}^\top
$
and its global spin-flipped state
$
\begin{bsmallmatrix}
    \bullet & \circ & \circ & \bullet & \bullet & \circ
\end{bsmallmatrix}^\top
$.
The two-ring system has several ground states
owing to the frustrated interaction involving spin 6, such as
$
\mathcal{C}
=
\left\{
    J_{6, 10},
    J_{10, 9},
    J_{9, 6}
\right\}
$
and
$
\mathcal{C}
=
\left\{
    J_{6, 3},
    J_{3, 2},
    J_{2, 6}
\right\}
$.
In the three-ring system,
there are more frustrated cycles containing spins 6, 9, and 10,
e.g.,
$
\mathcal{C}
=
\left\{
    J_{6, 10},
    J_{10, 9},
    J_{9, 6}
\right\}
$,
$
\mathcal{C}
=
\left\{
    J_{6, 3},
    J_{3, 2},
    J_{2, 6}
\right\}
$,
$
\mathcal{C}
=
\left\{
    J_{9, 8},
    J_{8, 12},
    J_{12, 9}
\right\}
$,
and
$
\mathcal{C}
=
\left\{
    J_{10, 13},
    J_{13, 11},
    J_{11, 10}
\right\}
$.
Thus, the one-ring system has an unfrustrated interaction network,
and the two- and three-ring systems have frustrated interaction networks.

\subsection{Projecting the hypercubic energy landscape
of the kagome spin-ice system}
For the projection of the hypercubic energy landscape,
we calculate the covariance matrix [Eq.~\eqref{eq:covariance-matrix}]
with the probability distribution of each state.
The probability distribution of each state (hypercubic vertex) is
determined by the canonical ensemble: Boltzmann distribution
\begin{equation}
    p\left(\bm{s}\right)
    =
    \frac{1}{\mathcal{Z}}
    \exp
    \left[
        -
        \beta
        \mathcal{H}\left(\bm{s}\right)
    \right]
    \label{eq:Boltzmann-distribution}
    ,
\end{equation}
where
$
\mathcal{Z}
\coloneqq
\sum_{\bm{s}}
\exp
\left[
    -
    \beta
    \mathcal{H}\left(\bm{s}\right)
\right]
$
is the partition function.
The temperature $T$ of the reservoir (bath) or inverse temperature
$\beta$ is assigned to be
$
k_\mathrm{B}T
=
\frac{1}{\beta}
= 0.3
$
in our projection.
Here, $k_\mathrm{B}$ is the Boltzmann constant.
Note that
the probability distribution is an even function,
$
p\left(-\bm{s}\right)
=
p\left(\bm{s}\right)
$
when the external field is zero, $\bm{h}=\bm{0}$,
because the Hamiltonian has symmetry under the global spin flip,
$\bm{s} \leftarrow -\bm{s}$, or $\mathbb{Z}_2$ symmetry,
$
\mathcal{H} \left(-\bm{s}\right)
=
-
\frac{1}{2}
\left(-\bm{s}\right)^\top
\bm{J}
\left(-\bm{s}\right)
=
-
\frac{1}{2}
\bm{s}^\top
\bm{J}
\bm{s}
=
\mathcal{H} \left(\bm{s}\right)
$.
This symmetry of the probability distribution makes
the mean state vector the zero vector,
$
\left<\bm{s}\right>
=
\bm{0}
$
because
$
\left<\bm{s}\right>
=
\sum_{\bm{s}}
p\left(\bm{s}\right)
\bm{s}
=
\sum_{-\bm{s}}
p\left(-\bm{s}\right)
\left(-\bm{s}\right)
=
\sum_{\bm{s}}
p\left(\bm{s}\right)
\left(-\bm{s}\right)
=
-\left<\bm{s}\right>
$.
Thus, the covariance matrix becomes
$
\bm{\varSigma}
=
\sum_{\bm{s}}
p\left(\bm{s}\right)
\bm{s}
\bm{s}^\top
$.
The off-diagonal element (covariance) is
$
\varSigma_{i,j}
=
\sum_{\bm{s}}
p\left(\bm{s}\right)
s_i s_j
$,
and the diagonal element (variance) is
$
\varSigma_{i,i}
=
\sum_{\bm{s}}
p\left(\bm{s}\right)
s_i^2
=
\sum_{\bm{s}}
p\left(\bm{s}\right)
=
1
$.
Therefore, the correlation between spin $i$ and $j$ is
the covariance between them,
$
\frac{
    \varSigma_{i,j}
}{
    \sqrt{\varSigma_{i,i}}
    \sqrt{\varSigma_{j,j}}
}
=
\varSigma_{i,j}
$.
The covariance matrix becomes the correlation matrix.
PCA finds the most correlated direction
in the hypercubic state space.

\begin{figure*}[p]
    \includegraphics{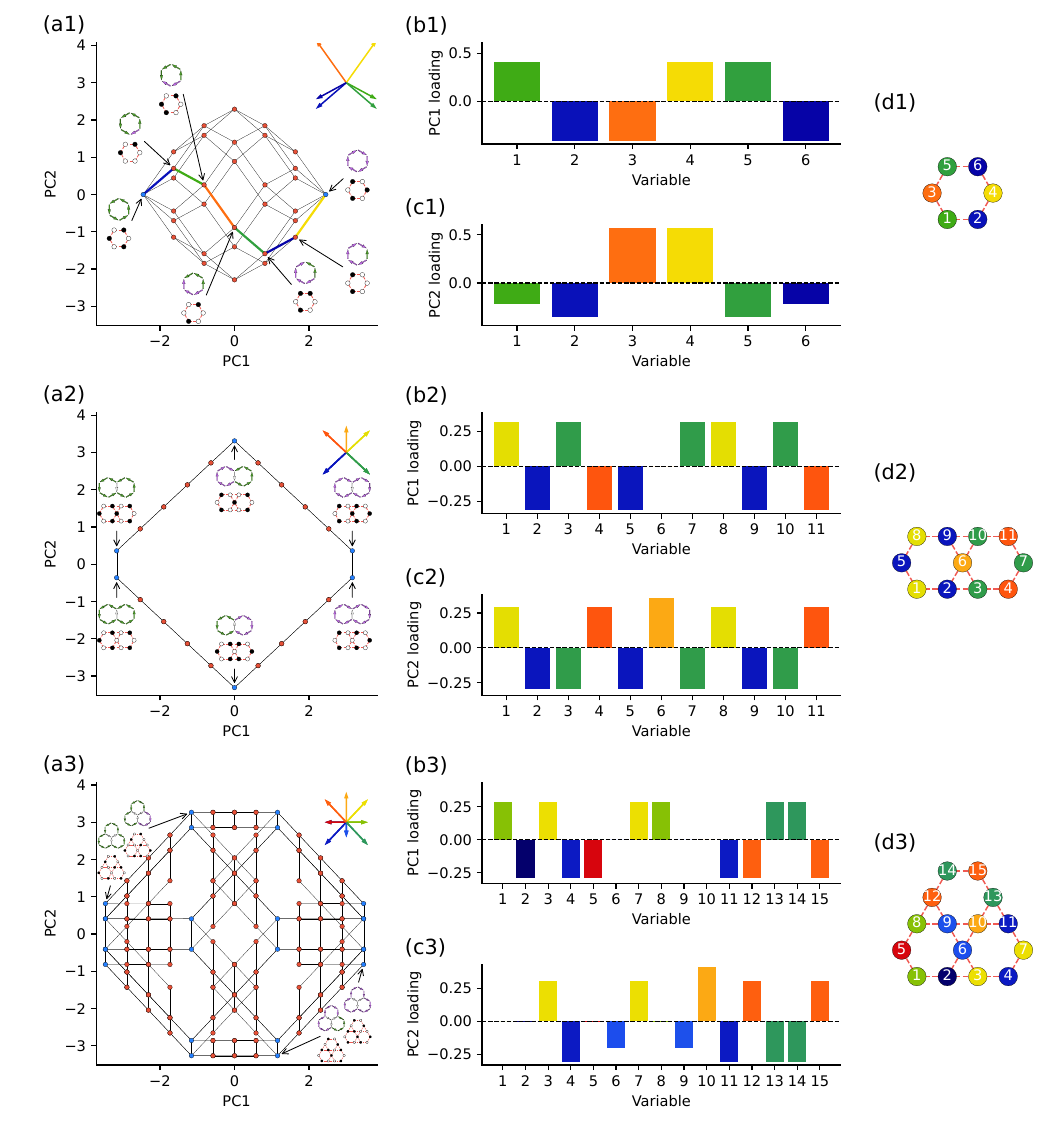}
    \caption{
        Hypercubic energy landscape of the finite artificial spin-ice
        systems and
        interaction networks colored by the angle of the biplot vectors.
        (a) The hypercubic energy landscape of one-ring (a1), two-ring (a2),
        and three-ring (a3) systems by PC1 and PC2.
        For clarity of the panel, only the states with the lowest
        and the second-lowest energy are shown.
        The energy of the state colors vertices
        (blue for the lowest energy and red for the second-lowest energy).
        Biplot vectors are shown on the top right, and
        their color indicates the angle of the biplot vectors.
        Notice that in (a2) and (a3),
        several biplot vectors are overlapped.
        One example pathway from a ground state to another ground state is
        shown as colored hypercubic edges in (a1).
        The colors of these lines are the same as the color of
        the corresponding
        biplot vectors.
        Examples of states are shown as insets in (a1), (a2), and (a3).
        See the caption of Fig.~\ref{fig:spin-ice} for
        the definition of state visualization on the interaction network.
        (b) PC1 loading of one-ring (b1), two-ring (b2),
        and three-ring (b3) systems.
        The color of the bar matches the color of
        the corresponding biplot vectors in (a).
        (c) Same as (b) but for PC2.
        (d) The Ising spin interaction networks of
        one-ring (d1), two-ring (d2), and three-ring (d3) systems,
        where the nodes are colored by the angle of
        the corresponding biplot vectors in (a).
    }
    \label{fig:spin-ice-pc-1-2}
\end{figure*}

In Figs.~\ref{fig:spin-ice-pc-1-2}(a1)--\ref{fig:spin-ice-pc-1-2}(a3),
the hypercubic energy landscapes of the finite artificial kagome
Ising spin systems
are projected using PCA\@.
Each state $\bm{s}$ corresponds to a vertex of the hypercube, and
a state transition with a single spin flip corresponds to
the edge of the hypercube.
Thus, a hypercubic energy landscape is a kind of state transition diagram.
For clarity of visualization,
we show only the vertices with the lowest and the second-lowest energy,
i.e., the ground states and the first-excited states,
out of $2^N$ vertices (states)
in Figs.~\ref{fig:spin-ice-pc-1-2}(a1)--\ref{fig:spin-ice-pc-1-2}(a3).
By visualizing hypercubic edges connecting
the ground states and first-excited states,
one can observe the pathways expected to be followed
in the dynamics at low temperatures.
Note also that the distributions discussed in
Sec.~\ref{sec:pc-loading-corresponding} appear here:
the one-, two-, and three-ring systems have probability distributions
that closely resemble the bipolar, quadripolar, and sexapolar cases,
respectively.

A glance at the hypercubic energy landscapes in
Figs.~\ref{fig:spin-ice-pc-1-2}(a1)--\ref{fig:spin-ice-pc-1-2}(a3)
reveals a qualitative difference between the unfrustrated (one-ring) system
and the frustrated (two- and three-ring) systems.
In the one-ring system, the ground states are not directly connected
by hypercubic edges; instead, they are always separated by excited states.
In contrast, in the two- and three-ring systems, some ground states are
directly connected by hypercubic edges.
This shows that, in unfrustrated systems, ground states are always
surrounded by excited states, whereas in frustrated systems, certain ground
states are adjacent to each other.
Thus, the degeneracy of ground states in frustrated systems qualitatively
changes the structure of the hypercubic energy landscape.

\subsection{The hypercubic energy landscape of the one-ring system}
\label{sec:one-ring}
We first examine the one-ring system of Figs.~\ref{fig:spin-ice}(a1)
and~\ref{fig:spin-ice}(b1).
In Fig.~\ref{fig:spin-ice-pc-1-2}(a1),
we show the possible pathway from a ground state
$
\begin{bsmallmatrix}
    \circ & \bullet & \bullet & \circ & \circ & \bullet
\end{bsmallmatrix}^\top
$
to the globally spin-flipped state
$
\begin{bsmallmatrix}
    \bullet & \circ & \circ & \bullet & \bullet & \circ
\end{bsmallmatrix}^\top
$
by connecting the vertices with the lowest energy and the
second-lowest energy
using hypercubic edges.
(See Figs.~\ref{fig:spin-ice}(b1) or~\ref{fig:spin-ice-pc-1-2}(d1)
for labeling of spins.)
In this example, the hypercubic edge corresponds to
the state transition with a single spin flip.
The biplot vectors are shown at the top right of
Fig.~\ref{fig:spin-ice-pc-1-2}(a1), and its color indicates the angle.
We emphasize one particular pathway between two ground states
by coloring the edges
following the biplot vectors in Fig.~\ref{fig:spin-ice-pc-1-2}(b1).
We found the state transition pathway is collective:
the flipping order is restricted to keep the energy low.
In a one-ring system, there are $6!=720$ possible pathways from
one ground state to another ground state,
but the number of most probable pathways is limited, as shown
in Fig.~\ref{fig:spin-ice-pc-1-2}(a1), and it is reduced to
$6 \times 2^4 \times 1 = 96$~\cite{Farhan2013}.
This reduction arises from the interaction network of the system and
the associated hypercubic energy landscape.

In addition, one can infer which spin is involved in which state transition
by the biplot vectors.
One can know which spin flips in the state transition by
comparing the length and angle of the hypercubic edge of interest with
that of the biplot vectors.
The direction of the arrow of the biplot vector indicates
the direction of the spin flip.
In other words, if the state transition is along the direction of
the arrow of the biplot vector, the spin flips from down to up.
Otherwise, the spin flips from up to down.
Notice that Fig.~\ref{fig:spin-ice-pc-1-2}(a1) is the Hamming
projection because
the probability distribution is roughly the same
as Eq.~\eqref{eq:bipolar-distribution}.

To interpret the relation between PC loading and the interaction network,
we visualize the loading of PC1 and PC2 in
Figs.~\ref{fig:spin-ice-pc-1-2}(b1)
and~\ref{fig:spin-ice-pc-1-2}(c1), respectively,
colored by the angle of the biplot vector in
Fig.~\ref{fig:spin-ice-pc-1-2}(a1).
We confirmed that the element-wise sign of PC1 corresponds to
the ground state [Fig.~\ref{fig:spin-ice-pc-1-2}(b1)].
The element-wise sign of PC2 loading corresponds to the state
perpendicular to
the ground state, which has a Hamming distance of $\frac{N}{2}$
from the ground
state [Fig.~\ref{fig:spin-ice-pc-1-2}(c1)].

The interaction network in Fig.~\ref{fig:spin-ice-pc-1-2}(d1),
where nodes are colored by the angle of the biplot vectors,
shows the correlation captured by the first two PC loadings.
Spins 1 and 5 have the same color,
and spin 3, interacting with both spins 1 and 5, has the color
complementary to
that of spins 1 and 5.
We can argue the same for spins 2, 4, and 6.
Hence, the resulting hypercubic energy landscape is drawn to
emphasize the correlation arising from the interaction.

\subsection{The hypercubic energy landscape of the two-ring system}
The hypercubic energy landscape of the two-ring system,
which has several ground states, is illustrated
in Fig.~\ref{fig:spin-ice-pc-1-2}(a2).
We show six ground states and pathways
between them in a two-ring system with a biplot vector on the top right.
Similar to the one-ring system in Fig.~\ref{fig:spin-ice-pc-1-2}(a1),
the most probable pathways between the ground states are shown by
connecting the vertices
with the lowest and the second-lowest energy.
We discuss two pathways connecting the ground states as examples.

The first example involves two ground states on the left side
(around
    $
    \begin{bsmallmatrix}
        -3 & 0
    \end{bsmallmatrix}^\top
$),
i.e.,
$
\begin{bsmallmatrix}
    \circ & \bullet & \circ & \bullet & \bullet & \bullet & \circ & \circ &
    \bullet & \circ & \bullet
\end{bsmallmatrix}^\top
$
and
$
\begin{bsmallmatrix}
    \circ & \bullet & \circ & \bullet & \bullet & \circ & \circ & \circ &
    \bullet & \circ & \bullet
\end{bsmallmatrix}^\top
$
which are connected by the hypercubic edge involving the flip of spin 6,
indicating that spin 6 is expected to fluctuate on
the left side of the landscape.
One of the interactions in pairs of $J_{6, 3}$ and $J_{6, 10}$, or
$J_{6, 2}$ and $J_{6, 9}$, is unsatisfied with any state of spin 6.
This frustration of the interaction network cancels
the local field to spin 6, i.e.,
$
\sum_{j=1}^{11}
J_{6, j} s_j
=
0
$.
This is the reason why spin 6 does not have an energetically favored state.

The second example is the transition pathway
from the ground state on the upper left side
(around
    $
    \begin{bsmallmatrix}
        -3 & 0.3
    \end{bsmallmatrix}^\top
$),
to the ground state on the top
(around
    $
    \begin{bsmallmatrix}
        0 & 3.2
    \end{bsmallmatrix}^\top
$),
i.e.,
state transition from
$
\begin{bsmallmatrix}
    \circ & \bullet & \circ & \bullet & \bullet & \bullet & \circ & \circ &
    \bullet & \circ & \bullet
\end{bsmallmatrix}^\top
$
to
$
\begin{bsmallmatrix}
    \bullet & \circ & \circ & \bullet & \circ & \bullet & \circ & \bullet &
    \circ & \circ & \bullet
\end{bsmallmatrix}^\top
$,
which requires the flipping of spins 1, 2, 5, 8, and 9
[see biplot vectors of Fig.~\ref{fig:spin-ice-pc-1-2}(a2)
    and color of
    Figs.~\ref{fig:spin-ice-pc-1-2}(b2),~\ref{fig:spin-ice-pc-1-2}(c2),
and~\ref{fig:spin-ice-pc-1-2}(d2)].
This projection of the hypercube indicates that state transition
happens with
the correlated spin flips on the left-half part of the system
(spins 1, 2, 5, 8, and 9).
Among the left-half part of the system, spins 2 and 9 are more
likely to flip
at the beginning of the pathway.
This difference arises from the unsatisfied interactions:
$J_{3, 9}$ and $J_{6, 2}$.
Spins 2 and 9 receive weaker local fields than spins 1, 5, and 8
because they have interactions involving spins 3, 6, and 10.
Thus, spins 2 and 9 are expected to flip more than spins 1, 5, and 8.
Because of the frustrated interactions involving spins 3, 6, and 10,
spins 2 or 9 are the most probable to flip first.
Then,
the flips happen in the order of spins 2, 1, 5, 8, and 9 if spin
2 flips first,
or in the order of spins 9, 8, 5, 1, and 2 if spin 9 flips first.
Similar to the one-ring system in Fig.~\ref{fig:spin-ice-pc-1-2}(a1),
the number of most probable pathways is limited,
but the constraints are stricter and fewer pathways are possible.

To deepen our understanding of PC loading and interaction networks,
we show the loading of PC1 and PC2 in
Figs.~\ref{fig:spin-ice-pc-1-2}(b2) and~\ref{fig:spin-ice-pc-1-2}(c2),
coloring them by the angles of the biplot vectors in
Fig.~\ref{fig:spin-ice-pc-1-2}(a2).
Again, we confirmed that the sign of
PC1 loading [Fig.~\ref{fig:spin-ice-pc-1-2}(b2)]
corresponds to the ground state with
the largest magnitude of the PC1 score, but spin 6 is not determined.
Thus, the frustrated dynamics of spin 6 are captured by PC1.
The element-wise sign of PC2 loading [Fig.~\ref{fig:spin-ice-pc-1-2}(c2)]
corresponds to the ground state with the largest magnitude of the PC2 score.
Notice that the sixth element of PC2 loading has a slightly
higher magnitude than the others,
indicating the importance of spin 6.
If spin 6 flips when the system is in the ground state
$\operatorname{sgn}\left(\bm{u}_2\right)$,
the energy of the system increases more than when another spin flips.

The interaction network with nodes colored by the angles of the
biplot vectors in
Fig.~\ref{fig:spin-ice-pc-1-2}(d2) shows the
inter-spin correlation captured by the biplot vectors.
The biplot vectors capture the (anti)correlation of spins 1, 2, 5, 8, and 9
[yellow and blue arrows in Fig.~\ref{fig:spin-ice-pc-1-2}(a2) top right]
which corresponds to the left ring of the interaction network
in Fig.~\ref{fig:spin-ice-pc-1-2}(d2).
The biplot vectors also capture the (anti)correlation of
spins 3, 4, 7, 10, and 11
[green and purple arrows in Fig.~\ref{fig:spin-ice-pc-1-2}(a2) top right]
which corresponds to the right ring of the interaction network
in Fig.~\ref{fig:spin-ice-pc-1-2}(d2).
The hypercubic energy landscape using PCA provides insight into the dynamics
through the correlation of spins.

\subsection{The hypercubic energy landscape of the three-ring system}
\label{sec:three-ring}
More ground states and probable pathways emerge in
the hypercubic energy landscape of a three-ring system
[Fig.~\ref{fig:spin-ice-pc-1-2}(a3)].
There are eight groups of connected ground states;
two of them around
$
\begin{bsmallmatrix}
    -3 & 0
\end{bsmallmatrix}^\top
$
and
$
\begin{bsmallmatrix}
    3 & 0
\end{bsmallmatrix}^\top
$
consist of
six ground states, and the rest consist of two ground states.

Unlike the one-ring and two-ring systems,
the structure of pathways is more complex, but
we can still infer the state transition from the projection.
For example, the pathway from the ground state
$
\begin{bsmallmatrix}
    \circ & \bullet & \circ & \bullet & \bullet & \circ & \circ & \circ &
    \circ & \bullet & \bullet & \bullet & \circ & \circ & \bullet
\end{bsmallmatrix}^\top
$
(around
    $
    \begin{bsmallmatrix}
        -3 & 0.9
    \end{bsmallmatrix}^\top
$)
to
$
\begin{bsmallmatrix}
    \circ & \bullet & \bullet & \circ & \bullet & \circ & \bullet & \circ &
    \circ & \bullet & \circ & \bullet & \circ & \circ & \bullet
\end{bsmallmatrix}^\top
$
(around
    $
    \begin{bsmallmatrix}
        -1.5 & 3.2
    \end{bsmallmatrix}^\top
$)
is a transition with the correlated spin flips of spins located
in the lower right hexagon (spins 3, 4, 7, and 11).
Pathways parallel to this pathway involve the same spin flips.
From the biplot vectors on the top right of
Fig.~\ref{fig:spin-ice-pc-1-2}(a3),
one can generally know that by following the angle of each biplot vector,
the transitions involve spins on the top hexagon
(spins 12, 13, 14, and 15) being flipped:
the pathways are lines from the left top to the right bottom,
which are parallel to the biplot vectors of spins 12, 13, 14, and 15.
Likewise, horizontal pathways are used
when spins in the lower left hexagon (spins 1, 2, 5, and 8) are flipped,
and vertical pathways are followed
when spins in the inverted triangle at the center (spins 6, 9, and 10) are
flipped.
Similar to the one-ring and two-ring systems,
the pathway is a collective flip of spins reflecting the
interaction network.
Notice that this projection of
the hypercubic energy landscape [Fig.~\ref{fig:spin-ice-pc-1-2}(a3)] is
a partial Hamming projection:
the PC1 score is equivalent to the Hamming distance from the ground state on
the left side
(around
    $
    \begin{bsmallmatrix}
        -3 & 0
    \end{bsmallmatrix}^\top
$)
or the right side
(around
    $
    \begin{bsmallmatrix}
        3 & 0
    \end{bsmallmatrix}^\top
$)
but without considering spins 6, 9, and 10.

To understand the relation between PC loading and the interaction network,
in particular,
the relation between PC loading and the frustration of the system,
we color the loading of PC1 and PC2 in
Figs.~\ref{fig:spin-ice-pc-1-2}(b3) and~\ref{fig:spin-ice-pc-1-2}(c3),
following the angle of the biplot vectors in
Fig.~\ref{fig:spin-ice-pc-1-2}(a3).
We confirmed that the element-wise sign of PC1
[Fig.~\ref{fig:spin-ice-pc-1-2}(b3)]
corresponds to the ground state group containing
six ground states, but the states of spins 6, 9, and 10 are not determined.
These spins involve a frustrated cycle of the interaction network.
PC2 loading in Fig.~\ref{fig:spin-ice-pc-1-2}(c3) shows that
spins 1, 2, 5, and 8 do not contribute, and those spins belong to
the lower left ring.
Moreover, PC2 loading of spins 6, 9, and 10 has
slightly higher or lower magnitudes than the others,
indicating the uncommon contribution of spins 6, 9, and 10,
suggesting the frustration-related correlation.

These insights are consistent with the interaction network with
colored nodes
in Fig.~\ref{fig:spin-ice-pc-1-2}(d3):
the PC loading captures the correlation arising from the
interaction network.
Nevertheless, spins 6 and 9 have exactly the same color:
their biplot vectors are the same,
but it can be another combination of spins such as 6 and 10.
This breaks the symmetry of the system—PC2 ignores the lower left hexagon of
the interaction network even though the system has the same other
structure—and
we examine how the PC loading reflects the symmetry of the system.

\subsection{Fraction of explained variance}
\begin{figure}[tb]
    \includegraphics{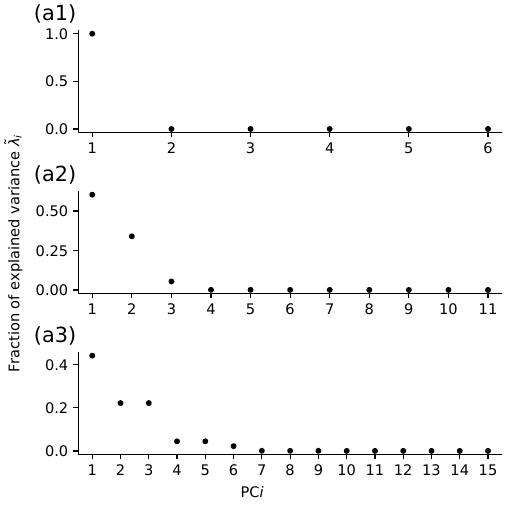}
    \caption{
        Fraction of explained variance by each PC of
        the hypercubic energy landscapes in Fig.~\ref{fig:spin-ice-pc-1-2}.
        (a1) One-ring system.
        (a2) Two-ring system.
        (a3) Three-ring system.
    }
    \label{fig:spin-ice-variance}
\end{figure}

We investigate the fraction of explained variance
by the PCs of the hypercubic energy landscape
in Fig.~\ref{fig:spin-ice-variance}
to see how the frustration of the system influences the PCs.
For the one-ring system [Fig.~\ref{fig:spin-ice-variance}(a1)],
PC1 dominates the explained variance by the PCs,
which is consistent with PC1 being proportional to the ground state.
Turning to the two-ring and three-ring systems
[Figs.~\ref{fig:spin-ice-variance}(a2) and~\ref{fig:spin-ice-variance}(a3)],
the first PCs explain less than 90\% of the variance.
This decrease in the explained variance
by the first component arises from the degeneracy of the ground states.
PC2 explains the variance of degenerate ground states.
For the three-ring system [Fig.~\ref{fig:spin-ice-variance}(a3)],
the first three PCs are required to explain more than 80\% of the variance,
and the second and third fractions of explained variance are the same,
indicating that PC2 and PC3 share the regularity:
the symmetry of the system.

\begin{figure*}[tb]
    \includegraphics{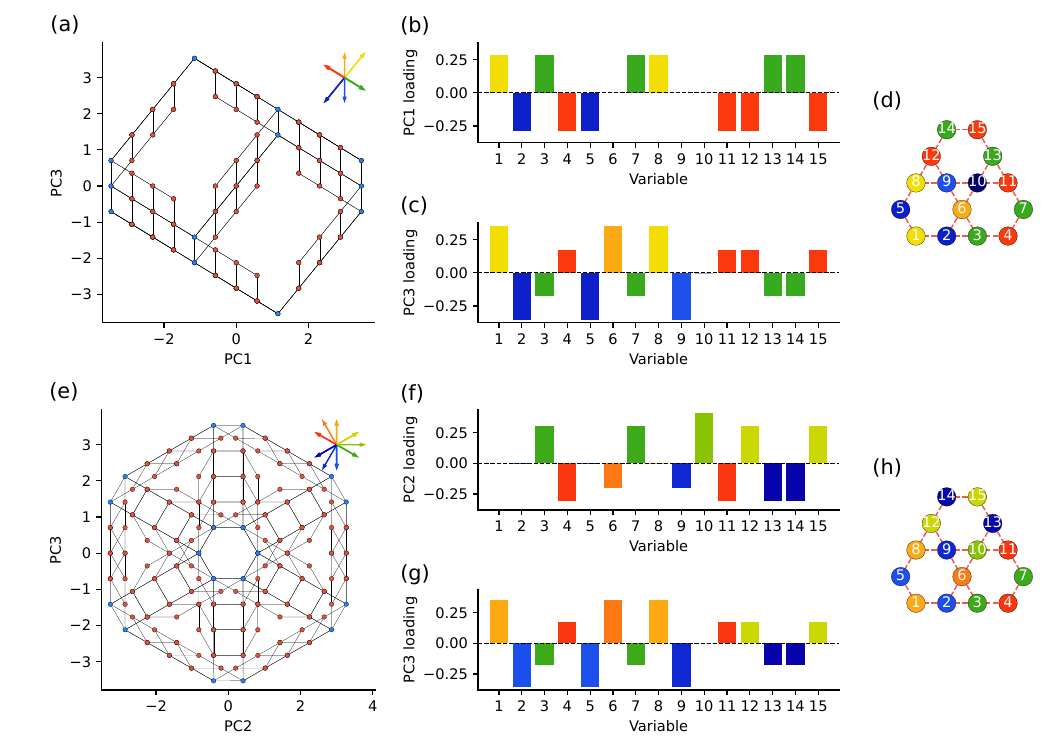}
    \caption{
        Same as
        Figs.~\ref{fig:spin-ice-pc-1-2}(a3),~\ref{fig:spin-ice-pc-1-2}(b3),~\ref{fig:spin-ice-pc-1-2}(c3),
        and~\ref{fig:spin-ice-pc-1-2}(d3) but by different PCs.
        (a) The hypercubic energy landscape of the three-ring system
        by PC1 and PC3.
        (b) PC1 loading of (a) and the color of the bar matches
        the color of the corresponding biplot vectors in (a).
        (c) Same as (b) but for PC3.
        (d) The Ising spin interaction networks of the three-ring systems,
        where the nodes are colored by the angle of the corresponding
        biplot vectors in (a).
        (e)--(h) Same as (a)--(d) but by PC2 and PC3.
        Notice that in (a) and (e),
        several biplot vectors are overlapped.
    }
    \label{fig:spin-ice-pc-2-3}
\end{figure*}

\subsection{The hypercubic energy landscape of
the three-ring system through PC3}
The fraction of explained variance of the three-ring system
[Fig.~\ref{fig:spin-ice-variance}(a3)]
leads us to visualize the hypercubic energy landscapes of
three-ring\footnote{
    See the Supplemental Material~\cite{sm}
    for hypercubic energy landscapes of two-ring systems using PC3.
}
systems using PC3.
In Fig.~\ref{fig:spin-ice-pc-2-3}(a),
the hypercubic energy landscape of three-ring systems,
the same as Fig.~\ref{fig:spin-ice-pc-1-2}(a3),
is shown but by PC1 and PC3.
We find that with PC1 and PC3, unlike the projection by PC1 and PC2,
the shape of the hypercubic energy landscape looks similar to a
parallelogram,
not a hexagon.
The biplot vectors on the top right of Fig.~\ref{fig:spin-ice-pc-2-3}(a) are
only six arrows despite the system having 15 spins,
indicating numerous overlaps of the biplot vectors.
To check which spins are overlapped,
we show the loading of PC1 and PC3 in
Figs.~\ref{fig:spin-ice-pc-2-3}(b) and~\ref{fig:spin-ice-pc-2-3}(c),
and the interaction network in Fig.~\ref{fig:spin-ice-pc-2-3}(d),
coloring them by the angles of the biplot vectors.
PC3 loading [Fig.~\ref{fig:spin-ice-pc-2-3}(c)], in particular,
has a nonuniform magnitude of the elements;
the magnitude of the elements of spins 1, 2, 5, 6, 8, and 9 are larger than
the others.
Those spins belong to the lower left hexagon of the interaction network
in Fig.~\ref{fig:spin-ice-pc-2-3}(d).
The interaction network colored by the angles of the biplot vectors
in Fig.~\ref{fig:spin-ice-pc-2-3}(d)
illustrates how the biplot vectors capture the correlation of spins.
As indicated by PC3 loading [Fig.~\ref{fig:spin-ice-pc-2-3}(c)],
the spins on the lower left hexagon (spins 1, 2, 5, 6, 8, and 9) form
the (anti)correlated group.
The rest of the two rings form a large (anti)correlated group.
PC3 loading creates the hypercubic energy landscape, emphasizing
the lower left hexagon of the interaction network.

We further investigate the hypercubic energy landscape of the
three-ring system
by PC2 and PC3
to understand the symmetry of the system.
The hypercubic energy landscape of the three-ring system by PC2 and PC3
[Fig.~\ref{fig:spin-ice-pc-2-3}(e)], looks like a hexagon,
similar to the projection by PC1 and PC2.
Nevertheless, the landscape is more symmetric and regular than
that by PC1 and PC2.
Around the origin, the connected six ground states also form a
regular hexagon,
but they are overlapping with the other group of six ground states,
cf. Fig.~\ref{fig:spin-ice-pc-1-2}(a3).
The angles of the biplot vectors are almost uniform (with some overlapping),
and the lengths of them are nearly the same,
resulting in a symmetric and regular hypercubic energy landscape;
this projection is the same kind as the isometric projection, but
that of a nine-dimensional hypercube.

The loading of PC2 and PC3 is shown in
Figs.~\ref{fig:spin-ice-pc-2-3}(f) and~\ref{fig:spin-ice-pc-2-3}(g),
with the node color indicating the angle of the biplot vectors.
As we mentioned in Sec.~\ref{sec:three-ring},
the PC2 loading emphasizes contributions from spins 6, 9, and 10
but ignores spins 1, 2, 5, and 8.
On the contrary, PC3 loading has a higher magnitude for spins 1, 2, 5, and 8
in addition to spins 6 and 9, but ignores spin 10.
Spins belonging to other hexagons contribute equally to the PC2
and PC3 loading.
Therefore, as shown in the biplot vectors of
Fig.~\ref{fig:spin-ice-pc-2-3}(a),
the contributions by PC2 and PC3 are complementary,
and the resulting hypercubic energy landscape
shows the symmetry of the state space reflecting the interaction network.

The interaction network colored by the angles of the biplot vectors in
Fig.~\ref{fig:spin-ice-pc-2-3}(h) supports our interpretation:
the coloring is symmetric on the interaction network.
For example, there is a stripe pattern along the outer interaction cycle
(spins 1, 2, 3, 4, 7, 11, 13, 15, 14, 12, 8, and 5).
The smaller interaction cycle consisting of six spins
(such as spins 1, 2, 6, 9, 8, and 5)
also exhibits the stripe pattern.
There is a red-green-blue stripe pattern in the interaction cycle with
three spins (such as spins 6, 10, and 9).
The stripe pattern in all sizes of interaction cycles indicates that
the three-ring system has a collective mode involving the whole system,
and those are hierarchical:
the stripe pattern of the outer cycle arises from
the stripe pattern of the smaller cycle.
This non-local correlation is captured by the PC2 and PC3 loading as
the symmetry of the interaction network.
With an appropriate combination of PC loading,
such as PCs sharing the same fraction of explained variance,
PCA can capture the symmetry of the system.

\begin{figure*}[tb]
    \includegraphics{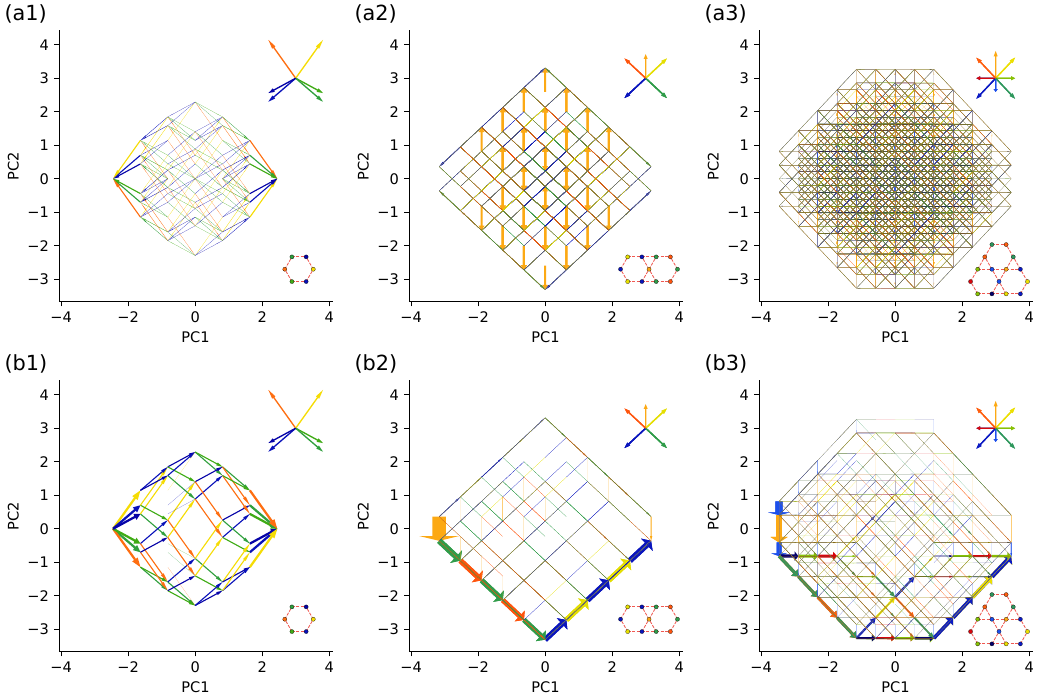}
    \caption{
        Visualization of time-integrated probability flux on the
        hypercubic energy landscape from Fig.~\ref{fig:spin-ice-pc-1-2}.
        (a1) The one-ring system.
        (a2) The two-ring system.
        (a3) The three-ring system.
        (b1) The one-ring system with an external field
        $
        \bm{h}
        =
        \begin{bsmallmatrix}
            + & - & - & + & + & -
        \end{bsmallmatrix}^\top
        $,
        which biases the state transition toward the state projected around
        $
        \begin{bsmallmatrix}
            2.5 & 0
        \end{bsmallmatrix}^\top
        $.
        (b2) The two-ring system with an external field
        $
        \bm{h}
        =
        \begin{bsmallmatrix}
            + & - & + & - & - & - & + & + & - & + & -
        \end{bsmallmatrix}^\top
        $,
        which biases the state transition toward the state projected around
        $
        \begin{bsmallmatrix}
            3.2 & -0.5
        \end{bsmallmatrix}^\top
        $.
        (b3) The three-ring system with an external field
        $
        \bm{h}
        =
        \begin{bsmallmatrix}
            + & - & + & - & - & + & + & + & + & - & - & - & + & + & -
        \end{bsmallmatrix}^\top
        $,
        which biases the state transition toward the state projected around
        $
        \begin{bsmallmatrix}
            3.7 & -0.8
        \end{bsmallmatrix}^\top
        $.
        Time-integrated flux is shown as hypercubic arrows.
        The width of each arrow is proportional to the magnitude of
        the time-integrated flux,
        $\left|\Delta \mathcal{J}_{\bm{s}, \bm{s}^\prime}\right|$.
        The direction indicates the sign of the time-integrated flux
        $
        \operatorname{sgn}
        \left(\Delta \mathcal{J}_{\bm{s}, \bm{s}^\prime}\right)
        $, i.e., it is positive if the corresponding state transition aligns
        with the biplot vector and negative if it is anti-aligned.
        Biplot vectors are shown in the top right, with their color
        indicating the angle of the biplot vectors.
        The width of the biplot vector corresponds to
        $
        \left|\Delta \mathcal{J}_{\bm{s}, \bm{s}^\prime}\right|
        =
        0.02
        $.
        We only visualize the arrow with
        $
        \left|\Delta \mathcal{J}_{\bm{s}, \bm{s}^\prime}\right|
        >
        0.0002
        $
        for clarity.
        The Ising spin interaction network is shown in the lower right,
        where nodes are colored by the angle of the corresponding
        biplot vectors.
        The initial probability distribution is set to be uniform for
        (a1)--(a3), i.e.,
        $p_{\bm{s}}\left(0\right) = \frac{1}{2^N}$, $\forall \bm{s}$.
        For (b1)--(b3),
        the initial probability distribution is set to be
        unipolar for the state
        $\bm{s} = -\bm{h}$, i.e.,
        $p_{\bm{s}}\left(0\right) = \delta_{-\bm{h}, \bm{s}}$.
        We use $A = 1$, $k_\mathrm{B} T = \frac{1}{\beta} = 0.3$,
        and $\tau=50$
        in all panels.
        Note that in (b2) and (b3), several arrows appear at the same
        location due to overlapping biplot vectors.
        See the Supplemental Material~\cite{sm} for visualization
        by other PCs.
    }
    \label{fig:probability-flux}
\end{figure*}

\subsection{Probability flux on the hypercubic energy landscape}
\label{sec:probability-flux}
We now validate the state transition pathways discussed in this
Sec.~\ref{sec:applications} by analyzing the probability flux,
which represents the ensemble of experimentally observable trajectories of
state transitions.
To this end,
we introduce the time dependence of the probability distribution.\footnote{
    In this context,
    the probability distribution is also called the
    \emph{statistical} state,
    which provides a probabilistic description of an ensemble of
    trajectories.
}
Assuming a Markov process,
the time evolution of the probability distribution is described by the
master equation~\cite{Glauber1963}:
\begin{equation}
    \frac{\mathrm{d}}{\mathrm{d}t}
    p_{\bm{s}}\left(t\right)
    =
    \sum_{\bm{s}^\prime}
    \big[
        w_{\bm{s}, \bm{s}^\prime}
        p_{\bm{s}^\prime}\left(t\right)
        -
        w_{\bm{s}^\prime, \bm{s}}
        p_{\bm{s}}\left(t\right)
    \big]
    .
    \label{eq:master-equation}
\end{equation}
Here, $p_{\bm{s}}\left(t\right)$ is the probability of finding the system
in state $\bm{s}$ at time $t$, and
$w_{\bm{s}^\prime, \bm{s}} \in \mathbb{R}_{\geq 0}$
is the transition rate from state
$
\bm{s}
=
\begin{bsmallmatrix}
    s_1 & \cdots & s_k & \cdots & s_N
\end{bsmallmatrix}^\top
$
to state
$
\bm{s}^\prime
=
\bm{F}_{\left(k\right)}\bm{s}
=
\begin{bsmallmatrix}
    s_1 & \cdots & -s_k & \cdots & s_N
\end{bsmallmatrix}^\top
$.
Here,
$
\bm{F}_{\left(k\right)}
\coloneqq
\bm{I}
-
2 \bm{e}_k \bm{e}_k^\top
$
is the spin-flip matrix that flips the $k$th spin.
Therefore, the transition rates
$\left\{w_{\bm{s}^\prime, \bm{s}}\right\}$
are nonzero if
$
\bm{s}^\prime = \bm{F}_{\left(k\right)} \bm{s}
$
and zero otherwise.
Thus, the transition rate connects states in the same way as
hypercubic edges.
The term
$
w_{\bm{s}^\prime, \bm{s}}
p_{\bm{s}}\left(t\right)
$
is the joint transition rate from state $\bm{s}$ to state $\bm{s}^\prime$
at time $t$,
which denotes the rate at which probability mass is transported from
state $\bm{s}$ to state $\bm{s}^\prime$.

The transition rates
$\left\{w_{\bm{s}^\prime, \bm{s}}\right\}$
satisfy the detailed balance condition as $t \to \infty$,
which corresponds to the equilibrium condition:
\begin{align}
    w_{\bm{s}, \bm{s}^\prime}
    p_{\bm{s}^\prime}
    =
    w_{\bm{s}^\prime, \bm{s}}
    p_{\bm{s}}
    ,
    \quad
    \forall \left(\bm{s}, \bm{s}^\prime\right)
    \label{eq:detailed-balance}
    ,
\end{align}
This expresses the microscopic reversibility of the forward and backward
joint transition rates along each hypercubic edge.
Here, $p_{\bm{s}} \coloneqq p_{\bm{s}}\left(t \to \infty\right)$.
At equilibrium, the probability distribution is given by the
Boltzmann distribution [Eq.~\eqref{eq:Boltzmann-distribution}],
$
p_{\bm{s}}\left(t \to \infty\right)
=
\frac{1}{\mathcal{Z}} \exp\left[-\beta
\mathcal{H}\left(\bm{s}\right)\right]
$.
The Arrhenius transition rate~\cite{Farhan2013}
\begin{equation}
    w_{\bm{s}^\prime, \bm{s}}
    =
    A
    \exp\left[
        -
        \beta
        \frac{
            \Delta E\left(\bm{s}^\prime, \bm{s}\right)
        }{2}
    \right]
\end{equation}
satisfies the detailed balance condition of Eq.~\eqref{eq:detailed-balance}.
Here, $A \in \mathbb{R}_{> 0}$ is a positive constant and
$
\Delta E\left(\bm{s}^\prime, \bm{s}\right)
\coloneqq
\mathcal{H} \left(\bm{s}^\prime\right) - \mathcal{H}\left(\bm{s}\right)
=
2 s_k
\left(
    \sum_{j=1}^{N}
    J_{k, j} s_j
    +
    h_k
\right)
$
is the energy difference from state $\bm{s}$ to state $\bm{s}^\prime$ for
the Hamiltonian [Eq.~\eqref{eq:ising-spin-hamiltonian}] of
the Ising spin system.
See the Supplemental Material~\cite{sm} for the derivation of
the Arrhenius transition rate from the detailed balance condition,
as well as the derivation of the energy difference from the Hamiltonian.

We quantify the state transition dynamics by examining the time evolution
of the probability distribution.
We define the probability flux\footnote{
    Also called probability current or probability flow.
} from state $\bm{s}^\prime$ to state $\bm{s}$ as
\begin{equation}
    \mathcal{J}_{\bm{s}, \bm{s}^\prime} \left(t\right)
    \coloneqq
    w_{\bm{s}, \bm{s}^\prime}
    p_{\bm{s}^\prime}\left(t\right)
    -
    w_{\bm{s}^\prime, \bm{s}}
    p_{\bm{s}}\left(t\right)
    ,
\end{equation}
which allows us to rewrite the master equation
[Eq.~\eqref{eq:master-equation}] as
\begin{align}
    \frac{\mathrm{d}}{\mathrm{d}t}
    p_{\bm{s}}\left(t\right)
    &=
    \sum_{\bm{s}^\prime}
    \mathcal{J}_{\bm{s}, \bm{s}^\prime} \left(t\right)
    \label{eq:master-equation-flux}
    .
\end{align}
Integrating Eq.~\eqref{eq:master-equation-flux} over the time interval
$\left[0, \tau\right]$, we obtain
\begin{align}
    p_{\bm{s}}\left(\tau\right)
    -
    p_{\bm{s}}\left(0\right)
    &=
    \sum_{\bm{s}^\prime}
    \int_0^\tau
    \mathrm{d}t
    \,
    \mathcal{J}_{\bm{s}, \bm{s}^\prime} \left(t\right)
    \\
    \Delta p_{\bm{s}}
    &=
    \sum_{\bm{s}^\prime}
    \Delta
    \mathcal{J}_{\bm{s}, \bm{s}^\prime}
    ,
    \label{eq:master-equation-flux-integral}
\end{align}
where the time-integrated probability flux is defined as
$
\Delta
\mathcal{J}_{\bm{s}, \bm{s}^\prime}
\coloneqq
\int_0^\tau
\mathrm{d}t
\,
\mathcal{J}_{\bm{s}, \bm{s}^\prime} \left(t\right)
$.
The time-integrated probability flux represents the net flow from the
initial probability distribution to the distribution at time $\tau$; that is,
$
\Delta p_{\bm{s}}
\coloneqq
p_{\bm{s}}\left(\tau\right)
-
p_{\bm{s}}\left(0\right)
$,
and characterizes the relaxation dynamics of state transitions in
the system.
Since the transition rates $\left\{w_{\bm{s}^\prime, \bm{s}}\right\}$
permit state transitions only between states connected by a
hypercubic edge,
the time-integrated probability flux reflects the net flux along these edges
during the interval $\left[0, \tau\right]$.
We numerically solve the master equation [Eq.~\eqref{eq:master-equation}]
with a specified initial probability $p_{\bm{s}}\left(0\right)$ up to
time $\tau$,
and compute the time-integrated probability flux
$\Delta \mathcal{J}_{\bm{s}, \bm{s}^\prime}$
from the time evolution of $p_{\bm{s}}\left(t\right)$.
In Fig.~\ref{fig:probability-flux},
we visualize these fluxes on the hypercubic energy landscape created in
Fig.~\ref{fig:spin-ice-pc-1-2}.
For clarity, we only display the time-integrated flux
$\Delta \mathcal{J}_{\bm{s}, \bm{s}^\prime}$
with magnitude greater than $0.0002$.

We first investigate the relaxation dynamics of state transitions starting
from a random state.
To do this, we consider relaxation from a high-temperature
probability distribution---that is, a uniform probability distribution,
$
p_{\bm{s}} \left(0\right)
=
\frac{1}{2^N}
$,
$
\forall \bm{s}
$,
to the low-temperature canonical ensemble distribution,
$
p_{\bm{s}} \left(\tau\right)
\approx
p_{\bm{s}}
=
\frac{1}{\mathcal{Z}}
\exp
\left[-\beta \mathcal{H} \left(\bm{s}\right)\right]
$.
In Fig.~\ref{fig:probability-flux}(a1), we show the time-integrated
probability flux
$
\Delta \mathcal{J}_{\bm{s}, \bm{s}^\prime}
$
of the one-ring system.
The magnitude of the flux increases as it approaches the ground
states around
$
\begin{bsmallmatrix}
    -2.5 & 0
\end{bsmallmatrix}^\top
$
and
$
\begin{bsmallmatrix}
    2.5 & 0
\end{bsmallmatrix}^\top
$.
In contrast, the two-ring system exhibits spin-dependent flux
magnitudes, as shown in
Fig.~\ref{fig:probability-flux}(a2).
We observe that the flux involving the flip of spin 6 is larger than
the others, reflecting the unique interactions of spin 6.
The other fluxes increase in magnitude as they approach the ground states.
The time-integrated flux for the three-ring system is shown in
Fig.~\ref{fig:probability-flux}(a3).
Because the three-ring system has more ground states than the other
systems, we do not observe clear convergence of the flux.
Our projection of the hypercubic energy landscape captures the
relaxation dynamics from random states to the ground states.

We then investigated the state transition---that is, the transport of
probability mass---from one ground state to the all-spin-flipped ground state.
To drive the system toward the target ground state,
we apply a magnetic field $\bm{h}$ corresponding to that state;
the probability distribution is then expected to converge to
a unipolar distribution centered on the target state,
$
p_{\bm{s}}\left(\tau\right)
\approx
p_{\bm{s}}
=
\delta_{\bm{h}, \bm{s}}
$.
We let the probability distribution evolve from the unipolar
distribution of the all-spin-flipped state, i.e., $\bm{s} = -\bm{h}$.
Thus, the initial probability distribution is set as $p_{\bm{s}}
\left(0\right) = \delta_{-\bm{h}, \bm{s}}$.
Naively, there are $N!$ possible orders in which the spins can flip,
but as we emphasized earlier in this Sec.~\ref{sec:applications},
the hypercubic energy landscape constrains state transitions to a limited
number of pathways.

The state transition dynamics from a specific initial state to its
all-spin-flipped counterpart are shown in Fig.~\ref{fig:probability-flux}(b).
In Fig.~\ref{fig:probability-flux}(b1),
we display the time-integrated flux of the one-ring system as it
evolves from one ground state to the other.
As discussed in Sec.~\ref{sec:one-ring}, the larger time-integrated
flux is confined to transitions between the ground states or first
excited states.

We then visualize, in Fig.~\ref{fig:probability-flux}(b2),
the time-integrated flux of the two-ring system as it evolves from
one ground state to the other.
Unlike the one-ring system, the two-ring system exhibits a dominant flux pathway
along with several minor pathways.
The dominant pathway involves correlated spin flips within spin domains:
state transitions that flip spin 6 carry the largest flux from
the initial state, followed by transitions involving correlated spin flips
in the right half of the system (spins 3, 4, 7, 10, 11),
and subsequently in the left half of the system (spins 1, 2, 5, 8, 9).
Similar to the dominant pathway, the minor pathways also display
correlated spin flips, but in a different sequence.

Turning to the three-ring system in Fig.~\ref{fig:probability-flux}(b3),
we observe multiple distinguished dominant pathways,
in contrast to the two-ring system.
At the beginning of the time evolution,
state transitions involving spins located in the inner part of
the interaction network (spins 6, 9, and 10) exhibit large flux.
Subsequently, the dominant pathway splits into two branches:
one involves correlated spin flips of spins in the lower left
(spins 1, 2, 5, and 8), which is less dominant,
while the other involves correlated flips of spins in the top region
(spins 12, 13, 14, and 15), which is more dominant.
The less dominant pathway diverges again around
$
\begin{bsmallmatrix}
    -1.8 & -0.9
\end{bsmallmatrix}^\top
$
and some flux then converges with the dominant pathway near
$
\begin{bsmallmatrix}
    0.5 & -3.2
\end{bsmallmatrix}^\top
$
and
$
\begin{bsmallmatrix}
    1.2 & -3.2
\end{bsmallmatrix}^\top
$.
Returning to the dominant pathway, it diverges at
$
\begin{bsmallmatrix}
    1.3 & -3.2
\end{bsmallmatrix}^\top
$,
with one branch involving correlated flips of spins in the lower
right (spins 3, 4, 7, and 11), and the other emerging from correlated
flips of spins in the lower left (spins 1, 2, 5, and 8).
Ultimately, the dominant pathway leads the system into the target state
through correlated flips of spins in the lower right (spins 3, 4, 7, and 11).
These results demonstrate how hypercubic energy landscapes projected
via PCA capture state transition dynamics,
including information about correlated spin flips.

\subsection{Energy landscape of the mean-field model}
In Fig.~\ref{fig:probability-flux} (b2) and (b3),
we observed that the dominant state transition pathways emerge around the
periphery of the hypercubic energy landscape.
To understand this center-avoiding behavior,
we consider the mean-field approximation of Ising spin systems.
We show that for Hopfield models~\cite{Hopfield1982} of
associative memory,\footnote{
    Also called Amari--Hopfield model~\cite{Amari1972,Hopfield1982}.
}
the energy is maximized at the center of the hypercubic energy landscape
and minimized at the periphery.
As a result, relaxational state transition pathways tend to avoid the
center of the hypercubic energy landscape.
We also point out that the bound of the inner-product error can be
expressed in terms of energy.
Although our analysis is entirely based on the mean-field model,
the qualitative behavior on the hypercubic energy landscape
is expected to be similar for other Ising spin systems.
See the Supplemental Material~\cite{sm} for detailed derivations
and additional figures related to the discussion below.

In general, it is possible to calculate
the energy of a state projected at the periphery of the projected space
because, as we have shown in Sec.~\ref{sec:upper-bound},
the inner-product error vanishes at the periphery of the
projected hypercube,
i.e.,
$
\varepsilon
=
\bm{s}^\top \bm{s}^\prime
-
\bm{r}^\top\left(\bm{s}\right) \bm{r}\left(\bm{s}^\prime\right)
= 0
$.
For example, the reconstructed state
$
r_1\left(\bm{s}\right) \bm{u}_1
$
with the bipolar distribution~[Eq.~\eqref{eq:bipolar-distribution}]
is identical to the original state
$
\bm{s}
$
if
$
\left| r_1\left(\bm{s}\right) \right| = \sqrt{N}
$.
Thus, using the projected coordinates
$\bm{r}\left(\bm{s}\right) \in \mathbb{R}^2$
at the periphery
$\left|\bm{r}\left(\bm{s}\right)\right| = \sqrt{N}$
of the projected space,
we can calculate the Hamiltonian as a function of the projected coordinates:
\begin{equation}
    \mathcal{H}\left(\bm{r}\right)
    =
    -
    \frac{1}{2}
    \left(\bm{V} \bm{r}\right)^\top
    \bm{J}
    \left(\bm{V} \bm{r}\right)
    =
    -
    \frac{1}{2}
    \bm{r}^\top
    \left(
        \bm{V}^\top
        \bm{J}
        \bm{V}
    \right)
    \bm{r}
    \label{eq:hamiltonian-projected-coordinates}
\end{equation}
where $\bm{V} \in \mathbb{R}^{N \times 2}$ is the
transformation matrix whose columns are the selected PC loadings
$\bm{u}_i \in \mathbb{R}^N$.
Although Eq.~\eqref{eq:hamiltonian-projected-coordinates} provides
a visual estimate of the energy of a state from its projected coordinates,
it is not straightforward to estimate the energy of a state that is not at the
periphery of the projected space.
To estimate the energy of such a state,
we need to consider how the structure of the interaction network
$\bm{J}$ is modified
when transformed by the two PC loadings, i.e., $\bm{V}^\top \bm{J} \bm{V}$.
This is not straightforward for a general interaction matrix.

Although it is generally difficult to infer the energy from the
projected coordinates, it is possible for mean-field models,
where the Hamiltonian is a function of the order parameter,
and the interaction matrix and covariance matrix share the same eigenvectors.
To begin, we introduce the Hopfield model,
where the interaction network (matrix) $\bm{J}_\mathrm{H}$
is given by the Hebbian rule~\cite{Hopfield1982}:
\begin{equation}
    \bm{J}_\mathrm{H}
    =
    \frac{J}{N}
    \sum_{\mu=1}^{P}
    c_\mu
    \bm{\xi}_\mu
    \bm{\xi}_\mu^\top
    \label{eq:hopfield-network}
\end{equation}
Here, $J \in \mathbb{R}_{> 0}$ is a positive constant,
$P \in \mathbb{N}$ is the number of patterns,
and $\bm{\xi}_\mu \in \left\{+1, -1\right\}^N$ is the $\mu$th
pattern vector.
The parameter $c_\mu \in \left[0, 1\right]$ is the weight of
the $\mu$th pattern,
satisfying $\sum_{\mu=1}^{P} c_\mu = 1$.
Note that the factor $\frac{1}{N}$ ensures that the Hamiltonian is extensive.
This interaction network is an all-to-all interaction network,
where each spin interacts with all other spins.
The order parameters $\left\{m_\mu\left(\bm{s}\right)\right\}_{\mu=1}^P$ of
the Hopfield model are the overlaps, or cosine similarities, between a state
and the $\mu$th pattern:
\begin{equation}
    m_\mu \left(\bm{s}\right)
    \coloneqq
    Q\left(\bm{\xi}_\mu, \bm{s}\right)
    =
    \frac{1}{N}
    \bm{\xi}_\mu^\top \bm{s}
    =
    \frac{1}{N}
    \sum_{i=1}^{N}
    \xi_{i; \mu} s_i
    ,
    \label{eq:overlap}
\end{equation}
where $\xi_{i; \mu}$ is the $i$th element of the $\mu$th pattern vector.
We now show that the Hopfield model is a generalization of the
mean-field (or infinite-range ferromagnetic) Ising spin system.
With the Hamiltonian-invariant transformation of variables,
\begin{equation}
    s_i
    \leftarrow
    \xi_{i; \mu} s_i
\end{equation}
and
\begin{equation}
    J_{i, j}
    \leftarrow
    \xi_{i; \mu} J_{i, j} \xi_{j; \mu}
    ,
\end{equation}
the Hamiltonian of the Hopfield model
$\mathcal{H}_\mathrm{H}\left(\bm{s}\right)$
without external field ($\bm{h} = \bm{0}$) is transformed to
\begin{align}
    \mathcal{H}_\mathrm{H}\left(\bm{s}\right)
    &=
    -
    \frac{1}{2}
    \frac{J}{N}
    \sum_{\mu=1}^{P}
    c_\mu
    \sum_{i=1}^{N}
    \sum_{j=1}^{N}
    s_i \xi_{i; \mu}
    \underbrace{
        \left(
            \xi_{i; \mu} \xi_{i; \mu} \xi_{j; \mu} \xi_{j; \mu}
        \right)
    }_{=1}
    \xi_{j; \mu} s_j
    \nonumber
    \\&=
    -
    N
    \frac{J}{2}
    \sum_{\mu=1}^{P}
    c_\mu
    m_\mu^2 \left(\bm{s}\right)
    \eqqcolon
    \mathcal{H}_\mathrm{H}\left(m_1, \ldots, m_P\right)
    \label{eq:hopfield-hamiltonian}
\end{align}
under zero external field, $\bm{h} = \bm{0}$.
Thus, the Hamiltonian of the Hopfield model is a function of the order
parameters $\left\{m_\mu\left(\bm{s}\right)\right\}_{\mu=1}^P$.

We now show that the mean-field Hamiltonian is a function of the projected
coordinates $\bm{r}\left(\bm{s}\right)$,
and that the upper bound of the inner-product error is determined by
the Hamiltonian.
For the case of a single pattern ($P=1$ and $c_\mu = \delta_{1, \mu}$),
the Hamiltonian is exactly mapped to the infinite-range ferromagnetic
Ising spin system (i.e., $\bm{J} = \frac{J}{N}\bm{1}\bm{1}^\top$),
and it becomes a function of the order parameter:
the PC1 score $m_1\left(\bm{s}\right) = \frac{1}{\sqrt{N}} r_1
\left(\bm{s}\right)$,
\begin{equation}
    \mathcal{H}_\mathrm{H}\left(m_1\right)
    =
    - N \frac{J}{2}  m_1^2
    =
    - \frac{J}{2} r_1^2
    =
    \mathcal{H}_\mathrm{H}\left(r_1\right)
    \label{eq:hamiltonian-one-pattern}
    ,
\end{equation}
because, at low temperature $k_\mathrm{B} T \ll 1$ and zero external field
$\bm{h} = \bm{0}$,
the probability distribution $p \left(\bm{s}\right)$ is the
bipolar distribution
[Eq.~\eqref{eq:bipolar-distribution}],
and the projected coordinate is equivalent to the order parameter
[Eq.~\eqref{eq:PC1-score-order-parameter}].
The Hamming projection emerges as a result.
Therefore, the energy is highest at the origin of the projected space.
If the order parameter is treated as a continuous variable,
Eq.~\eqref{eq:hamiltonian-one-pattern} gives the continuous
energy landscape.
In addition, the upper bound of the inner-product error
[Eq.~\eqref{eq:inner-product-abs-bound-normalized}]
is given by the squared order parameter, i.e., the Hamiltonian,
\begin{equation}
    \left(
        1 - \frac{l^2}{N}
    \right)^2
    =
    \left(
        1 - m_1^2
    \right)^2
    =
    \left[
        1 + \frac{2}{NJ} \mathcal{H}_\mathrm{H}\left(m_1\right)
    \right]^2
    ,
\end{equation}
because the coordinate is, again, given by the order parameter,
$
l = \sqrt{N} m_1
$.
Thus,
the upper bound of the absolute normalized inner-product error
$
\frac{\left|\varepsilon\right|}{N}
=
\frac{
    \left|
    \bm{s}^\top \bm{s}^\prime
    -
    \bm{r}^\top\left(\bm{s}\right)
    \bm{r}\left(\bm{s}^\prime\right)
    \right|
}{N}
$
coincides with the scaled Hamiltonian relative to the ground state energy
$
1
+
\frac{2}{NJ}
\mathcal{H}_\mathrm{H}\left(m_1\right)
\in
\left[0, 1\right]
$.
As the energy increases,
the inner-product error bound also increases,
and the inner-product error is minimized when the system is at the ground state.

\begin{figure}[tb]
    \includegraphics{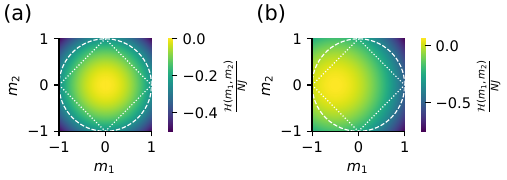}
    \caption{
        Energy landscape of the Hopfield model with two orthogonal patterns,
        $\bm{\xi}_1$ and $\bm{\xi}_2$.
        (a) The Hamiltonian
        $
        \frac{\mathcal{H}_\mathrm{H}\left(m_1, m_2\right)}{NJ}
        =
        -\frac{1}{2}\left(c_1 m_1^2 + c_2 m_2^2\right)
        $
        from Eq.~\eqref{eq:hamiltonian-two-patterns}
        plotted on the projected space spanned by the scaled first two PCs,
        $\frac{r_1}{\sqrt{N}} = m_1$ and $\frac{r_2}{\sqrt{N}} = m_2$.
        (b) The Hamiltonian
        $
        \frac{\mathcal{H}_\mathrm{H}\left(m_1, m_2\right)}{NJ}
        =
        - \frac{1}{2}\left(c_1 m_1^2 + c_2 m_2^2\right)
        - \frac{h}{J} m_1
        $
        from Eq.~\eqref{eq:hamiltonian-two-patterns-external-field}
        on the same projected space as in (a).
        We assume the system size is large ($N \gg 1$) and
        treat $m_1$ and $m_2$ as continuous variables.
        For both (a) and (b), $c_1 = c_2 = \frac{1}{2}$.
        The external field for (b) is $\frac{h}{J} = \frac{1}{4}$.
        In both panels, hypercubic vertices are projected inside the
        dashed circle
        [Eq.~\eqref{eq:possible-region-general}] in general,
        and, in the case of the Hamming projection,
        inside the dotted square
        [Eqs.~\eqref{eq:bound-r1plusr2} and~\eqref{eq:bound-r1minusr2}].
    }
    \label{fig:energy-landscape-Hopfield}
\end{figure}

We then extend our result to the case of two patterns
($P=2$, $c_1 \in \left[\frac{1}{2}, 1\right]$, $c_2 = 1 - c_1$,
and $c_\mu = 0$ for $\mu \geq 3$).
The Hamiltonian is a function of the two order parameters
$\left\{m_\mu\left(\bm{s}\right)\right\}_{\mu=1}^2$
or the projected coordinates
$
\left\{r_\mu \left(\bm{s}\right)
    =
\sqrt{N} m_\mu \left(\bm{s}\right)\right\}_{\mu=1}^2
$,
\begin{align}
    \mathcal{H}_\mathrm{H}\left(m_1, m_2\right)
    &=
    - N \frac{J}{2}
    \left(
        c_1 m_1^2 + c_2 m_2^2
    \right)
    \nonumber
    \\&=
    - \frac{J}{2} \left( c_1 r_1^2 + c_2 r_2^2 \right)
    =
    \mathcal{H}_\mathrm{H}\left(r_1, r_2\right)
    \label{eq:hamiltonian-two-patterns}
    ,
\end{align}
because the probability distribution is quadripolar:
$
p \left(\bm{s}\right)
=
\frac{c_1}{2}
\left(
    \delta_{+\bm{\xi}_1, \bm{s}}
    +
    \delta_{-\bm{\xi}_1, \bm{s}}
\right)
+
\frac{c_2}{2}
\left(
    \delta_{+\bm{\xi}_2, \bm{s}}
    +
    \delta_{-\bm{\xi}_2, \bm{s}}
\right)
$,
and we assume that the two patterns are orthogonal,
$\bm{\xi}_1^\top \bm{\xi}_2 = 0$.
Thus, the equi-energy contours are ellipses in the projected space.
The vertex projected at the origin of the projected space has the highest
energy.
Vertices projected at the periphery of the projected space tend
to have lower energy than those at the origin.
In Fig.~\ref{fig:energy-landscape-Hopfield}(a),
we visualize an example of this Hamiltonian as an energy landscape,
a function of the projected coordinates.
Note that the possible region of the projected coordinates is limited,
as shown in Figs.~\ref{fig:inner-product-error-bounds}(b)
and~\ref{fig:energy-landscape-Hopfield}(a).
See also Appendix~\ref{sec:possible-region}.
Similar to the single-pattern case discussed above,
the upper bound of the inner-product error
[Eq.~\eqref{eq:inner-product-abs-bound-normalized-2D}]
is given by the Hamiltonian,
\begin{align}
    \left(1 - \frac{l^2}{N}\right)^2
    &=
    \left[
        1 -
        \left(
            m_1^2 + m_2^2
        \right)
    \right]^2
    \label{eq:inner-product-abs-bound-two-patterns-order-parameters}
    \\&=
    \left[
        1 + \frac{4}{NJ} \mathcal{H}_\mathrm{H}\left(m_1, m_2\right)
    \right]^2
    \label{eq:inner-product-abs-bound-two-patterns-hamiltonian}
    ,
\end{align}
if the two patterns have the same weight, i.e.,
$c_1 = c_2 = \frac{1}{2}$.
Note that
Eq.~\eqref{eq:inner-product-abs-bound-two-patterns-order-parameters}
is valid for any $c_1 \in \left[\frac{1}{2}, 1\right]$ and $c_2 =
1 - c_1$, but
Eq.~\eqref{eq:inner-product-abs-bound-two-patterns-hamiltonian} is
valid only for
$c_1 = c_2 = \frac{1}{2}$.
Consequently, the upper bound of the two-dimensional inner-product error
increases as the system becomes more disordered ($m_\mu \to 0$),
and it is a quadratic function of the Hamiltonian in this special case.

We continue our discussion of the two-pattern Hamiltonian,
now considering the case with a nonzero external field as discussed
in Sec.~\ref{sec:probability-flux}.
If the external field is proportional to the first pattern,
i.e., $\bm{h} = h \bm{\xi}_1$, $h \in \mathbb{R}_{\neq 0}$,
the Hamiltonian becomes
\begin{align}
    \mathcal{H}_\mathrm{H}\left(m_1, m_2\right)
    &=
    -
    N
    \frac{J}{2}
    \left(
        c_1 m_1^2 + c_2 m_2^2
    \right)
    -
    N
    h m_1
    \\&=
    -
    N
    \frac{J}{2}
    \left[
        c_1
        \left(
            m_1 + \frac{h}{J c_1}
        \right)^2
        +
        c_2
        m_2^2
    \right]
    +
    N
    \frac{h^2}{2Jc_1}
    \label{eq:hamiltonian-two-patterns-external-field}
    ,
\end{align}
which describes a shifted ellipse in the projected space if
$r_1 \left(\bm{s}\right) = \sqrt{N} m_1 \left(\bm{s}\right)$
and
$r_2 \left(\bm{s}\right) = \sqrt{N} m_2 \left(\bm{s}\right)$.
We visualize an example of this Hamiltonian in
Fig.~\ref{fig:energy-landscape-Hopfield}(b) on the $m_1$--$m_2$ plane.
The qualitative shape of the energy landscape described by this
Hamiltonian changes depending on the value of $h$.
If $h \leq J c_1$, the Hamiltonian reaches its maximum within the allowed range
$\left[-1, 1\right]^2$, so trajectories starting from
$
\begin{bsmallmatrix}
    -1 & 0
\end{bsmallmatrix}^\top
$
are expected to avoid the maximum at
$
\begin{bsmallmatrix}
    -\frac{h}{J c_1} & 0
\end{bsmallmatrix}^\top
$
and instead evolve along the periphery of the projected space.
If $h > J c_1$, the Hamiltonian has its maximum outside the allowed range
$\left[-1, 1\right]^2$, and trajectories starting from
$
\begin{bsmallmatrix}
    -1 & 0
\end{bsmallmatrix}^\top
$
are expected to cross the center of the projected space.
Although this discussion is based on the mean-field model, we can
expect that the center-avoiding behavior of state transition pathways
is a general feature, as observed in
Figs.~\ref{fig:probability-flux}(b2) and~\ref{fig:probability-flux}(b3).
Note that as the temperature increases, the entropy landscape (or
free energy landscape)---rather than the energy landscape---dominates
the state transition pathways.
In such situations, center-accumulating pathways are expected because
of the general tendency for a high density of states around the
origin of the projected space in the Hamming projection, as shown in
Sec.~\ref{sec:number-of-vertices-along-pc1}.

\section{Conclusions}
\label{sec:conclusions}
The hypercubic representation of binary state space is a powerful tool to
reveal the high-dimensional structures of binary data in various sciences,
yet as R.~P.~Feynman once pointed out~\cite{Feynman1996},
``unfortunately our brains can't visualize'' the high-dimensional hypercube.
Conventional methods of projection have both advantages and disadvantages,
and there is a demand for informative and practical methods.
In this study, we suggest several orthogonal projection methods to obtain
reproducible, interpretable, and automatic visualizations of hypercubes,
concluding that PCA is a suitable method for this purpose.
We merge ideas from high-dimensional geometry and statistics
(unsupervised machine learning),
and apply the projections of hypercubes to understand
statistical mechanical models.
Extending our work to other linear dimensionality reduction methods is
straightforward if one obtains a biplot.

Our contributions are summarized as follows:
(1) We introduce the Hamming and fractal projection of hypercubes
in Sec.~\ref{sec:orthogonal-projection-hypercubes}.
(2) We propose interpreting the biplot vectors of PCA as
the vectors of the orthogonal projection of hypercubes in
Sec.~\ref{sec:interpreting-pca}.
(3) We find that the PC1 loading of PCA on hypercubes is
equivalent to the weighted hypercubic vertices in
Sec.~\ref{sec:hypercubic-pca},
and validate it analytically and numerically in
Sec.~\ref{sec:pc-loading-corresponding}.
(4) We reveal numerically and analytically
that the quality of the orthogonal projection of hypercubes tends
to be worse around the origin of the projected space in
Sec.~\ref{sec:quality-centrality-upper-bound}.
(5) We visualize the hypercubic energy landscapes of
the finite artificial spin-ice systems
and extract physical interpretations---particularly focusing on correlated spin
flips---in Sec.~\ref{sec:applications}.

Although we have demonstrated that PCA yields reproducible, interpretable, and
automatic projections of hypercubes,
there are two remaining challenges:
(1) analytically predicting the eigenvectors of the covariance matrix
for non-bipolar probability distributions, and
(2) the overlap of hypercubic vertices
and hypercubic edges in the projected space.
The first challenge is not yet resolved in this study,
but we believe that the symmetric outer product matrix
[the second term of Eq.~\eqref{eq:covariance-matrix-generalized}]
is key to addressing this problem.
We present some nontrivial examples
in Fig.~\ref{fig:pca-hypercubes-random-pair-weighted}(a) and
Supplemental Material~\cite{sm},
where the PC loading is not proportional to
the most heavily weighted hypercubic vertex.
Perturbation theory (Appendix~\ref{sec:pca-with-perturbation}) is
promising to address this challenge.
Turning to the second challenge,
many hypercubic vertices and edges tend to overlap
in the projected space, especially near the origin.
As we show numerically and analytically
in Sec.~\ref{sec:quality-centrality-upper-bound},
PCA locates numerous hypercubic vertices around
the origin of the projected space,
and those overlapped hypercubic vertices can increase the error of
the projection.
We, however, analytically know these limitations of PCA beforehand:
we can interpret the resulting plot while grasping the limitations,
unlike other nonlinear dimensionality reduction methods,
where such analytical properties are not obvious and might be
more difficult to understand.
When constructing the hypercubic energy landscape and probability flux in
Sec.~\ref{sec:applications},
we mitigate the overlap problem by projecting only a limited subset of
hypercubic vertices and edges.
A similar strategy may be employed to bypass this issue in other
applications.

H.~S.~M.~Coxeter mentioned that illustrations of high-dimensional objects
have psychological and artistic merit~\cite{Coxeter1973}.
In this study,
we initiate unveiling the new merit---or possibility---of projecting
high-dimensional hypercubes:
interpreting binary data through \emph{data-driven} visualization.
As Anscombe's quartet~\cite{Anscombe1973} shows hidden patterns
not in statistics but in graphs~\cite{Matejka2017},
informative orthogonal projection using PCA might lead to fresh
understanding, interpretation, and discovery~\cite{Tukey1977,
Rudin2022, Allen2024}
in high-dimensional binary data across sciences.

\bigskip
Calculations and visualizations of this work were performed using
open-source \textsc{python}~\cite{Rossum2010} libraries:
\textsc{matplotlib}~\cite{Hunter2007},
\textsc{networkx}~\cite{Hagberg2008},
\textsc{numpy}~\cite{Harris2020},
and
\textsc{scipy}~\cite{Virtanen2020}.
The color map of some figures are generated by
\textsc{colorcet}~\cite{Kovesi2015}.
All data and code are available online from Zenodo~\cite{codes}.

\begin{acknowledgments}
    Special thanks go to
    Takashi Horiike, Kennichi Hagiwara, Nozomu Imai, Takumi Nishina,
    and Hiroto Murata
    for fruitful discussions at the early stage of this research.
    Y.H.~thank
    Mizuho Aoki, Johanna von Drachenfels, Anton G. Golles, and Henri Schmidt
    for their valuable comments on this manuscript.
    This work was supported by KAKENHI Grant Number 22H00406 and 24H00061 of
    the Japan Society for the Promotion of Science.
    Y.H.~is supported by JST SPRING, Grant Number
    JPMJSP2125
    ``THERS Make New Standards Program for the Next Generation
    Researchers''.
\end{acknowledgments}

\bigskip
Y.H.~performed
conceptualization, data curation, formal analysis, investigation,
software development, validation, and visualization;
Y.H.~partially contributed to the funding acquisition;
Y.H.~and S.F.~developed the methodology;
S.F.~provided computational resources;
Y.H.~wrote the original draft;
and Y.H.~and S.F.~reviewed and edited the manuscript.

\appendix
\section{Constructing hypercubes}
\label{sec:constructing-hypercube}
A hypercube is constructed as follows~\cite{Coxeter1973, Bragdon2005,
Gardner1975}.
Suppose one has a zero-dimensional point,
one obtains a one-dimensional line by moving the point in a direction.
By moving the line in a direction but not along its line,
one can obtain a two-dimensional parallelogram.
Again, moving the parallelogram in a direction but not along its plane,
one creates a three-dimensional parallelepiped.
By repeating this translation process $N$ times,
one obtains an $N$-dimensional parallelepiped or a parallelotope
with $2^N$ vertices and $\frac{2^N N}{2}$ edges.\footnote{
    Each of the $2^N$ vertices has $N$ edges.
    To avoid double counting, one needs to divide them by two.
}
The unit vectors of each dimension correspond to
the direction of the translation,
i.e., edges created by the translation are parallel to
the directions of unit vectors.

Some special objects of parallelotopes are known~\cite{Coxeter1973}.
An orthotope is a special parallelotope where the unit vectors of
each dimension are mutually perpendicular.
A hypercube (also written hyper-cube) or measure polytope is a type of
orthotope where
all unit vectors of each dimension share the same length.
An $N$-dimensional hypercube is labeled as $\gamma_N$.

\section{Formulations of PCA}
\label{sec:formulations-pca}
In this Appendix, we briefly review four formulations of
PCA~\cite{Jolliffe2002, Bishop2006, Cox2001, Borg2005, Ghojogh2023}.
Our review includes the well-known two formulations:
the maximum projection variance formulation~\cite{Hotelling1933}
and the minimum reconstruction error formulation~\cite{Pearson1901}.
Then, the equivalence of classical MDS and PCA is
reintroduced~\cite{Gower1966},
and we present the minimum inner-product formulation of
PCA~\cite{Ghojogh2023}.

Throughout this Appendix~\ref{sec:formulations-pca},
we use an $N$-dimensional vector with real elements
$\bm{x} \in \mathbb{R}^N$
as the data point in the original space.
In our main text, we use the Ising state vector
$\bm{s} \in \left\{+1, -1\right\}^N$ as
hypercubic vertices (states or data points),
so replace $\bm{x}$ with $\bm{s}$ if readers consider specifically
the projections of hypercubic vertices.

\subsection{Maximum projection variance formulation}
\label{sec:maximum-projection-variance-formulation}
Here, we reintroduce the maximum variance formulation of
PCA~\cite{Hotelling1933, Bishop2006, Jolliffe2002}.
Each data point $\bm{x}$ is projected to
$r_1\left(\bm{x}\right) = \bm{u}_1^\top\bm{x}$
by unit vector $\bm{u}_1$.
The mean of the data set is given by
$
\left<\bm{x}\right>
\coloneqq
\sum_{\bm{x}}
p\left(\bm{x}\right)
\bm{x}
,
$
and the mean of the projected data point is
$
\left<r_1\left(\bm{x}\right)\right>
=
\bm{u}_1^\top
\left<\bm{x}\right>
$.
The variance of the projected data set is the target function to maximize:
\begin{align}
    L
    &\coloneqq
    \sum_{\bm{x}}
    p\left(\bm{x}\right)
    \left[
        r_1\left(\bm{x}\right)
        -
        \left<
        r_1\left(\bm{x}\right)
        \right>
    \right]^2
    \label{eq:variance-func}
    \\&=
    \nonumber
    \sum_{\bm{x}}
    p\left(\bm{x}\right)
    \left[
        \left(
            \bm{x}
            -
            \left<\bm{x}\right>
        \right)^\top
        \bm{u}_1
    \right]^2
    \\&=
    \nonumber
    \sum_{\bm{x}}
    p\left(\bm{x}\right)
    \bm{u}_1^\top
    \left(
        \bm{x}
        -
        \left<\bm{x}\right>
    \right)
    \left(
        \bm{x}
        -
        \left<\bm{x}\right>
    \right)^\top
    \bm{u}_1
    \\&=
    \bm{u}_1^\top
    \bm{\varSigma}
    \bm{u}_1
    ,
\end{align}
where
$
\bm{\varSigma}
\coloneqq
\sum_{\bm{x}}
p\left(\bm{x}\right)
\left(
    \bm{x}
    -
    \left<\bm{x}\right>
\right)
\left(
    \bm{x}
    -
    \left<\bm{x}\right>
\right)^\top
$
is the covariance matrix of the data set.
One can maximize the variance of the projected data
while keeping the normalization $\bm{u}_1^\top\bm{u}_1 = 1$
by the method of Lagrange multipliers.
The Lagrange function is
\begin{equation}
    \mathcal{L}
    \coloneqq
    \bm{u}_1^\top
    \bm{\varSigma}
    \bm{u}_1
    +
    \lambda_1
    \left(
        1
        -
        \bm{u}_1^\top\bm{u}_1
    \right)
    ,
\end{equation}
where $\lambda_1$ is the Lagrange multiplier.
Deriving $\mathcal{L}$ with respect to $\bm{u}_1$ and setting it to be
the zero vector,
\begin{equation}
    \bm{0}
    =
    \frac{\partial}{\partial \bm{u}_1}
    \mathcal{L}
    =
    2
    \bm{\varSigma}
    \bm{u}_1
    -
    2
    \lambda_1
    \bm{u}_1
    ,
\end{equation}
one obtains
$
\bm{\varSigma}
\bm{u}_1
=
\lambda_1
\bm{u}_1
$
and the variance of the projected data set is
$
\bm{u}_1^\top
\bm{\varSigma}
\bm{u}_1
=
\lambda_1
.
$
The eigenvector $\bm{u}_1$ is the PC1 explaining the variance
$\lambda_1$ which
is maximized.

The further PCs are obtained by an incremental procedure while keeping
the orthogonality to all the previous PCs.
For example, PC2 is obtained by the same procedure but with
the normalization constraint
$
\bm{u}_2^\top\bm{u}_2
=
1
$
and the orthogonality constraint to PC1
$
\bm{u}_1^\top\bm{u}_2
=
0
$.
The Lagrange function is
\begin{equation}
    \mathcal{L}
    \coloneqq
    \bm{u}_2^\top
    \bm{\varSigma}
    \bm{u}_2
    +
    \lambda_2
    \left(
        1
        -
        \bm{u}_2^\top\bm{u}_2
    \right)
    +
    \eta
    \bm{u}_1^\top\bm{u}_2
    ,
\end{equation}
where $\lambda_2$ and $\eta$ are Lagrange multipliers.
Similarly to PC1,
deriving $\mathcal{L}$ with respect to $\bm{u}_2$ and setting it to be
the zero vector,
one obtains
\begin{equation}
    \bm{0}
    =
    \frac{\partial}{\partial \bm{u}_2}
    \mathcal{L}
    =
    2
    \bm{\varSigma}
    \bm{u}_2
    -
    2
    \lambda_2
    \bm{u}_2
    +
    \eta
    \bm{u}_1
    .
\end{equation}
Multiplying $\bm{u}_1^\top$ to both sides of the equation
from the left side yields\footnote{
    $
    \bm{u}_1^\top
    \bm{\varSigma}
    \bm{u}_2
    =
    \bm{u}_2^\top
    \bm{\varSigma}
    \bm{u}_1
    =
    \bm{u}_2^\top
    \lambda_1
    \bm{u}_1
    =
    \lambda_1
    \bm{u}_2^\top
    \bm{u}_1
    =
    0
    $
    because of the orthogonality between $\bm{u}_1$ and $\bm{u}_2$,
    $\bm{u}_1^\top\bm{u}_2 = 0$~\cite{Jolliffe2002}.
}
$
0
=
\bm{u}_1^\top
\bm{\varSigma}
\bm{u}_2
-
\lambda_2
\bm{u}_1^\top
\bm{u}_2
+
\eta
\bm{u}_1^\top
\bm{u}_1
=
\eta
\bm{u}_1^\top
\bm{u}_1
=
\eta
$
resulting in $\eta = 0$.
Thus, we obtain
$
\bm{\varSigma}
\bm{u}_2
=
\lambda_2
\bm{u}_2
$
and $\bm{u}_2$ is also the eigenvector of
the covariance matrix $\bm{\varSigma}$,
explaining variance
$
\bm{u}_2^\top
\bm{\varSigma}
\bm{u}_2
=
\lambda_2
.
$
The same procedure is applied to obtain further eigenvectors (PCs).

\subsection{Minimum reconstruction error formulation}
We review the minimum reconstruction error formulation of
PCA~\cite{Pearson1901, Bishop2006} in this part.
Suppose we have a set of vectors
$\left\{\bm{u}_i\right\}_{i=1}^N$, which
are complete and orthonormal, satisfying
$\sum_{i=1}^{N} \bm{u}_i \bm{u}_i^\top = \bm{I}$
and
$\bm{u}_i^\top\bm{u}_j = \delta_{i,j}$.
Here,
$\bm{I}\coloneqq \operatorname{diag}\left(1, \ldots, 1\right)$
is the identity matrix,
and $\delta_{i,j}$ is the Kronecker delta.
Each data point $\bm{x}$ is projected to
$r_i\left(\bm{x}\right) = \bm{u}_i^\top \bm{x}$
and using these projected coordinates,
one can reconstruct the data point exactly as
$
\sum_{i=1}^N
r_i\left(\bm{x}\right)
\bm{u}_i
=
\sum_{i=1}^N
\bm{u}_i
\bm{u}_i^\top
\bm{x}
=
\bm{I}
\bm{x}
=
\bm{x}
,
$
We then want to approximate the data point by using only
$
n
\in
\left\{
    n
    \in
    \mathbb{Z}
    \, \mid \,
    0 \leq n < N
\right\}
$
projected coordinates
and corresponding orthonormal vectors, i.e.,
express original data in fewer $n$-dimensional space.
We approximate the data point by,
$
\tilde{\bm{x}}
=
\sum_{i=1}^n
f_i\left(\bm{x}\right)
\bm{u}_i
+
\sum_{i=n+1}^N
g_i
\bm{u}_i
\approx
\bm{x}
,
$
where $\left\{f_i\left(\bm{x}\right)\right\}_{i=1}^n$ depend on
data point $\bm{x}$
but $\left\{g_i\right\}_{i=n+1}^N$ are constant for all data
points $\bm{x}$.
The reconstruction error $L$ is defined as
\begin{align}
    L
    &\coloneqq
    \sum_{\bm{x}}
    p\left(\bm{x}\right)
    \left|
    \bm{x}
    -
    \tilde{\bm{x}}
    \right|^2
    \label{eq:reconstruction-error-func}
    \\&=
    \nonumber
    \sum_{\bm{x}}
    p\left(\bm{x}\right)
    \left|
    \sum_{i=1}^n
    \left[
        r_i\left(\bm{x}\right)
        -
        f_i\left(\bm{x}\right)
    \right]
    \bm{u}_i
    +
    \sum_{i=n+1}^N
    \left[
        r_i\left(\bm{x}\right)
        -
        g_i
    \right]
    \bm{u}_i
    \right|^2
    \\&=
    \nonumber
    \sum_{\bm{x}}
    p\left(\bm{x}\right)
    \sum_{i=1}^n
    \left[
        r_i\left(\bm{x}\right)
        -
        f_i\left(\bm{x}\right)
    \right]^2
    \\&\quad +
    \sum_{\bm{x}}
    p\left(\bm{x}\right)
    \sum_{i=n+1}^N
    \left[
        r_i\left(\bm{x}\right)
        -
        g_i
    \right]^2
    ,
\end{align}
which is minimized by choosing the appropriate
$\left\{f_i\left(\bm{x}\right)\right\}_{i=1}^n$,
$\left\{g_i\right\}_{i=n+1}^N$,
and $\left\{\bm{u}_i\right\}_{i=1}^N$.

We first derive $L$ with respect to $f_i\left(\bm{x}\right)$, resulting in
\begin{equation}
    0
    =
    \frac{\partial}{\partial f_i\left(\bm{x}\right)}
    L
    =
    -2
    \sum_{\bm{x}}
    p\left(\bm{x}\right)
    \left[
        r_i\left(\bm{x}\right)
        -
        f_i\left(\bm{x}\right)
    \right]
    .
\end{equation}
We obtain
$
\sum_{\bm{x}}
p\left(\bm{x}\right)
f_i\left(\bm{x}\right)
=
\sum_{\bm{x}}
p\left(\bm{x}\right)
r_i\left(\bm{x}\right)
$
which is equivalent to
$f_i\left(\bm{x}\right)=r_i\left(\bm{x}\right)$.
Then, we derive $L$ with respect to $g_i$, getting
\begin{equation}
    0
    =
    \frac{\partial}{\partial g_i}
    L
    =
    -2
    \sum_{\bm{x}}
    p\left(\bm{x}\right)
    \left[
        r_i\left(\bm{x}\right)
        -
        g_i
    \right]
    .
\end{equation}
The solution is
$
\sum_{\bm{x}}
p\left(\bm{x}\right)
g_i
=
\sum_{\bm{x}}
p\left(\bm{x}\right)
r_i\left(\bm{x}\right)
$
which can further be simplified as
$
g_i
=
\left<
r_i\left(\bm{x}\right)
\right>
.
$
Then, the reconstruction error becomes
\begin{align}
    L
    &=
    \nonumber
    \sum_{\bm{x}}
    p\left(\bm{x}\right)
    \sum_{i=n+1}^N
    \left[
        r_i\left(\bm{x}\right)
        -
        \left<
        r_i\left(\bm{x}\right)
        \right>
    \right]^2
    \\&=
    \nonumber
    \sum_{\bm{x}}
    p\left(\bm{x}\right)
    \sum_{i=n+1}^N
    \left[
        \left(
            \bm{x}
            -
            \left<\bm{x}\right>
        \right)^\top
        \bm{u}_i
    \right]^2
    \\&=
    \nonumber
    \sum_{\bm{x}}
    p\left(\bm{x}\right)
    \sum_{i=n+1}^N
    \bm{u}_i^\top
    \left(
        \bm{x}
        -
        \left<\bm{x}\right>
    \right)
    \left(
        \bm{x}
        -
        \left<\bm{x}\right>
    \right)^\top
    \bm{u}_i
    \\&=
    \sum_{i=n+1}^N
    \bm{u}_i^\top
    \bm{\varSigma}
    \bm{u}_i
    .
\end{align}
We then minimize $L$ by choosing appropriate
$\left\{\bm{u}_i\right\}_{i=n+1}^N$
by the method of Lagrange multipliers.
The Lagrange function is
\begin{equation}
    \mathcal{L}
    \coloneqq
    \underbrace{
        \sum_{i=n+1}^N
        \bm{u}_i^\top
        \bm{\varSigma}
        \bm{u}_i
    }_{=L}
    +
    \sum_{i=n+1}^N
    \lambda_i
    \left(
        1
        -
        \bm{u}_i^\top\bm{u}_i
    \right)
    .
\end{equation}
Differentiating $\mathcal{L}$ with respect to $\bm{u}_i$ and
setting it to be the zero vector,
one obtains
$
\bm{\varSigma}
\bm{u}_i
=
\lambda_i
\bm{u}_i
$
for
$
i
\in
\left\{
    i \in \mathbb{Z}
    \, \mid \,
    n+1 \leq i \leq N
\right\}
$.
Notice that this should be valid for all
$
n
\in
\left\{
    n \in \mathbb{Z}
    \, \mid \,
    0 \leq n < N
\right\}
$.
Then, the eigenvectors $\left\{\bm{u}_i\right\}_{i=1}^N$ are solutions of
$
\bm{\varSigma}
\bm{u}_i
=
\lambda_i
\bm{u}_i
$
for
$
i
\in
\left\{
    i \in \mathbb{Z}
    \, \mid \,
    1 \leq i \leq N
\right\}
$.
When we sort the eigenvectors following the descending order of
the eigenvalues
$\left\{\lambda_i\right\}_{i=1}^N$, PCA is performed.

\subsection{Distance, similarity, or overlap preserving formulation}
In this subsection,
we briefly review the pairwise distance-preserving formulation of PCA
via classical MDS~\cite{Gower1966, Cox2001, Ghojogh2023}.
The goal of classical MDS is to
find a set of points in a lower-dimensional space
that preserves the original pairwise distances (or similarity).
To achieve this, we rewrite the squared Euclidean distance
using the inner product.
Then, by performing eigendecomposition of the inner product matrix,
we obtain the eigenvectors, which are the desired
lower-dimensional coordinates.
Finally, we show that this classical MDS is equivalent to PCA\@.
We first consider the classical MDS with the squared Euclidean distance
then try to formulate it with the probability of data points.

\subsubsection{Classical MDS}
A popular distance measure is the squared Euclidean distance.
Suppose we have $M$ data points
$\left\{\bm{x}_i\right\}_{i=1}^M$.
We assume $M > N$, i.e.,
the number of data points is larger than the dimension of the data.
We consider the Euclidean distance matrix
$\bm{D} \in \mathbb{R}_{\geq 0}^{M\times M}$
but with the squared elements
$D_{i, j} \coloneqq D_\mathrm{E}^2 \left(\bm{x}_i, \bm{x}_j\right)$
between the $i$th and $j$th data points.
The squared Euclidean distance is
\begin{align}
    D_{i,j}
    &=
    \nonumber
    \left|
    \bm{x}_i
    -
    \bm{x}_j
    \right|^2
    \\&=
    \nonumber
    \left(
        \bm{x}_i
        -
        \bm{x}_j
    \right)^\top
    \left(
        \bm{x}_i
        -
        \bm{x}_j
    \right)
    \\&=
    \nonumber
    \bm{x}_i^\top\bm{x}_i
    -
    2
    \bm{x}_i^\top\bm{x}_j
    +
    \bm{x}_j^\top\bm{x}_j
    \\&=
    G_{i,i}
    -
    2
    G_{i,j}
    +
    G_{j,j}
    ,
    \label{eq:squared-euclidean-distance}
\end{align}
where we introduce the similarity (inner product) matrix or
the Gram matrix $\bm{G} \in \mathbb{R}^{M\times M}$ which has the
inner product
as its element $G_{i, j} \coloneqq \bm{x}_i^\top\bm{x}_j$.
Notice that the element of the Gram matrix is proportional to the
overlap, i.e.,
$G_{i, j} = N Q\left(\bm{s}_i, \bm{s}_j\right)$ if the data points are
Ising spin states $\bm{x} \rightarrow \bm{s}$.
We can write the element as
$
D_{i,j}
=
g_i
-
2
G_{i,j}
+
g_j
.
$
Here, $g_i\coloneqq G_{i,i}$ is the $i$th diagonal element of
the Gram matrix $\bm{G}$ and
$\bm{g} = \operatorname{diag} \left(\bm{G}\right)$
is the vector of diagonal elements of the Gram matrix.
Then, Eq.~\eqref{eq:squared-euclidean-distance} becomes, in matrix form,
\begin{equation}
    \bm{D}
    =
    \bm{g}
    \bm{1}^\top
    -
    2
    \bm{G}
    +
    \bm{1}
    \bm{g}^\top
    \label{eq:squared-euclidean-distance-matrix}
    ,
\end{equation}
where
$
\bm{1}
\coloneqq
\begin{bsmallmatrix}1 & \cdots & 1
\end{bsmallmatrix}^\top
\in \left\{1\right\}^M
$
is the vector of $M$ ones.
We then introduce the centering matrix as
$
\bm{C}
\coloneqq
\bm{I}
-
\frac{1}{M}
\bm{1}
\bm{1}^\top
,
$
where $\bm{I} = \operatorname{diag}\left(\bm{1}\right)$ is
the $M \times M$ identity matrix.
Notice that the centering matrix is symmetric
$
\bm{C}^\top
=
\bm{C}
$
and idempotent
$
\bm{C}^2
=
\bm{I} - \frac{2}{M} \bm{1} \bm{1}^\top + \frac{1}{M} \bm{1} \bm{1}^\top
=
\bm{C}
$.
The centering matrix subtracts the mean of the column (row) of a matrix when
it is multiplied from the left (right) side of the matrix.
Next, we reduce the mean of both row and column of
the squared Euclidean distance matrix,
i.e., we double-center the squared Euclidean distance matrix, resulting in
\begin{align}
    \bm{C}
    \bm{D}
    \bm{C}
    &=
    \bm{C}
    \left(
        \bm{g}
        \bm{1}^\top
        -
        2
        \bm{G}
        +
        \bm{1}
        \bm{g}^\top
    \right)
    \bm{C}
    \nonumber
    \\&=
    -
    2
    \bm{C}
    \bm{G}
    \bm{C}
    \label{eq:double-centering}
    ,
\end{align}
because
$\bm{1}^\top \bm{C} = \bm{0}^\top$
and
$\bm{C}\bm{1} = \bm{0}$.
Here,
$
\bm{0}
\coloneqq
\begin{bsmallmatrix}
    0 & \cdots & 0
\end{bsmallmatrix}^\top
\in
\left\{0\right\}^M
$
is the vector of $M$ zeros.
Thus, when we double-center the Euclidean distance matrix,
we obtain the double-centered Gram matrix,
i.e., the \emph{distance} matrix becomes the \emph{similarity} matrix.

We perform eigendecomposition of the Gram matrix
\begin{equation}
    \bm{G}
    =
    \bm{Y}
    \bm{\varOmega}
    \bm{Y}^\top
    \label{eq:gram-matrix-decomposition}
    ,
\end{equation}
where
$
\bm{Y}
\coloneqq
\begin{bsmallmatrix}
    \bm{y}_1 & \cdots & \bm{y}_M
\end{bsmallmatrix}
\in
\mathbb{R}^{M\times M}
$
is the matrix with columns being eigenvectors $\{\bm{y}_i\}$ and
$
\bm{\varOmega}
\coloneqq
\operatorname{diag}\left(\omega_1, \ldots, \omega_M\right)
\in
\mathbb{R}_{\geq 0}^{M\times M}
$
is the diagonal matrix of eigenvalues $\left\{\omega_i\right\}$
in descending order
$\omega_1 \geq \cdots \geq \omega_M \geq 0$.
Noticing that the Gram matrix is the inner product matrix,
we can express the Gram matrix as a multiplication of matrices
\begin{equation}
    \bm{G}
    \coloneqq
    \bm{X}^\top
    \bm{X}
    \label{eq:gram-matrix-as-inner-product}
    ,
\end{equation}
where
$
\bm{X}
\coloneqq
\begin{bsmallmatrix}
    \bm{x}_1 & \cdots & \bm{x}_M
\end{bsmallmatrix}
\in
\mathbb{R}^{N\times M}
$
is the data matrix, with each column being a data point.\footnote{
    Note that our definition of the data matrix is the transpose of
    the convention commonly used in the statistics literature.
}
Then, the rank of the Gram matrix becomes
$
\operatorname{rank}\left(\bm{G}\right)
=
\operatorname{rank}
\left(
    \bm{X}^\top
    \bm{X}
\right)
=
\operatorname{rank}\left(\bm{X}\right)
=
N
$.
Because the Gram matrix $\bm{G}$ is symmetric and
positive semi-definite of rank $N$,
it has $N$ non-negative eigenvalues and $M-N$ zero eigenvalues:
$
\omega_1
\geq
\cdots
\geq
\omega_N
>
\omega_{N+1}
=
\cdots
=
\omega_M
=
0
$.
We then rewrite the eigendecomposition of
the Gram matrix [Eq.~\eqref{eq:gram-matrix-decomposition}] as
\begin{align}
    \bm{G}
    & =
    \tilde{\bm{Y}}
    \tilde{\bm{\varOmega}}
    \tilde{\bm{Y}}^\top
    \label{eq:gram-matrix-decomposition-new}
    \\&=
    \tilde{\bm{Y}}
    \tilde{\bm{\varOmega}}^{\frac{1}{2}}
    \tilde{\bm{\varOmega}}^{\frac{1}{2}}
    \tilde{\bm{Y}}^\top
    \label{eq:gram-matrix-decomposition-new-splitted}
    ,
\end{align}
with
reduced eigenvalue diagonal matrix
$
\tilde{\bm{\varOmega}}
\coloneqq
\operatorname{diag}\left(\omega_1, \ldots, \omega_N\right)
\in
\mathbb{R}_{\geq 0}^{N\times N}
$
and
reduced eigenvector matrix
$
\tilde{\bm{Y}}
\coloneqq
\begin{bsmallmatrix}
    \bm{y}_1 & \cdots & \bm{y}_N
\end{bsmallmatrix}
\in
\mathbb{R}^{M\times N}
$
which exclude the zero eigenvalues and corresponding
eigenvectors, respectively.

Calculating the Gram matrix with the centered data matrix,
$
\bm{X}
\leftarrow
\bm{X}
\bm{C}
$,
corresponds to the double centering of the distance matrix
\begin{align}
    \bm{G}
    = &
    \bm{C}
    \bm{X}^\top
    \bm{X}
    \bm{C}
    \label{eq:gram-matrix-transformation}
    \\=&
    -\frac{1}{2}
    \bm{C}
    \bm{D}
    \bm{C}
    \nonumber
    .
\end{align}
By comparing
Eqs.~\eqref{eq:gram-matrix-decomposition-new-splitted}
and~\eqref{eq:gram-matrix-transformation},
one sees that the centered data matrix is expressed as
\begin{align}
    \bm{X}
    \bm{C}
    =
    \tilde{\bm{\varOmega}}^{\frac{1}{2}}
    \tilde{\bm{Y}}^\top
    .
\end{align}
To approximate the original data points by the lower-dimensional coordinates
while maximally preserving the pairwise distances,
one can use the desired number of the first rows of
$
\tilde{\bm{\varOmega}}^{\frac{1}{2}}
\tilde{\bm{Y}}^\top
$.
For example, if one wants to approximate the original
$N$-dimensional data point
by two-dimensional coordinates,
classical MDS provides a set of scaled coordinates
$
\bm{R}
=
\begin{bsmallmatrix}
    \bm{r}_1 & \cdots & \bm{r}_M
\end{bsmallmatrix}
=
\begin{bsmallmatrix}
    \sqrt{\omega_1} \bm{y}_1 &
    \sqrt{\omega_2} \bm{y}_2
\end{bsmallmatrix}^\top
\in \mathbb{R}^{2\times M}
,
$
where $\bm{r}_i$ is the two-dimensional scaled coordinate of the $i$th data
point.
Replacing original coordinates with principal coordinates is also called
principal \emph{coordinate} analysis (PCO or PCoA)~\cite{Gower1966}.

Numerically, the classical MDS is performed by four steps:
(1) calculating the squared Euclidean distance matrix,
(2) performing double centering of the distance matrix,
(3) eigendecomposition of the double-centered distance matrix,
and (4) obtaining the scaled (principal) coordinates
from the eigenvalues and eigenvectors.

\subsubsection{Classical MDS is equivalent to PCA}
\label{sec:mds-pca}
Here, we show that
obtaining the scaled coordinates by classical MDS is equivalent to PCA\@.
From Eq.~\eqref{eq:gram-matrix-decomposition-new},
the eigenvalues and eigenvectors of the double-centered Gram matrix satisfy
\begin{equation}
    \underbrace{
        \bm{C}
        \bm{X}^\top
        \bm{X}
        \bm{C}
    }_{=\bm{G}}
    \bm{y}_i
    =
    \omega_i
    \bm{y}_i
    \label{eq:gram-matrix-eigenequation}
    ,
\end{equation}
for
$
i
\in
\left\{
    i \in \mathbb{Z}
    \, \mid \,
    1 \leq i \leq N
\right\}
$.
Multiplying $\frac{1}{M}\bm{X}\bm{C}$ from the left of both sides of
Eq.~\eqref{eq:gram-matrix-eigenequation},
we obtain
\begin{equation}
    \underbrace{
        \frac{1}{M}
        \bm{X}
        \bm{C}
        \bm{X}^\top
    }_{=\bm{\varSigma}}
    \left(
        \bm{X}
        \bm{C}
        \bm{y}_i
    \right)
    =
    \frac{\omega_i}{M}
    \left(
        \bm{X}
        \bm{C}
        \bm{y}_i
    \right)
    \label{eq:gram-to-covariance}
    .
\end{equation}
Notice $\bm{C}^\top = \bm{C}$ and $\bm{C}^2 = \bm{C}$.
We introduce the covariance matrix in the matrix form
\begin{equation}
    \bm{\varSigma}
    \coloneqq
    \frac{1}{M}
    \left(
        \bm{X}
        \bm{C}
    \right)
    \left(
        \bm{X}
        \bm{C}
    \right)^\top
    =
    \frac{1}{M}
    \bm{X}
    \bm{C}
    \bm{X}^\top
    .
\end{equation}
Note that PCA performs eigendecomposition of the covariance matrix
$
\bm{\varSigma}
\bm{u}_i
=
\lambda_i
\bm{u}_i
$.
Thus, from Eq.~\eqref{eq:gram-to-covariance},
the eigenvalues of the covariance matrix are proportional to
those of the Gram matrix,
\begin{equation}
    \lambda_i
    =
    \frac{\omega_i}{M}
    ,
\end{equation}
and the eigenvector of the covariance matrix is expressed with
the eigenvector of the Gram matrix as
\begin{equation}
    \bm{u}_i
    =
    \frac{1}{\sqrt{\omega_i}}
    \bm{X}
    \bm{C}
    \bm{y}_i
    .
\end{equation}
Following Eq.~\eqref{eq:gram-matrix-eigenequation}, or
$
\bm{y}_i^\top
\bm{C}
\bm{X}^\top
\bm{X}
\bm{C}
\bm{y}_i
=
\omega_i
$,
the normalization factor $\frac{1}{\sqrt{\omega_i}}$ is introduced.
The PC score by PC loading $\bm{u}_i$
and the $i$th element of principal coordinates
$\sqrt{\omega_i}\bm{y}_i^\top$ are equivalent:
\begin{align}
    \bm{u}_i^\top
    \bm{X}
    \bm{C}
    &=
    \frac{1}{\sqrt{\omega_i}}
    \bm{y}_i^\top
    \bm{C}
    \bm{X}^\top
    \bm{X}
    \bm{C}
    \nonumber
    \\&=
    \sqrt{\omega_i}
    \bm{y}_i^\top
    .
\end{align}
Thus, the results of classical MDS are equivalent to those of PCA\@.
This derivation of PCA provides us with a new understanding of PCA
through pairwise distance or inner product.

\subsubsection{Weighted classical MDS as PCA}
We then extend classical MDS with the probability of data points.
Our goal is to find the covariance matrix,
\begin{equation}
    \bm{\varSigma}
    \coloneqq
    \bm{X}
    \acute{\bm{C}}
    \bm{P}
    \acute{\bm{C}}^\top
    \bm{X}^\top
    =
    \left(
        \bm{X}
        \acute{\bm{C}}
        \bm{P}^\frac{1}{2}
    \right)
    \left(
        \bm{X}
        \acute{\bm{C}}
        \bm{P}^\frac{1}{2}
    \right)^\top
    ,
\end{equation}
from the Gram matrix.
Here,
$
\bm{P}
\coloneqq
\operatorname{diag} \left(\bm{p}\right)
\in
\mathbb{R}_{\geq 0}^{M\times M}
$
is
the probability matrix and
$\bm{p} \in \left[0, 1\right]^M$ is the vector of probabilities,
satisfying $\operatorname{tr} \left(\bm{P}\right) = 1$.
Because of the probabilities, the centering matrix is modified as
$
\acute{\bm{C}}
\coloneqq
\bm{I}
-
\bm{1}
\bm{p}^\top
=
\bm{I}
-
\bm{1}
\bm{1}^\top
\bm{P}
$,
which is not symmetric $\acute{\bm{C}}^\top \neq \acute{\bm{C}}$
nor idempotent $\acute{\bm{C}}^2 \neq \acute{\bm{C}}$.
We define the Gram matrix with the probability of data points as
\begin{equation}
    \bm{G}
    \coloneqq
    \bm{P}^\frac{1}{2}
    \acute{\bm{C}}^\top
    \bm{X}^\top
    \bm{X}
    \acute{\bm{C}}
    \bm{P}^\frac{1}{2}
    =
    \left(
        \bm{X}
        \acute{\bm{C}}
        \bm{P}^\frac{1}{2}
    \right)^\top
    \left(
        \bm{X}
        \acute{\bm{C}}
        \bm{P}^\frac{1}{2}
    \right)
    \label{eq:gram-matrix-probability}
    .
\end{equation}
The Gram matrix of Eq.~\eqref{eq:gram-matrix-probability} has an element
$
G_{i,j}
\coloneqq
\sqrt{p_i}
\left(
    \bm{x}_i
    -
    \left<\bm{x}\right>
\right)^\top
\left(
    \bm{x}_j
    -
    \left<\bm{x}\right>
\right)
\sqrt{p_j}
$,
which is the inner product of the centered data points
weighted by the probability of each data point.

With the same procedure as in Appendix~\ref{sec:mds-pca},
we can show that the eigendecomposition of the Gram matrix is
equivalent to PCA\@.
The eigenvalue equation of Eq.~\eqref{eq:gram-matrix-probability} becomes
\begin{equation*}
    \underbrace{
        \bm{P}^\frac{1}{2}
        \acute{\bm{C}}^\top
        \bm{X}^\top
        \bm{X}
        \acute{\bm{C}}
        \bm{P}^\frac{1}{2}
    }_{=\bm{G}}
    \bm{y}_i
    =
    \omega_i
    \bm{y}_i
    .
\end{equation*}
Multiplying $\bm{X}\acute{\bm{C}}\bm{P}^\frac{1}{2}$ from
the left of both sides,
we obtain
\begin{equation*}
    \underbrace{
        \bm{X}
        \acute{\bm{C}}
        \bm{P}
        \acute{\bm{C}}^\top
        \bm{X}^\top
    }_{=\bm{\varSigma}}
    \left(
        \bm{X}
        \acute{\bm{C}}
        \bm{P}^\frac{1}{2}
        \bm{y}_i
    \right)
    =
    \omega_i
    \left(
        \bm{X}
        \acute{\bm{C}}
        \bm{P}^\frac{1}{2}
        \bm{y}_i
    \right)
    .
\end{equation*}
Then, the eigenvalues and the eigenvectors of the covariance matrix are
\begin{gather*}
    \lambda_i
    =
    \omega_i
    ,
    \\
    \bm{u}_i
    =
    \frac{1}{\sqrt{\omega_i}}
    \bm{X}
    \acute{\bm{C}}
    \bm{P}^\frac{1}{2}
    \bm{y}_i
    ,
\end{gather*}
and PC$i$ score is
\begin{align*}
    \bm{u}_i^\top
    \bm{X}
    \acute{\bm{C}}
    \bm{P}^\frac{1}{2}
    &=
    \frac{1}{\sqrt{\omega_i}}
    \bm{y}_i^\top
    \bm{P}^\frac{1}{2}
    \acute{\bm{C}}^\top
    \bm{X}^\top
    \bm{X}
    \acute{\bm{C}}
    \bm{P}^\frac{1}{2}
    \\&=
    \sqrt{\omega_i}
    \bm{y}_i^\top
    .
\end{align*}

Nevertheless, Eq.~\eqref{eq:gram-matrix-probability} is unavailable from
double centering [Eq.~\eqref{eq:double-centering}] of the distance matrix
[Eq.~\eqref{eq:squared-euclidean-distance-matrix}]
by $\acute{\bm{C}}\bm{P}^\frac{1}{2}$ and
$\bm{P}^\frac{1}{2}\acute{\bm{C}}^\top$
because the centering matrix does not erase the vector $\bm{1}$, i.e.,
$
\bm{1}^\top
\acute{\bm{C}}
\neq
\bm{0}^\top
$ and
$
\acute{\bm{C}}^\top
\bm{1}
\neq
\bm{0}
$.
Therefore, the connection between the weighted Gram matrix
[Eq.~\eqref{eq:gram-matrix-probability}] and distance matrix is vague.
In the following Appendix~\ref{sec:minimum-inner-product-error-formulation},
we instead introduce an alternative derivation of PCA
starting from pairwise inner-product similarity, not distance.

\subsection{Minimum inner-product error formulation}
\label{sec:minimum-inner-product-error-formulation}
Here, we introduce the minimum inner product error formulation of PCA\@.
By minimizing the inner product error,
we obtain the same eigenvalues and eigenvectors as PCA\@.
First, we introduce the function to minimize
the mean squared inner product error $L$,
\begin{equation}
    L
    \coloneqq
    \left<
    \varepsilon^2
    \right>
    =
    \sum_{\bm{x}, \bm{x}^\prime}
    p\left(\bm{x}\right)
    p\left(\bm{x}^\prime\right)
    \varepsilon^2\left(\bm{x}, \bm{x}^\prime\right)
    ,
    \label{eq:inner-product-error-func}
\end{equation}
where
\begin{equation}
    \varepsilon^2\left(\bm{x}, \bm{x}^\prime\right)
    \coloneqq
    \frac{1}{2}
    \left[
        \bm{x}^\top
        \bm{x}^\prime
        -
        r
        \left(
            \bm{x}
        \right)
        r
        \left(
            \bm{x}^{\prime}
        \right)
    \right]^2
\end{equation}
is the inner product error of two data points $\bm{x}$ and $\bm{x}^\prime$
between the original high-dimensional and projected one-dimensional space.
The direction of projection is $\bm{u}$, the unit vector,
and the projected coordinate is
$r\left(\bm{x}\right) \coloneqq \bm{u}^\top\bm{x}$.
The factor $\frac{1}{2}$ is introduced for convenience of calculation and
does not change the result.
We would like to find the vector $\bm{u}$ that minimizes $L$
under the normalization constraint $\bm{u}^\top\bm{u} = 1$.
We introduce the Lagrangian function
\begin{equation}
    \mathcal{L}
    \coloneqq
    L
    +
    \alpha
    \left(
        1
        -
        \bm{u}^\top\bm{u}
    \right)
    ,
\end{equation}
where $\alpha$ is the Lagrange multiplier.
The differentiation of
$\varepsilon^2\left(\bm{x}, \bm{x}^\prime\right)$
with respect to $\bm{u}$ is
\begin{align*}
    \frac{\partial}{\partial \bm{u}}
    \varepsilon^2\left(\bm{x}, \bm{x}^\prime\right)
    &=
    \left(
        \bm{x}^\top
        \bm{x}^\prime
        -
        \bm{u}^\top
        \bm{x}
        \bm{x}^{\prime\top}
        \bm{u}
    \right)
    \left(
        -2
        \bm{x}
        \bm{x}^{\prime\top}
        \bm{u}
    \right)
    \\&=
    2
    \left[
        \left(
            \bm{u}^\top
            \bm{x}
        \right)
        \left(
            \bm{x}^{\prime\top}
            \bm{u}
        \right)
        \bm{x}
        \left(
            \bm{x}^{\prime\top}
            \bm{u}
        \right)
        -
        \left(
            \bm{x}^\top
            \bm{x}^\prime
        \right)
        \bm{x}
        \left(
            \bm{x}^{\prime\top}
            \bm{u}
        \right)
    \right]
    \\&=
    2
    \left[
        \bm{x}
        \left(
            \bm{x}^\top
            \bm{u}
        \right)
        \left(
            \bm{u}^\top
            \bm{x}^{\prime}
        \right)
        \left(
            \bm{x}^{\prime\top}
            \bm{u}
        \right)
        -
        \bm{x}
        \left(
            \bm{x}^\top
            \bm{x}^\prime
        \right)
        \left(
            \bm{x}^{\prime\top}
            \bm{u}
        \right)
    \right]
\end{align*}
Thus, the differentiation of $L$ with respect to $\bm{u}$ is
\begin{align*}
    \frac{\partial}{\partial \bm{u}}
    L
    &=
    \sum_{\bm{x}, \bm{x}^\prime}
    p\left(\bm{x}\right)
    p\left(\bm{x}^\prime\right)
    \frac{\partial}{\partial \bm{u}}
    \varepsilon^2\left(\bm{x}, \bm{x}^\prime\right)
    \\&=
    2
    \sum_{\bm{x}, \bm{x}^\prime}
    p\left(\bm{x}\right)
    p\left(\bm{x}^\prime\right)
    \left(
        \bm{x}
        \bm{x}^\top
        \bm{u}
        \bm{u}^\top
        \bm{x}^{\prime}
        \bm{x}^{\prime\top}
        \bm{u}
        -
        \bm{x}
        \bm{x}^\top
        \bm{x}^\prime
        \bm{x}^{\prime\top}
        \bm{u}
    \right)
    \\&=
    2
    \left(
        \acute{\bm{\varSigma}}
        \bm{u}
        \bm{u}^\top
        \acute{\bm{\varSigma}}
        \bm{u}
        -
        \acute{\bm{\varSigma}}
        \acute{\bm{\varSigma}}
        \bm{u}
    \right)
    .
\end{align*}
Here,
$
\acute{\bm{\varSigma}}
\coloneqq
\sum_{\bm{x}}
p\left(\bm{x}\right)
\bm{x}
\bm{x}^\top
$
is the \emph{uncentered} covariance matrix.
Then, the method of Lagrange multipliers gives
$
\bm{0}
=
\frac{\partial}{\partial \bm{u}}
\mathcal{L}
\nonumber
=
2
\left(
    \acute{\bm{\varSigma}}
    \bm{u}
    \bm{u}^\top
    \acute{\bm{\varSigma}}
    \bm{u}
    -
    \acute{\bm{\varSigma}}
    \acute{\bm{\varSigma}}
    \bm{u}
\right)
-
2
\alpha
\bm{u}
$
or
\begin{equation}
    \acute{\bm{\varSigma}}
    \bm{u}
    \bm{u}^\top
    \acute{\bm{\varSigma}}
    \bm{u}
    -
    \acute{\bm{\varSigma}}
    \acute{\bm{\varSigma}}
    \bm{u}
    =
    \alpha
    \bm{u}
    .
    \label{eq:MIPE-equation}
\end{equation}

We will prove by contradiction that every solution
$\bm{u}$ to Eq.~\eqref{eq:MIPE-equation}
is an eigenvector of the uncentered covariance matrix
$\acute{\bm{\varSigma}}$.
Now, assume that $\bm{u}$ is \emph{not}
an eigenvector of $\acute{\bm{\varSigma}}$.
Let $\acute{\bm{u}}_i$ be the eigenvector of
$\acute{\bm{\varSigma}}$ corresponding to eigenvalue $\lambda_i$, i.e.,
$\acute{\bm{\varSigma}}\acute{\bm{u}}_i = \lambda_i \acute{\bm{u}}_i$.
Without loss of generality,
assuming that $\left\{\lambda_i\right\}_{i=1}^n$ are distinct eigenvalues
for $n \geq 2$,
we write the solution $\bm{u}$ as
a linear combination of the corresponding eigenvectors
\begin{equation}
    \bm{u}
    =
    \sum_{i=1}^n
    \kappa_i
    \acute{\bm{u}}_i
    ,
\end{equation}
where $\kappa_i$ is a coefficient satisfying
$\kappa_i^2 \in \left[0, 1\right)$ and $\sum_{i=1}^n \kappa_i^2 = 1$.
With this expansion,
$
\acute{\bm{\varSigma}}
\bm{u}
=
\sum_{i=1}^n
\kappa_i
\lambda_i
\acute{\bm{u}}_i
$,
$
\bm{u}^\top
\acute{\bm{\varSigma}}
\bm{u}
=
\sum_{i=1}^n
\kappa_i^2
\lambda_i
$,
and
$
\acute{\bm{\varSigma}}
\acute{\bm{\varSigma}}
\bm{u}
=
\sum_{i=1}^n
\kappa_i
\lambda_i^2
\acute{\bm{u}}_i
$.
Plugging these into Eq.~\eqref{eq:MIPE-equation},
we obtain
\begin{equation}
    \left(
        \sum_{i=1}^n
        \kappa_i
        \lambda_i
        \acute{\bm{u}}_i
    \right)
    \left(
        \sum_{i=1}^n
        \kappa_i^2
        \lambda_i
    \right)
    -
    \sum_{i=1}^n
    \kappa_i
    \lambda_i^2
    \acute{\bm{u}}_i
    =
    \alpha
    \sum_{i=1}^n
    \kappa_i
    \acute{\bm{u}}_i
    ,
\end{equation}
or
\begin{equation}
    \sum_{i=1}^n
    \left[
        \left(
            \sum_{j=1}^n
            \kappa_j^2
            \lambda_j
        \right)
        \kappa_i
        \lambda_i
        -
        \kappa_i
        \lambda_i^2
        -
        \alpha
        \kappa_i
    \right]
    \acute{\bm{u}}_i
    =
    \bm{0}
    .
\end{equation}
Since $\left\{\acute{\bm{u}}_i\right\}$ are linearly independent,
we get a series of $n$ equations
\begin{equation}
    \alpha
    =
    \left(
        \sum_{j=1}^n
        \kappa_j^2
        \lambda_j
        -
        \lambda_i
    \right)
    \lambda_i
    ,
    \label{eq:proof-by-contradiction}
\end{equation}
for
$
i
\in
\left\{
    i \in \mathbb{Z}
    \, \mid \,
    1 \leq i \leq n
\right\}
$.
Without loss of generality, we assume that
$\lambda_1 > \lambda_2 > \lambda_3 > \cdots > \lambda_n \geq 0$.
Then, we have the upper and lower bounds for the summation,
$
\sum_{j=1}^n
\kappa_j^2
\lambda_j
$,
namely
\begin{equation}
    \lambda_1
    >
    \sum_{j=1}^n
    \kappa_j^2
    \lambda_j
    >
    \lambda_n
    ,
\end{equation}
because $\kappa_i \neq 0$ and $\sum_{i=1}^n \kappa_i^2 = 1$.
Thus, the first equation of Eq.~\eqref{eq:proof-by-contradiction} becomes
\begin{equation}
    \alpha
    =
    \left(
        \sum_{j=1}^n
        \kappa_j^2
        \lambda_j
        -
        \lambda_1
    \right)
    \lambda_1
    < 0
\end{equation}
and that of the $n$th becomes
\begin{equation}
    \alpha
    =
    \left(
        \sum_{j=1}^n
        \kappa_j^2
        \lambda_j
        -
        \lambda_n
    \right)
    \lambda_n
    \geq 0
    .
\end{equation}
These two inequalities contradict each other (i.e., $\alpha < 0
\leq \alpha$).
Therefore, by contradiction,
the solution $\bm{u}$ to Eq.~\eqref{eq:MIPE-equation}
must be an eigenvector of the uncentered covariance matrix
$\acute{\bm{\varSigma}}$,
i.e., $\bm{u} = \acute{\bm{u}}_i$.
Conversely, when $\bm{u}$ is a normalized eigenvector of
$\acute{\bm{\varSigma}}$,
or $\acute{\bm{\varSigma}}\bm{u} = \lambda \bm{u}$
and $\bm{u}^\top\bm{u} = 1$,
we have
$
\acute{\bm{\varSigma}}
\bm{u}
\bm{u}^\top
\acute{\bm{\varSigma}}
\bm{u}
-
\acute{\bm{\varSigma}}
\acute{\bm{\varSigma}}
\bm{u}
=
\acute{\bm{\varSigma}}
\left(
    \bm{u}
    \bm{u}^\top
    -
    \bm{I}
\right)
\acute{\bm{\varSigma}}
\bm{u}
=
\lambda
\acute{\bm{\varSigma}}
\left(
    \bm{u}
    -
    \bm{u}
\right)
=
\bm{0}
,
$
so it is a solution to Eq.~\eqref{eq:MIPE-equation} with $\alpha = 0$.

Finally, we determine the optimal solution of the eigenvector.
Let us expand the data vector,
$
\bm{x}
=
\sum_{i=1}^N
x_i
\acute{\bm{u}}_i
,
$
with
$
\bm{u} = \acute{\bm{u}}_1
.
$
Then, the mean squared inner product error is
\begin{align}
    L
    &=
    \frac{1}{2}
    \sum_{\bm{x}, \bm{x}^\prime}
    p\left(\bm{x}\right)
    p\left(\bm{x}^\prime\right)
    \left(
        \sum_{i=1}^N
        x_i x_i^\prime
        -
        x_1 x_1^\prime
    \right)^2
    \nonumber
    \\&=
    \frac{1}{2}
    \sum_{\bm{x}, \bm{x}^\prime}
    p\left(\bm{x}\right)
    p\left(\bm{x}^\prime\right)
    \left(
        \sum_{i=2}^N
        x_i x_i^\prime
    \right)^2
    \nonumber
    \\&=
    \frac{1}{2}
    \sum_{\bm{x}, \bm{x}^\prime}
    p\left(\bm{x}\right)
    p\left(\bm{x}^\prime\right)
    \sum_{i = 2}^{N}
    \sum_{j = 2}^{N}
    x_i x_j
    x_i^\prime x_j^\prime
    \nonumber
    \\&=
    \frac{1}{2}
    \sum_{i = 2}^{N}
    \sum_{j = 2}^{N}
    \left(
        \sum_{\bm{x}}
        p\left(\bm{x}\right)
        x_i x_j
    \right)
    \left(
        \sum_{\bm{x}^\prime}
        p\left(\bm{x}^\prime\right)
        x_i^\prime x_j^\prime
    \right)
    \nonumber
    \\&=
    \frac{1}{2}
    \sum_{i = 2}^{N}
    \sum_{j = 2}^{N}
    \left(
        \left<x_i x_j\right>
        \delta_{i, j}
    \right)^2
    \nonumber
    \\&=
    \frac{1}{2}
    \sum_{i = 2}^{N}
    \lambda_i^2
    \nonumber
    \\&=
    \frac{1}{2}
    \left[
        \operatorname{tr}
        \left(
            \acute{\bm{\varSigma}}^2
        \right)
        -
        \lambda_1^2
    \right]
    ,
\end{align}
which is minimized when $\lambda_1$ is the largest eigenvalue of
$\acute{\bm{\varSigma}}$.
When the data points and the projected coordinates are centered,
$
\bm{x}
\leftarrow
\bm{x}
-
\left<\bm{x}\right>
$,
$
r\left(\bm{x}\right)
\leftarrow
r\left(\bm{x}\right)
-
\left<r\left(\bm{x}\right)\right>
$
the uncentered covariance matrix becomes the covariance matrix,
and the optimal solution corresponds to the first PC loading.

One can arrive at the same conclusion starting
from Eq.~\eqref{eq:inner-product-error-func}
but through different calculations and proofs~\cite{Ghojogh2023,
Cox2001, Eckart1936}.

\section{PCA with perturbed distribution}
\label{sec:pca-with-perturbation}
\subsection{General formulation}
To expand our discussion in Sec.~\ref{sec:pc-loading-corresponding},
we consider the perturbation to the ideal bipolar distribution as
\begin{equation}
    p\left(\bm{s}\right)
    =
    \frac{1 - \epsilon}{2}
    \left(
        \delta_{+\bm{\xi}_1, \bm{s}}
        +
        \delta_{-\bm{\xi}_1, \bm{s}}
    \right)
    +
    \epsilon
    \rho\left(\bm{s}\right)
    ,
\end{equation}
where $\bm{\xi}_1 \in \left\{+1, -1\right\}^N$ is
the most weighted vertex,
$0 \leq \epsilon \ll 1$ is the perturbation parameter, and
$\rho\left(\bm{s}\right) \in \left[0, 1\right]$ is
any perturbation or noise distribution, satisfying
$\sum_{\bm{s}} \rho\left(\bm{s}\right) = 1$.
We define the perturbation-mean of the vertex as
$
\left<\bm{s}\right>_\rho
\coloneqq
\sum_{\bm{s}}
\rho\left(\bm{s}\right)
\bm{s}
$.
Then, the mean of the vertex becomes
\begin{equation}
    \left<\bm{s}\right>
    =
    \frac{1-\epsilon}{2}
    \left(\bm{\xi}_1 - \bm{\xi}_1\right)
    +
    \epsilon
    \left<\bm{s}\right>_\rho
    =
    \epsilon
    \left<\bm{s}\right>_\rho
\end{equation}
and the perturbed covariance matrix becomes
\begin{align}
    \bm{\varSigma}
    &=
    \left<
    \left(
        \bm{s}
        -
        \left<\bm{s}\right>
    \right)
    \left(
        \bm{s}
        -
        \left<\bm{s}\right>
    \right)^\top
    \right>
    \nonumber
    \\&=
    \left<
    \bm{s}
    \bm{s}^\top
    \right>
    -
    \left<
    \bm{s}
    \right>
    \left<
    \bm{s}
    \right>^\top
    \nonumber
    \\&=
    \frac{1-\epsilon}{2}
    \left(
        \bm{\xi}_1
        \bm{\xi}_1^\top
        +
        \bm{\xi}_1
        \bm{\xi}_1^\top
    \right)
    +
    \epsilon
    \sum_{\bm{s}}
    \rho \left(\bm{s}\right)
    \bm{s}
    \bm{s}^\top
    -
    \epsilon^2
    \left<\bm{s}\right>_\rho
    \left<\bm{s}\right>_\rho^\top
    \nonumber
    \\&=
    \grave{\bm{\varSigma}}
    +
    \epsilon
    \bm{B}
    ,
\end{align}
where
\begin{equation}
    \grave{\bm{\varSigma}}
    \coloneqq
    \bm{\xi}_1
    \bm{\xi}_1^\top
\end{equation}
is the unperturbed covariance matrix and
\begin{equation}
    \bm{B}
    \coloneqq
    -
    \bm{\xi}_1
    \bm{\xi}_1^\top
    +
    \sum_{\bm{s}}
    \rho \left(\bm{s}\right)
    \bm{s}
    \bm{s}^\top
    -
    \epsilon
    \left<\bm{s}\right>_\rho
    \left<\bm{s}\right>_\rho^\top
\end{equation}
is the perturbation matrix.
The rank of the unperturbed covariance matrix
$\grave{\bm{\varSigma}}$ is one.
As we show in Sec.~\ref{sec:pc-loading-corresponding},
the non-zero eigenvalue is
$
\grave{\lambda}_1
=
\bm{\xi}_1^\top
\bm{\xi}_1
=
N
,
$
and the corresponding eigenvector is
$
\grave{\bm{u}}_1
=
\frac{1}{\sqrt{N}}
\bm{\xi}_1
$.
The remaining eigenvalues $\grave{\lambda}_{i\neq 1} = 0$ are all zero and
$(N-1)$-fold degenerate,
and the eigenvectors $\left\{\grave{\bm{u}}_i\right\}_{i=2}^N$ form
a complete orthonormal basis.

Now, we consider the eigenvalue equation of the perturbed covariance matrix,
\begin{equation}
    \bm{\varSigma}
    \bm{u}_i
    =
    \lambda_i
    \bm{u}_i,
\end{equation}
for
$
i
\in
\left\{
    i \in \mathbb{Z}
    \, \mid \,
    1 \leq i \leq N
\right\}
$.
Following the perturbation theory of quantum mechanics,
we define the projection matrix onto the degenerate eigenspace as
\begin{equation}
    \bm{\varPi}
    \coloneqq
    \sum_{i=2}^N
    \grave{\bm{u}}_i
    \grave{\bm{u}}_i^\top
    =
    \bm{I}
    -
    \grave{\bm{u}}_1
    \grave{\bm{u}}_1^\top
\end{equation}
and solve the eigenvalue equation for the projected perturbation matrix
\begin{equation}
    \bm{\varPi}
    \bm{B}
    \bm{\varPi}^\top
    \grave{\bm{u}}_i
    =
    \chi_i
    \grave{\bm{u}}_i
    \label{eq:perturbation-eigenvalue-equation}
    ,
\end{equation}
for
$
i
\in
\left\{
    i \in \mathbb{Z}
    \, \mid \,
    2 \leq i \leq N
\right\}
$.
Those vectors which satisfy
Eq.~\eqref{eq:perturbation-eigenvalue-equation},
along with the eigenvector of
the unperturbed covariance matrix,
define the orthonormal basis $\left\{\bm{u}_i\right\}_{i=1}^N$ for
the expression of the eigenvalues and
eigenvectors of the perturbed covariance matrix $\bm{\varSigma}$.
The first-order approximation is given by
\begin{equation}
    \lambda_1
    \approx
    \grave{\lambda}_1
    +
    \epsilon
    \grave{\bm{u}}_1^\top
    \bm{B}
    \grave{\bm{u}}_1
\end{equation}
and
\begin{equation}
    \bm{u}_1
    \approx
    \grave{\bm{u}}_1
    -
    \epsilon
    \sum_{i=2}^N
    \grave{\bm{u}}_i
    \frac{
        \grave{\bm{u}}_i^\top \bm{B} \grave{\bm{u}}_1
    }{
        \grave{\lambda}_{i} - \grave{\lambda}_1
    }
    =
    \grave{\bm{u}}_1
    +
    \epsilon
    \frac{
        \bm{\varPi} \bm{B} \grave{\bm{u}}_1
    }{
        \grave{\lambda}_1
    }
\end{equation}
for $i=1$.
For
$
i
\in
\left\{
    i \in \mathbb{Z}
    \, \mid \,
    2 \leq i \leq N
\right\}
$,
\begin{equation}
    \lambda_i
    \approx
    \epsilon \chi_i
\end{equation}
and
\begin{equation}
    \bm{u}_i
    \approx
    \grave{\bm{u}}_i
    -
    \epsilon
    \grave{\bm{u}}_1
    \frac{
        \grave{\bm{u}}_1^\top
        \bm{B}
        \grave{\bm{u}}_i
    }{
        \grave{\lambda}_1 - \grave{\lambda}_{i}
    }
    =
    \grave{\bm{u}}_i
    -
    \epsilon
    \grave{\bm{u}}_1
    \frac{
        \grave{\bm{u}}_1^\top
        \bm{B}
        \grave{\bm{u}}_i
    }{
        \grave{\lambda}_1
    }
    .
\end{equation}

\subsection{Quadripolar distribution}
We then examine the quadripolar distribution by considering
the bipolar distribution as the perturbation:
\begin{equation}
    \rho \left(\bm{s}\right)
    =
    \frac{1}{2}
    \left(
        \delta_{+\bm{\xi}_2, \bm{s}}
        +
        \delta_{-\bm{\xi}_2, \bm{s}}
    \right)
    \label{eq:perturbation-distribution}
    .
\end{equation}
The perturbation matrix becomes
\begin{equation}
    \bm{B}
    =
    -
    \bm{\xi}_1
    \bm{\xi}_1^\top
    +
    \bm{\xi}_2
    \bm{\xi}_2^\top
\end{equation}
because $\left<\bm{s}\right>_\rho = \bm{0}$,
and the projected perturbation matrix becomes
\begin{align}
    \bm{\varPi}
    \bm{B}
    \bm{\varPi}^\top
    &=
    \left(
        \bm{I}
        -
        \grave{\bm{u}}_1
        \grave{\bm{u}}_1^\top
    \right)
    \left(
        -
        \bm{\xi}_1
        \bm{\xi}_1^\top
        +
        \bm{\xi}_2
        \bm{\xi}_2^\top
    \right)
    \left(
        \bm{I}
        -
        \grave{\bm{u}}_1
        \grave{\bm{u}}_1^\top
    \right)^\top
    \nonumber
    \\&=
    \left(
        \bm{I}
        -
        \frac{1}{N}
        \bm{\xi}_1
        \bm{\xi}_1^\top
    \right)
    \left(
        -
        \bm{\xi}_1
        \bm{\xi}_1^\top
        +
        \bm{\xi}_2
        \bm{\xi}_2^\top
    \right)
    \left(
        \bm{I}
        -
        \frac{1}{N}
        \bm{\xi}_1
        \bm{\xi}_1^\top
    \right)^\top
    \nonumber
    \\&=
    \bm{\xi}_2
    \bm{\xi}_2^\top
    -
    \frac{
        \bm{\xi}_1^\top \bm{\xi}_2
    }{
        N
    }
    \left(
        \bm{\xi}_2 \bm{\xi}_1^\top
        +
        \bm{\xi}_1 \bm{\xi}_2^\top
    \right)
    \nonumber
    +
    \left(
        \frac{
            \bm{\xi}_1^\top \bm{\xi}_2
        }{
            N
        }
    \right)^2
    \bm{\xi}_1 \bm{\xi}_1^\top
    \nonumber
    \\&=
    \left(
        \bm{\xi}_2 - \frac{
            \bm{\xi}_1^\top \bm{\xi}_2
        }{
            N
        }
        \bm{\xi}_1
    \right)
    \left(
        \bm{\xi}_2 - \frac{
            \bm{\xi}_1^\top \bm{\xi}_2
        }{
            N
        }
        \bm{\xi}_1
    \right)^\top
    ,
\end{align}
which is the rank-one matrix with the nonzero eigenvalue,
\begin{align}
    \chi_2
    &=
    \left(
        \bm{\xi}_2 - \frac{
            \bm{\xi}_1^\top \bm{\xi}_2
        }{
            N
        }
        \bm{\xi}_1
    \right)^\top
    \left(
        \bm{\xi}_2 - \frac{
            \bm{\xi}_1^\top \bm{\xi}_2
        }{
            N
        }
        \bm{\xi}_1
    \right)
    \nonumber
    \\&=
    \bm{\xi}_2^\top \bm{\xi}_2
    -
    2
    \frac{
        \bm{\xi}_1^\top \bm{\xi}_2
    }{
        N
    }
    \bm{\xi}_1^\top \bm{\xi}_2
    +
    \left(
        \frac{
            \bm{\xi}_1^\top \bm{\xi}_2
        }{
            N
        }
    \right)^2
    \bm{\xi}_1^\top \bm{\xi}_1
    \nonumber
    \\&=
    N
    \left[
        1
        -
        \left(
            \frac{
                \bm{\xi}_1^\top \bm{\xi}_2
            }{
                N
            }
        \right)^2
    \right]
    ,
\end{align}
and the corresponding eigenvector
\begin{equation}
    \grave{\bm{u}}_2
    =
    \frac{
        1
    }{
        \sqrt{
            N
            \left[
                1
                -
                \left(
                    \frac{
                        \bm{\xi}_1^\top \bm{\xi}_2
                    }{
                        N
                    }
                \right)^2
            \right]
        }
    }
    \left(
        \bm{\xi}_2
        -
        \frac{
            \bm{\xi}_1^\top \bm{\xi}_2
        }{N}
        \bm{\xi}_1
    \right)
    .
\end{equation}
With straightforward calculations,
we derive the first-order approximation of the eigenvalues and
eigenvectors of the first two PCs.
The eigenvalue and eigenvector of PC1 are
\begin{equation}
    \lambda_1
    \approx
    N
    +
    \epsilon
    N
    \left[
        -1
        +
        \left(
            \frac{
                \bm{\xi}_1^\top \bm{\xi}_2
            }{
                N
            }
        \right)^2
    \right]
\end{equation}
and
\begin{equation}
    \bm{u}_1
    \approx
    \grave{\bm{u}}_1
    -
    \epsilon
    \frac{
        \bm{\xi}_1^\top \bm{\xi}_2
    }{
        N
    }
    \left(
        \frac{
            \bm{\xi}_1^\top \bm{\xi}_2
        }{
            N
        }
        \frac{\bm{\xi}_1}{\sqrt{N}}
        -
        \frac{\bm{\xi}_2}{\sqrt{N}}
    \right)
    .
\end{equation}
Those of PC2 are
\begin{equation}
    \lambda_2
    \approx
    \epsilon
    N
    \left[
        1
        -
        \left(
            \frac{
                \bm{\xi}_1^\top \bm{\xi}_2
            }{
                N
            }
        \right)^2
    \right]
\end{equation}
and
\begin{equation}
    \bm{u}_2
    \approx
    \grave{\bm{u}}_2
    -
    \epsilon
    \frac{
        \bm{\xi}_1^\top \bm{\xi}_2
    }{
        N
    }
    \sqrt{
        N
        \left[
            1
            -
            \left(
                \frac{
                    \bm{\xi}_1^\top \bm{\xi}_2
                }{
                    N
                }
            \right)^2
        \right]
    }
    \bm{\xi}_1
    .
\end{equation}

\subsection{Relation with Hamming distance}
We define the angle $\vartheta$ between the two weighted vertices as
$
\vartheta
\coloneqq
\arccos
\left(
    \frac{
        \bm{\xi}_1^\top \bm{\xi}_2
    }{
        N
    }
\right)
$,
i.e.,
$
\frac{
    \bm{\xi}_1^\top \bm{\xi}_2
}{
    N
}
=
\cos
\left(
    \vartheta
\right)
$,
and consider the projected coordinates of vertex $\bm{s}$
on PC1 and PC2 with zeroth-order approximation, obtaining
\begin{equation}
    r_1
    =
    \bm{u}_1^\top
    \bm{s}
    =
    \frac{1}{\sqrt{N}}
    \bm{\xi}_1^\top
    \bm{s}
    \label{eq:perturbed-pc1-projection}
\end{equation}
and
\begin{equation}
    r_2
    =
    \bm{u}_2^\top
    \bm{s}
    =
    \frac{1}{
        \sqrt{N}
        \sin
        \left(
            \vartheta
        \right)
    }
    \left[
        \bm{\xi}_2^\top
        \bm{s}
        -
        \bm{\xi}_1^\top
        \bm{s}
        \cos \left(\vartheta\right)
    \right]
    \label{eq:perturbed-pc2-projection}
    .
\end{equation}
We then reveal the relationship between the projected coordinates and
Hamming distance.
Using Eqs.~\eqref{eq:perturbed-pc1-projection}
and~\eqref{eq:perturbed-pc2-projection}, we have
\begin{align}
    \bm{\xi}_2^\top
    \bm{s}
    &=
    \nonumber
    \sqrt{N} r_2
    \sin
    \left(
        \vartheta
    \right)
    +
    \bm{\xi}_1^\top
    \bm{s}
    \cos
    \left(
        \vartheta
    \right)
    \\&=
    \sqrt{N}
    \left[
        r_1 \cos \left(\vartheta\right)
        +
        r_2 \sin \left(\vartheta\right)
    \right]
    .
\end{align}
As seen in Sec.~\ref{sec:pc-loading-corresponding},
the Hamming distance
is a function of the inner product,
$
D_\mathrm{H} \left(\bm{\xi}, \bm{s}\right)
=
\frac{
    N
    -
    \bm{\xi}^\top
    \bm{s}
}{
    2
}
.
$
Therefore, we express the Hamming distance as a function of
the projected coordinates
\begin{equation}
    D_\mathrm{H}
    \left(\bm{\xi}_1, \bm{s}\right)
    =
    \frac{
        N
        -
        \sqrt{N}
        r_1
    }{
        2
    }
\end{equation}
and
\begin{equation}
    D_\mathrm{H} \left(\bm{\xi}_2, \bm{s}\right)
    =
    \frac{
        N
        -
        \sqrt{N}
        \left[
            r_1 \cos \left(\vartheta\right)
            +
            r_2 \sin \left(\vartheta\right)
        \right]
    }{
        2
    }
    .
\end{equation}
If the two weighted vertices are orthogonal,
$\bm{\xi}_1^\top \bm{\xi}_2 = 0$ or $\vartheta = \frac{\uppi}{2}$,
the Hamming distance from the second weighted vertex is
$
D_\mathrm{H} \left(\bm{\xi}_2, \bm{s}\right)
=
\frac{
    N
    -
    \sqrt{N}
    r_2
}{
    2
}
$
but otherwise, the Hamming distance is
a function of both $r_1$ and $r_2$.
The angle $\vartheta$ might not be known \emph{a priori},
but it can be estimated by measuring the angle between the
weighted vertices in the projected space.
Indeed, the projected coordinates of the second weighted vertex
on PC1 is
\begin{equation}
    \bm{u}_1^\top
    \bm{\xi}_2
    =
    \frac{1}{\sqrt{N}}
    \bm{\xi}_1^\top
    \bm{\xi}_2
    =
    \sqrt{N}
    \cos \left(\vartheta\right)
\end{equation}
and on PC2 is
\begin{equation}
    \bm{u}_2^\top
    \bm{\xi}_2
    =
    \frac{1}{\sqrt{N} \sin \left(\vartheta\right)}
    \left[
        \bm{\xi}_2^\top \bm{\xi}_2
        -
        \bm{\xi}_1^\top \bm{\xi}_2
        \cos
        \left(
            \vartheta
        \right)
    \right]
    =
    \sqrt{N}
    \sin \left(\vartheta\right)
    .
\end{equation}

In this Appendix~\ref{sec:pca-with-perturbation},
we consider the bipolar distribution [Eq.~\eqref{eq:PC1-score-distribution}]
as the distribution of perturbation.
In the framework of perturbation theory,
one can consider any distribution as the perturbation.
Thus, the extension of this discussion to
any distribution is straightforward.

\section{Possible region of projected coordinates of hypercubic vertices}
\label{sec:possible-region}
\subsection{General bound of projected coordinates}
In general, the orthogonally projected coordinates of hypercubic
vertices are
bounded by a circle with radius $\sqrt{N}$.
This is because we have
$
\left|\bm{r}\left(\bm{s}\right)\right|
\leq
\sqrt{N}
$
[Eq.~\eqref{eq:possible-region-general}] for any vertex.
To show this, we consider the upper bound of
$\left|\bm{r}\left(\bm{s}\right)\right|^2$,
\begin{align}
    \left|\bm{r}\left(\bm{s}\right)\right|^2
    &=
    \bm{r}^\top\left(\bm{s}\right)
    \bm{r}\left(\bm{s}\right)
    =
    \sum_{i=1}^{2}
    \left(\bm{u}_i^\top \bm{s}\right)^2
    \nonumber
    \\&\leq
    \sum_{i=1}^{N}
    \left(\bm{u}_i^\top \bm{s}\right)^2
    =
    \sum_{i=1}^{N}
    \bm{s}^\top \bm{u}_i
    \bm{u}_i^\top \bm{s}
    =
    \bm{s}^\top
    \left(
        \sum_{i=1}^{N}
        \bm{u}_i
        \bm{u}_i^\top
    \right)
    \bm{s}
    \nonumber
    \\&=
    \bm{s}^\top
    \bm{s}
    =
    N
    ,
\end{align}
which is
\begin{equation}
    \left|\bm{r}\left(\bm{s}\right)\right|
    \leq
    \sqrt{N}.
\end{equation}

\subsection{The bound of projected coordinates with quadripolar
distribution}
In Sec.~\ref{sec:upper-bound}, we consider the projected
coordinates assuming
the quadripolar distribution of Eq.~\eqref{eq:perturbation-distribution}.
We show the possible region of the projected coordinates
in Fig.~\ref{fig:inner-product-error-bounds}(b) and
we derive the boundary of this region in this Appendix.

We have the triangular inequalities
\begin{align}
    &
    D_\mathrm{H}\left(\bm{\xi}_1, \bm{\xi}_2\right)
    \leq
    D_\mathrm{H}\left(+\bm{s}, \bm{\xi}_1\right)
    +
    D_\mathrm{H}\left(+\bm{s}, \bm{\xi}_2\right)
    \\&
    D_\mathrm{H}\left(\bm{\xi}_1, \bm{\xi}_2\right)
    \leq
    D_\mathrm{H}\left(-\bm{s}, \bm{\xi}_1\right)
    +
    D_\mathrm{H}\left(-\bm{s}, \bm{\xi}_2\right)
\end{align}
which are similar to Eqs.~\eqref{eq:triangular-inequality-1}
and~\eqref{eq:triangular-inequality-2} but different inequalities.
Because weighted states are orthogonal to each other,
$D_\mathrm{H}\left(\bm{\xi}_1, \bm{\xi}_2\right) = \frac{N}{2}$
with the assumption that $N$ is even.
With sufficiently large $N$,
the results are qualitatively the same when $N$ is odd.
Following the same procedure to derive
Eq.~\eqref{eq:Hamming-distance-inequalities},
we obtain
\begin{equation}
    \frac{N}{2}
    \leq
    N
    -
    \frac{\sqrt{N}}{2}
    \left|
    r_1\left(\bm{s}\right)
    +
    r_2\left(\bm{s}\right)
    \right|
    ,
\end{equation}
which is equivalent to Eq.~\eqref{eq:bound-r1plusr2}.
Similarly, we have
\begin{align}
    &
    D_\mathrm{H}\left(\bm{\xi}_1, -\bm{\xi}_2\right)
    \leq
    D_\mathrm{H}\left(+\bm{s}, \bm{\xi}_1\right)
    +
    D_\mathrm{H}\left(+\bm{s}, -\bm{\xi}_2\right)
    \\&
    D_\mathrm{H}\left(\bm{\xi}_1, -\bm{\xi}_2\right)
    \leq
    D_\mathrm{H}\left(-\bm{s}, \bm{\xi}_1\right)
    +
    D_\mathrm{H}\left(-\bm{s}, -\bm{\xi}_2\right)
\end{align}
and
$
D_\mathrm{H}\left(\bm{\xi}_1, -\bm{\xi}_2\right)
=
N - D_\mathrm{H}\left(\bm{\xi}_1, \bm{\xi}_2\right)
=
\frac{N}{2}
$.
Repeating the same procedure to derive
Eq.~\eqref{eq:Hamming-distance-inequalities},
we have
\begin{equation}
    \frac{N}{2}
    \leq
    N
    -
    \frac{\sqrt{N}}{2}
    \left|
    r_1\left(\bm{s}\right)
    -
    r_2\left(\bm{s}\right)
    \right|
    ,
\end{equation}
which is Eq.~\eqref{eq:bound-r1minusr2}.
Therefore, the possible region of the projected coordinates is the square
bounded by Eqs.~\eqref{eq:bound-r1plusr2} and~\eqref{eq:bound-r1minusr2},
which excludes the corners of Fig.~\ref{fig:inner-product-error-bounds}(b).
We show the boundary of this region
in Fig.~\ref{fig:inner-product-error-bounds}(b).

\bibliography{references}

\end{document}



\title{
    Supplemental Material for\\``Orthogonal projections of hypercubes''
}

\author{Yoshiaki Horiike\,\orcidlink{0009-0000-2010-2598}}
\email{yoshi.h@nagoya-u.jp}
\affiliation{
    Department of Applied Physics,
    \href{https://ror.org/04chrp450}{Nagoya University},
    Nagoya, Japan
}
\affiliation{
    Department of Neuroscience,
    \href{https://ror.org/035b05819}{University of Copenhagen},
    Copenhagen, Denmark
}

\author{Shin Fujishiro\,\orcidlink{0000-0002-0127-0761}}
\affiliation{
    Fukui Institute for Fundamental Chemistry,
    \href{https://ror.org/02kpeqv85}{Kyoto University},
    Kyoto, Japan
}

\date{August 19, 2025}

\maketitle


\renewcommand{\thesection}{S\Roman{section}}
\renewcommand{\theequation}{S\arabic{equation}}
\renewcommand{\thefigure}{S\arabic{figure}}
\renewcommand{\thetable}{S\Roman{table}}

\section{Historical background of higher dimensions}
Human interest in higher dimensions, particularly the fourth dimension, has been
emerging since the 19th century~\cite{Henderson2013, Blacklock2013,
Throesch2016, Blacklock2018}---before the
theory of relativity by A.~Einstein~\cite{Einstein1905, Einstein1911,
Einstein1916},
which is well known as the theory of four-dimensional
spacetime~\cite{Misner2017}.
Unlike the theory of relativity,
19th-century mathematicians, physicists, and philosophers focused on
the \emph{spatial} fourth dimension.
Among them, C.~H.~Hinton has been influential
because his works~\cite{Hinton1980} indicate that the tesseract,
or four-dimensional cube, can embody four-dimensional space.
Since then high-dimensional space has been explored with hypercubes.

\section{Arts and hypercube}
Hypercubes have been applied
in a wide range of interdisciplinary arts~\cite{White2018}.
In various arts,
e.g., literature~\cite{Heinlein1940, LEngle1962, Gardner1987},
visual art~\cite{Brisson2019, Henderson2013, Dali1954, Kemp1998},
architecture~\cite{Brisson2019},
ornament~\cite{Bragdon1992},
and film~\cite{Favreau2010, Nolan2014, Thorne2014, DuVernay2018},
hypercubes have repeatedly inspired human imagination
and have become an embodiment of high-dimensional space.

\section{The gallery of orthogonal projections of hypercubes}
\label{sec:gallery}
In this Sec.~\ref{sec:gallery},
we visualize hypercubes through several methods of orthogonal projections.
In Fig.~\ref{figs:s-isometric-gallery}, we show isometric projections of
hypercubes up to eight dimensions.
In Fig.~\ref{fig:s-hamming-gallery}, we show Hamming projections of
hypercubes up to eight dimensions.
In Fig.~\ref{fig:s-fractal-decaract}, we show fractal projections of
decaract.

\section{Hypercubic PCA with random weights}
Here, we show additional figures of the PCA of hypercubes with
random weights.
In Fig.~\ref{fig:s-hpca-random}, we show the orthogonal projections of
four-dimensional hypercubic vertices by PCA with random weights.
We show two orthogonal projections of the hypercubes,
each with a different realization of random weights.

\section{Hypercubic energy landscape of two-ring system by PC3}
\label{sec:two-rings-pc3}
In this Sec.~\ref{sec:two-rings-pc3},
we show the hypercubic energy landscape of the two-ring system
by the third principal component (PC3).
In Fig.~\ref{fig:s-two-rings-pc3}, we show the hypercubic energy landscapes,
PC loading, and interaction networks of the two-ring system by PC3.

With PC1 and PC3, we find a strong emphasis on spin 6
in Fig.~\ref{fig:s-two-rings-pc3}(a).
Unlike the PC2 loading in Fig.~\ref{fig:s-two-rings-pc3}(b),
spin (variable) 6 dominates the PC3 loading
in Fig.~\ref{fig:s-two-rings-pc3}(c).
This is consistent with the biplot vector in Fig.~\ref{fig:s-two-rings-pc3}(a),
where spin 6 has the longest vector.
The interaction network in Fig.~\ref{fig:s-two-rings-pc3}(d)
shows that the angle of the biplot vector corresponds to
the correlation arising from interaction.

The hypercubic energy landscape by PC2 and PC3 in
Fig.~\ref{fig:s-two-rings-pc3}(e)
draws attention to spin 6,
which is expected from the PC2 and PC3 loadings in
Figs.~\ref{fig:s-two-rings-pc3}(f)
and~\ref{fig:s-two-rings-pc3}(g).
PC2 has a slightly larger contribution from spin 6,
and PC3 has the largest contribution from spin 6.
The interaction network in Fig.~\ref{fig:s-two-rings-pc3}(h)
reveals the correlation between the spins except for spin 6.

\section{Deriving covariance matrix of generalized polar distribution}
\label{sec:deriving-covariance-matrix-generalized}
Here, we derive the covariance matrix of the generalized polar distribution.
Our goal is to derive the covariance matrix
\begin{equation}
    \bm{\varSigma}
    \coloneqq
    \sum_{\bm{s}}
    p\left(\bm{s}\right)
    \left(
        \bm{s} - \left<\bm{s}\right>
    \right)
    \left(
        \bm{s} - \left<\bm{s}\right>
    \right)^\top
    \label{eq:cov-matrix}
\end{equation}
with the probability distribution
\begin{equation}
    p\left(\bm{s}\right)
    =
    \sum_{\mu=1}^{M}
    \left(
        c_{+, \mu}
        \delta_{+\bm{\xi}_\mu, \bm{s}}
        +
        c_{-, \mu}
        \delta_{-\bm{\xi}_\mu, \bm{s}}
    \right)
    ,
\end{equation}
where $c_{+, \mu}, c_{-, \mu} \in [0, 1]$ are the weights of the
selected states,
and satisfying
$
\sum_{\bm{s}} p\left(\bm{s}\right)
=
\sum_{\mu=1}^{M} \left(c_{+, \mu} + c_{-, \mu}\right)
=
1
$.
We define the mean of coefficient
\begin{equation}
    \overline{c}_\mu \coloneqq \frac{1}{2}\left(c_{+, \mu} + c_{-, \mu}\right)
\end{equation}
and the difference of coefficient
\begin{equation}
    \Delta c_\mu \coloneqq \frac{1}{2}\left(c_{+, \mu} - c_{-, \mu}\right)
\end{equation}
for each $\mu$.
Thus,
$c_{+, \mu} = \overline{c}_\mu + \Delta c_\mu$ and
$c_{-, \mu} = \overline{c}_\mu - \Delta c_\mu$.
The mean vector is given by
\begin{align}
    \left<\bm{s}\right>
    &\coloneqq
    \sum_{\bm{s}} p\left(\bm{s}\right) \bm{s}
    \nonumber
    \\&=
    \sum_{\mu=1}^{M}
    \left[
        \left(
            \overline{c}_\mu + \Delta c_\mu
        \right)
        -
        \left(
            \overline{c}_\mu - \Delta c_\mu
        \right)
    \right]
    \bm{\xi}_\mu
    \nonumber
    \\&=
    \sum_{\mu=1}^{M}
    \left(
        2 \Delta c_\mu
    \right)
    \bm{\xi}_\mu
    \label{eq:mean-vector-generalized}
    .
\end{align}
\newpage
The detailed derivation of covariance matrix is as follows.
\begin{align}
    \bm{\varSigma}
    &=
    \sum_{\mu=1}^{M}
    \left[
        \left(
            \overline{c}_\mu + \Delta c_\mu
        \right)
        \left(
            +\bm{\xi}_\mu
            -
            \sum_{\nu=1}^{M}
            2
            \Delta c_\nu
            \bm{\xi}_\nu
        \right)
        \left(
            +\bm{\xi}_\mu
            -
            \sum_{\nu^\prime=1}^{M}
            2
            \Delta c_{\nu^\prime}
            \bm{\xi}_{\nu^\prime}
        \right)^\top
        +
        \left(
            \overline{c}_\mu - \Delta c_\mu
        \right)
        \left(
            -\bm{\xi}_\mu
            -
            \sum_{\nu=1}^{M}
            2
            \Delta c_\nu
            \bm{\xi}_\nu
        \right)
        \left(
            -\bm{\xi}_\mu
            -
            \sum_{\nu^\prime=1}^{M}
            2
            \Delta c_{\nu^\prime}
            \bm{\xi}_{\nu^\prime}
        \right)^\top
    \right]
    \nonumber
    \\&=
    \sum_{\mu=1}^{M}
    \left\{
        \left(
            \overline{c}_\mu + \Delta c_\mu
        \right)
        \left[
            \bm{\xi}_\mu
            \bm{\xi}_\mu^\top
            -
            \sum_{\nu=1}^{M}
            2
            \Delta c_\nu
            \left(
                \bm{\xi}_\mu
                \bm{\xi}_\nu^\top
                +
                \bm{\xi}_\nu
                \bm{\xi}_\mu^\top
            \right)
            +
            \sum_{\nu=1}^{M}
            \sum_{\nu^\prime=1}^{M}
            \left(2\Delta c_\nu\right)
            \left(2\Delta c_{\nu^\prime}\right)
            \bm{\xi}_\nu
            \bm{\xi}_{\nu^\prime}^\top
        \right]
        \right.
        \nonumber
        \\&\quad\quad\quad+
        \left.
        \left(
            \overline{c}_\mu - \Delta c_\mu
        \right)
        \left[
            \bm{\xi}_\mu
            \bm{\xi}_\mu^\top
            +
            \sum_{\nu=1}^{M}
            2
            \Delta c_\nu
            \left(
                \bm{\xi}_\mu
                \bm{\xi}_\nu^\top
                +
                \bm{\xi}_\nu
                \bm{\xi}_\mu^\top
            \right)
            +
            \sum_{\nu=1}^{M}
            \sum_{\nu^\prime=1}^{M}
            \left(2\Delta c_\nu\right)
            \left(2\Delta c_{\nu^\prime}\right)
            \bm{\xi}_\nu
            \bm{\xi}_{\nu^\prime}^\top
        \right]
    \right\}
    \nonumber
    \\&=
    \sum_{\mu=1}^{M}
    \left\{
        2
        \overline{c}_\mu
        \left[
            \bm{\xi}_\mu
            \bm{\xi}_\mu^\top
            +
            \sum_{\nu=1}^{M}
            \sum_{\nu^\prime=1}^{M}
            \left(2\Delta c_\nu\right)
            \left(2\Delta c_{\nu^\prime}\right)
            \bm{\xi}_\nu
            \bm{\xi}_{\nu^\prime}^\top
        \right]
        -
        2
        \Delta c_\mu
        \sum_{\nu=1}^{M}
        \left(
            2
            \Delta c_\nu
        \right)
        \left(
            \bm{\xi}_\mu
            \bm{\xi}_\nu^\top
            +
            \bm{\xi}_\nu
            \bm{\xi}_\mu^\top
        \right)
    \right\}
    \nonumber
    \\&=
    \sum_{\mu=1}^{M}
    2
    \overline{c}_\mu
    \bm{\xi}_\mu
    \bm{\xi}_\mu^\top
    +
    \underbrace{
        \sum_{\mu=1}^{M}
        2
        \overline{c}_\mu
    }_{=1}
    \sum_{\nu=1}^{M}
    \sum_{\nu^\prime=1}^{M}
    \left(2\Delta c_\nu\right)
    \left(2\Delta c_{\nu^\prime}\right)
    \bm{\xi}_\nu
    \bm{\xi}_{\nu^\prime}^\top
    -
    \sum_{\mu=1}^{M}
    \sum_{\nu=1}^{M}
    \left(
        2
        \Delta c_\mu
    \right)
    \left(
        2
        \Delta c_\nu
    \right)
    \left(
        \bm{\xi}_\mu
        \bm{\xi}_\nu^\top
        +
        \bm{\xi}_\nu
        \bm{\xi}_\mu^\top
    \right)
    \nonumber
    \\&=
    \sum_{\mu=1}^{M}
    2
    \overline{c}_\mu
    \bm{\xi}_\mu
    \bm{\xi}_\mu^\top
    +
    \sum_{\mu=1}^{M}
    \sum_{\nu=1}^{M}
    \left(2\Delta c_\mu\right)
    \left(2\Delta c_\nu\right)
    \bm{\xi}_\mu
    \bm{\xi}_\nu^\top
    -
    \sum_{\mu=1}^{M}
    \sum_{\nu=1}^{M}
    \left(
        2
        \Delta c_\mu
    \right)
    \left(
        2
        \Delta c_\nu
    \right)
    \left(
        \bm{\xi}_\mu
        \bm{\xi}_\nu^\top
        +
        \bm{\xi}_\nu
        \bm{\xi}_\mu^\top
    \right)
    \nonumber
    \\&=
    \sum_{\mu=1}^{M}
    2
    \overline{c}_\mu
    \bm{\xi}_\mu
    \bm{\xi}_\mu^\top
    \nonumber
    \\&\quad+
    \sum_{\mu=1}^{M}
    \left(2\Delta c_\mu\right)
    \left(2\Delta c_\mu\right)
    \bm{\xi}_\mu
    \bm{\xi}_\mu^\top
    +
    \sum_{\mu=1}^{M}
    \sum_{\nu=\mu+1}^{M}
    \left(2\Delta c_\mu\right)
    \left(2\Delta c_\nu\right)
    \bm{\xi}_\mu
    \bm{\xi}_\nu^\top
    +
    \sum_{\nu=1}^{M}
    \sum_{\mu=\nu+1}^{M}
    \left(2\Delta c_\nu\right)
    \left(2\Delta c_\mu\right)
    \bm{\xi}_\nu
    \bm{\xi}_\mu^\top
    \nonumber
    \\&\quad-
    \sum_{\mu=1}^{M}
    \left(2\Delta c_\mu\right)
    \left(2\Delta c_\mu\right)
    \left(
        \bm{\xi}_\mu
        \bm{\xi}_\mu^\top
        +
        \bm{\xi}_\mu
        \bm{\xi}_\mu^\top
    \right)
    -
    \sum_{\mu=1}^{M}
    \sum_{\nu=\mu+1}^{M}
    \left(2\Delta c_\mu\right)
    \left(2\Delta c_\nu\right)
    \left(
        \bm{\xi}_\mu
        \bm{\xi}_\nu^\top
        +
        \bm{\xi}_\nu
        \bm{\xi}_\mu^\top
    \right)
    -
    \sum_{\nu=1}^{M}
    \sum_{\mu=\nu+1}^{M}
    \left(2\Delta c_\nu\right)
    \left(2\Delta c_\mu\right)
    \left(
        \bm{\xi}_\nu
        \bm{\xi}_\mu^\top
        +
        \bm{\xi}_\mu
        \bm{\xi}_\nu^\top
    \right)
    \nonumber
    \\&=
    \sum_{\mu=1}^{M}
    2
    \overline{c}_\mu
    \bm{\xi}_\mu
    \bm{\xi}_\mu^\top
    \nonumber
    \\&\quad+
    \sum_{\mu=1}^{M}
    \left(2\Delta c_\mu\right)^2
    \bm{\xi}_\mu
    \bm{\xi}_\mu^\top
    +
    \sum_{\mu=1}^{M}
    \sum_{\nu=\mu+1}^{M}
    \left(2\Delta c_\mu\right)
    \left(2\Delta c_\nu\right)
    \left(
        \bm{\xi}_\mu
        \bm{\xi}_\nu^\top
        +
        \bm{\xi}_\nu
        \bm{\xi}_\mu^\top
    \right)
    \nonumber
    \\&\quad-
    \sum_{\mu=1}^{M}
    2
    \left(
        2
        \Delta c_\mu
    \right)^2
    \bm{\xi}_\mu
    \bm{\xi}_\mu^\top
    -
    \sum_{\mu=1}^{M}
    \sum_{\nu=\mu+1}^{M}
    2
    \left(2\Delta c_\mu\right)
    \left(2\Delta c_\nu\right)
    \left(
        \bm{\xi}_\mu
        \bm{\xi}_\nu^\top
        +
        \bm{\xi}_\nu
        \bm{\xi}_\mu^\top
    \right)
    \nonumber
    \\&=
    \sum_{\mu=1}^{M}
    \left[
        2
        \overline{c}_\mu
        -
        \left(
            2
            \Delta c_\mu
        \right)^2
    \right]
    \bm{\xi}_\mu
    \bm{\xi}_\mu^\top
    -
    \sum_{\mu=1}^{M}
    \sum_{\nu=\mu+1}^{M}
    \left(2\Delta c_\mu\right)
    \left(2\Delta c_\nu\right)
    \left(
        \bm{\xi}_\mu
        \bm{\xi}_\nu^\top
        +
        \bm{\xi}_\nu
        \bm{\xi}_\mu^\top
    \right)
    \label{eq:covariance-matrix-generalized}
\end{align}

\newpage
\section{Deriving the transition rate of the master equation of an Ising
spin system}
We show the detailed derivation of the transition rate of the master
equation introduced in the main text.
From the detailed balance condition, we derive the transition rate as
a function of the energy difference,
$
\Delta E\left(\bm{s}^\prime, \bm{s}\right)
\coloneqq
\mathcal{H} \left(\bm{s}^\prime\right) - \mathcal{H}\left(\bm{s}\right)
=
2 s_k
\sum_{j=1}^{N}
J_{k, j} s_j
+
2 s_k h_k
$.
From the detailed balance condition, we have
\begin{align}
    \frac{
        w_{\bm{s}^\prime, \bm{s}}
    }{
        w_{\bm{s}, \bm{s}^\prime}
    }
    &=
    \frac{
        p_{\bm{s}^\prime}
    }{
        p_{\bm{s}}
    }
    \nonumber
    \\&=
    \frac{
        \frac{1}{Z}
        \exp\left[-\beta \mathcal{H}\left(\bm{s}^\prime\right)\right]
    }{
        \frac{1}{Z}
        \exp\left[-\beta \mathcal{H}\left(\bm{s}\right)\right]
    }
    \nonumber
    \\&=
    \exp\left[-\beta \, \Delta E \left(\bm{s}^\prime, \bm{s}\right)\right]
    \nonumber
    \\&=
    \frac{
        \exp
        \left[
            -
            \beta
            \frac{
                \Delta E \left(\bm{s}^\prime, \bm{s}\right)
            }{2}
        \right]
    }{
        \exp
        \left[
            -
            \beta
            \frac{
                \Delta E \left(\bm{s}, \bm{s}^\prime\right)
            }{2}
        \right]
    }
    .
\end{align}
Here, $\beta \coloneqq 1/k_\mathrm{B} T$ is the inverse temperature
with the Boltzmann constant $k_\mathrm{B}$ and the temperature $T$.
Thus, we assign
\begin{equation}
    w_{\bm{s}^\prime, \bm{s}}
    =
    A
    \exp
    \left[
        -
        \beta
        \frac{
            \Delta E \left(\bm{s}^\prime, \bm{s}\right)
        }{2}
    \right]
\end{equation}
with a constant $A \in \mathbb{R}_{> 0}$.

We then derive the energy difference from state $\bm{s}$ to $\bm{s}^\prime$,
$
\Delta E \left(\bm{s}^\prime, \bm{s}\right)
\coloneqq
\mathcal{H} \left(\bm{s}^\prime\right) - \mathcal{H}\left(\bm{s}\right)
$,
for the Hamiltonian of an Ising spin system
with a single spin flip of spin $k$.
Using the Hamiltonian of an Ising spin system,
\begin{align}
    \Delta E \left(\bm{s}^\prime, \bm{s}\right)
    &=
    \mathcal{H} \left(\bm{s}^\prime\right) - \mathcal{H}\left(\bm{s}\right)
    \nonumber
    \\&=
    \Bigg[
        -
        \frac{1}{2}
        \sum_{i=1,\, i\neq k}^{N}
        \sum_{j=1,\, j\neq k}^{N}
        s_i J_{i, j} s_j
        -
        \frac{1}{2}
        \left(-s_k\right)
        \sum_{j=1}^{N}
        J_{k, j} s_j
        -
        \frac{1}{2}
        \sum_{i=1}^{N}
        s_i J_{i, k} \left(-s_k\right)
        -
        \sum_{i=1,\, i\neq k}^{N}
        s_i h_i
        -
        \left(-s_k\right) h_k
    \Bigg]
    \nonumber
    \\&\quad -
    \Bigg(
        -
        \frac{1}{2}
        \sum_{i=1,\, i\neq k}^{N}
        \sum_{j=1,\, j\neq k}^{N}
        s_i J_{i, j} s_j
        -
        \frac{1}{2}
        s_k
        \sum_{j=1}^{N}
        J_{k, j} s_j
        -
        \frac{1}{2}
        \sum_{i=1}^{N}
        s_i J_{i, k} s_k
        -
        \sum_{i=1,\, i\neq k}^{N}
        s_i h_i
        -
        s_k h_k
    \Bigg)
    \nonumber
    \\&=
    s_k
    \sum_{j=1}^{N}
    J_{k, j} s_j
    +
    \sum_{i=1}^{N}
    s_i J_{i, k} s_k
    +
    2s_k h_k
    \nonumber
    \\&=
    s_k
    \sum_{j=1}^{N}
    J_{k, j} s_j
    +
    s_k
    \sum_{i=1}^{N}
    J_{i, k} s_i
    +
    2s_k h_k
    \label{eq:calculation-1}
    \\&=
    s_k
    \sum_{j=1}^{N}
    J_{k, j} s_j
    +
    s_k
    \sum_{i=1}^{N}
    J_{k, i} s_i
    +
    2s_k h_k
    \label{eq:calculation-2}
    \\&=
    2s_k
    \sum_{j=1}^{N}
    J_{k, j} s_j
    +
    2s_k h_k
    \nonumber
    \\&=
    2s_k
    \left(
        \sum_{j=1}^{N}
        J_{k, j} s_j
        +
        h_k
    \right)
    .
\end{align}
From Eq.~\eqref{eq:calculation-1} to Eq.~\eqref{eq:calculation-2},
we used the symmetry of the interaction matrix
$\bm{J} = \bm{J}^\top$.

\section{Probability flux on the hypercubic energy landscapes}
\label{sec:probability-flux}
In this Sec.~\ref{sec:probability-flux},
we show the time-integrated probability flux on the hypercubic
energy landscapes
which we do not show in the main text.

We show, in Fig.~\ref{fig:probability-flux-pc13}, the time-integrated
probability flux on the hypercubic energy landscape projected by
PC1 and PC3.
The time-integrated probability flux of the one-ring system on
PC1--PC3 space looks similar to that of PC1--PC2 space.
Nevertheless, the time-integrated flux of the two- and three-ring
system on PC1--PC3 space exhibits the center-crossing structure
which is not observed in the PC1--PC2 space.
The state transition dynamics arising from our choice of the initial
state and the external field is well captured by the first two PCs.

Same visualization of time-integrated flux but by PC2 and PC3 are
shown in Fig.~\ref{fig:probability-flux-pc23}.
As expected from the degeneracy of the explained variance of PC2 and PC3,
the time-integrated flux on the PC2--PC3 space exhibits a symmetry.

The time-integrated probability flux projected by PC1 and PC2
at high temperature is shown in
Fig.~\ref{fig:probability-flux-pc12-high-temperature}.
Our choice of the temperature is $k_\mathrm{B}T = 8.0$,
which is high enough to let the entropy dominate the free energy.
We show, in Figs.~\ref{fig:probability-flux-pc12-high-temperature}(a1)--(a3),
that the uniform initial probability distribution does not exhibit
the high-magnitude of time-integrated probability flux, as expected.
If we start the state transition dynamics from the unipolar
probability distribution
$p_{\bm{s}} \left(0\right) = \delta_{-\bm{h}, \bm{s}}$
with external field $\bm{h} \in \left\{+1, -1\right\}^N$
[Figs.~\ref{fig:probability-flux-pc12-high-temperature}(b1)--(b3)],
the time-integrated probability fluxes become smaller as they reach
to the destination state $\bm{h}$.

\section{Review of the mean field approximation of Ising spin system}
\label{sec:mean-field-review}
In this Sec.~\ref{sec:mean-field-review},
we review the mean field approximation\footnote{
    The mean field approximation is also called molecular field approximation,
    Weiss field approximation, or Bragg--Williams approximation.
    The Landau theory is also a mean field theory.
}
of an Ising spin system~\cite{Kubo1988}.
The mean field approximation is a method to replace the interaction
by the mean magnetization.
The model is physically corresponds to the infinite-range model,\footnote{
    The model with such all-to-all interaction is also called
    Husimi--Temperley model or Curie--Weiss model.
    In the context of quantum Ising spin system,
    it is Lipkin--Meshkov--Glick model.
}
where a spin interacts with all other spins.

\subsection{The mean field approximation of ferromagnetic Ising spin system}
We first show that the mean-field Hamiltonian is a function of the magnetization
or order parameter.
We begin with the Hamiltonian
\begin{align}
    \mathcal{H}
    \left(\bm{s}\right)
    &=
    -
    \frac{1}{2}
    \bm{s}^\top
    \bm{J}
    \bm{s}
    -
    \bm{s}^\top
    \bm{h}
    \\&=
    -
    \frac{1}{2}
    \sum_{i=1}^{N}
    \sum_{j=1}^{N}
    s_i
    J_{i, j}
    s_j
    -
    \sum_{i=1}^{N}
    s_i
    h_i
    \label{eq:ising-hamiltonian}
    ,
\end{align}
where
$
\bm{s}
\coloneqq
\begin{bsmallmatrix}
    s_1 & \cdots & s_N
\end{bsmallmatrix}^\top
\in
\left\{\,\uparrow\,\coloneqq +1, \,\downarrow\,\coloneqq -1\right\}^N
$
is a state vector of $N$-Ising-spin system,
$\bm{J} \in \mathbb{R}^{N\times N}$ is an interaction matrix,
$
\bm{h}
\coloneqq
\begin{bsmallmatrix}
    h_1 & \cdots & h_N
\end{bsmallmatrix}^\top
\in
\mathbb{R}^N
$
is the external field vector,
and $N$ is the number of spins.
In the mean field model,
the interaction matrix is that of all-to-all ferromagnetic interaction matrix
\begin{equation}
    \bm{J} = \frac{J}{N} \bm{1} \bm{1}^\top,
\end{equation}
where $J \in \mathbb{R}_{> 0}$ is the interaction strength and
the factor $\frac{1}{N}$ is to ensure that the energy is extensive.
Applying this interaction matrix to the Hamiltonian,
we have the mean-field Hamiltonian
\begin{equation}
    \mathcal{H}_\mathrm{mf} \left(\bm{s}\right)
    =
    -
    \frac{1}{2}
    \frac{J}{N}
    \bm{s}^\top
    \bm{1} \bm{1}^\top
    \bm{s}
    -
    \bm{s}^\top
    \bm{h}
    .
\end{equation}
When the external field is uniform, $\bm{h} = h \bm{1}$,
the Hamiltonian becomes the function of the order parameter
\begin{align}
    \mathcal{H}_\mathrm{mf} \left(m\right)
    &=
    -
    \frac{1}{2}
    \frac{J}{N}
    \left(Nm\right)^2
    -
    h\left(Nm\right)
    \\&=
    -
    \frac{1}{2}
    \frac{J}{N}M^2
    -
    h M
    \\&=
    \mathcal{H}_\mathrm{mf} \left(M\right)
    .
\end{align}
Here, we introduce the order parameter (magnetization per spin)
\begin{align}
    m
    &\coloneqq
    \frac{1}{N}
    \bm{1}^\top \bm{s}
    \\&=
    \frac{1}{N}
    \sum_{i=1}^{N}
    s_i
    ,
\end{align}
and define the total magnetization as
\begin{equation}
    M
    \coloneqq
    N m
    .
\end{equation}

We then calculate the partition function and free energy of the mean-field
Ising spin system.
With the inverse temperature $\beta \coloneqq \frac{1}{k_\mathrm{B}T}$,
the partition function is given by
\begin{align}
    \mathcal{Z}
    &=
    \sum_{\bm{s}}
    \exp\left[-\beta \mathcal{H}\left(\bm{s}\right)\right]
    \\&\approx
    \sum_{M=-N}^{N}
    W\left(M\right) \,
    \exp\left[-\beta \mathcal{H}_\mathrm{mf}\left(M\right)\right]
    \\&=
    \sum_{N_\uparrow=0}^{N}
    \binom{N}{N_\uparrow}
    \exp\left[
        \beta
        \frac{1}{2}
        \frac{J}{N}
        \left(2N_\uparrow - N\right)^2
        +
        \beta h
        \left(2N_\uparrow - N\right)
    \right]
    ,
\end{align}
where
$W\left(M\right) = \binom{N}{N_\uparrow}$
is the number of states $\bm{s}$ with total magnetization $M$.
We express the total magnetization $M=Nm$ in terms of the number of
up spins $N_\uparrow$ and down spins $N_\downarrow$,
\begin{align}
    M
    &=
    N_\uparrow - N_\downarrow
    \\&=
    2N_\uparrow - N
    \\&=
    N - 2N_\downarrow
    .
\end{align}
Note that $N = N_\uparrow + N_\downarrow$.
In the limit of large $N$ ($N \gg 1$),
we can approximate the number of states $W\left(M\right)$
using Stirling's approximation,
$
\ln \left(x!\right) \approx x \ln \left(x\right) - x
$.
After some algebra,\footnote{
    $
    W\left(M\right)
    =
    \binom{N}{N_\uparrow}
    =
    \frac{N!}{N_\uparrow! \left(N-N_\uparrow\right)!}
    =
    \frac{N!}{N_\uparrow! N_\downarrow!}
    =
    \exp \left[\ln \left(\frac{N!}{N_\uparrow! N_\downarrow!}\right)\right]
    \approx
    \exp
    \left[
        N\ln \left(N\right) - N
        -N_\uparrow \ln \left(N_\uparrow\right) + N_\uparrow
        -N_\downarrow \ln \left(N_\downarrow\right) + N_\downarrow
    \right]
    =
    \exp
    \left[
        N_\uparrow \ln \left(N\right)
        +N_\downarrow \ln \left(N\right)
        -N_\uparrow \ln \left(N_\uparrow\right)
        -N_\downarrow \ln \left(N_\downarrow\right)
    \right]
    =
    \exp
    \left[
        -N_\uparrow \ln \left(\frac{N_\uparrow}{N}\right)
        +N_\downarrow \ln \left(\frac{N_\downarrow}{N}\right)
    \right]
    =
    \exp
    \left\{
        -N
        \left[
            \frac{N_\uparrow}{N}\ln \left(\frac{N_\uparrow}{N}\right)
            +
            \frac{N_\downarrow}{N}\ln \left(\frac{N_\downarrow}{N}\right)
        \right]
    \right\}
    $
}
we have
\begin{align}
    W\left(M\right)
    &=
    \binom{N}{N_\uparrow}
    \\&\approx
    \exp
    \left\{
        -N
        \left[
            \frac{N_\uparrow}{N}\ln \left(\frac{N_\uparrow}{N}\right)
            +
            \frac{N_\downarrow}{N}\ln \left(\frac{N_\downarrow}{N}\right)
        \right]
    \right\}
    .
\end{align}
With $N \gg 1$,
we approximate the discrete sum over the total magnetization by
a continuous integral over the order parameter $m$,
$
\sum_{M=-N}^{N}
W\left(M\right) \,
\exp\left[-\beta \mathcal{H}\left(M\right)\right]
\approx
\int_{-1}^{+1}
\mathrm{d} m \,
W\left(m\right) \,
\exp\left[-\beta \mathcal{H}\left(m\right)\right]
$.
Applying these approximations, we obtain
\begin{align}
    \mathcal{Z}
    &\approx
    \sum_{N_\downarrow=0}^{N}
    \exp\left\{
        \beta
        \frac{1}{2}
        \frac{J}{N}
        \left(2N_\uparrow - N\right)^2
        +
        \beta h
        \left(2N_\uparrow - N\right)
        -N
        \left[
            \frac{N_\uparrow}{N}\ln \left(\frac{N_\uparrow}{N}\right)
            +
            \frac{N_\downarrow}{N}\ln \left(\frac{N_\downarrow}{N}\right)
        \right]
    \right\}
    \nonumber
    \\&\approx
    \int_{-1}^{+1}
    \mathrm{d} m \,
    \exp\left\{
        \beta
        \frac{1}{2}
        \frac{J}{N}
        \left(Nm\right)^2
        +
        \beta h
        \left(Nm\right)
        -N
        \left[
            \frac{1+m}{2}\ln \left(\frac{1+m}{2}\right)
            +
            \frac{1-m}{2}\ln \left(\frac{1-m}{2}\right)
        \right]
    \right\}
    \nonumber
    \\&=
    \int_{-1}^{+1}
    \mathrm{d} m \,
    \exp\left[
        - N \beta f\left(m\right)
    \right]
    ,
\end{align}
where we use
$
\frac{N_\uparrow}{N}
=
\frac{N + M}{2N}
=
\frac{1 + m}{2}
$
and
$
\frac{N_\downarrow}{N}
=
\frac{N - M}{2N}
=
\frac{1 - m}{2}
$,
and we define the effective free energy\footnote{
    Also called Bragg--Williams free energy, pseudo free energy, or
    Landau free energy.
} as
\begin{align}
    F\left(m\right)
    =
    N
    f\left(m\right)
    &\coloneqq
    -
    \frac{1}{\beta}
    \ln
    \left[
        Z \left(m\right)
    \right]
    \\&=
    N\left\{
        -\frac{J}{2}
        m^2
        -
        h m
        +
        \frac{1}{\beta}
        \left[
            \frac{1+m}{2}\ln \left(\frac{1+m}{2}\right)
            +
            \frac{1-m}{2}\ln \left(\frac{1-m}{2}\right)
        \right]
    \right\}
    \label{eq:effective-free-energy-per-spin}
\end{align}
with effective free energy per spin $f\left(m\right)$ and
effective partition function\footnote{
    Notice that
    $
    \mathcal{Z}
    =
    \int_{-1}^{+1}
    \mathrm{d} m \,
    W\left(m\right) \,
    \exp\left[- \beta \mathcal{H}_\mathrm{mf}\left(m\right)\right]
    =
    \int_{-1}^{+1}
    \mathrm{d} m \,
    \exp\left\{
        -
        \beta
        \left[
            \mathcal{H}_\mathrm{mf}\left(m\right)
            -
            \frac{1}{\beta}
            \ln \left[W\left(m\right)\right]
        \right]
    \right\}
    =
    \int_{-1}^{+1}
    \mathrm{d} m \,
    \exp\left[
        -
        \beta
        F \left(m\right)
    \right]
    =
    \int_{-1}^{+1}
    \mathrm{d} m \,
    Z\left(m\right)
    $.
}
$Z\left(m\right)
\coloneqq
\int_{-1}^{+1}
\mathrm{d} m^\prime \,
\delta\left(m^\prime - m\right) \,
\exp\left[-N \beta f\left(m^\prime\right)\right]
=
\exp\left[-N \beta f\left(m\right)\right]
$.
Here, $\delta\left(x\right)$ is the Dirac delta function.
See Fig.~\ref{fig:free-energy-landscape-mean-field} for the plot of
Eq.~\eqref{eq:effective-free-energy-per-spin} with various parameters.
Since $N \gg 1$, we apply the Laplace's method\footnote{
    Also referred as saddle-point method or method of the steepest descent.
    In our case, we first approximate the effective free energy at
    the stationary point $m^\ast$,
    $
    f \left(m\right)
    \approx
    f \left(m^\ast\right)
    +
    \frac{1}{2}
    \left.
    \frac{\mathrm{d}^2 f\left(m\right)}{\mathrm{d} m^2}
    \right|_{m=m^\ast}
    \left(m - m^\ast\right)^2
    $.
    Then, the integral is
    $
    \mathcal{Z}
    \approx
    \int_{-1}^{+1}
    \mathrm{d} m \,
    \exp
    \left\{
        - N \beta
        \left[
            f \left(m^\ast\right)
            +
            \frac{1}{2}
            \left.
            \frac{\mathrm{d}^2 f\left(m\right)}{\mathrm{d} m^2}
            \right|_{m=m^\ast}
            \left(m - m^\ast\right)^2
        \right]
    \right\}
    =
    \sqrt{
        \frac{
            2 \uppi
        }{
            N \beta
        }
        \left[
            \left.
            \frac{\mathrm{d}^2 f\left(m\right)}{\mathrm{d} m^2}
            \right|_{m=m^\ast}
        \right]^{-1}
    }
    \exp
    \left[
        - N \beta f\left(m^\ast\right)
    \right]
    $.
    The second derivative is given by
    $
    \frac{\mathrm{d}^2 f\left(m\right)}{\mathrm{d} m^2}
    =
    \frac{1 - \beta J \left(1 - m^2\right)}{\beta \left(1 - m^2\right)}
    $
    and should be positive $\beta J \left(1 - m^2\right) < 1$.
}
to the integral over the order parameter $m$,
\begin{align}
    \mathcal{Z}
    &\approx
    \sum_{m^\ast}
    \exp\left[
        - N \beta f\left(m^*\right)
    \right]
    \label{eq:mean-field-partition-function}
    ,
\end{align}
with ignoring the factor of the integral.\footnote{
    The factor of the integral disappears in the free energy with
    large $N$ limit because it is proportional to $\frac{1}{\sqrt{N}}$.
}
Here, $m^*$ is the value of the order parameter at the stationary point
of the effective free energy per spin, satisfying
\begin{equation}
    \left.
    \frac{\mathrm{d} f\left(m\right)}{\mathrm{d} m}
    \right|_{m=m^\ast}
    =
    -Jm^\ast - h
    +
    \frac{1}{2 \beta}
    \ln
    \left(
        \frac{1+m^\ast}{1-m^\ast}
    \right)
    =
    0
    .
\end{equation}
After some algebra,\footnote{
    $
    \ln
    \left(
        \frac{1+m^\ast}{1-m^\ast}
    \right)
    =
    2 \beta \left(
        J m^\ast + h
    \right)
    \Rightarrow
    \frac{1+m^\ast}{1-m^\ast}
    =
    \exp\left[
        2 \beta \left(
            J m^\ast + h
        \right)
    \right]
    \Rightarrow
    1 + m^\ast
    =
    \left(1 - m^\ast\right)
    \exp\left[
        2 \beta \left(
            J m^\ast + h
        \right)
    \right]
    \Rightarrow
    m^\ast
    \left\{
        \exp\left[
            2 \beta \left(
                J m^\ast + h
            \right)
        \right]
        +1
    \right\}
    =
    \exp\left[
        2 \beta \left(
            J m^\ast + h
        \right)
    \right]
    -1
    \Rightarrow
    m^\ast
    =
    \frac{
        \exp\left[
            2 \beta \left(
                J m^\ast + h
            \right)
        \right]
        -1
    }{
        \exp\left[
            2 \beta \left(
                J m^\ast + h
            \right)
        \right]
        +1
    }
    \Rightarrow
    m^\ast
    =
    \frac{
        \exp\left[
            \beta \left(
                J m^\ast + h
            \right)
        \right]
        -
        \exp\left[
            -\beta \left(
                J m^\ast + h
            \right)
        \right]
    }{
        \exp\left[
            \beta \left(
                J m^\ast + h
            \right)
        \right]
        +
        \exp\left[
            -\beta \left(
                J m^\ast + h
            \right)
        \right]
    }
    \Rightarrow
    m^\ast
    =
    \frac{
        2
        \sinh\left[
            \beta \left(
                J m^\ast + h
            \right)
        \right]
    }{
        2
        \cosh\left[
            \beta \left(
                J m^\ast + h
            \right)
        \right]
    }
    \Rightarrow
    m^\ast
    =
    \tanh\left[
        \beta \left(
            J m^\ast + h
        \right)
    \right]
    $
}
we obtain the self-consistent equation\footnote{
    Also called the equation of state.
}
of $m^\ast$ for the saddle-point approximation of the partition
function of Eq.~\eqref{eq:mean-field-partition-function}:
\begin{equation}
    m^\ast
    =
    \tanh\left[
        \beta \left(J m^\ast + h\right)
    \right]
    \label{eq:self-consistent-equation}
    .
\end{equation}
From the partition function in Eq.~\eqref{eq:mean-field-partition-function},
we obtain the (equilibrium) free energy
$
\mathcal{F}
=
-\frac{1}{\beta}
\ln \left(\mathcal{Z}\right)
$,
energy
$
\mathcal{E}
=
-
\frac{\partial}{\partial \beta}
\ln \left(\mathcal{Z}\right)
$,
and entropy
$
\mathcal{S}
=
-
\frac{\partial}{\partial T}
\mathcal{F}
=
k_\mathrm{B}
\beta^2
\frac{\partial}{\partial \beta}
\mathcal{F}
$.
See, for example, the Appendix~\ref{sec:self-consistent-equation}.

To understand the effective free energy of
Eq.~\eqref{eq:effective-free-energy-per-spin},
we decomposed into an energy term and an entropy term:
\begin{equation}
    F\left(m\right)
    =
    E\left(m\right)
    -
    T S\left(m\right)
    ,
\end{equation}
where the energy term is given by the mean-field Hamiltonian
\begin{align}
    E\left(m\right)
    &=
    \mathcal{H}_\mathrm{mf}\left(m\right)
    \\&=
    N
    \left(
        -\frac{J}{2}
        m^2
        -
        h m
    \right)
    ,
\end{align}
and the entropy term corresponds to the Boltzmann entropy,
\begin{align}
    S\left(m\right)
    &=
    N
    k_\mathrm{B}
    \ln
    \left[
        W \left(m\right)
    \right]
    \\&=
    -
    N
    k_\mathrm{B}
    \left[
        \frac{1+m}{2}\ln \left(\frac{1+m}{2}\right)
        +
        \frac{1-m}{2}\ln \left(\frac{1-m}{2}\right)
    \right]
    \label{eq:effective-entropy}
    .
\end{align}
See the Appendix~\ref{sec:effective-entropy} for
derivation of this Eq.~\eqref{eq:effective-entropy} from the
Shannon entropy.

To show the connection to the Landau theory,
we further approximate the effective free energy per spin
[Eq.~\eqref{eq:effective-free-energy-per-spin}]
using the Newton--Mercator series expansion,
$
\ln \left(1+x\right) \approx x - \frac{x^2}{2} + \frac{x^3}{3} -
\frac{x^4}{4}
$.
After some algebra,\footnote{
    $
    \frac{1+m}{2}\ln \left(\frac{1+m}{2}\right)
    +
    \frac{1-m}{2}\ln \left(\frac{1-m}{2}\right)
    \approx
    \frac{1+m}{2}
    \left[
        -\ln \left(2\right)
        + m - \frac{m^2}{2} + \frac{m^3}{3} - \frac{m^4}{4}
    \right]
    +
    \frac{1-m}{2}
    \left[
        -\ln \left(2\right)
        - m - \frac{m^2}{2} - \frac{m^3}{3} - \frac{m^4}{4}
    \right]
    =
    \frac{1}{2}
    \left[
        - 2 \ln \left(2\right)
        - m^2 - \frac{m^4}{2}
    \right]
    +
    \frac{m}{2}
    \left[
        2m + \frac{2}{3} m^3
    \right]
    =
    -\ln \left(2\right)
    - \frac{1}{2} m^2
    - \frac{1}{4} m^4
    + m^2
    + \frac{1}{3} m^4
    =
    -\ln \left(2\right)
    + \frac{1}{2} m^2
    + \frac{1}{12} m^4
    $
}
we have
\begin{equation}
    S\left(m\right)
    \approx
    -
    N k_\mathrm{B}
    \left[
        -
        \ln \left(2\right)
        +
        \frac{1}{2}
        m^2
        +
        \frac{1}{12}
        m^4
    \right]
    .
\end{equation}
Then the effective free energy per spin becomes
\begin{align}
    F\left(m\right)
    &\approx
    N
    \left\{
        -\frac{J}{2}
        m^2
        -
        h m
        +
        \frac{1}{\beta}
        \left[
            -\ln \left(2\right)
            +
            \frac{1}{2}
            m^2
            +
            \frac{1}{12}
            m^4
        \right]
    \right\}
    \nonumber
    \\&=
    N
    \left[
        -
        \frac{1}{\beta}
        \ln \left(2\right)
        -
        h m
        +
        \frac{1}{2}
        \left(\frac{1 - \beta J}{\beta}\right)
        m^2
        +
        \frac{1}{12 \beta}
        m^4
    \right]
    ,
\end{align}
which is the forth-order polynomial in order parameter as the Landau theory,
and the sign of the second order coefficient determines the
number of extrema.
Thus, when $\beta J < 1$, i.e., the temperature is high enough,
$k_\mathrm{B}T > J$,
the free energy has a single minimum,
and when $\beta J > 1$, i.e., the temperature is low enough,
$k_\mathrm{B}T < J$,
the free energy has two minima, corresponding to the ordered phase
and the disordered phase, respectively.
We define the critical temperature as
$
T_\mathrm{c} \coloneqq \frac{J}{k_\mathrm{B}}
$
and its inverse as
$
\beta_\mathrm{c} \coloneqq \frac{1}{k_\mathrm{B}T_\mathrm{c}}
$.

\subsection{The covariance matrix of the mean-field model}
\label{sec:mean-field-covariance}
We consider the covariance matrix of the mean-field model.
The element of the covariance matrix is given by
$
\varSigma_{i, j}
=
\left<s_i - \left<s_i\right>\right>
\left<s_j - \left<s_j\right>\right>
=
\left<s_i s_j\right>
-\left<s_i\right> \left<s_j\right>
$,
thus we derive the two-spin mean (correlation) and single-spin mean to obtain
the covariance matrix.
In general, the spin correlation is given by
\begin{equation}
    \left<s_i s_j\right>
    -
    \left<s_i\right>
    \left<s_j\right>
    =
    \frac{1}{\mathcal{Z}}
    \sum_{\bm{s}}
    \exp\left[-\beta \mathcal{H}\left(\bm{s}\right)\right]
    s_i s_j
    -
    \left\{
        \frac{1}{\mathcal{Z}}
        \sum_{\bm{s}}
        \exp\left[-\beta \mathcal{H}\left(\bm{s}\right)\right]
        s_i
    \right\}
    \left\{
        \frac{1}{\mathcal{Z}}
        \sum_{\bm{s}}
        \exp\left[-\beta \mathcal{H}\left(\bm{s}\right)\right]
        s_j
    \right\}
    .
\end{equation}
The two-spin mean is given by
\begin{align}
    \left<s_i s_j\right>
    &=
    \frac{1}{\mathcal{Z}}
    \sum_{\bm{s}}
    \exp
    \left[
        -
        \beta
        \left(
            -
            \frac{1}{2}
            \sum_{i=1}^{N}
            \sum_{j=1}^{N}
            s_i
            J_{i, j}
            s_j
            -
            \sum_{i=1}^{N}
            s_i
            h_i
        \right)
    \right]
    s_i s_j
    \nonumber
    \\&=
    \frac{2}{\beta}
    \frac{\partial}{\partial J_{i,j}}
    \ln \left(\mathcal{Z}\right)
    \nonumber
    \\&=
    -
    2
    \frac{\partial}{\partial J_{i,j}}
    \mathcal{F}
    ,
\end{align}
and the single spin mean is given by
\begin{align}
    \left<s_i\right>
    &=
    \frac{1}{\mathcal{Z}}
    \sum_{\bm{s}}
    \exp
    \left[
        -
        \beta
        \left(
            -
            \frac{1}{2}
            \sum_{i=1}^{N}
            \sum_{j=1}^{N}
            s_i
            J_{i, j}
            s_j
            -
            \sum_{i=1}^{N}
            s_i
            h_i
        \right)
    \right]
    s_i
    \nonumber
    \\&=
    \frac{1}{\beta}
    \frac{\partial}{\partial h_i} \ln \left(\mathcal{Z}\right)
    \nonumber
    \\&=
    -
    \frac{\partial}{\partial h_i} \mathcal{F}
    .
\end{align}
With the mean-field Hamiltonian and effective free energy per spin,
these are calculated as functions of order parameter $m$,\footnote{
    Strictly speaking, the diagonal element is determined as
    $
    \left<s_i s_j\right> \left(m\right)
    =
    -
    2
    \frac{\partial}{\partial J}
    f \left(m\right)
    =
    \begin{cases}
        m^2 \quad &\text{if} \quad i \neq j \\
        1   \quad &\text{if} \quad i = j \\
    \end{cases}
    $
    because $s_i^2 = 1$.
}
\begin{align}
    \left<s_i s_j\right> \left(m\right)
    &=
    -
    2
    \frac{\partial}{\partial J}
    f \left(m\right)
    \\&=
    m^2
    ,
\end{align}
and
\begin{align}
    \left<s_i\right> \left(m\right)
    &=
    -
    \frac{\partial}{\partial h}
    f \left(m\right)
    \\&=
    \begin{cases}
        m \quad &\text{if} \quad h \neq 0 \\
        0 \quad &\text{if} \quad h = 0
    \end{cases}
    .
\end{align}
Thus, the element of covariance matrix $\varSigma_{i, j}\left(m\right)$
is\footnote{
    If we consider that the diagonal element is
    $\left<s_i s_j\right> = 1$ for $i = j$,
    the correlation is
    $
    \varSigma_{i, j} \left(m\right)
    =
    \left<s_i s_j\right> \left(m\right)
    -
    \left<s_i\right> \left(m\right)
    \left<s_j\right> \left(m\right)
    =
    \begin{cases}
        0   \quad &\text{if} \quad i \neq j \quad \text{and}
        \quad h \neq 0 \\
        m^2 \quad &\text{if} \quad i \neq j \quad \text{and} \quad h = 0 \\
        1 - m^2 \quad &\text{if} \quad i = j \quad \text{and}
        \quad h \neq 0 \\
        1  \quad &\text{if} \quad i = j \quad \text{and} \quad h = 0
    \end{cases}
    $.
}
\begin{align}
    \varSigma_{i, j} \left(m\right)
    &=
    \left<s_i s_j\right> \left(m\right)
    -
    \left<s_i\right> \left(m\right)
    \left<s_j\right> \left(m\right)
    \\&=
    \begin{cases}
        0   \quad &\text{if} \quad h \neq 0 \\
        m^2 \quad &\text{if} \quad h = 0 \\
    \end{cases}
    .
\end{align}
Thus, the covariance matrix $\bm{\varSigma}\left(m\right)$ is\footnote{
    With the knowledge of
    $\left<s_i s_j\right> = 1$ for $i = j$,
    $
    \bm{\varSigma} \left(m\right)
    =
    \begin{cases}
        \bm{0} \bm{0}^\top + \left(1 - m^2\right) \bm{I}
        \quad &\text{if} \quad h \neq 0 \\
        m^2 \bm{1} \bm{1}^\top + \left(1 - m^2\right) \bm{I}
        \quad &\text{if} \quad h = 0
    \end{cases}
    $,
    where
    $\bm{I} \coloneqq \operatorname{diag} \left(1, 1, \ldots, 1\right)$
    is the identity matrix.
}
\begin{equation}
    \bm{\varSigma} \left(m\right)
    =
    \begin{cases}
        \bm{0} \bm{0}^\top     \quad &\text{if} \quad h \neq 0 \\
        m^2 \bm{1} \bm{1}^\top \quad &\text{if} \quad h = 0
    \end{cases}
    ,
\end{equation}
where
$\bm{0} \in \left\{0\right\}^N$
and
$\bm{1} \in \left\{1\right\}^N$
are the vector of zeros and ones, respectively.
If $h = 0$, the covariance matrix
$
m^2 \bm{1} \bm{1}^\top
=
N m^2
\left(\frac{1}{\sqrt{N}}\bm{1}\right)
\left(\frac{1}{\sqrt{N}}\bm{1}\right)^\top
$
satisfies the eigenvalue equation\footnote{
    In case of $\left<s_i s_j\right> = 1$,
    $
    \bm{\varSigma} \left(m\right)
    \left(
        \frac{1}{\sqrt{N}}
        \bm{1}
    \right)
    =
    \left(
        N m^2
        +
        1 - m^2
    \right)
    \left(
        \frac{1}{\sqrt{N}}
        \bm{1}
    \right)
    $.
}
\begin{equation}
    \bm{\varSigma} \left(m\right)
    \left(
        \frac{1}{\sqrt{N}}
        \bm{1}
    \right)
    =
    N m^2
    \left(
        \frac{1}{\sqrt{N}}
        \bm{1}
    \right)
    .
\end{equation}
Therefore, the eigenvalue is $N m^2$ with the eigenvector
$\frac{1}{\sqrt{N}} \bm{1}$.
These covariance matrices $\bm{0}\bm{0}^\top$ and $m^2 \bm{1}\bm{1}^\top$ are
derived from the unipolar\footnote{
    If $\left<s_i s_j\right> = 1$ for $i = j$,
    $
    p_m \left(\bm{s}\right)
    \propto
    \delta_{m\bm{1}, \bm{s}}
    +
    \left(
        \frac{1}{2^N}
        \sum_{\bm{s}^\prime} \delta_{\sqrt{1 - m^2} \bm{s}^\prime, \bm{s}}
    \right)
    $.
}
\begin{equation}
    p_m \left(\bm{s}\right)
    =
    \delta_{m\bm{1}, \bm{s}}
\end{equation}
and the bipolar distribution\footnote{
    If $\left<s_i s_j\right> = 1$ for $i = j$,
    $
    p_m \left(\bm{s}\right)
    \propto
    \left(
        \frac{1}{2}
        \delta_{+m\bm{1}, \bm{s}}
        +
        \frac{1}{2}
        \delta_{-m\bm{1}, \bm{s}}
    \right)
    +
    \left(
        \frac{1}{2^N}
        \sum_{\bm{s}^\prime} \delta_{\sqrt{1 - m^2} \bm{s}^\prime, \bm{s}}
    \right)
    $.
}
\begin{equation}
    p_m \left(\bm{s}\right)
    =
    \frac{1}{2}
    \left(
        \delta_{+m\bm{1}, \bm{s}}
        +
        \delta_{-m\bm{1}, \bm{s}}
    \right)
    ,
\end{equation}
respectively, using Eq.~\eqref{eq:cov-matrix}.
Note that the mean vector is
$
\left<\bm{s}\right>
=
m\bm{1}
$
and
$
\left<\bm{s}\right>
=
\bm{0}
$, respectively.
Also, normalization condition is satisfied for both distributions.
The mean-field approximation is rough because it assumes that
$\left<s_i s_i\right> = m^2$.
In real, it is $\left<s_i s_i\right> = 1$, but the qualitative behavior
is the same.
Note that, if $h = 0$, the covariance matrix
$\bm{\varSigma} \left(m\right) = m^2 \bm{1} \bm{1}^\top$
is proportional to the interaction matrix
$\bm{J} = \frac{J}{N} \bm{1}\bm{1}^\top$,
\begin{equation}
    \bm{\varSigma} \left(m\right)
    \propto
    \bm{J}
    .
\end{equation}
Thus, the correlation is causation in the mean-field model.

\subsection{Hopfield model is the generalized mean field model}
Next, we show that the Hopfield model is equivalent to the mean field model.
We consider the Hopfield model,\footnote{
    Also called the Amari--Hopfield model.
}
in which the interaction matrix is given by the Hebbian rule~\cite{Hopfield1982}
\begin{equation}
    \bm{J}_\mathrm{H}
    =
    \frac{J}{N}
    \sum_{\mu=1}^{P}
    c_\mu
    \bm{\xi}_\mu
    \bm{\xi}_\mu^\top
\end{equation}
with the element
\begin{equation}
    J_{i, j}
    =
    \frac{J}{N}
    \sum_{\mu=1}^{P}
    c_\mu
    \xi_{i; \mu} \xi_{j; \mu}
    ,
\end{equation}
and the corresponding Hamiltonian is
\begin{align}
    \mathcal{H}_\mathrm{H} \left(\bm{s}\right)
    &=
    -
    \frac{1}{2}
    \frac{J}{N}
    \bm{s}^\top
    \left(
        \sum_{\mu=1}^{P}
        c_\mu
        \bm{\xi}_\mu
        \bm{\xi}_\mu^\top
    \right)
    \bm{s}
    -
    \bm{s}^\top \bm{h}
    \\&=
    -
    \frac{1}{2}
    \frac{J}{N}
    \sum_{\mu=1}^{P}
    c_\mu
    \sum_{i=1}^{N}
    \sum_{j=1}^{N}
    s_i \xi_{i; \mu} \xi_{j; \mu} s_j
    -
    \sum_{i=1}^{N}
    s_i h_i
    .
\end{align}
Here, $\bm{\xi}_\mu \in \left\{+1, -1\right\}^N$ denotes the $\mu$th
pattern vector, $c_\mu \in \left[0, 1\right]$
(with $\sum_{\mu=1}^{P} c_\mu = 1$) is the weight
of the $\mu$th pattern, and $P \in \mathbb{N}$ is the number
of patterns.
Note that $J \in \mathbb{R}_{> 0}$ is the interaction strength and
the factor $\frac{1}{N}$ is to ensure that the energy is extensive.

For the case of a single pattern ($P=1$, $c_\mu = \delta_{1, \mu}$),
\begin{equation}
    \bm{J}_\mathrm{H}
    =
    \frac{J}{N}
    \bm{\xi}
    \bm{\xi}^\top
    ,
\end{equation}
the Hopfield interaction network is exactly reduced to the mean
field model by
introducing transformations\footnote{
    This transformation is known as the Mattis (gauge) transformation.
}
\begin{equation}
    s_i^\prime
    \coloneqq
    \xi_i s_i
\end{equation}
with a transformed state vector
\begin{equation}
    \bm{s}^\prime
    \coloneqq
    \bm{\xi}
    \odot
    \bm{s}
    ,
\end{equation}
transformed interaction
\begin{equation}
    J_{i, j}^\prime
    \coloneqq
    \xi_i J_{i, j} \xi_j
\end{equation}
with a transformed interaction matrix
\begin{equation}
    \bm{J}^\prime
    \coloneqq
    \bm{\xi}
    \odot
    \bm{J}
    \odot
    \bm{\xi}^\top
    ,
\end{equation}
and transformed external field
\begin{equation}
    h_i^\prime
    \coloneqq
    \xi_i h_i
\end{equation}
with a transformed external field vector
\begin{equation}
    \bm{h}^\prime
    \coloneqq
    \bm{\xi} \odot \bm{h}
    .
\end{equation}
Here, $\odot$ denotes the element-wise (Hadamard) product.
In general, the Hamiltonian is invariant under this transformation,
\begin{align}
    \mathcal{H}^\prime
    \left(\bm{s}^\prime\right)
    &=
    -
    \frac{1}{2}
    \left(\bm{\xi} \odot \bm{s}\right)^\top
    \left(\bm{\xi} \odot \bm{J} \odot \bm{\xi}^\top\right)
    \left(\bm{\xi} \odot \bm{s}\right)
    -
    \left(\bm{\xi} \odot \bm{s}\right)^\top
    \left(\bm{\xi} \odot \bm{h}\right)
    \nonumber
    \\&=
    -
    \frac{1}{2}
    \bm{s}^\top
    \bm{J}
    \bm{s}
    -
    \bm{s}^\top
    \bm{h}
    \nonumber
    \\&=
    -
    \frac{1}{2}
    \sum_{i=1}^{N}
    \sum_{j=1}^{N}
    \left(s_i \xi_i\right)
    \left(\xi_i J_{i, j} \xi_j\right)
    \left(\xi_j s_j\right)
    -
    \sum_{i=1}^{N}
    \left(s_i \xi_i\right)
    \left(\xi_i h_i\right)
    \nonumber
    \\&=
    -
    \frac{1}{2}
    \sum_{i=1}^{N}
    \sum_{j=1}^{N}
    s_i
    J_{i, j}
    s_j
    -
    \sum_{i=1}^{N}
    s_i
    h_i
    \nonumber
    \\&=
    \mathcal{H}\left(\bm{s}\right)
    .
\end{align}
Notice that $\xi_i \xi_i = 1$, $\forall i$.
Using the invariance of the Hamiltonian under this transformation,
we rewrite the Hamiltonian of the Hopfield model as
\begin{align}
    \mathcal{H}_\mathrm{H} \left(\bm{s}\right)
    &=
    -
    \frac{1}{2}
    \frac{J}{N}
    \sum_{i=1}^{N}
    \sum_{j=1}^{N}
    \left(s_i \xi_i\right)
    \underbrace{
        \left[\xi_i \left( \xi_i \xi_j\right) \xi_j\right]
    }_{=1}
    \left(\xi_j s_j\right)
    -
    \sum_{i=1}^{N}
    \left(s_i \xi_i\right)
    \left(\xi_i h_i\right)
    \nonumber
    \\&=
    -
    \frac{1}{2}
    \frac{J}{N}
    \sum_{i=1}^{N}
    \sum_{j=1}^{N}
    \left(s_i \xi_i\right)
    \left(\xi_j s_j\right)
    -
    \sum_{i=1}^{N}
    \left(s_i \xi_i\right)
    \left(\xi_i h_i\right)
    \nonumber
    \\&=
    -
    \frac{1}{2}
    \frac{J}{N}
    \left(\bm{s}^\prime\right)^\top
    \bm{1} \bm{1}^\top
    \bm{s}^\prime
    -
    \left(\bm{s}^\prime\right)^\top
    \bm{h}^\prime
\end{align}
When the transformed external field is uniform,
$\bm{h}^\prime = h^\prime \bm{1}$, i.e., $\bm{h} \propto \bm{\xi}$,
the Hamiltonian is rewritten as
\begin{align}
    \mathcal{H}_\mathrm{H} \left(m^\prime\right)
    &=
    -
    \frac{1}{2}
    \frac{J}{N}
    \left( Nm^\prime \right)^2
    -
    h^\prime \left( Nm^\prime\right)
    \\&=
    -
    \frac{1}{2}
    \frac{J}{N}
    \left(M^\prime\right)^2
    -
    h^\prime
    M^\prime
    \\&=
    \mathcal{H}_\mathrm{H} \left(M^\prime\right)
    \\&=
    \mathcal{H}_\mathrm{mf} \left(m^\prime\right)
    .
\end{align}
Here, we define the transformed order parameter as
\begin{align}
    m^\prime
    &\coloneqq
    \frac{1}{N}
    \bm{1}^\top \bm{s}^\prime
    \\&=
    \frac{1}{N}
    \bm{1}^\top \left(\bm{\xi} \odot \bm{s}\right)
    \\&=
    \frac{1}{N}
    \bm{\xi}^\top \bm{s}
    \\&=
    \frac{1}{N}
    \sum_{i=1}^{N}
    \xi_i s_i
\end{align}
which represents the overlap, or cosine similarity, between
the pattern vector $\bm{\xi}$ and the state vector $\bm{s}$.
The total overlap or similarity is defined as
\begin{equation}
    M^\prime
    \coloneqq
    N m^\prime
    .
\end{equation}
Therefore, the Hopfield model is thus equivalent to the mean field model,
but with the order parameter replaced by the transformed order parameter
$m^\prime$ and the external field replaced by transformed external
field $h^\prime$.
All results and discussions for the mean field model can be applied
directly to the Hopfield model by substituting
$m \leftarrow m^\prime$ and $h \leftarrow h^\prime$.

We then consider the Hopfield model with multiple patterns ($P \geq 2$)
to show that the Hopfield model is a generalized mean field model.
We introduce the transformed order parameters
$\left\{m_\mu\right\}_{\mu=1}^{P}$,
\begin{align}
    m_\mu
    &\coloneqq
    \frac{1}{N}
    \bm{\xi}_\mu^\top \bm{s}
    \\&=
    \frac{1}{N}
    \sum_{i=1}^{N}
    \xi_{i; \mu} s_i
    ,
\end{align}
the Hopfield model Hamiltonian is rewritten as
\begin{align}
    \mathcal{H}_\mathrm{H} \left(\bm{s}\right)
    &=
    -
    \frac{1}{2}
    \frac{J}{N}
    \sum_{\mu=1}^{P}
    c_\mu
    \sum_{i=1}^{N}
    \sum_{j=1}^{N}
    \left(s_i \xi_{i; \mu}\right)
    \underbrace{
        \left[
            \xi_{i; \mu} \left(\xi_{i; \mu} \xi_{j; \mu}\right) \xi_{j; \mu}
        \right]
    }_{=1}
    \left(\xi_{j; \mu} s_j\right)
    -
    \sum_{i=1}^{N}
    s_i h_i
    \nonumber
    \\&=
    -
    \frac{1}{2}
    \frac{J}{N}
    \sum_{\mu=1}^{P}
    c_\mu
    \left(
        \bm{\xi}_\mu^\top \bm{s}
    \right)^2
    -
    \bm{s}^\top
    \bm{h}
    \nonumber
    \\&=
    -
    \frac{1}{2}
    \frac{J}{N}
    \sum_{\mu=1}^{P}
    c_\mu
    \left(
        N m_\mu
    \right)^2
    -
    \bm{s}^\top
    \bm{h}
    .
\end{align}
Thus, the interaction term of the Hamiltonian of Hopfield model
is a function of the order parameters $\left\{m_\mu\right\}_{\mu=1}^{P}$.

We then consider some examples of external fields.
If the external field is the sum of the weighted patterns with the
same coefficients $\left\{c_\mu\right\}_{\mu=1}^{P}$,
\begin{equation}
    \bm{h}
    =
    h
    \sum_{\mu=1}^{P}
    c_\mu
    \bm{\xi}_\mu
    \label{eq:hopfield-external-field}
    ,
\end{equation}
the Hamiltonian becomes
\begin{align}
    \mathcal{H}_\mathrm{H} \left(\bm{s}\right)
    &=
    -
    \frac{1}{2}
    \frac{J}{N}
    \sum_{\mu=1}^{P}
    c_\mu
    \left(
        N m_\mu
    \right)^2
    -
    h
    \sum_{\mu=1}^{P}
    c_\mu
    \left(
        N m_\mu
    \right)
    \nonumber
    \\&=
    \sum_{\mu=1}^{P}
    c_\mu
    \left[
        -
        \frac{1}{2}
        \frac{J}{N}
        \left(N m_\mu\right)^2
        -
        N h m_\mu
    \right]
    \nonumber
    \\&=
    \sum_{\mu=1}^{P}
    c_\mu
    \mathcal{H}_\mathrm{mf} \left(m_\mu\right)
    \\&\eqqcolon
    \mathcal{H}_\mathrm{H} \left(m_1, \ldots, m_P\right)
    .
\end{align}
Thus, the Hopfield model is equivalent to a sum of mean field Hamiltonian
for each pattern.
If the external field is zero, $\bm{h} = \bm{0}$,
the Hamiltonian is rewritten as,
\begin{align}
    \mathcal{H}_\mathrm{H} \left(\bm{s}\right)
    &=
    -
    N
    \frac{J}{2}
    \sum_{\mu=1}^{P}
    c_\mu
    m_\mu^2
    .
\end{align}
For example, for two patterns ($P=2$), we have
\begin{equation}
    \mathcal{H}_\mathrm{H} \left(m_1, m_2\right)
    =
    -
    N
    \frac{J}{2}
    \left(
        c_1
        m_1^2
        +
        c_2
        m_2^2
    \right)
    .
\end{equation}
If the patterns are orthogonal to each other,
$
\bm{\xi}_1^\top \bm{\xi}_2 = 0
$,
the equi-energy contours in the two-dimensional order parameter space
spanned by $m_1$ and $m_2$ are ellipses.

\subsection{The covariance matrix of the Hopfield model}
Following the same procedure as in
Sec.~\ref{sec:mean-field-covariance},
we derive the covariance matrix of the Hopfield model.
The two-spin mean is given by
\begin{equation}
    \left<s_i s_j\right> \left(m_1, \ldots, m_P\right)
    =
    \sum_{\mu = 1}^{P}
    c_\mu
    m_\mu^2
    .
\end{equation}
Assuming the external field of Eq.~\eqref{eq:hopfield-external-field},
the single spin mean is given by
\begin{equation}
    \left<s_i\right> \left(m_1, \ldots, m_P\right)
    =
    \begin{cases}
        \sum_{\mu = 1}^{P} c_\mu m_\mu \quad &\text{if} \quad h \neq 0\\
        0 \quad &\text{if} \quad h = 0\\
    \end{cases}
    .
\end{equation}
With these, the element of covariance matrix
is given by
\begin{align}
    \varSigma_{i, j} \left(m_1, \ldots, m_P\right)
    &=
    \left<s_i s_j\right> \left(m_1, \ldots, m_P\right)
    -
    \left<s_i\right> \left(m_1, \ldots, m_P\right)
    \left<s_j\right> \left(m_1, \ldots, m_P\right)
    \\&=
    \begin{cases}
        \sum_{\mu = 1}^{P} c_\mu m_\mu^2
        -
        \left(\sum_{\mu=1}^{P} c_\mu m_\mu\right)^2
        =
        \sum_{\mu = 1}^{P} \left(c_\mu - c_\mu^2\right) m_\mu^2
        -
        2
        \sum_{\mu = 1}^{P} \sum_{\nu = \mu + 1}^{P} c_\mu c_\nu m_\mu m_\nu
        \quad &\text{if} \quad h \neq 0 \\
        \sum_{\mu = 1}^{P} c_\mu m_\mu^2 \quad &\text{if} \quad h = 0 \\
    \end{cases}
    \label{eq:hopfield-covariance-matrix}
    .
\end{align}
The case of $h \neq 0$ of Eq.~\eqref{eq:hopfield-covariance-matrix} is
reminiscent to Eq.~\eqref{eq:covariance-matrix-generalized}.
If external field is zero vector ($h = 0$),
the covariance matrix is
\begin{align}
    \bm{\varSigma} \left(m_1, \ldots, m_P\right)
    &=
    \sum_{\mu = 1}^{P} c_\mu m_\mu^2
    \bm{1} \bm{1}^\top
    \\&=
    N
    \sum_{\mu = 1}^{P} c_\mu m_\mu^2
    \left(
        \frac{1}{\sqrt{N}}
        \bm{1}
    \right)
    \left(
        \frac{1}{\sqrt{N}}
        \bm{1}
    \right)^\top
    ,
\end{align}
which has the eigenvalue
$
N
\sum_{\mu = 1}^{P} c_\mu m_\mu^2
$
and the eigenvector
$
\frac{1}{\sqrt{N}} \bm{1}
$.
With Eq.~\eqref{eq:cov-matrix}, this covariance matrix is derived from
the distribution
\begin{equation}
    p_m \left(\bm{s}\right)
    =
    \sum_{\mu=1}^{P}
    c_\mu
    \frac{1}{2}
    \left(
        \delta_{+m_\mu \bm{1}, \bm{s}}
        +
        \delta_{-m_\mu \bm{1}, \bm{s}}
    \right)
    .
\end{equation}
The mean vector is zero vector with this distribution.
Note that, as we show in Sec.~\ref{sec:mean-field-covariance},
the covariance matrix of $h = 0$ case
$
\bm{\varSigma} \left(m_1, \ldots, m_P\right)
=
\sum_{\mu = 1}^{P}
c_\mu
m_\mu^2
\bm{1} \bm{1}^\top
$
is proportional to the transformed interaction matrix of the Hopfield model
$
\bm{J}_\mathrm{H}^\prime
=
\frac{J}{N}
\sum_{\mu=1}^{P}
c_\mu
\bm{1}
\bm{1}^\top
\leftarrow
\bm{J}_\mathrm{H}
=
\frac{J}{N}
\sum_{\mu=1}^{P}
c_\mu
\bm{\xi}_\mu
\bm{\xi}_\mu^\top
$,
\begin{equation}
    \bm{\varSigma} \left(m_1, \ldots, m_P\right)
    \propto
    \bm{J}_\mathrm{H}
    .
\end{equation}
Thus, the correlation is causation in the Hopfield model as well.

\appendix

\renewcommand{\thesection}{S\Alph{section}}
\renewcommand{\theequation}{S\Alph{section}\arabic{equation}}
\renewcommand{\thefigure}{S\Alph{section}\arabic{figure}}
\renewcommand{\thetable}{S\Alph{section}\Roman{table}}

\section{Solving the self-consistent equation}
\label{sec:self-consistent-equation}
The self-consistent equation generally does not admit an
analytical solution, but its behavior can be analyzed in the high-
and low-temperature limits.
Our scope is limited in the absence of an external field ($h=0$),
which gives the self-consistent equation
\begin{equation}
    m^\ast
    =
    \tanh\left(\beta J m^\ast\right)
    .
\end{equation}

In the high-temperature limit ($\beta J \ll 1$),
expanding the hyperbolic tangent function to first order,
$\tanh\left(x\right) \approx x$,
yields
\begin{equation}
    m^\ast \approx \beta J m^\ast
\end{equation}
which gives
\begin{equation}
    m^\ast
    =
    0
    ,
\end{equation}
corresponding to the disordered phase.
The effective free energy per spin in the disordered phase is
\begin{equation}
    f\left(0\right)
    =
    -
    \frac{1}{\beta}
    \ln \left(2\right)
    ,
\end{equation}
and the partition function is approximated as
\begin{equation}
    \mathcal{Z}
    \approx
    2^N
    .
\end{equation}
Thus, the (equilibrium) free energy is
\begin{equation}
    \mathcal{F}
    =
    -
    \frac{N}{\beta}
    \ln \left(2\right)
    .
\end{equation}
Here, the effective free energy per spin is dominated by the entropy
term, and spontaneous magnetization does not occur.

In the low-temperature limit ($\beta J \gg 1$),
since the hyperbolic tangent function approaches
$
\lim_{\left|x\right| \to \infty}\tanh\left(x\right)
=
\operatorname{sgn}\left(x\right)
$,
we obtain
\begin{align}
    m^\ast
    &\approx
    \operatorname{sgn}\left(m^\ast\right)
    \\&=
    \pm 1
    ,
\end{align}
in addition to the trivial solution $m^\ast = 0$.
Because the second derivative at the trivial stationary point
$m^\ast = 0$
is negative, it does not contribute during the Laplace's method.
Therefore, we only consider the non-trivial stationary points
$m^\ast = \pm 1$.
The new stationary points correspond to the ordered phase.
The effective free energy per spin in the ordered phase is
\begin{equation}
    f\left(\pm 1\right)
    =
    -
    \frac{J}{2}
    ,
\end{equation}
and the partition function is approximated as
\begin{align}
    \mathcal{Z}
    &\approx
    \exp\left(
        N \beta \frac{J}{2}
    \right)
    +
    \exp\left(
        N \beta \frac{J}{2}
    \right)
    \\&\approx
    2
    \exp\left(
        N \beta \frac{J}{2}
    \right)
    .
\end{align}
Thus, the (equilibrium) free energy is
\begin{align}
    \mathcal{F}
    &=
    -
    N
    \frac{J}{2}
    -
    \frac{1}{\beta}
    \ln \left(2\right)
    \\&\approx
    -
    N\frac{J}{2}
    .
\end{align}
In this regime, the effective free energy per spin is dominated
by the interaction
energy term, and the system exhibits spontaneous magnetization.
Note that $m^\ast = 0$ is also a solution in the ordered phase,
but the effective free energy per spin
$f\left(0\right)$ is higher than that of the ordered phase
$f\left(\pm 1\right)$, and the stationary point $m^\ast = 0$
contribute less.

\section{Deriving the effective entropy of mean-field model from
the Shannon entropy}
\label{sec:effective-entropy}
Because of the all-to-all interaction,
the probability distribution is approximated by
the product of the individual spin probability distributions,
\begin{equation}
    p \left(\bm{s}\right)
    =
    \prod_{i=1}^{N}
    p \left(s_i\right)
    .
\end{equation}
With this assumption, the Shannon entropy is given by
\begin{align}
    S
    &=
    k_\mathrm{B}
    \sum_{\bm{s}}
    p \left(\bm{s}\right)
    \left\{
        -
        \ln
        \left[p \left(\bm{s}\right)\right]
    \right\}
    \nonumber
    \\&=
    - k_\mathrm{B}
    \sum_{\bm{s}}
    p \left(\bm{s}\right)
    \ln
    \left[p \left(\bm{s}\right)\right]
    \nonumber
    \\&=
    - k_\mathrm{B}
    \sum_{\bm{s}}
    \left[
        \prod_{i=1}^{N}
        p \left(s_i\right)
    \right]
    \ln
    \left[
        \prod_{i=1}^{N}
        p \left(s_i\right)
    \right]
    \nonumber
    \\&=
    - k_\mathrm{B}
    \sum_{\bm{s}}
    \left[
        \prod_{j=1}^{N}
        p \left(s_j\right)
    \right]
    \sum_{i=1}^{N}
    \ln
    \left[p \left(s_i\right)\right]
    \nonumber
    \\&=
    - k_\mathrm{B}
    \sum_{\bm{s}}
    \sum_{i=1}^{N}
    p \left(s_i\right)
    \ln
    \left[p \left(s_i\right)\right]
    \prod_{j=1,\, j\neq i}^{N}
    p \left(s_j\right)
    \nonumber
    \\&=
    - k_\mathrm{B}
    \sum_{s_1 \in \left\{+1, -1\right\}}
    \sum_{s_2 \in \left\{+1, -1\right\}}
    \cdots
    \sum_{s_N \in \left\{+1, -1\right\}}
    \sum_{i=1}^{N}
    p \left(s_i\right)
    \ln
    \left[p \left(s_i\right)\right]
    \prod_{j=1,\, j\neq i}^{N}
    p \left(s_j\right)
    \nonumber
    \\&=
    - k_\mathrm{B}
    \sum_{i=1}^{N}
    \sum_{s_i \in \left\{+1, -1\right\}}
    p \left(s_i\right)
    \ln
    \left[p \left(s_i\right)\right]
    \prod_{j=1,\, j\neq i}^{N}
    \underbrace{
        \sum_{s_j \in \left\{+1, -1\right\}}
        p \left(s_j\right)
    }_{=1}
    \nonumber
    \\&=
    - k_\mathrm{B}
    \sum_{i=1}^{N}
    \sum_{s_i \in \left\{+1, -1\right\}}
    p \left(s_i\right)
    \ln
    \left[p \left(s_i\right)\right]
    .
\end{align}
Because of the all-to-all interaction,
the probability distribution of each spin is expected to be uniform,
i.e., $p \left(s_i\right)$ is independent of $i$ but rather depends on
the order parameter $m$.
We assume that the phenomenological probability distribution
of each spin,
which depends on the order parameter $m$ as
\begin{equation}
    p_m \left(s_i\right)
    =
    \frac{
        1 + s_i m
    }{2}
    ,
    \quad
    \forall i,
\end{equation}
which satisfies the normalization condition
\begin{equation}
    \sum_{s_i \in \left\{+1, -1\right\}} p_m \left(s_i\right) = 1
    ,
    \quad
    \forall i,
\end{equation}
and the single-spin mean is given by
\begin{equation}
    \left<s_i\right>
    =
    \sum_{s_i \in \left\{+1, -1\right\}} p_m \left(s_i\right) s_i
    = m
    ,
    \quad
    \forall i,
\end{equation}
which is consistent with the mean field model.
Then, the Shannon entropy is rewritten as a function of the order
parameter $m$,
\begin{align}
    S \left(m\right)
    &=
    - N k_\mathrm{B}
    \sum_{s_i \in \left\{+1, -1\right\}}
    p_m \left(s_i\right)
    \ln
    \left[p_m \left(s_i\right)\right]
    \\&=
    - N k_\mathrm{B}
    \left[
        \frac{1+m}{2}\ln \left(\frac{1+m}{2}\right)
        +
        \frac{1-m}{2}\ln \left(\frac{1-m}{2}\right)
    \right]
    ,
\end{align}
which is the same as Eq.~\eqref{eq:effective-entropy}.

\bibliography{references}

\renewcommand{\thesection}{S\Roman{section}}
\renewcommand{\theequation}{S\arabic{equation}}
\renewcommand{\thefigure}{S\arabic{figure}}
\renewcommand{\thetable}{S\Roman{table}}

\begin{figure}[p]
    \includegraphics{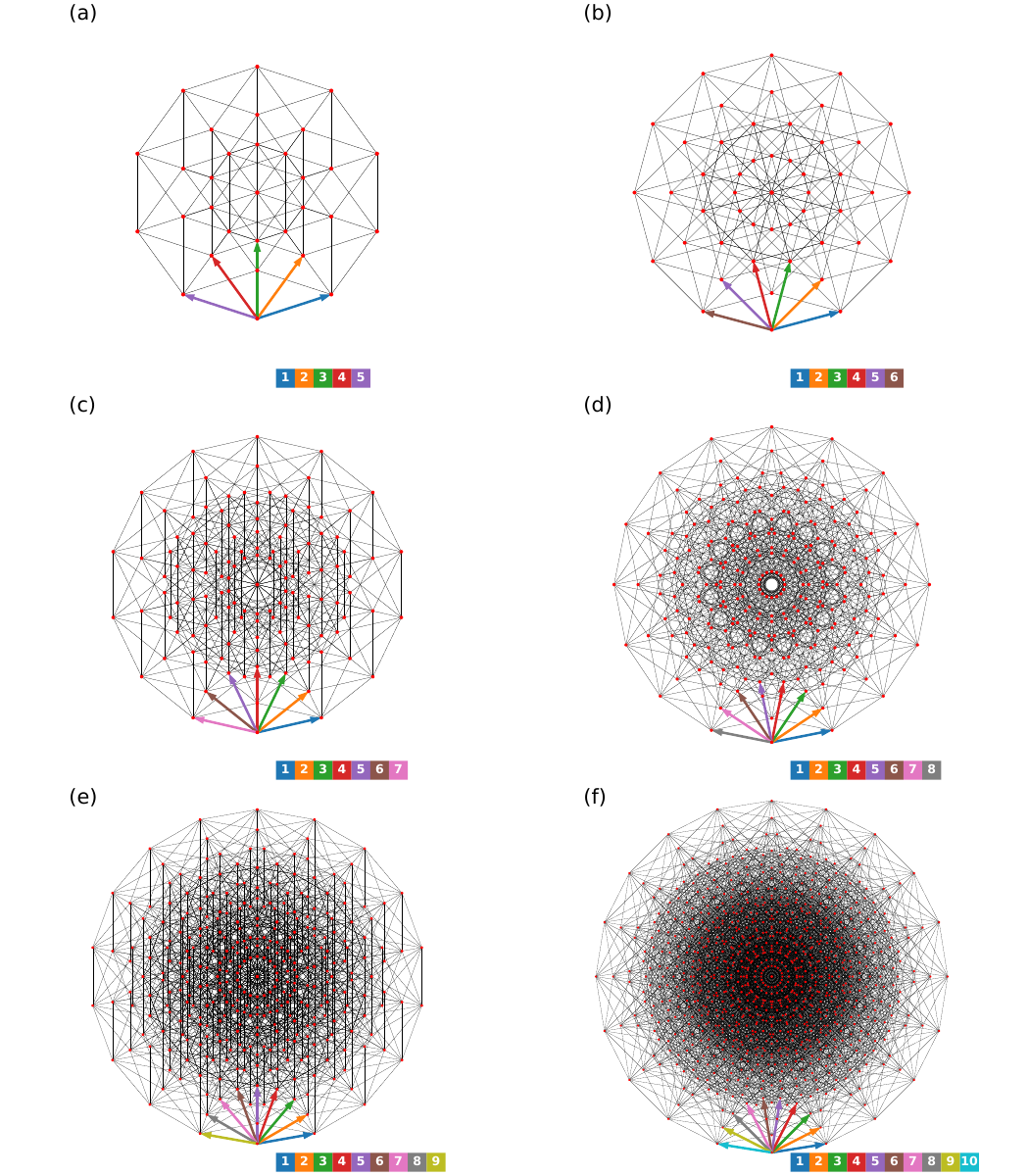}
    \caption{
        Isometric projections of
        (a) five-dimensional cube or penteract,
        (b) six-dimensional cube or hexeract,
        (c) seven-dimensional cube or hepteract,
        (d) eight-dimensional cube or octeract,
        (e) nine-dimensional cube or enneract,
        and
        (f) ten-dimensional cube or decaract.
        Colored arrows represent the contribution basis of
        each dimension.
        The boxes on the bottom right indicate
        the correspondence between the colors and the dimensions.
    }
    \label{figs:s-isometric-gallery}
\end{figure}

\begin{figure}[p]
    \includegraphics{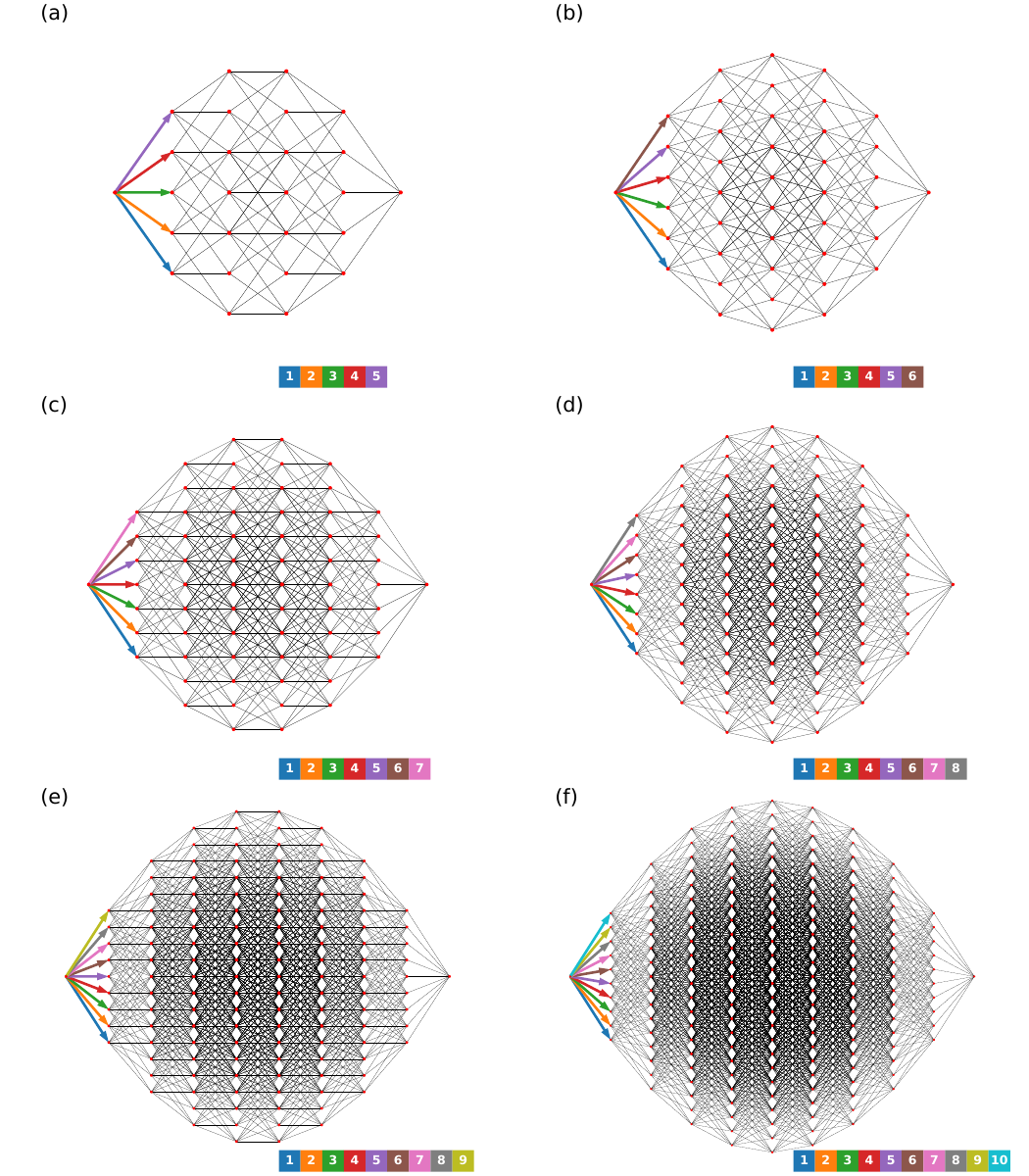}
    \caption{
        Hamming projections of
        (a) five-dimensional cube or penteract,
        (b) six-dimensional cube or hexeract,
        (c) seven-dimensional cube or hepteract,
        (d) eight-dimensional cube or octeract,
        (e) nine-dimensional cube or enneract,
        and
        (f) ten-dimensional cube or decaract.
        Colored arrows represent the contribution basis of
        each dimension.
        The boxes on the bottom right indicate
        the correspondence between the colors and the dimensions.
    }
    \label{fig:s-hamming-gallery}
\end{figure}

\begin{figure}[p]
    \includegraphics{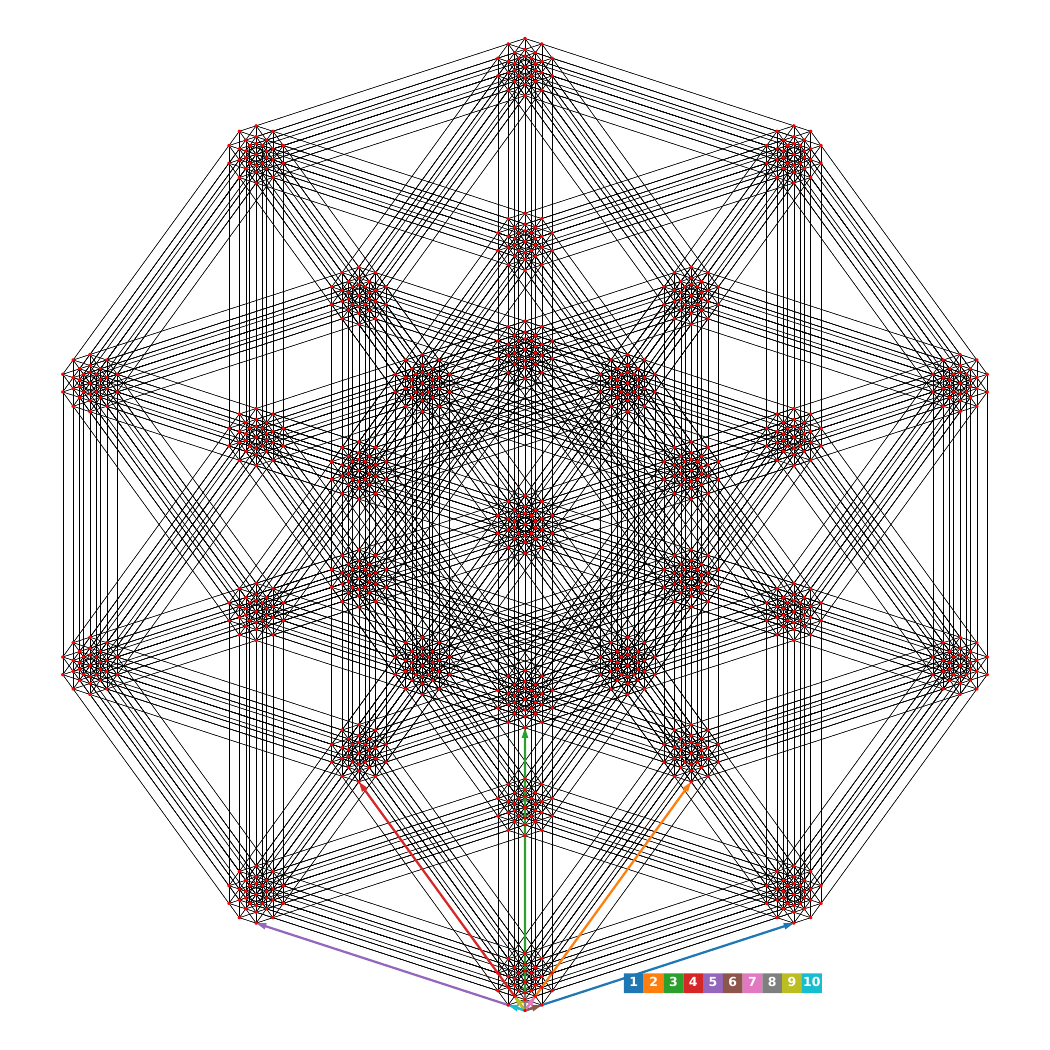}
    \caption{
        Fractal projections of decaract.
        Colored arrows represent the contribution basis of
        each dimension.
        The boxes on the bottom right indicate
        the correspondence between the colors and the dimensions.
    }
    \label{fig:s-fractal-decaract}
\end{figure}

\begin{figure}[p]
    \includegraphics{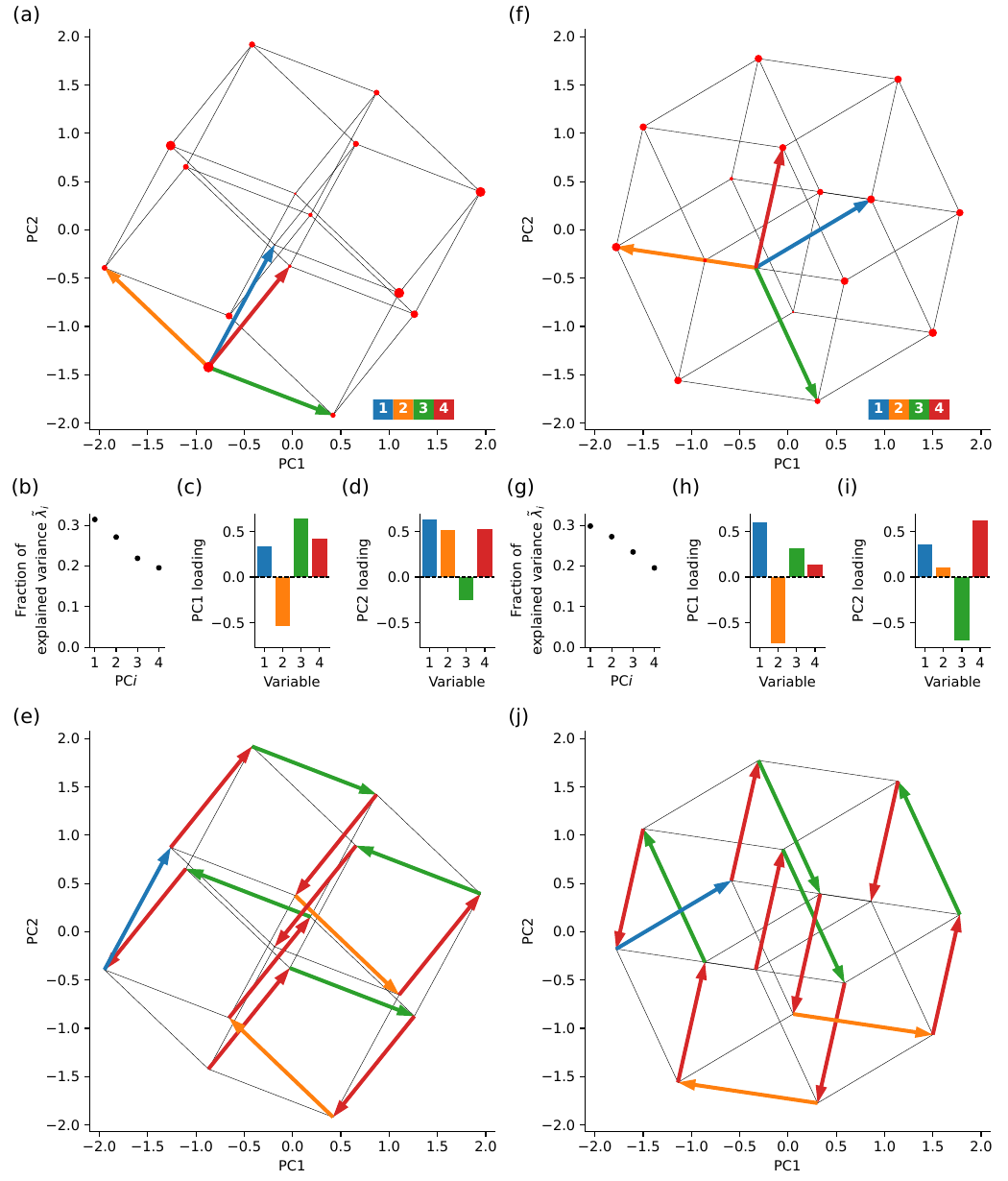}
    \caption{
        Orthogonal projections of four-dimensional hypercubic
        vertices by PCA\@.
        This figure is the same as Fig.~4 of the main text, but
        with a different
        realization of random weights.
        (a) A projection of a four-dimensional hypercube
        where vertices are
        weighted randomly.
        Red filled circles are the vertices and lines are the edges of
        the hypercube.
        The magnitude of weight is proportional to the area
        of the vertex.
        Arrows are biplot vectors originating from
        $
        \begin{bsmallmatrix}
            - & - & - & -
        \end{bsmallmatrix}^\top
        $
        and the boxes on the bottom right indicate
        the correspondence between the colors of the arrows and
        the original dimensions.
        (b) Fraction of explained variance by each PC of (a).
        (c) PC1 loading,
        and (d) PC2 loading
        of random weighted hypercubic vertices of (a).
        (e) Hamiltonian path on a four-dimensional hypercube in (a).
        (f)--(j) Same as (a)--(e) but with a different realization of
        random weights.
    }
    \label{fig:s-hpca-random}
\end{figure}

\begin{figure}[p]
    \includegraphics{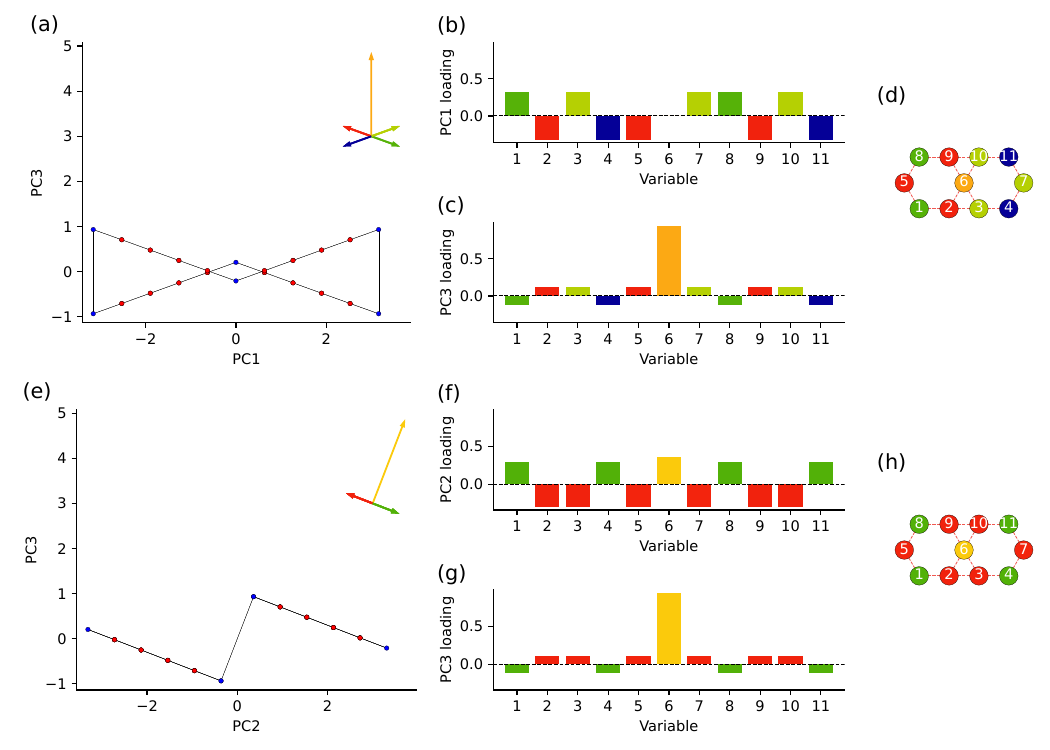}
    \caption{
        Hypercubic energy landscape of two-ring systems and
        the interaction network colored by the angle of the
        biplot vector.
        (a) The hypercubic energy landscape of the two-ring system by
        PC1 and PC3.
        (b) PC1 loading of (a), with the color of the bar matching
        the color of the corresponding biplot vector in (a).
        (c) Same as (b) but for PC3.
        (d) The Ising spin interaction network of the two-ring system,
        where the nodes are colored by the angle of the corresponding
        biplot vector.
        (e)--(h) Same as (a)--(d) but by PC2 and PC3.
        Notice that in (a) and (e),
        several biplot vectors are overlapped.
    }
    \label{fig:s-two-rings-pc3}
\end{figure}

\begin{figure*}[p]
    \includegraphics{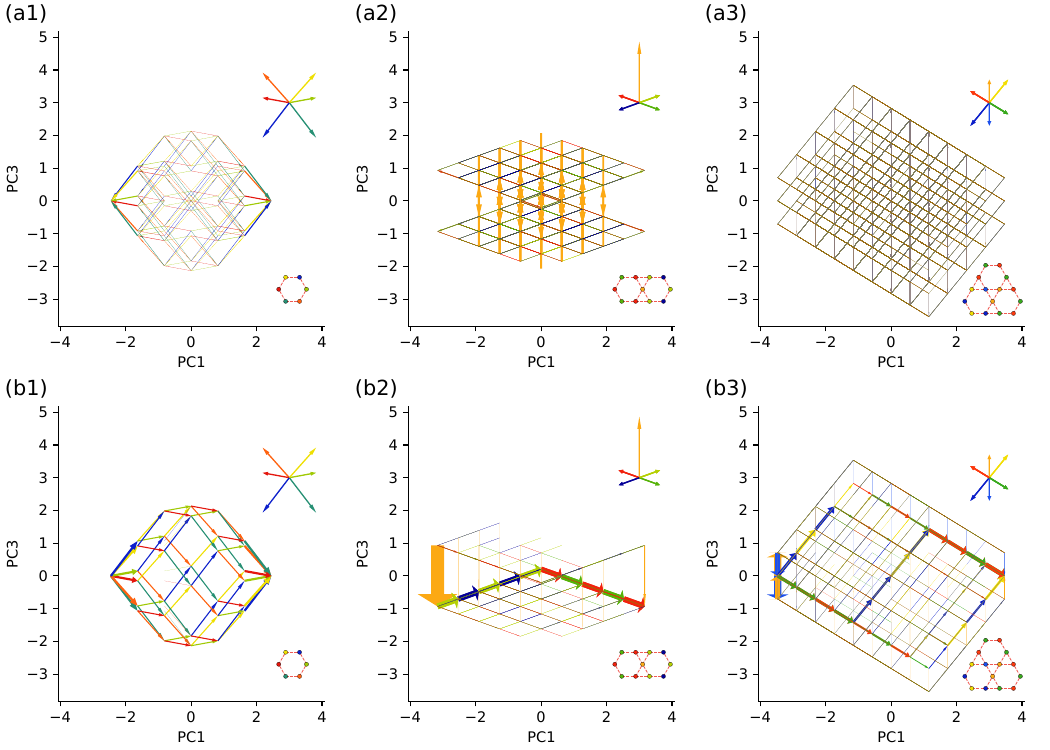}
    \caption{
        Visualization of time-integrated probability flux on the
        hypercubic energy landscape projected by PC1 and PC3.
        (a1) The one-ring system.
        (a2) The two-ring system.
        (a3) The three-ring system.
        (b1) The one-ring system with an external field
        $
        \bm{h}
        =
        \begin{bsmallmatrix}
            + & - & - & + & + & -
        \end{bsmallmatrix}^\top
        $,
        which biases the state transition toward the state
        projected around
        $
        \begin{bsmallmatrix}
            2.5 & 0
        \end{bsmallmatrix}^\top
        $.
        (b2) The two-ring system with an external field
        $
        \bm{h}
        =
        \begin{bsmallmatrix}
            + & - & + & - & - & - & + & + & - & + & -
        \end{bsmallmatrix}^\top
        $,
        which biases the state transition toward the state
        projected around
        $
        \begin{bsmallmatrix}
            3.2 & -0.5
        \end{bsmallmatrix}^\top
        $.
        (b3) The three-ring system with an external field
        $
        \bm{h}
        =
        \begin{bsmallmatrix}
            + & - & + & - & - & + & + & + & + & - & - & - & + & + & -
        \end{bsmallmatrix}^\top
        $,
        which biases the state transition toward the state
        projected around
        $
        \begin{bsmallmatrix}
            3.7 & -0.8
        \end{bsmallmatrix}^\top
        $.
        Time-integrated flux is shown as hypercubic arrows.
        The width of each arrow is proportional to the magnitude of
        the time-integrated flux,
        $\left|\Delta \mathcal{J}_{\bm{s}, \bm{s}^\prime}\right|$.
        The direction indicates the sign of the time-integrated flux
        $
        \operatorname{sgn}
        \left(\Delta \mathcal{J}_{\bm{s}, \bm{s}^\prime}\right)
        $, i.e., it is positive if the corresponding state
        transition aligns
        with the biplot vector and negative if it is anti-aligned.
        Biplot vectors are shown in the top right, with their color
        indicating the angle of the biplot vectors.
        The width of the biplot vector corresponds to
        $
        \left|\Delta \mathcal{J}_{\bm{s}, \bm{s}^\prime}\right|
        =
        0.02
        $.
        We only visualize the arrow with
        $
        \left|\Delta \mathcal{J}_{\bm{s}, \bm{s}^\prime}\right|
        >
        0.0002
        $
        for clarity.
        The Ising spin interaction network is shown in the lower right,
        where nodes are colored by the angle of the corresponding
        biplot vectors.
        The initial probability distribution is set to be uniform for
        (a1)--(a3), i.e.,
        $p_{\bm{s}}\left(0\right) = \frac{1}{2^N}$, $\forall \bm{s}$.
        For (b1)--(b3),
        the initial probability distribution is set to be
        unipolar for the state
        $\bm{s} = -\bm{h}$, i.e.,
        $p_{\bm{s}}\left(0\right) = \delta_{-\bm{h}, \bm{s}}$.
        We use $A = 1$, $k_\mathrm{B} T = \frac{1}{\beta} = 0.3$,
        and $\tau=50$
        in all panels.
        Note that in (b2) and (b3), several arrows appear at the same
        location due to overlapping biplot vectors.
    }
    \label{fig:probability-flux-pc13}
\end{figure*}

\begin{figure*}[p]
    \includegraphics{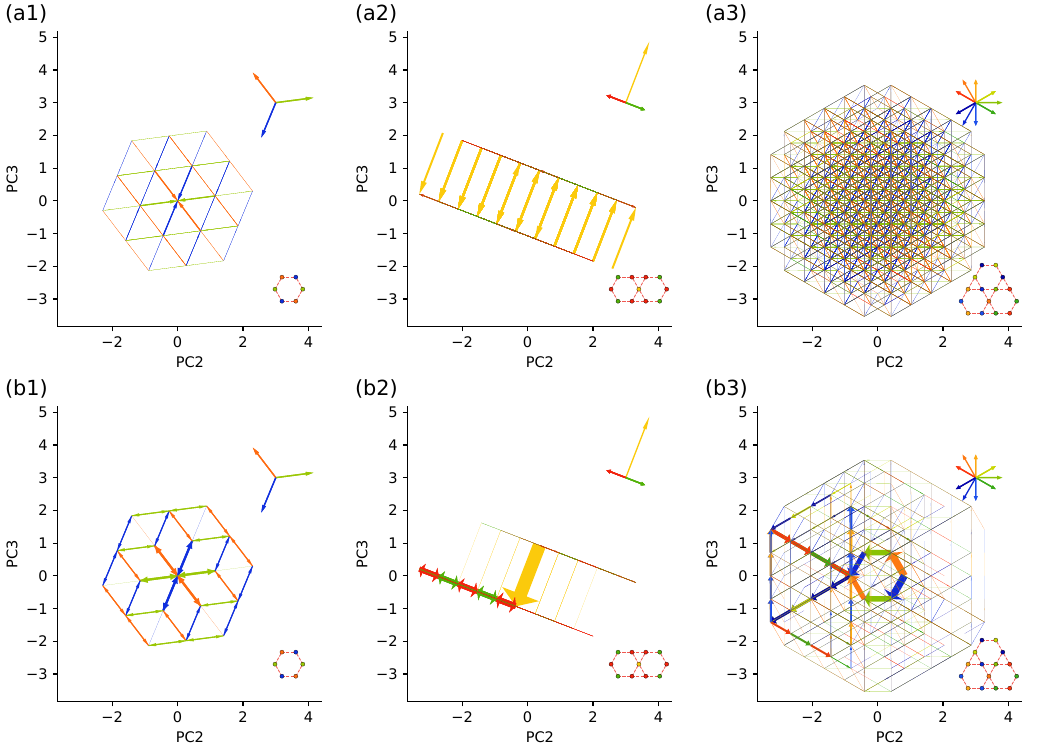}
    \caption{
        Visualization of time-integrated probability flux on the
        hypercubic energy landscape projected by PC2 and PC3.
        (a1) The one-ring system.
        (a2) The two-ring system.
        (a3) The three-ring system.
        (b1) The one-ring system with an external field
        $
        \bm{h}
        =
        \begin{bsmallmatrix}
            + & - & - & + & + & -
        \end{bsmallmatrix}^\top
        $,
        which biases the state transition toward the state
        projected around
        $
        \begin{bsmallmatrix}
            2.5 & 0
        \end{bsmallmatrix}^\top
        $.
        (b2) The two-ring system with an external field
        $
        \bm{h}
        =
        \begin{bsmallmatrix}
            + & - & + & - & - & - & + & + & - & + & -
        \end{bsmallmatrix}^\top
        $,
        which biases the state transition toward the state
        projected around
        $
        \begin{bsmallmatrix}
            3.2 & -0.5
        \end{bsmallmatrix}^\top
        $.
        (b3) The three-ring system with an external field
        $
        \bm{h}
        =
        \begin{bsmallmatrix}
            + & - & + & - & - & + & + & + & + & - & - & - & + & + & -
        \end{bsmallmatrix}^\top
        $,
        which biases the state transition toward the state
        projected around
        $
        \begin{bsmallmatrix}
            3.7 & -0.8
        \end{bsmallmatrix}^\top
        $.
        Time-integrated flux is shown as hypercubic arrows.
        The width of each arrow is proportional to the magnitude of
        the time-integrated flux,
        $\left|\Delta \mathcal{J}_{\bm{s}, \bm{s}^\prime}\right|$.
        The direction indicates the sign of the time-integrated flux
        $
        \operatorname{sgn}
        \left(\Delta \mathcal{J}_{\bm{s}, \bm{s}^\prime}\right)
        $, i.e., it is positive if the corresponding state
        transition aligns
        with the biplot vector and negative if it is anti-aligned.
        Biplot vectors are shown in the top right, with their color
        indicating the angle of the biplot vectors.
        The width of the biplot vector corresponds to
        $
        \left|\Delta \mathcal{J}_{\bm{s}, \bm{s}^\prime}\right|
        =
        0.02
        $.
        We only visualize the arrow with
        $
        \left|\Delta \mathcal{J}_{\bm{s}, \bm{s}^\prime}\right|
        >
        0.0002
        $
        for clarity.
        The Ising spin interaction network is shown in the lower right,
        where nodes are colored by the angle of the corresponding
        biplot vectors.
        The initial probability distribution is set to be uniform for
        (a1)--(a3), i.e.,
        $p_{\bm{s}}\left(0\right) = \frac{1}{2^N}$, $\forall \bm{s}$.
        For (b1)--(b3),
        the initial probability distribution is set to be
        unipolar for the state
        $\bm{s} = -\bm{h}$, i.e.,
        $p_{\bm{s}}\left(0\right) = \delta_{-\bm{h}, \bm{s}}$.
        We use $A = 1$, $k_\mathrm{B} T = \frac{1}{\beta} = 0.3$,
        and $\tau=50$
        in all panels.
        Note that in (b2) and (b3), several arrows appear at the same
        location due to overlapping biplot vectors.
    }
    \label{fig:probability-flux-pc23}
\end{figure*}

\begin{figure*}[p]
    \includegraphics{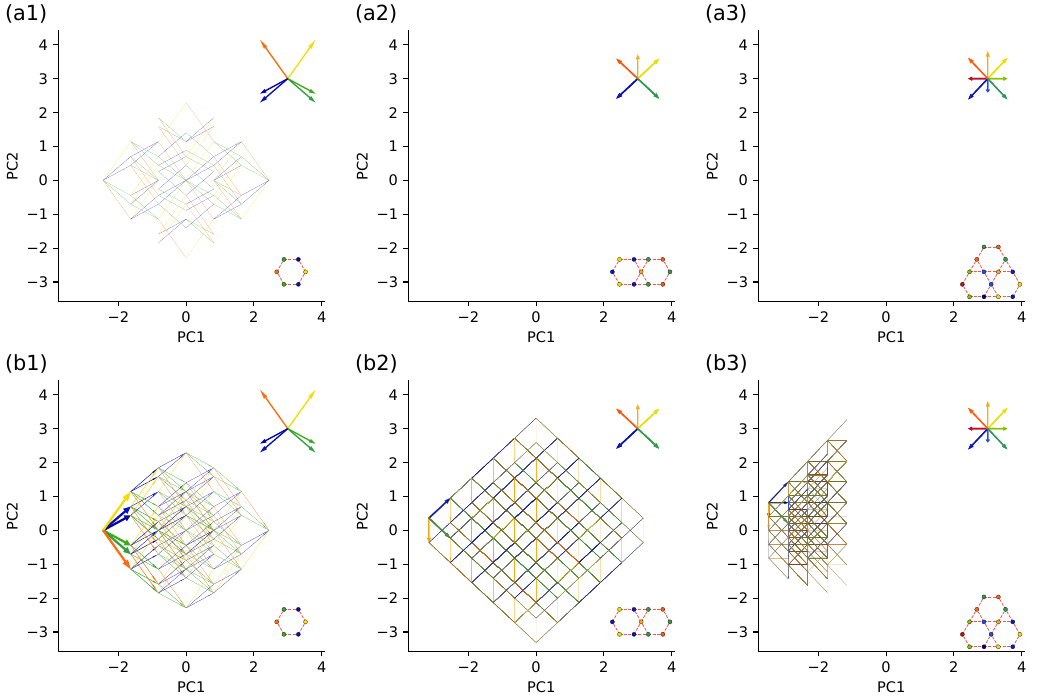}
    \caption{
        Visualization of time-integrated probability flux on the
        hypercubic energy landscape projected by PC1 and PC2.
        (a1) The one-ring system.
        (a2) The two-ring system.
        (a3) The three-ring system.
        (b1) The one-ring system with an external field
        $
        \bm{h}
        =
        \begin{bsmallmatrix}
            + & - & - & + & + & -
        \end{bsmallmatrix}^\top
        $,
        which biases the state transition toward the state
        projected around
        $
        \begin{bsmallmatrix}
            2.5 & 0
        \end{bsmallmatrix}^\top
        $.
        (b2) The two-ring system with an external field
        $
        \bm{h}
        =
        \begin{bsmallmatrix}
            + & - & + & - & - & - & + & + & - & + & -
        \end{bsmallmatrix}^\top
        $,
        which biases the state transition toward the state
        projected around
        $
        \begin{bsmallmatrix}
            3.2 & -0.5
        \end{bsmallmatrix}^\top
        $.
        (b3) The three-ring system with an external field
        $
        \bm{h}
        =
        \begin{bsmallmatrix}
            + & - & + & - & - & + & + & + & + & - & - & - & + & + & -
        \end{bsmallmatrix}^\top
        $,
        which biases the state transition toward the state
        projected around
        $
        \begin{bsmallmatrix}
            3.7 & -0.8
        \end{bsmallmatrix}^\top
        $.
        Time-integrated flux is shown as hypercubic arrows.
        The width of each arrow is proportional to the magnitude of
        the time-integrated flux,
        $\left|\Delta \mathcal{J}_{\bm{s}, \bm{s}^\prime}\right|$.
        The direction indicates the sign of the time-integrated flux
        $
        \operatorname{sgn}
        \left(\Delta \mathcal{J}_{\bm{s}, \bm{s}^\prime}\right)
        $, i.e., it is positive if the corresponding state
        transition aligns
        with the biplot vector and negative if it is anti-aligned.
        Biplot vectors are shown in the top right, with their color
        indicating the angle of the biplot vectors.
        The width of the biplot vector corresponds to
        $
        \left|\Delta \mathcal{J}_{\bm{s}, \bm{s}^\prime}\right|
        =
        0.02
        $.
        We only visualize the arrow with
        $
        \left|\Delta \mathcal{J}_{\bm{s}, \bm{s}^\prime}\right|
        >
        0.0002
        $
        for clarity.
        The Ising spin interaction network is shown in the lower right,
        where nodes are colored by the angle of the corresponding
        biplot vectors.
        The initial probability distribution is set to be uniform for
        (a1)--(a3), i.e.,
        $p_{\bm{s}}\left(0\right) = \frac{1}{2^N}$, $\forall \bm{s}$.
        For (b1)--(b3),
        the initial probability distribution is set to be
        unipolar for the state
        $\bm{s} = -\bm{h}$, i.e.,
        $p_{\bm{s}}\left(0\right) = \delta_{-\bm{h}, \bm{s}}$.
        We use $A = 1$, $k_\mathrm{B} T = \frac{1}{\beta} = 8.0$,
        and $\tau=50$
        in all panels.
        Note that in (b2) and (b3), several arrows appear at the same
        location due to overlapping biplot vectors.
        Note also that there is no clearly visible flux in (a1)--(a3).
    }
    \label{fig:probability-flux-pc12-high-temperature}
\end{figure*}

\begin{figure*}[p]
    \includegraphics{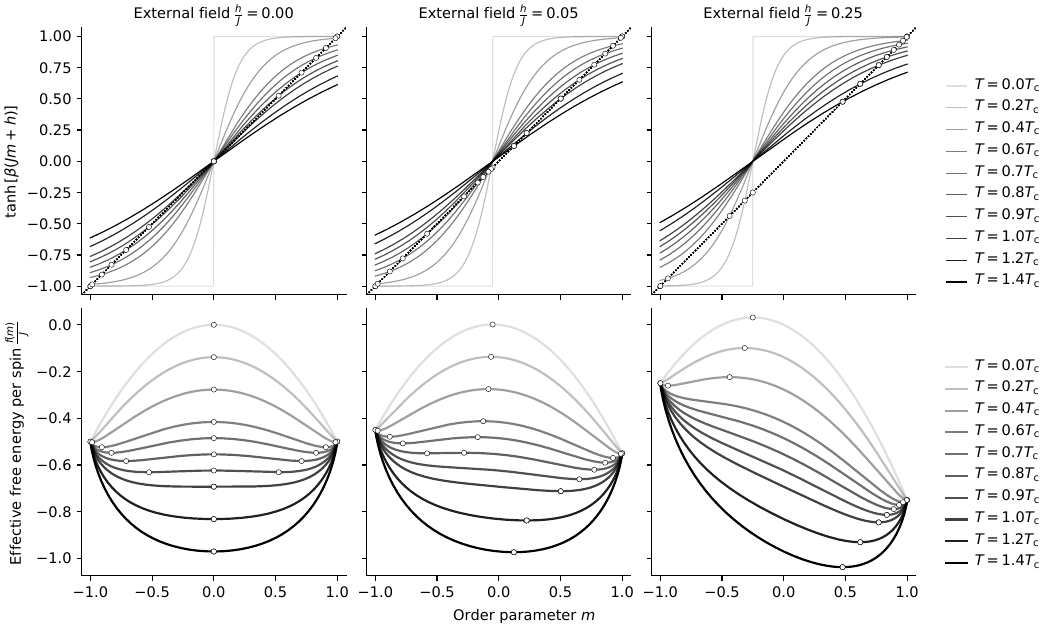}
    \caption{
        Visualizing self-consistent equation of the mean-field model
        (upper panel) [Eq.~\eqref{eq:self-consistent-equation}], and
        free energy landscapes of the mean-field model (lower panel).
        The effective free energy per spin as a function of the order
        parameter $m$ [Eq.~\eqref{eq:effective-free-energy-per-spin}] is shown.
        The color of line indicates the temperature in the unit of
        the critical temperature
        $T_\mathrm{c} = \frac{J}{k_\mathrm{B}}$.
        The white circles indicate stationary points $\left\{m^\ast\right\}$ of
        the effective free energy per spin.
        From left to right, the external field $\frac{h}{J}$ is set to be
        $0.00$, $0.05$, $0.25$.
    }
    \label{fig:free-energy-landscape-mean-field}
\end{figure*}